\documentclass[a4paper,11pt,oneside]{amsart} 

\usepackage[english]{babel}

\usepackage{fullpage}

\usepackage[utf8]{inputenc} 
\usepackage{lmodern} 
\usepackage[T1]{fontenc}

\usepackage{amssymb}
\usepackage{mathtools} 
	\numberwithin{equation}{section}

\usepackage[dvipsnames]{xcolor} 

\usepackage{graphicx}

\usepackage{tikz}
	\usetikzlibrary{matrix,arrows,decorations.pathmorphing} 
	\usetikzlibrary{patterns} 

\DeclarePairedDelimiter{\ket}{\lvert}{\rangle}

\DeclarePairedDelimiter{\cbra}{\langle\!\langle}{\rvert}
\DeclarePairedDelimiter{\cket}{\lvert}{\rangle\!\rangle}
\DeclarePairedDelimiterX{\cbraket}[2]{\langle\!\langle}{\rangle}{#1\vert#2}
\DeclarePairedDelimiterX{\bracket}[2]{\langle}{\rangle\!\rangle}{#1\vert#2}
\DeclarePairedDelimiterX{\cbracket}[2]{\langle\!\langle}{\rangle\!\rangle}{#1\vert#2}

\usepackage{bm} 
\newcommand{\vect}[1]{\bm{{#1}}}

\newcommand{\I}{\mathrm{i}}
\newcommand{\E}{\mathrm{e}}

 \let\Im\undefined \DeclareMathOperator{\Im}{Im}
 \let\Re\undefined \DeclareMathOperator{\Re}{Re}

\newcommand{\modulo}[1]{\overline{\vphantom{t}#1}}

\newcommand{\checkedMma}{}

\usepackage[
	ocgcolorlinks, 
	linkcolor=BlueViolet, citecolor=OliveGreen, urlcolor=RawSienna, filecolor=Sepia,
	hyperfootnotes=false,
	]{hyperref} 

\usepackage{multirow} 
\usepackage{booktabs} 

\usepackage[alphabetic,short-journals]{amsrefs}

\DeclareRobustCommand{\SkipTocEntry}[5]{} 

\newcommand{\mom}[1]{\bar{#1}^{\vee}}
\newcommand{\momL}[1]{{#1}^{\vee}}

\begin{document}

\title{How coordinate Bethe ansatz works \\ for Inozemtsev model}

\author{Rob Klabbers}
\address{\textsc{Nordita}, KTH Royal Institute of Technology and Stockholm University, Roslagstullsbacken~23, \indent 106 91 Stockholm, Sweden}
\email{rob.klabbers@su.se}

\author{Jules Lamers}
\address{School of Mathematics and Statistics, University of Melbourne, Vic 3010, Australia} 
\email{jules.l@unimelb.edu.au}

\date{\today}

\begin{abstract}
Three decades ago, Inozemtsev discovered an isotropic long-range spin chain with elliptic pair potential that interpolates between the Heisenberg and Haldane--Shastry spin chains while admitting an exact solution throughout, based on a connection with the elliptic quantum Calogero--Sutherland model.
Though Inozemtsev's spin chain is widely believed to be quantum integrable, the underlying algebraic reason for its exact solvability is not yet well understood. As a step in this direction we refine Inozemtsev's `extended coordinate Bethe ansatz' and clarify various aspects of the model's exact spectrum and its limits. 

We identify quasimomenta in terms of which the $M$-particle energy is close to being (functionally) additive, as one would expect from the limiting models. This moreover makes it possible to rewrite the energy and Bethe-ansatz equations on the elliptic curve, turning the spectral problem into a rational
problem as might be expected for an isotropic spin chain. 

We treat the $M=2$ particle sector and its limits in detail. We identify an \textit{S}-matrix that is independent of positions despite the more complicated form of the extended coordinate Bethe ansatz. We show that the Bethe-ansatz equations reduce to those of Heisenberg in one limit and give rise to the `motifs' of Haldane--Shastry in the other limit. We show that, as the interpolation parameter changes, the `scattering states' from Heisenberg become Yangian highest-weight states for Haldane--Shastry, while bound states become ($\mathfrak{sl}_2$-highest weight versions of) affine descendants of the magnons from $M=1$. We are able to treat this at the level of the wave function and quasimomenta. For bound states we find an equation that, for given Bethe integers, relates the `critical' values of the spin-chain length and the interpolation parameter for which the two complex quasimomenta collide; it reduces to the known equation for the `critical length' in the limit of the Heisenberg spin chain. We also elaborate on Inozemtsev's proof of the completeness for $M=2$ by passing to the elliptic curve. 

Our review of the two-particle sectors of the Heisenberg and Haldane--Shastry spin chains may be of independent interest.

\bigskip

\end{abstract}

\begin{flushright}
	\footnotesize
	\textsc{nordita} 2020-007
\end{flushright}
\bigskip

{\let\newpage\relax\maketitle} 


\setcounter{tocdepth}{3}
\tableofcontents

\section{Introduction} \label{sec:intro}
\noindent
One of the big open problems in the field of quantum integrability is the question whether Ino\-zemtsev's elliptic spin chain \cite{Inozemtsev:1989yq} is quantum integrable, cf.~e.g.~\cite[\textsection5]{Hal_94}. 
This (isotropic) spin chain interpolates between the well-understood Heisenberg (\textsc{xxx}) \cite{heisenberg1928ferromagnetismus} and Haldane--Shastry spin chains \cite{haldane1988exact,shastry1988exact} while being exact solvable throughout. It famously made a guest appearance in the context of the AdS/CFT correspondence, where it was found to describe the scaling dimensions of $\mathcal{N}=4$ super Yang-Mills theory up to three loops~\cite{Serban:2004jf}. Unlike the limiting models, no underlying quantum-integrable structure has been identified for Inozemtsev's spin chain yet. 

We focus on the spin-1/2 case, so `isotropy' means invariance under $\mathfrak{sl}_2 = (\mathfrak{su}_2)_{\mathbb{C}}$. This is the simplest instance of the Heisenberg model, for which the exact characterisation of the spectrum goes back to Bethe~\cite{bethe1931theorie} and has been refined many times since. One milestone is its algebraic incarnation in the framework of the quantum inverse-scattering method, see e.g.~\cite{Fad_95u}, which uncovered a deep relation with quantum algebra. This established the model's \emph{quantum integrability}: exact solvability by virtue of an underlying quantum group (\mbox{\textit{R}-matrix}, Yang--Baxter equation). The relevant quantum group, the \emph{Yangian} $Y(\mathfrak{gl}_2)$, plays a two-fold role for the Heisenberg spin chain:
\begin{itemize}
	\item it has an abelian subalgebra, known as the `Bethe subalgebra', whose generating function, the transfer matrix, generates a hierarchy of commuting operators that includes the Hamiltonian;
	\item its (nilpotent) Borel part, which does \emph{not} commute with the Bethe subalgebra, generates the model's eigenvectors: this is the algebraic Bethe ansatz.
\end{itemize}
These two type of `symmetries' are sometimes called \emph{symmetry generating} and \emph{spectrum generating}, respectively. Only in the asymptotic regime of infinite system size the full Yangian commutes with the Heisenberg Hamiltonian~\cite{Ber_92}. 
The Heisenberg spin chain admits various deformations, 
including (partial of fully) anisotropic counterparts, that preserve its quantum integrability. In this work we focus on the isotropic case.

The Haldane--Shastry (HS) spin chain has many striking properties. By construction \cite{haldane1988exact,shastry1988exact} its (antiferromagnetic) ground-state wave function is of Jastrow-type (Vandermonde squared) resembling the Laughlin state at filling fraction $\nu=1/2$ of the fractional quantum Hall effect.
It is one of the simplest models exhibiting fractional (exclusion) statistics, with `semionic' quasiparticles with statistics halfway between that of bosons and fermions~\cite{Hal_91a,Hal_91b}. 
The exact solution of the HS spin chain is closely related to quantum algebra. This time the action of $Y\mspace{-1mu}(\mathfrak{gl}_2)$ is a little more complicated \cite{HH+_92,bernard1993yang} and is such that
\begin{itemize}
	\item its centre, which is generated by the quantum determinant, provides the commuting operators including the Hamiltonian: these are the model's \emph{abelian symmetries};
	\item as a consequence, the whole Yangian commutes with the Hamiltonian, providing \emph{nonabelian symmetries};
	\item in particular, its Borel part only produces \emph{descendants} of (highest-weight) eigenvectors.
\end{itemize}
The model's highly degenerate eigenspaces \cite{haldane1988exact} are in part \cite{FG_15} due to these enhanced symmetries. Its exact eigenvectors are known in closed form, with wave functions that feature a special case (viz.\ zonal spherical) of Jack polynomials~\cite{bernard1993yang}. Their construction hinges on the presence of a second quantum algebra \cite{bernard1993yang}: the \emph{degenerate affine Hecke algebra}, which is Schur--Weyl dual to the Yangian \cite{drinfel1986degenerate}. Here
\begin{itemize}
	\item its centre coincides with that of $Y\mspace{-1mu}(\mathfrak{gl}_2)$, and is contained in
	\item its abelian subalgebra, generated by Dunkl operators, can be simultaneously diagonalised to produce the eigenvectors.
\end{itemize}
More precisely, one first constructs a more general quantum many-body model, the spin-analogue of the (trigonometric, quantum) Calogero--Sutherland model, with spin-1/2 particles moving on a circle while interacting in pairs. This model already has Yangian invariance~\cite{bernard1993yang}, where the Yangian action differs from that for Heisenberg. The Haldane--Shastry spin chain is obtained by \emph{freezing}, i.e.\ taking the limit in which the potential dominates the kinetic energy so that the particles come to a halt at their equally spaced (classical) equilibrium positions \cite{shastry1988exact,Pol_93,bernard1993yang,TH_95}. Through a rather subtle argument the `frozen' Dunkl operators (i.e.\ no differential part) allow one to obtain the spin-chain eigenvectors; this relies on the identification of a suitable subspace of polynomials and exploits the special equidistant positions of the spin chain~\cite{bernard1993yang,LPS_20u}.

For Inozemtsev's spin chain there are many clues that all point towards quantum integrability. Inozemtsev already proposed a quantum Lax pair in \cite{Inozemtsev:1989yq} as well as a set of commuting Hamiltonians~\cite{Inozemtsev:1989yq,inozemtsev1996invariants}, but so far only the commutativity of the first few of these have been shown \cite{Inozemtsev:1989yq,dittrich2008commutativity}.\,\footnote{\ Note that the existence of a family of commuting Hamiltonians does not follow from Inozemtsev's quantum Lax pair $(L,M)$, as the entries of $L$ and $M$ are operators that do not commute. Moreover, the remedy proposed in \cite{ujino1992,Shastry1993} cannot be applied here because the sum condition $\sum_k M_{ik} = 0$ is not satisfied.} Finkel and Gonz\'{a}lez-L\'{o}pez obtained the (first) Hamiltonian by `freezing' the integrable elliptic Calogero-Sutherland model with spin \cite{Finkel_2014}. Additional evidence comes from the large-$L$ regime; Inozemtsev's spin chain is at least `asymptotically quantum integrable': its thermodynamics in large volume can be studied exactly using the thermodynamic Bethe ansatz \cite{dittrich1997two,klabbers2015inozemtsev} and the model exhibits Yangian symmetry in the infinite-length limit \cite{ha1993squeezed,Gomez:2016zcb}. Such an explicit link to quantum algebra is lacking  at finite size. Nevertheless, Inozemtsev showed in a series of works that, using an extended version of the coordinate Bethe ansatz, the model is \emph{exactly solvable}. The spectral problem was solved exactly for the two-magnon sector both at finite and infinite length in \cite{Inozemtsev:1989yq} using an ansatz inspired by Hermite's solution of the Lam\'{e} equation. Not much later the infinite-length case was solved in full generality \cite{inozemtsev1992extended}, gradually followed by the finite-length case \cites{Inozemtsev_1995,Inozemtsev_2000}. A crucial insight in this endeavour was that the eigenfunctions of the spin-$1/2$ chain are closely related \cite{Inozemtsev_1995} to the wave functions of the corresponding (elliptic or hyperbolic) \emph{spinless} Calogero--Sutherland models.
Armed with this solution, some further progress was made in understanding the model by analysing its critical point~\cite{inozemtsev1993ground} and computing spin-spin correlation functions~\cite{dittrich1997second}.

Some elementary questions are yet to be answered: the relation between the solution to the Inozemtsev model and its limiting cases is currently unclear, partly because in its present formulation these limits are tricky. There is no (numerical) analysis of the spectrum known beyond the relatively simple two-particle sector, which is due to the fact that solving the constraints from \cite{Inozemtsev_2000} becomes quite involved for more than two particles: the many trivial solutions that are to be discarded from the solution set seem to act as attractors for any numerical algorithm, making it difficult to find true solutions. It is therefore not yet known whether the ansatz is complete, so strictly speaking it is not known whether the spectrum is completely understood.   

In the present paper we work towards solving these questions by proposing a different parametrisation of the solution. This allows us to study the limits explicitly and uncover some new relations between the parameters in the respective solutions. Moreover, in this parametrisation the $M$-particle energy becomes additive up to a potential term and we show what would be needed for full additivity. Our parametrisation furthermore allows for a good definition of the two-particle $S$-matrix at finite length and it has manifest Bethe-ansatz equations (\textsc{bae}). Finally, for $M=2$ the accompanying constraints (extended \textsc{bae}) can be put on an elliptic curve, yielding a set of algebraic equations instead of transcendental ones. We show that also the two-particle energy admits such a lift, turning the full spectral problem into a rational one. 

\bigskip 
\noindent
In \textsection\ref{sec:model} we introduce the model, its limits and other necessary definitions. In \textsection\ref{sec:Bethe_analysis} we present a full solution of the spectral problem in our new parametrisation, indicate the relation to the elliptic Calogero--Sutherland model and introduce the momentum lattices that simplify our further analysis considerably. In \textsection\ref{sec:M=2} we discuss the consequences of our new parametrisation for the two-particle case: we show how to take the various limits of our solution and compare with the known solutions in the limits, in particular proving completeness. We show that both the constraints and the energy can be put on an elliptic curve, turning the spectral problem into a rational one. In \textsection\ref{sec:conclusions} we discuss future directions. 

There are various appendices. In \textsection\ref{app:notation_comparison} we summarise how our notations for key quantities are related to those used in the literature. The elliptic functions that we need are reviewed in \textsection\ref{app:elliptic_fns}. In \textsection\ref{app:Inozemtsev_trick} we discuss how to compute certain sums of elliptic functions that appear in our analysis. Finally, \textsection\ref{app:further_proofs} contains some computational details for \textsection\ref{sec:Bethe_analysis}, a discussion of the proof of completeness given by Inozemtsev \cite{inozemtsev1993Hermitelike}, as well as a proof that the wave function vanishes on trivial roots.

\addtocontents{toc}{\SkipTocEntry}
\subsection*{Acknowledgements}
\noindent
The work of RK was supported by the grant “Exact Results in Gauge and String Theories” from the Knut and Alice Wallenberg foundation (KAW). JL acknowledges the Australian Research Council Centre of Excellence for Mathematical and Statistical Frontiers (\textsc{acems}) for financial support. We are indebted to G.~Arutyunov for motivating us to study the Inozemtsev spin chain back when we were MSc students. We are grateful to J.~Br\"{o}del, \mbox{J.-S.}~Caux, A.~Kaderli, E.~Langmann, D.~Serban, I.\ M.\ Sz\'{e}cs\'{e}nyi and V.~Pasquier for their interest and stimulating discussions. We thank the organisers of \textit{Elliptic integrable systems, special functions and quantum field theory} (Nordita, 2019) and \textit{Baxter 2020: Frontiers in Integrability} (ANU Canberra, 2020) for opportunities to present part of this work.
We thank the referees for their feedback on the manuscript.

\section{Inozemtsev's spin chain}
\label{sec:model}

\subsection{Hamiltonians} Consider a chain with $L$ spin-$1/2$ sites, represented by the Hilbert space $\mathcal{H} \coloneqq (\mathbb{C}\ket{\uparrow} \oplus \mathbb{C}\ket{\downarrow})^{\otimes L}$. Inozemtsev's elliptic spin chain has
(unnormalised) Hamiltonian 
\cite{Inozemtsev:1989yq}
\begin{equation*}
	H_\text{u} = \sum_{j< k}^L \wp(j-k) \, \frac{1 - \vec{\sigma}_j \cdot \vec{\sigma}_k}{2} \, ,
\checkedMma
\end{equation*}
where the sum runs over all pairs of sites. As usual $\vec{\sigma}_j$ is a vector of Pauli matrices acting nontrivially at the $j$th site, and $(1 - \vec{\sigma}_j\cdot \vec{\sigma}_k )/2 = 1 - P_{jk}${\checkedMma}
essentially antisymmetrises the spins at sites $j$ and~$k$. The long-range nature is characterised by the pair potential $\wp(z)$, which denotes the Weierstrass elliptic function, which is even, has periods $(\omega_1,\omega_2) \coloneqq (L,\omega) \in \mathbb{N} \times \I \,\mathbb{R}_{>0}$ and a double pole at $z=0$ with residue one. The real period sets the chain's length (periodic boundary conditions) while the imaginary period (ensuring that $\wp$ is real on $\mathbb{R}$) controls the interaction range as we will see at the end of \textsection\ref{sec:limits_combs}.

The two periods of $\wp$ allow us to take various limits. Three of these are of physical interest: $L\to \infty$ and $\omega \to 0, \I\,\infty$. To make the limit $\omega\to 0$ nonsingular we need to shift the ground-state energy a little. As we will see momentarily (\textsection\ref{sec:limits_combs}) this can be done in a way that also removes an irrelevant additive constant in the two other limits. Consider the Weierstrass $\zeta$-function with quasiperiods $(L,\omega)$, so that $\wp(z) = -\zeta'(z)$.{\checkedMma} 
Define the constants $\eta_b \coloneqq 2\, \zeta(\omega_b/2)$ for $b=1,2$. With this notation we set
\begin{equation} \label{eq:ham_shifted}
	H_\text{s} \coloneqq \sum_{j< k}^L \biggl( \wp(j-k) + \frac{\eta_2}{\omega} \biggr) \frac{1 - \vec{\sigma}_j \cdot \vec{\sigma}_k}{2} 
		\, .
\checkedMma
\end{equation}
This Hamiltonian does no longer diverge if $\omega\to 0$, but instead vanishes. This can be dealt with by a further rescaling, yielding the properly normalised Hamiltonian\,%
\footnote{\ The above Hamiltonians are ferromagnetic: the potentials are positive (for real arguments) while $1-P$ is a positive operator, yielding energies $\geq 0$, with the ferromagnetic (pseudo)vacuum~$\ket{\uparrow\cdots\uparrow}$ at zero energy. Importantly, this choice of overall sign will yield a relative sign in the expression \eqref{eq:M_particle_energy'} for the energy of $H_\text{s}$, and eventually give energies that are elliptic in the sense of \textsection\ref{sec:M=2_energy_rat}, \textsection\ref{sec:conclusions}. Indeed, in \eqref{eq:M_particle_energy'} the dispersion $\varepsilon_\text{s}$, see \eqref{eq:dispersion}, inherits the sign of $H_\text{s}$ while the potential part $\widetilde{U}$ originates from the elliptic Calogero--Sutherland (eCS) operator~\eqref{eq:ham_eCS}, which should be a positive operator. Only with the relative signs as in \eqref{eq:M_particle_energy'} a shift of the energy in $\varphi$ (\textsection\ref{sec:M=2_energy_rat}) or any $t_\alpha$ (\textsection\ref{sec:conclusions}) is proportional to the eCS Bethe-ansatz equations, vanishing on shell.} 
\begin{subequations} \label{eq:ham_normalised}
	\begin{gather}
	H_\text{n} \coloneqq \, n_H 
	H_\text{s} \, ,
\intertext{where $n_H = n_H(\omega)$ is a function that diverges as $n_H \sim (\omega\,\E^{\pi\I/\omega})^2$ for $\omega\to 0$ without affecting the other limit, $\lim_{\omega\to\I\infty} n_H = 1$. 
The hyperbolic limit $L\to\infty$ suggests taking~\cite{Inozemtsev_1995}}
	\text{e.g.} \quad n_H = \frac{\sin^2(\pi/\omega)}{(\pi/\omega)^2} = \frac{\sinh^2 \kappa}{\kappa^2} \, ,
	\end{gather}
\checkedMma
\end{subequations}
where, following Inozemtsev, we parametrise the imaginary period~$\omega$ by
\begin{equation*}
\kappa \coloneqq \frac{\I\,\pi}{\omega} \, \in \, \mathbb{R}_{>0} \, .
\end{equation*} 
We will use \eqref{eq:ham_shifted} for computations to avoid the normalising prefactor floating around.

\subsection{Limits and combs} \label{sec:limits_combs}
The different limits of the normalised pair potential are 
\begin{equation} \label{eq:limits_potential}
\tikz[baseline={([yshift=-.5*10pt*0.6+5pt]current bounding box.center)},scale=0.6]{
	\matrix (m) [matrix of math nodes,row sep=0em,column sep=5em]{
		& \substack{\displaystyle \text{elliptic} \\[.2em] \displaystyle 
		n_H(\kappa) \, \biggl(\wp(z) + \frac{\eta_2}{\omega} \biggr)} & \\
		\substack{\displaystyle \text{contact} \\[.7em] \displaystyle \delta_{d_L(z),1} } & & \substack{\displaystyle \text{trigonometric} \\[.5em] \displaystyle \frac{(\pi/L)^2}{\sin^2(\pi\,z/L)} } \\
		& \substack{\displaystyle
		\frac{n_H(\kappa) \, \kappa^2}{\sinh^2(\kappa\,z)} 
		\\[.7em] \displaystyle \text{hyperbolic}} & \\
		\delta_{|z|,1} & & \displaystyle \frac{1}{\textit{z}^2} \rlap{\ \text{rational}} \\
	};
	\path[->] 
	(m-1-2) edge node [above left,xshift=1cm,yshift=.2cm] {\footnotesize $\begin{array}{c} \kappa\to\infty \\ (\omega \to 0) \\ z \in\mathbb{R}, d_L(z)\geq 1 \end{array}$} ([yshift=.1cm]m-2-1.east)
	(m-1-2) edge node [left] {\footnotesize $L\to\infty$ \ } (m-3-2)
	(m-1-2) edge node [above right,xshift=-1cm,yshift=.25cm] {\footnotesize $\begin{array}{c} \kappa\to0 \\ (\omega \to \I\,\infty) \end{array}$} ([yshift=.2cm]m-2-3.west)
	(m-2-1) edge node [left] {\footnotesize $L\to\infty$ \ } (m-4-1)
	(m-2-3) edge node [right] {\ \footnotesize $L\to\infty$} ([yshift=.1cm]m-4-3.north)
	([yshift=.1cm]m-3-2.west) edge node [below right,xshift=-.9cm] {\footnotesize $\begin{array}{c} \kappa\to\infty \\ (\omega \to 0) \\ z \in\mathbb{R}, |z|\geq 1 \end{array}$} (m-4-1)
	([yshift=.1cm]m-3-2.east) edge node [below left,xshift=.6cm,yshift=0cm] {\footnotesize $\begin{array}{c} \kappa\to0 \\ (\omega \to \I\,\infty) \end{array}$\ } (m-4-3);
	\path[<->,dotted] (m-3-2) edge node [fill=white,inner sep=-.1cm] {\footnotesize $\begin{array}{c} \I\,\pi/\kappa \leftrightarrow L \\ \text{\& rescale} \end{array}$} (m-2-3);
}
\checkedMma
\end{equation}
On the left $\delta$ denotes a Kronecker delta function, and in the top left $d_L(z) \coloneqq \min_{n \in \mathbb{Z}} |z+n\,L|$ the distance function on $\mathbb{Z}_L \coloneqq \mathbb{Z}/L\,\mathbb{Z}$. See also the plots in Figures \ref{fig:potential_kappa} and~\ref{fig:potential_L} below.

The arrows in the right half of \eqref{eq:limits_potential} admit inverses, as is well known in the area of integrable quantum-many body systems of Calogero--Sutherland type, see e.g.~\cite{sutherland2004beautiful}. Indeed, one can introduce a period~$a$ in a (suitable) function $f$ via
\begin{equation} \label{eq:comb_def}
	\mathrm{comb}_a \, f(z) \coloneqq \sum_{n\in\mathbb{Z}} f(z+n\,a) \, .
\end{equation}
Physically one can think of this as a `comb of particles', that is, infinitely many copies of a particle on a line, equally spaced at distance $a$ of each other and moving in exactly the same way. Starting with the rational case we obtain
\begin{equation} \label{eq:combs}
\tikz[baseline={([yshift=-.5*10pt*0.6+5pt]current bounding box.center)},scale=0.6]{
	\matrix (m) [matrix of math nodes,row sep=0em,column sep=5em]{
		\substack{\displaystyle \text{elliptic} \\[.2em] \displaystyle \wp(z) + \frac{\eta_2}{\omega} \qquad\quad \wp(z) + \frac{\eta_1}{L} } & \\
		& \substack{\displaystyle \text{trigonometric} \\[.5em] \displaystyle \frac{(\pi/L)^2}{\sin^2(\pi \, z/L)} } \\
		\substack{\displaystyle \frac{\kappa^2}{\sinh^2(\kappa \, z)} = \frac{(\pi/\omega)^2}{\sin^2(\pi \, z/\omega)} \\[.7em] \displaystyle \text{hyperbolic}} & \\
		& \displaystyle \frac{1}{z^2} \rlap{\ \text{rational}} \\
	};
	\path[->] 
	([xshift=-2.1cm]m-3-1.north) edge node [left] {\footnotesize $\mathrm{comb}_L$ \ } ([xshift=-2.1cm]m-1-1.south)
	([yshift=.3cm]m-2-2.west) edge node [above right,xshift=-.7cm,yshift=.2cm] {\footnotesize \ $\mathrm{comb}_\omega$} ([yshift=-.6cm]m-1-1.east)
	([yshift=.1cm]m-4-2.north) edge node [right] {\ \footnotesize $\mathrm{comb}_L$} (m-2-2)
	(m-4-2) edge node [below left,xshift=.5cm,yshift=-.1cm] {\footnotesize $\mathrm{comb}_\omega$\ } ([yshift=.3cm]m-3-1.east);
	\path[<->,dotted] (m-2-2) edge node [fill=white,inner sep=0.05cm] {\footnotesize $\omega \leftrightarrow L$} (m-3-1);
}
\checkedMma
\end{equation}
The functions at the intermediate stage are related by `Wick rotating' $L = \omega_1 \leftrightarrow \omega_2 = \omega = \I \, \pi/\kappa$. Observe that the elliptic function depends on the route taken. By the Legendre relation the difference is $\eta_1/L - \eta_2/\omega = 2\pi\I/(L\,\omega) = 2\,\kappa/L$; in particular both elliptic functions have the same limits when $L\to\infty$ or $\omega\to\I\,\infty$ ($\kappa\to 0$). Either shift removes an (irrelevant but unwanted) additive constant: when $|\omega_a| \to \infty$ the Weierstrass elliptic function (without the shift) tends to $\wp(z) \to (\pi/\omega_b)^2 \,[\sin^{-2}(\pi\,z/\omega_b) - 1/3]$ for $b\neq a$.{\checkedMma}

Before we return to the spin chain let us give another way to understand the shift. Let $\vartheta(z\,|\,\tau)$ denote the odd Jacobi theta function with nome $\E^{\I \pi \tau}$ (see \textsection\ref{app:elliptic_fns}).
We will need two variants of this odd Jacobi theta function, which we denote by
\begin{equation} \label{eq:theta}
	\theta_1(z) \coloneqq \vartheta\bigl(\pi \,z/L\bigm|\omega/L\bigr) \, , \qquad \theta_2(z)\coloneqq \vartheta\bigl(\pi\,z/\omega\bigm|{-L}/\omega\bigr) \, ,
\checkedMma
\end{equation}
and are related by Jacobi's imaginary transformation as $\theta_1(z) = \I \, (L \, \kappa/\pi)^{1/2} \, \E^{-\kappa \, z^2/L} \, \theta_2(z)$.{\checkedMma}
Next define the (odd){\checkedMma}
`prepotential' functions as the logarithmic derivatives\,%
\footnote{\ \label{fn:theta_prime}Note the prefactor in $\theta_b'(z) = (\pi/\omega_b)\,\vartheta'(\pi\,z/\omega_b|\tau_b)$, where $\tau_b$ is the nome from \eqref{eq:theta}.{\checkedMma}}
\begin{equation} \label{eq:rho_def}
	\rho_b(z) \coloneqq \frac{\mathrm{d}}{\mathrm{d}z} \log \theta_b(z) = \frac{\theta_b'(z)}{\theta_b(z)} =  \zeta(z)-\frac{\eta_b}{\omega_b}\,z \, , \qquad b=1,2 \, .
\checkedMma
\end{equation}
Here the shift of $\zeta$ by a linear term ensures periodicity along the $b$th lattice constant, $\rho_b(z+\omega_b) = \rho_b(z)$.{\checkedMma} 
The two prepotentials differ by $\rho_2(z) - \rho_1(z) = (2\pi\I)/(L\,\omega)\,z${\checkedMma}
due to the Legendre relation (or Jacobi's imaginary transformation). As a consequence $\rho_b$ is (arithmetically) quasiperiodic in the other direction, $\rho_b(z+\omega_a) = \rho_b(z) + (-1)^b\, 2\pi\I/\omega_b$ for $a\neq b$.{\checkedMma}
The limits of $\rho_b$ will be given in \textsection\ref{sec:mtm_lattices}.
The important point here is that, in terms of these functions, the elliptic functions from \eqref{eq:combs} are $-\rho_b'(z) = \wp(z) + \eta_b/\omega_b$.{\checkedMma}

The spin-chain Hamiltonian \eqref{eq:ham_shifted} features the result of the route on the left in \eqref{eq:comb_def}, which is the inverse of the `central' route in \eqref{eq:limits_potential}. This allows one to determine the appropriate `renormalisation' of the potential already at the hyperbolic level; the factor of $\kappa^{-2} \sinh^2 \kappa$ in \eqref{eq:ham_normalised} and \eqref{eq:limits_potential} ensures finite limits for $\kappa\to\infty$ as well as $\kappa\to 0$.

From \eqref{eq:limits_potential} it follows that the limits of Inozemtsev's Hamiltonian are, in terms of the short-hand $h_{jk} \coloneqq (1 - \vec{\sigma}_j \cdot \vec{\sigma}_k)/2 = 1 - P_{jk}$, given by
\begin{equation} \label{eq:limits_ham}
\tikz[baseline={([yshift=-.5*10pt*0.6+5pt]current bounding box.center)},scale=0.6]{
	\matrix (m) [matrix of math nodes,row sep=0em,column sep=4.5em]{
		& \substack{\displaystyle \text{Inozemtsev} \\[.3em] \displaystyle H_\text{n} \ \eqref{eq:ham_normalised} } & \\
		\substack{\displaystyle \text{Heisenberg} \\[.2em] \displaystyle \sum_{j\in\mathbb{Z}_L} \! h_{j,j+1} } & & \substack{\displaystyle \text{Haldane--Shastry} \\[.2em] \displaystyle \sum_{j< k}^L \frac{(\pi/L)^2}{\sin^2[\pi\,(j-k)/L]} \, h_{jk} } \\
		& \displaystyle \sum_{j< k} \frac{\sinh^2 \kappa}{\sinh^2[\kappa\,(j-k)]} \, h_{jk} & \\
		\displaystyle \sum_{j\in\mathbb{Z}} h_{j,j+1} & & \displaystyle \sum_{j< k} \frac{1}{(j-k)^2} \, h_{jk} \\
		\text{short range} & \qquad\text{intermediate range} & \text{long range} \\
	};
	\path[->] 
	(m-1-2) edge node [above left,xshift=.7cm,yshift=.25cm] {\footnotesize $\begin{array}{c} \kappa\to\infty \\ (\omega \to 0) \end{array}$} ([yshift=.1cm]m-2-1.east)
	(m-1-2) edge node [left] {\footnotesize $L\to\infty$ \ } (m-3-2)
	(m-1-2) edge node [above right,xshift=-.7cm,yshift=.25cm] {\footnotesize $\begin{array}{c} \kappa\to0 \\ (\omega \to \I\,\infty) \end{array}$} ([yshift=.2cm]m-2-3.west)
	(m-2-1) edge node [right] {\footnotesize $L\to\infty$ \ } (m-4-1)
	(m-2-3) edge node [right] {\ \footnotesize $L\to\infty$} ([yshift=.1cm]m-4-3.north)
	([yshift=.1cm]m-3-2.west) edge node [below right,xshift=-.6cm] {\footnotesize $\begin{array}{c} \kappa\to\infty \\ (\omega \to 0) \end{array}$} ([yshift=1.1cm]m-4-1.east)
	([yshift=.1cm]m-3-2.east) edge node [below left,xshift=.5cm,yshift=0cm] {\footnotesize $\begin{array}{c} \kappa\to0 \\ (\omega \to \I\,\infty) \end{array}$\ } (m-4-3);
}
\end{equation}
The limits for finite size are the Heisenberg~\cite{heisenberg1928ferromagnetismus} and Haldane--Shastry~\cites{haldane1988exact,shastry1988exact} spin chains. Accordingly we will call $\kappa\to\infty$ ($\kappa\to0$) the \emph{Heisenberg (Haldane--Shastry) limit}, respectively. We see that from the spin-chain viewpoint the parameter $\kappa$ sets the interaction range. The nearest-neighbour Heisenberg model has \emph{(ultra) short-range} interactions. At the other extreme the \emph{long-range} HS spin chain has an inverse-square potential, $1/r^2$ with $r=(L/\pi) \, |\sin(\pi \,k/L)|$ the chord distance between the chain's sites, which are equally spaced on a circle of circumference~$L$; at least in classical one-dimensional systems $n=2$ is the maximal value for $1/r^n$-interactions to exhibit a phase transition \cite{thouless1969long}.
It is in this sense that the Inozemtsev spin chain can in general be considered to have \emph{intermediate-range} interactions. 

In the infinite-length limit $L \to \infty$ the Inozemtsev spin chain becomes hyperbolic~\cite{Inozemtsev:1989yq} and HS becomes rational. The focus of this work lies on $L<\infty$, but see \textsection\ref{sec:large_L} for $L\to\infty$.

\begin{figure}[h]
	\centering
	\begin{tikzpicture}
		\node at (0,0) {\includegraphics[width=.95\textwidth]{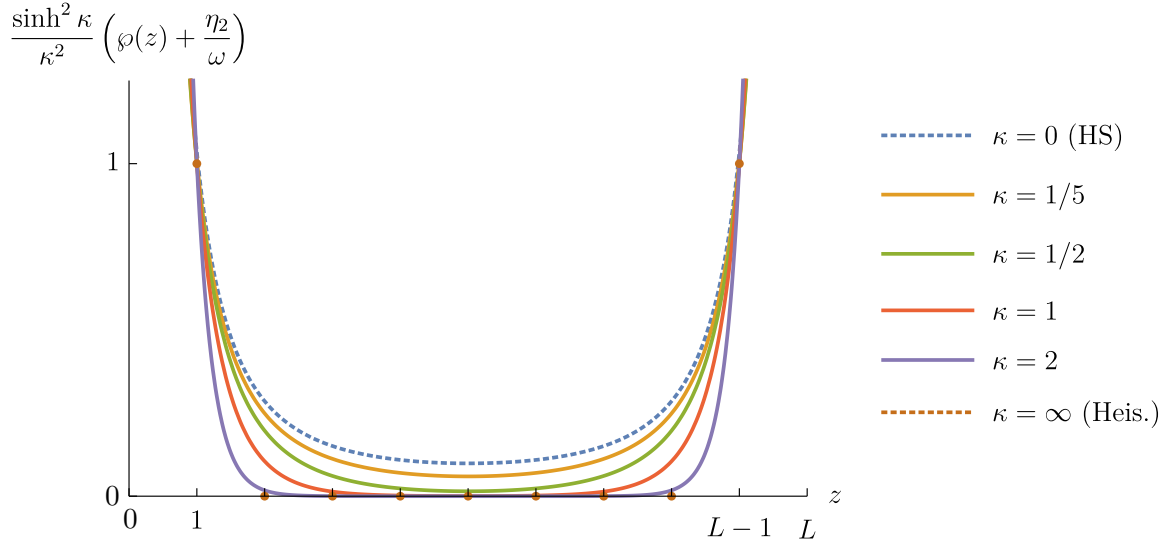}};
		\node at (-5.98,-3.3) {$0$};
		\node at (-6.22,-3.0) {$0$};
	\end{tikzpicture}
	\caption{The normalised potential at fixed~$L$ for values of $\kappa$ interpolating between the trigonometric ($\kappa=0$, dotted curve) and contact ($\kappa=\infty$; values shown by dots) limits. Note from $(\pi/L)^2/\sin^2(\pi/L) = 1 + \pi^2/(3L^2) + \mathcal{O}(L^{-4})$ that the potential equals one at $z=1$ only as $\kappa\to\infty$.}
	\label{fig:potential_kappa}
\end{figure}

\begin{figure}[h]
	\centering
	\begin{tikzpicture}
	\node at (0,0) {\includegraphics[width=.95\textwidth]{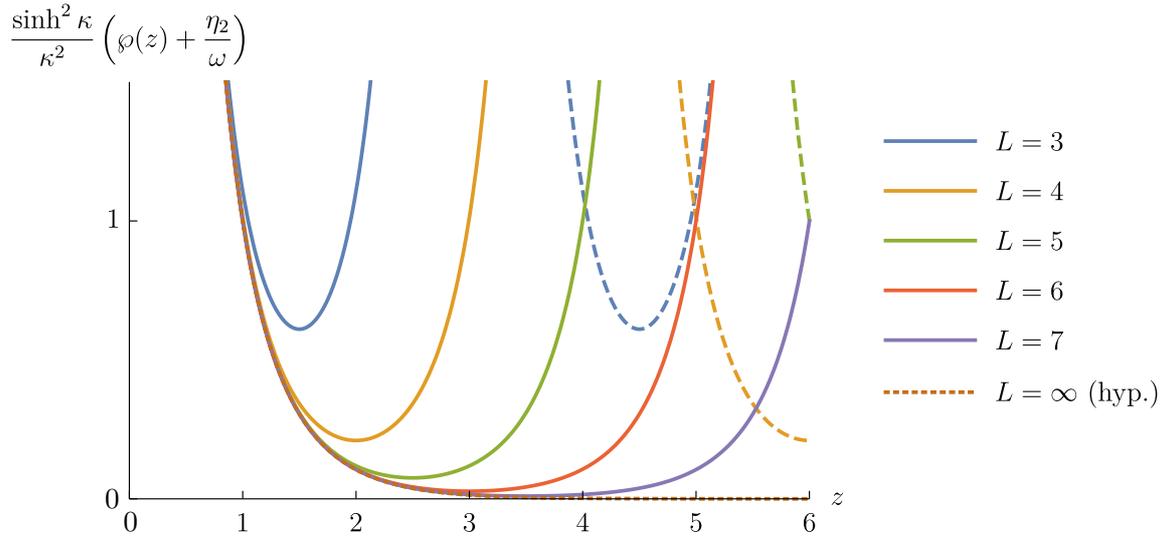}};
	\node at (-5.98,-3.35) {$0$};
	\node at (-6.22,-3.05) {$0$};
	\end{tikzpicture}
	\caption{The normalised potential at fixed imaginary period ($\kappa=1$) for increasing values of $L$. The graphs are dashed beyond $z=L$. Observe the rapid convergence to the hyperbolic limit (dotted curve).
	}
	\label{fig:potential_L}
\end{figure}

\section{Bethe-type analysis} \label{sec:Bethe_analysis}

\noindent One could cook up many models that interpolate between the Heisenberg and Haldane--Shastry spin chains. The crucial property of the Inozemtsev spin chain is that it does so in a way that is exactly solvable throughout. Our goal is to understand this better and study the solution to the eigenvalue problem
\begin{equation}
\label{eq:eigenvalue_equation}
	H_\text{s} \, \ket{\Psi} = E_\text{s} \, \ket{\Psi}
\end{equation} 
of the unnormalised but shifted Hamiltonian~\eqref{eq:ham_shifted}.

\subsection{General considerations}

Let $V$ be an even and $L$-periodic function, $V(-z) = V(z)$ and $V(z+L)=V(z)$, so that on $\mathbb{Z}_L$ it only depends on the distance~$d_L$ defined following~\eqref{eq:limits_potential}. Consider a general \mbox{spin-1/2} chain that is isotropic (invariant under $\mathfrak{sl}_2 = (\mathfrak{su}_2)_\mathbb{C}$) and homogeneous (translationally invariant) with pairwise interactions with pair potential $V$:
\begin{equation}
	H = \sum_{j < k}^L V(j-k) \, \frac{1 - \vec{\sigma}_j \cdot \vec{\sigma}_k}{2} 
	= \sum_{j \neq k}^L V(j-k) \, \frac{1 - P_{jk}}{2} \, ,
\checkedMma
\end{equation}
where the second form is convenient in calculations. This includes the Inozemtsev spin chain \eqref{eq:ham_shifted}, \eqref{eq:ham_normalised} and its limiting cases.

Consider the usual local raising and lowering operators $\sigma^\pm \coloneqq (\sigma^x \pm \I \, \sigma^y)/2$. The isotropy means that $[S^\alpha,H] = 0${\checkedMma} for each of the $\mathfrak{sl}_2$ generators
\begin{equation*}
	S^\pm \coloneqq \sum_{k \in \mathbb{Z}_L} \! \sigma^\pm_k \, , \quad 
	S^z \coloneqq \frac{1}{2} \sum_{k \in \mathbb{Z}_L} \! \sigma^z_k \, , \qquad 
	[S^z,S^\pm] = S^\pm \, , \quad [S^+ , S^- ] = 2 \, S^z \,.
	\checkedMma
\end{equation*}
The total spin-$z$ conservation allows us to diagonalise $H$ per $S^z$-eigenspace (weight space). The coordinate basis of the \emph{\textit{M}-particle sector} $\ker[S^z - (\tfrac12 L-M)]$
is labelled the positions of the $M$ excited spins. Let us denote these basis vectors by
\begin{equation} \label{eq:coord_basis}
\cket{\vect{n}} \coloneqq \sigma^-_{n_1} \cdots \sigma^-_{n_M} \ket{ \uparrow \uparrow \cdots \uparrow} \, , \qquad  1\leq n_1 < \dots < n_M \leq L \, .
\end{equation}
The wave function is $\Psi(\vect{n}) \coloneqq \cbraket{\vect{n}}{\Psi}$. As always when working with the coordinate basis for a translationally-invariant spin chain of finite size (so periodic boundaries) the cyclicity demands 
\begin{equation} \label{eq:cyclicity_gen}
\Psi(n_2,\dots\mspace{-1mu},n_M,n_1+L) = \Psi(\vect{n}) \, , \qquad  1\leq n_1 < \dots < n_M \leq L \, .
\end{equation}
By the isotropy it suffices to find the $\binom{L}{M} - \binom{L}{M-1}$ highest-weight vectors $\ket{\Psi}$ with $S^+ \ket{\Psi} = 0$ in each $M$-particle sector for $M \leq \lfloor L/2 \rfloor$.

\subsubsection{Low \textit{M}} \label{sec:M=1}
The operators $(1-\vec{\sigma}_j \cdot \vec{\sigma}_k)/2 = 1-P_{jk}$ kill aligned spins. The ($M=0$) vector with all spins up thus has $E = 0$, as do all its descendants $(S^-)^M \, \ket{\uparrow\uparrow\cdots\uparrow}$, $1\leq M \leq L$. 

Homogeneity fixes the $M=1$ sector. Indeed, $H$ commutes with the cyclic (left){\checkedMma}
shift 
\begin{equation} \label{eq:shift_op}
	G \coloneqq P_{(L\cdots 21)} = P_{L-1,L} \cdots P_{12} \, .
\checkedMma
\end{equation}
This operator has eigenvalues of the form $\E^{\I \mspace{1mu} p_\text{tot}}$, where the \emph{total (lattice) momentum}~$p_\text{tot}$ (defined mod~$2\pi$) is quantised as $p_\text{tot} \in (2\pi/L) \, \mathbb{Z}_L$ by the periodicity $G^L = 1$.{\checkedMma}

The (unnormalised) vector $\sum_n \E^{2\pi\I \, n\,I/L} \cket{n}$ has $p_\text{tot} = p = 2\pi\,I/L$ for $I \in \mathbb{Z}_L$,{\checkedMma} 
giving all $L$ (orthogonal){\checkedMma}
eigenvectors for $M=1$, including the descendant $S^- \, \ket{\uparrow\uparrow\cdots\uparrow}$ at $I=0$. 
To find the corresponding energy we act with $H$. Projecting the eigenvalue equation, with $\varepsilon(p) \coloneqq E^{M=1}$, onto $\cbra{n}$ gives
\begin{equation*}
	\begin{aligned}
	\E^{\I \, p \, n} \, \varepsilon(p) & = \frac{1}{2} \sum_{j\neq k}^L V(j-k) \sum_{n'=1}^L \E^{\I \, p \, n'} \cbra{n} (1-P_{jk}) \cket{n'} \\
	& = \sum_{k(\neq n)}^L \! (\E^{\I \, p \, n} - \E^{\I \, p \, k}) \, V(n-k) \, ,
	\end{aligned}
\checkedMma
\end{equation*}
where the symmetry $j\leftrightarrow k$ yields a factor of two. (Our notation means that the first sum in the first line ranges over pairs $j,k \in \{1,\cdots\mspace{-1mu},L\}$ with $j\neq k$, while the sum in the second line only runs over $k \in \{1,\cdots\mspace{-1mu},L\} \setminus \{n\}$.) Viewed as a function of the momentum~$p$ the \emph{dispersion} therefore assumes the form
\begin{equation} \label{eq:dispersion_gen}
	\varepsilon(p)	= \sum_{j=1}^{L-1} (1 - \E^{\I \, p \, j}) \, V(j) \, .
\checkedMma
\end{equation}

For the elliptic case, $V_\text{s}(z) = \wp(z) + \eta_2/\omega$, the sums for \eqref{eq:dispersion_gen} with $\wp$ are evaluated in \textsection\ref{app:dispersion}. The result for the (unnormalised) dispersion for the Inozemtsev spin chain is 
\begin{equation} \label{eq:dispersion}
	\varepsilon_\text{s}(p) 
	= \frac{1}{2} \, \bar{F}_2\Bigl(\frac{\omega \, p}{2\pi}\Bigr) = \frac{1}{2} \, \bar{F}_2\Bigl(\frac{\I \, p}{2\,\kappa}\Bigr) \, .
\checkedMma
\end{equation}
Here we define (even, pole-free on the real axis){\checkedMma}
functions that vanish at the origin
\begin{equation}\label{eq:F_def}
	\begin{aligned}
	\bar{F}_b(z) \coloneqq \frac{\bar{\theta}_b''(z)}{\bar{\theta}_b(z)} + \frac{3\,\bar{\eta}_b}{\bar{\omega}_b} & = \bar{\rho}_b'(z) + \bar{\rho}_b(z)^2 + \frac{3\,\bar{\eta}_b}{\bar{\omega}_b} \\
	& = {-}\bar{\wp}(z) + \Bigl( \bar{\zeta}(z) - \frac{\bar{\eta}_b}{\bar{\omega}_b} \, z\Bigr)^{\!2} + \frac{2 \, \bar{\eta}_b}{\bar{\omega}_b} \, , 
	\end{aligned}
	\quad\qquad b=1,2 \, ,
\checkedMma
\end{equation}
associated to a second (quasi)period lattice $\!\!\bar{\,\,\mathbb{L}} = \mathbb{Z} \oplus \omega \, \mathbb{Z}$ with $(\bar{\omega}_1,\bar{\omega}_2) \coloneqq (1,\omega)$, which also appears for the hyperbolic Inozemtsev spin chain at $L\to  \infty$. The function $p \mapsto \bar{F}_b(\I p/(2\kappa))$ is real, as follows from the real analyticity of the Weierstrass functions, and, although it might not be obvious, positive for $0\leq p \leq 2\pi$.\,%
\footnote{\ We will prove positivity of the dispersion in Footnote~\ref{fn:positive_dispersion} on p.\,\pageref{fn:positive_dispersion}.}

\begin{figure}[h]
	\centering
	\includegraphics[width=\textwidth]{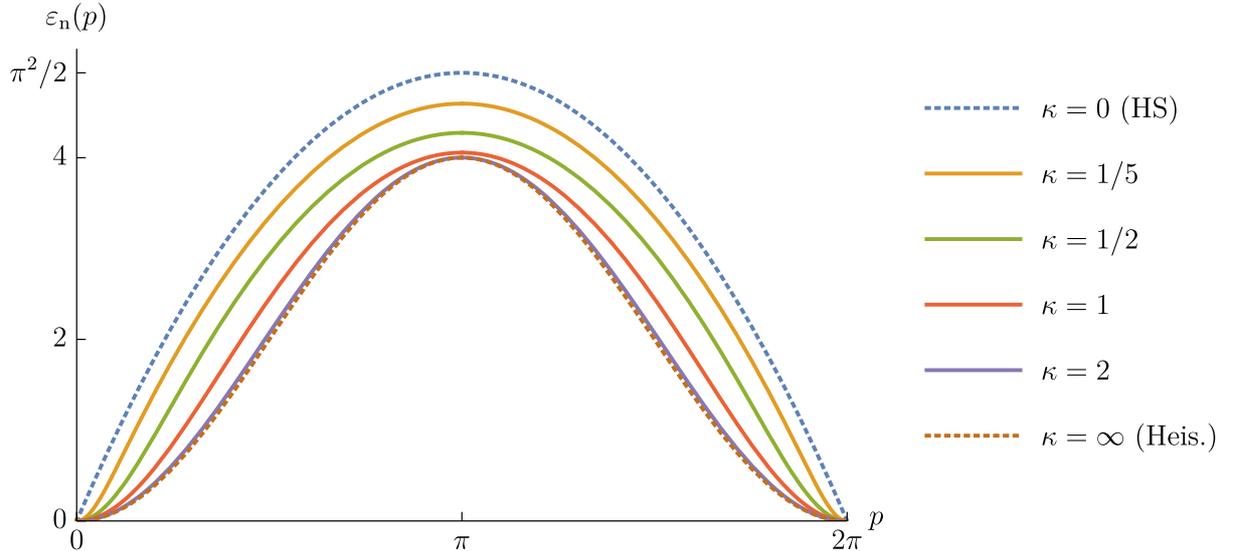}
	\caption{The normalised dispersion~\eqref{eq:dispersion_normalised}--\eqref{eq:dispersion_limits} for values of $\kappa$ interpolating between Haldane--Shastry and Heisenberg.}
	\label{fig:dispersion_kappa}
\end{figure}

In \textsection\ref{sec:mtm_lattices} we will compute the limits of the normalised dispersion,
\begin{subequations} \label{eq:dispersion_normalised}
	\begin{gather}
	\varepsilon_\text{n}(p) = \frac{n_H(\kappa)}{2} \, \bar{F}_2\Bigl(\frac{\I \, p}{2\,\kappa}\Bigr) \, ,
\intertext{where we recall from \eqref{eq:ham_normalised} that $n_H(\kappa)$ should diverge as $n_H \sim \kappa^{-2}\,\E^{2\kappa}$ for $\kappa\to \infty$ without affecting the other limit, $\lim_{\kappa\to 0} n_H = 1$,}
	\text{e.g.} \quad n_H(\kappa) = \frac{\sinh^2 \kappa}{\kappa^2} \, .
	\end{gather}
\checkedMma
\end{subequations}
To evaluate the Heisenberg limit of the dispersion we will further need the subleading term to behave as $n_H \sim \kappa^{-2}\,(\E^{2\kappa}-1)$ for $\kappa\to \infty$, which holds for the preceding choice.
The result is as follows. Let us denote the representative of $z \in \mathbb{C}/2\pi \mathbb{Z}$ in $[0,2\pi) \times \I \,\mathbb{R}$ by
\begin{equation} \label{eq:notation_mod2pi}
\modulo{z} \coloneqq z \ \mathrm{mod} \, 2\pi = z - 2\pi \lfloor \mathrm{Re}\,z/2\pi\rfloor \, .
\end{equation}
This notation is not to be confused with the complex conjugate $z^*$, nor with the bar that we use for the $L=1$ versions of the lattices and the associated functions. 
Then the limits yield the well-known Heisenberg and Haldane-Shastry dispersions
\begin{equation} \label{eq:dispersion_limits}
\tikz[baseline={([yshift=-.5*10pt*0.6+5pt]current bounding box.center)},scale=0.6]{
	\matrix (m) [matrix of math nodes,row sep=0em,column sep=4.5em]{
		& \substack{ \displaystyle \text{Inozemtsev} \\[.4em] \displaystyle \varepsilon_\text{n}(p) \ \eqref{eq:dispersion_normalised} } & \\
		\substack{ \displaystyle \text{Heisenberg} \\[.2em] \displaystyle 4 \, \sin^2(p/2) } & &  \substack{ \displaystyle \text{Haldane--Shastry} \\[.2em] \displaystyle \frac{\modulo{p} \,(2\pi-\modulo{p})}{2} \hfill } \\
	};
	\path[->] 
	(m-1-2) edge node [above left,xshift=.7cm,yshift=.25cm] {\footnotesize $\begin{array}{c} \kappa\to\infty \\ (\omega \to 0) \end{array}$} ([yshift=.1cm]m-2-1.east)
	(m-1-2) edge node [above right,xshift=-.7cm,yshift=.25cm] {\footnotesize $\begin{array}{c} \kappa\to0 \\ (\omega \to \I\,\infty) \end{array}$} ([yshift=.1cm]m-2-3.west);
}
\checkedMma
\end{equation}
Note that these functions are independent of the system size; $L$ only enters through the quantisation condition $p\in (2\pi/L)\,\mathbb{Z}_L$ for the allowed momenta. See Figure~\ref{fig:dispersion_kappa} for a plot.

\subsubsection{\textit{M}-particle difference equation} \label{sec:M-part_diff_eq} The real challenge is to find the highest-weight eigenvectors for multiple magnons. Projecting the eigenvalue equation~\eqref{eq:eigenvalue_equation} onto the coordinate-basis vector~\eqref{eq:coord_basis} like before yields the lattice Schrödinger equation for the $M$-magnon sector
\begin{equation*}
	\begin{aligned}
	E \, \Psi(\vect{n}) & = \frac{1}{2} \sum_{j\neq k}^L V(j-k) \, \cbra{\vect{n}} (1 - P_{jk}) \ket{\Psi} \\
	& = \sum_{m=1}^M \sum_{k(\notin \vect{n})}^L \! V(n_m - k) \, \bigl( \Psi(\vect{n}) - \Psi(n_1,\dots\mspace{-1mu},n_{m-1},k,n_{m+1},\dots\mspace{-1mu},n_M) \bigr) \, .
	\end{aligned}
\checkedMma
\end{equation*}
For one magnon $\Psi(n) = \E^{\I \mspace{1mu} p \mspace{1mu} n}$ gives \eqref{eq:dispersion_gen}. In general we obtain the \emph{\textit{M}-particle difference equation}
\begin{equation} \label{eq:M_particle_diff_eq_gen}
	\begin{aligned}
	\sum_{m=1}^M \sum_{k(\notin \vect{n})}^L \!\!V(n_m - k) \, [\Psi(\vect{n})]_{n_m \mapsto k} & = \Biggl( {-E} + \sum_{m=1}^M \sum_{k(\notin \vect{n})}^L \!\! V(n_m - k) \Biggr) \, \Psi(\vect{n}) \\
	& = \Biggl( {-E} + M \sum_{j=1}^{L-1} V(j) - \! \sum_{m\neq m'}^M \!\! V(n_m - n_{m'}) \Biggr) \, \Psi(\vect{n}) \, .
	\end{aligned}
\checkedMma
\end{equation}
Observe that, in order to avoid the need to distinguish various special cases, the left-hand side involves \emph{non}physical values of the wave function in the sense that $\Psi(\vect{n})$ only appears as a component $\Psi(\vect{n}) = \cbraket{\vect{n}}{\Psi}$ of $\ket{\Psi}$ when $n_1 < \dots < n_M$, cf.~\eqref{eq:coord_basis}. In writing \eqref{eq:M_particle_diff_eq_gen} we assume that the domain of the wave function can be extended to such nonphysical values. This assumption is standard, albeit often tacit, in the coordinate Bethe-ansatz analysis of the Heisenberg spin chain, see e.g.~\cite{Lam_14}.
The coordinate-basis vectors~\eqref{eq:coord_basis} are (formally) symmetric in the $n_m$, so the extended wave function may be taken to be symmetric without loss of generality. 

For Heisenberg $V(z) = \delta_{d_L(z),1}$ and \eqref{eq:M_particle_diff_eq_gen} becomes
\begin{equation*}
	\sum_{m=1}^M  \sum_{k(\notin \vect{n})}^L \!\! \delta_{k, n_m \pm 1} \, [\Psi(\vect{n})]_{n_m \mapsto k} = \Bigl( {-E} + 2 \, (M - N_{\vect{n}}) \Bigr) \, \Psi(\vect{n}) \, .
\checkedMma
\end{equation*}
The sums on the left-hand side give all possible ways of hopping any single excited spin from $n_m \in \vect{n}$ to a neighbouring site~$k$ that is not excited. On the right-hand side $N_{\vect{n}}$ denotes the number of pairs of neighbours among the coordinates~$\vect{n}$ of the excitations. This equation is familiar from the coordinate Bethe ansatz, cf.~e.g.~Appendix~B in \cite{Lam_14}.

For the HS spin chain it is possible \cites{Hal_91a,bernard1993yang} to recognise in the $M$-particle difference equation the action of the (trigonometric) Calogero--Sutherland Hamiltonian for $M$ particles moving on a circle with coupling $\beta\,(\beta-1)$ for $\beta=2$ (zonal spherical case). This approach might be feasible for Inozemtsev's elliptic spin chain too, but would lead in a different direction than we wish to take here. We plan to return to this in the future. 

For the elliptic case we wish to find the $M$-particle difference equation for $H_s$ from \eqref{eq:ham_shifted}. 
As for any potential shifted by a constant, $V_\text{s}(z) = V_\text{u}(z) + c$, the corresponding Hamiltonians are simply related by
\begin{equation*}
	H_\text{s} = H_\text{u} - c \, (\Omega - \Omega_0) \, ,
\checkedMma
\end{equation*}
where the difference is a multiple of the quadratic Casimir of $\mathfrak{sl}_2$ shifted by its value on $\ket{\uparrow\uparrow\cdots\uparrow}$,
\begin{equation*}
	\begin{aligned}
	\Omega \coloneqq {} & \Omega_0 - \sum_{j < k}^L (1 - P_{jk})
	\qquad\qquad\qquad\qquad\quad 
	\Omega_0 \coloneqq \frac{L}{2} \, \biggl(\frac{L}{2}+1\biggr) \\
	= {}& S^z \, S^z + \frac{1}{2} \, (S^+ \, S^- + S^- \, S^+) \\
	= {}& S^z \, (S^z \pm 1) + S^\mp \, S^\pm \, .
	\end{aligned}
\checkedMma 
\end{equation*}
Isotropy implies $[\Omega,H_\text{u}] = 0$.{\checkedMma}
The condition $\Omega = s\,(s+1)$, cf.\ the final expression for $\Omega$ on a highest-weight vector ($S^+ \, \ket{\Psi} = 0$), characterises the spin-$s$ (dimension $2\,s{+}1$) irrep of the Hilbert space. A highest-weight vector in the $M$-particle sector by definition has $s = \tfrac{L}{2} - M$, whence $E_\text{s} = E_\text{u} + M \, (L-M+1) \, c$ for such vectors. The upshot is that the $M$-particle difference equation for $H_\text{s}$ may, for highest-weight vectors, be obtained from that for $H_\text{u}$ by taking $E_\text{u} = E_\text{s} - M \, (L-M+1) \, c$.{\checkedMma}
With the aid of this observation we find the $M$-particle difference equation~\eqref{eq:M_particle_diff_eq_gen} for a highest-weight wave function of the shifted elliptic Hamiltonian~\eqref{eq:ham_shifted},\,%
\footnote{\ Our equation differs in the prefactor of $\eta_2$ from (59) in \cite{Inozemtsev:2002vb}: we find that it is independent of $L$.}
\begin{equation} \label{eq:M_particle_diff_eq_ell}
\begin{aligned}
\sum_{m=1}^M 
\sum_{k(\notin \vect{n})}^L
\! \wp(n_{m}-k) \, [\Psi(\vect{n})]_{n_m \mapsto k}
= \Biggl( {-E'_\text{s}} -\!\sum_{m \neq m'}^M \!\! \wp(n_{m}-n_{m'}) \Biggr) \Psi(\vect{n}) \, ,
 \\
E'_\text{s} \coloneqq E_\text{s} + M\,(M-1)\,\frac{\eta_2}{\omega} - M\frac{\bar{\eta}_2}{\omega} \, .
\end{aligned}
\checkedMma
\end{equation}
The left-hand side describes long-range hopping. For the right-hand side we used $\sum_{j = 1}^{L-1} \wp(j) = (\bar{\eta}_2 - L \, \eta_2)/\omega$, cf.~the elliptic dispersion relation~\eqref{eq:dispersion} and \textsection\ref{app:dispersion}.

\subsection{Extended coordinate Bethe ansatz} \label{sec:eCBA}
To solve the $M$-particle difference equation~\eqref{eq:M_particle_diff_eq_ell}, valid for highest-weight eigenvectors, Inozemtsev~\cites{Inozemtsev:1989yq,Inozemtsev_1995} sought for $M$-magnon wave functions in the form
\begin{equation} \label{eq:ansatz}
\Psi_{\tilde{\vect{p}},\vect{p}}(\vect{n}) \coloneqq \sum_{w\in S_M} \!\! \widetilde{\Psi}_{\tilde{\vect{p}}}(\vect{n}_w) \, \E^{\I \mspace{1mu} (\vect{p} - \tilde{\vect{p}}) \cdot \vect{n}_w} \, .
\end{equation}
Here $\vect{n}_w$ denotes the (right) action $(\vect{n}_w)_i = n_{w(i)}$ of $S_M$ and the symmetrisation is in accordance with the discussion following~\eqref{eq:M_particle_diff_eq_gen}. At a first glance the summand might be strange, featuring some function~$\widetilde{\Psi}$, a plane wave, and two sets of parameters $\tilde{\vect{p}}$ and~$\vect{p}$ that we wish to determine. Before we fix the meaning of these quantities through further assumptions let us provide some motivation for the form of \eqref{eq:ansatz}.

In the Heisenberg limit $\kappa\to\infty$ Bethe found long ago \cite{bethe1931theorie} that one may look for wave functions obtained from \eqref{eq:ansatz} with coefficients $\widetilde{\Psi}_{\tilde{\vect{p}}}(\vect{n}_w) \, \E^{-\I \mspace{1mu} \tilde{\vect{p}} \cdot \vect{n}_w}\rightsquigarrow A_w(\vect{p})$, to be supplemented by some identification between $\tilde{\vect{p}}$ and $\vect{p}$ (see \textsection\ref{sec:rapidity}). This captures the motion of (free) well-separated magnons with quasimomenta~$\vect{p}$, with (rational) coefficients that account for the nearest-neighbour interactions. The ansatz \eqref{eq:ansatz} looks a little like this, except that there are two sets of parameters and the coefficients are allowed to depend on $\vect{n}$ to reflect that the magnons interact at longer distances for finite $\kappa$. 

One might object that the exponentials in \eqref{eq:ansatz} may be absorbed in these coefficients, so that the extended coordinate Bethe ansatz seems to contain no information. It is even less clear how it can be reconciled with the very explicit eigenfunctions for the (Yangian highest-weight) vectors of the HS limit $\kappa\to 0$.\,%
\footnote{\ We will review these exact wave functions in \textsection\ref{sec:HS_recap_M=2} for $M=2$. In short: the HS wave function is the square of the Vandermonde polynomial times (a special case of) a Jack polynomial, where the variables are `evaluated' at roots of unity that represent the coordinates of the magnons on the chain. This has one similarity with \eqref{eq:ansatz} that is worth pointing out: Jack polynomials are naturally constructed by \emph{symmetrising} so-called nonsymmetric Jack polynomials. We plan to pursue this line of thought in future work.}
Nevertheless, the ansatz~\eqref{eq:ansatz} can be motivated in the infinite-length limit. There the hyperbolic pair potential, cf.~\eqref{eq:limits_potential}, allows for an asymptotic regime where all magnons are well separated and one may expect plane waves to be a reasonable ansatz. For finite separation one needs to correct these plane waves. By allowing their coefficients to be $\vect{n}$-dependent as in \eqref{eq:ansatz} Inozemtsev was able to determine the eigenfunctions~\cite{inozemtsev1992extended} via a connection with the hyperbolic Calogero--Sutherland model. In our notation $\vect{p}$ can then be identified as the quasimomenta, parametrising $p_\text{tot} = \sum_m p_m$ and $E = \sum_m \varepsilon_{\text{s}}(p_m)$ where the dispersion~\eqref{eq:dispersion} is as for finite~$L$. Although the coefficients depend on $\vect{n}$ their ratio becomes independent of $\vect{n}$ in the asymptotic regime with well-separated magnons and factorise into a product of infinite-length scattering matrices. The latter as just as for the Heisenberg spin chain, but with the usual rapidity replaced by its elliptic generalisation, given in~\eqref{eq:p_requirement} below \cites{dittrich1997two,klabbers2015inozemtsev}.

Rather than trying to provide any further physical motivation for the ansatz~\eqref{eq:ansatz} we will let the results speak for themselves. The merit of the explicit plane-wave factor will become clear in \textsection\ref{sec:connection}, where we will see that the coefficients $\widetilde{\Psi}_{\tilde{\vect{p}}}(\vect{n})$ have a neat physical interpretation. Our ansatz differs from Inozemtsev's~\cite{Inozemtsev_1995} in the characterisation (and meanings) of~$\tilde{\vect{p}}$ and $\vect{p}$, see \eqref{eq:quasiperiods}; see \textsection\ref{app:notation_comparison} for an overview of the changes with respect to the literature. This will allow us to identify $\vect{p}$ as the spin-chain quasimomenta and to streamline the Bethe-ansatz computations. This will yield more transparent results\,---\,see \eqref{eq:periodicitygenM}, \eqref{eq:M_particle_energy'}, \eqref{eq:p_requirement_mom}, \eqref{eq:M=2_constraint}\,---\,that remain finite in the Heisenberg and HS limits (\textsection\ref{sec:mtm_lattices} onwards).
The form \eqref{eq:ansatz} moreover helps consolidating the different parametrisations used by Inozemtsev in the analysis for $M=2$ \cite{Inozemtsev:1989yq}, $M=3$ \cite{inozemtsev1996solution}, and general $M$ \cites{Inozemtsev_1995,Inozemtsev_2000}.

\subsubsection{Assumptions} \label{sec:assumptions} To be able to analyse~\eqref{eq:M_particle_diff_eq_ell} we will need to look for the coefficients~$\widetilde{\Psi}_{\tilde{\vect{p}}}$ in \eqref{eq:ansatz}, with extended domain in view of the discussion following \eqref{eq:M_particle_diff_eq_gen}, in a space of reasonable functions. The following assumptions are quite natural in the elliptic setting.

\textit{Double quasiperiodicity.} Although the spin chain is only $L$-periodic, we will assume that the function $\widetilde{\Psi}_{\tilde{\vect{p}}}$ is \emph{doubly} (geometrically) quasiperiodic in each argument,
\begin{equation} 
\label{eq:quasiperiods}
\begin{aligned}
\widetilde{\Psi}_{\tilde{\vect{p}}}(n_1,\dots\mspace{-1mu},n_m + L,\dots\mspace{-1mu}, n_M) &= 
\E^{\I \mspace{1mu}\tilde{p}_m \mspace{1mu} L - \I \mspace{1mu} \varphi_m } \, \widetilde{\Psi}_{\tilde{\vect{p}}}(n_1,\dots\mspace{-1mu},n_M) \, , \\
\widetilde{\Psi}_{\tilde{\vect{p}}}(n_1,\dots\mspace{-1mu},n_m + \omega,\dots\mspace{-1mu}, n_M) &=\E^{\I\mspace{1mu}\tilde{p}_m \mspace{1mu}\omega} \, \widetilde{\Psi}_{\tilde{\vect{p}}}(n_1,\dots\mspace{-1mu},n_M) \, ,
\end{aligned}
\end{equation}
for some $\tilde{\vect{p}}$, defined (mod $2\pi/\omega = -2\,\I\,\kappa$) by the second equation, and $\vect{\varphi}$, which then parametrises the extent by which $\widetilde{\Psi}_{\tilde{\vect{p}}}$ differs from $\E^{\I \, \tilde{\vect{p}} \cdot \vect{n}}$. (We will see momentarily that these two sets of parameters must be related in our setting, which is why we only indicate the dependence of $\widetilde{\Psi}$ on one of them.)
An imaginary (quasi)period is not required by the spin-chain Hamiltonian, but can be motivated as follows. Consider \eqref{eq:M_particle_diff_eq_ell} and divide by $\Psi$, assuming that it does not vanish. Then the right-hand side is doubly periodic. For the resulting left-hand side to be so too it seems reasonable to expect $\widetilde{\Psi}_{\tilde{\vect{p}}}$, like the exponent, to have an imaginary quasiperiod.

\textit{Analytic structure.} We certainly want $\widetilde{\Psi}_{\tilde{\vect{p}}}$ to be meromorphic --- in the elliptic case any interesting wave function must have poles. Let us make this more precise. Like for the coordinate Bethe ansatz of the Heisenberg spin chain it will be convenient to (formally) allow for coinciding $n_m = n_{m'}$ in the extended domain, cf.~the discussion following \eqref{eq:M_particle_diff_eq_gen}. In the $M$-particle difference equation~\eqref{eq:M_particle_diff_eq_ell} the double pole of $\wp(z)$ at $z=0$ then makes an appearance. Towards the end of \textsection\ref{sec:connection} it will become clear that this pole can be balanced by asking $\widetilde{\Psi}_{\tilde{\vect{p}}}$ for coinciding arguments to have either a double zero (as in the HS limit) or a single pole. The derivation below crucially depends on the presence of poles, and will be simpler if we ask for them to lie on the real line. Following Inozemtsev~\cite{Inozemtsev_1995} we therefore demand simple poles at coinciding arguments:
\begin{equation}\label{eq:psi_decomposition}
	\widetilde{\Psi}_{\tilde{\vect{p}}}(\vect{n}) 
	= \frac{\widetilde{\Phi}_{\tilde{\vect{p}}}(\vect{n})}{\Delta(\vect{n})} \, ,  \qquad \Delta(\vect{n}) \coloneqq \prod_{m<m'}^M \!\!\! \sigma(n_m -n_{m'}) 
	\, .
\end{equation}
Here $\widetilde{\Phi}_{\tilde{\vect{p}}}$ is an analytic function and $\Delta$ is the elliptic Vandermonde factor in terms of the Weierstrass $\sigma$ function, see \textsection\ref{sec:Weierstrass}. The (geometric) quasiperiodicity of the latter ensures that $\Delta$ has quasiperiods $(L,\omega)$ too, so \eqref{eq:psi_decomposition} is compatible with \eqref{eq:quasiperiods}.

\subsubsection{Periodic boundary conditions} 
Besides the two assumptions on $\widetilde{\Psi}_{\tilde{\vect{p}}}$ we need to ensure that the ansatz $\Psi_{\tilde{\vect{p}},\vect{p}}$ has the correct periodicity properties, relating $\vect{p}$ from \eqref{eq:ansatz} with the quasiperiodicity parameters in \eqref{eq:quasiperiods}. The cyclicity condition~\eqref{eq:cyclicity_gen} demands from $\Psi_{\tilde{\vect{p}},\vect{p}}$ with the assumption \eqref{eq:quasiperiods} for $\widetilde{\Psi}_{\tilde{\vect{p}}}$ that 
\begin{equation} \label{eq:cyclicity}
	\sum_{w \in S_M} \!\! \widetilde{\Psi}_{\tilde{\vect{p}}}(\vect{n}_w) \, \E^{\I \mspace{1mu} (\vect{p} - \tilde{\vect{p}}) \cdot \vect{n}_w} \, \exp \, \I \left(L \, p_{w^{-1}(1)} - 	\varphi_{w^{-1}(1)} \right) = \! \sum_{w \in S_M} \!\!  \widetilde{\Psi}_{\tilde{\vect{p}}}(\vect{n}_w) \, \E^{\I \mspace{1mu} (\vect{p} - \tilde{\vect{p}})  \cdot \vect{n}_w} \, .
\end{equation}
This certainly holds\,%
\footnote{An argument for the passage from \eqref{eq:cyclicity} to \eqref{eq:periodicitygenM} goes as follows. Thinking of $\vect{n}$ in the extended domain as fixed let us view $\Psi_{\tilde{\vect{p}},\vect{p}}$ as a function of $\vect{r} \coloneqq \vect{p} - \tilde{\vect{p}}$. As long as $\vect{n}$ has all entries distinct there exists an $1\leq m^* \leq M$ for which $n_{m^{\!*}}$ is the (strict) largest among $\vect{n}$. For $1\leq m \leq M$ consider the limit $r_m \to -\I\, \infty$. On both sides of \eqref{eq:cyclicity} the terms for $w$ for which $w(m) = m^*$ then dominate all others. However, the additional exponential factor on the left-hand side contains $r_m = p_m - \tilde{p}_m$ too if $w(m)=1$. Provided $\Psi_{\tilde{\vect{p}},\vect{p}}(\vect{n}) \neq 0$ the only way that \eqref{eq:cyclicity} can hold in this limit is when $-\tilde{p}_m + \varphi_m/L$ grows like $r_m$, i.e.\ \eqref{eq:periodicitygenM} must hold asymptotically. Since we want all functions to depend meromorphically on all parameters this extend from a neighbourhood of infinity to the whole complex plane.}
if $\Psi_{\tilde{\vect{p}},\vect{p}}$ is fully $L$-periodic in each argument,
\begin{subequations} \label{eq:periodicitygenM}
\begin{gather}
\E^{\I \mspace{1mu} L \mspace{1mu} p_m} = \E^{\I \mspace{1mu} \varphi_m} 
\, , \qquad 1\leq m\leq M \, ,
\shortintertext{or in logarithmic form}
L \, p_m 
= 2\pi \mspace{1mu} I_m + \varphi_m
\, , \quad I_m \in \mathbb{Z}_L \, , \qquad 1\leq m\leq M \, .
\end{gather}
\end{subequations}
Given a choice of branch~$I_m$ for the complex logarithm this condition relates the parameters $\tilde{p}_m$ and $\varphi_m$ introduced in \eqref{eq:quasiperiods}. Inozemtsev's original derivation \cite{Inozemtsev_1995} hides \eqref{eq:periodicitygenM} by implementing it from the outset in \eqref{eq:ansatz}. However, it pays to make these equations explicit: these are nothing but the \emph{Bethe-ansatz equations} (\textsc{bae}).

In due course (\textsection\ref{sec:result}, \ref{sec:rapidity}, \ref{sec:S-matrix}) we will be able to give a physical interpretation to the parameters in \eqref{eq:periodicitygenM}. Recall that, by definition, the \emph{total (lattice) momentum}~$p_\text{tot}$ determines the eigenvalue of the (unitary) shift operator~$G$ from~\eqref{eq:shift_op} via $G \, \ket{\Psi} = \E^{\I \,p_\text{tot}} \, \ket{\Psi}$. Comparing the periodicity condition $G^L = 1$ with the preceding we can identify the combination
\begin{equation} \label{eq:tot_mtm}
p_\text{tot} = 
\sum_{m=1}^M \! p_m - \frac{1}{L} \sum_{m=1}^M \! \varphi_m
= \frac{2\pi}{L} \sum_{m=1}^M \! I_m \mod 2\pi
 \end{equation}
as the total momentum, with quantisation $p_\text{tot} \in (2\pi/L) \, \mathbb{Z}_L$ guaranteed by the \textsc{bae}~\eqref{eq:periodicitygenM}. 
In \textsection\ref{sec:result} we will see that $\sum_m \varphi_m = 0$ and that the $p_m$ remain finite in all limits. From the Heisenberg limit of \eqref{eq:periodicitygenM} one expects each $\varphi_m$ to be a function of $\vect{p}$. Indeed, in that limit $\varphi_m$ will be a sum of pairwise scattering phases with all other magnons, interpreted as the total scattering phase that the $m$th magnon picks up as it moves once around the chain. For general~$\kappa$ the long-range interactions spoil such an interpretation for \eqref{eq:periodicitygenM}. 
Nevertheless we will find further equations relating $\vect{p}$ and $\vect{\varphi}$, which together with \eqref{eq:periodicitygenM} characterise the `allowed' (`on shell') values of all parameters for which the extended \textsc{cba} gives solutions to the eigenvalue problem.

\subsection{Connection with elliptic Calogero--Sutherland} \label{sec:connection}

Armed with the ansatz~\eqref{eq:ansatz} we return to the $M$-particle difference equation~\eqref{eq:M_particle_diff_eq_ell}.
 
\subsubsection{Into the difference equation} Let us abbreviate $\vect{r} \coloneqq \vect{p} - \tilde{\vect{p}}$. Viewed as operators acting on $\Psi_{\tilde{\vect{p}},\vect{p}}(\vect{n})$ both sides of \eqref{eq:M_particle_diff_eq_ell} are symmetric in the $n_m$. We thus obtain an equality of sums over $w\in S_M$, coming from \eqref{eq:ansatz}, with summands depending only on the permuted coordinates~$\vect{n}_w$: 
\begin{equation} \label{eq:M_particle_general_M}
\sum_{w\in S_M} \sum_{m=1}^M \Sigma_m(\vect{n}_w) = \sum_{w\in S_M} \Biggl( {-E'_\text{s}} -\!\sum_{m \neq m'}^M \!\! \wp\bigl(n_{w(m)}-n_{w(m')}\bigr) \Biggr) \widetilde{\Psi}_{\tilde{\vect{p}}}(\vect{n}_w) \, \E^{\I \vect{r}\cdot \vect{n}_w} \, ,
\end{equation}
where we have defined
\begin{equation} \label{eq:M_particle_general_M_sum}
\Sigma_m(\vect{n}) \coloneqq \sum_{k(\notin \vect{n})}^L
\! \wp(n_m - k) \, \Bigl[\widetilde{\Psi}_{\tilde{\vect{p}}}(\vect{n}) \, \E^{\I \vect{r}\cdot \vect{n}} \Bigr]_{n_m \mapsto k} \, .
\end{equation}
The quasiperiodicity and poles from the assumptions for $\widetilde{\Psi}_{\tilde{\vect{p}}}$ make it possible to evaluate this sum following Inozemtsev~\cite{Inozemtsev_1995}. Since the calculations are somewhat involved we defer them to \textsection\ref{app:general_M_derivation}. In \textsection\ref{app:simplification} we show that plugging the result into the left-hand side of \eqref{eq:M_particle_general_M} gives
\begin{equation}
\label{eq:M_particle_plugged_in}
\begin{aligned}
& {-\frac{1}{2}} \sum_{w \in S_M} \sum_{m=1}^M \! \Bigl( \bigl( \partial_{\tilde{n}_m} - X_m \bigr)^{\!2} - X_m^2 -2 \, Y_m \Bigr) 
\widetilde{\Psi}_{\tilde{\vect{p}}}(\tilde{\vect{n}}) \, \E^{\I \vect{r}\cdot \tilde{\vect{n}}} \\
& \qquad - \! \sum_{m\neq m'}^M \mspace{1mu} \sum_{w \in S_M} \!\! \frac{\Pi_{m\mspace{1mu}m'}(\tilde{\vect{n}})}{2 \, \Delta(\tilde{\vect{n}})} \, \Bigl[ \Bigl(  \bigl(\partial_{\tilde{n}_m} - X_m \bigr)  - \bigl(\partial_{\tilde{n}_{m'}} - X_{m'} \bigr) \Bigr) 
\widetilde{\Phi}_{\tilde{\vect{p}}}(\tilde{\vect{n}}) \, \E^{\I \vect{r}\cdot \tilde{\vect{n}}} \Bigr]_{\tilde{n}_m = \tilde{n}_{m'} } \, .
\end{aligned}
\end{equation}
Here we abbreviate $\tilde{\vect{n}} \coloneqq \vect{n}_w$ and we have defined 
\begin{equation*}
X_m \coloneqq -\bar{\rho}_1\Bigl(\frac{\omega \, p_m}{2\,\pi} \Bigr) \, , \qquad Y_m \coloneqq \frac{1}{2} \, \bar{\wp}\Bigl(\frac{\omega \, p_m}{2\,\pi}\Bigr) - \frac{1}{2} \, \bar{\rho}_1\Bigl(\frac{\omega \, p_m}{2\,\pi} \Bigr)^2 \, .
\end{equation*}
Though it will not be relevant for us at the moment, the function $\Pi_{m,m'}$ in \eqref{eq:M_particle_plugged_in} is given by
\begin{equation*}
\Pi_{m m'}(\vect{n}) \coloneqq \wp(n_m-n_{m'}) \, \sigma(n_{m}-n_{m'}) \! \prod_{m''(\neq m,m')}^M \frac{\sigma(n_{m''} - n_{m})}{\sigma(n_{m''} - n_{m'})} \, .
\end{equation*}
Finally we recall that $\Delta$ is the (elliptic Vandermonde) denominator in our assumption~\eqref{eq:psi_decomposition}.

Observe that all derivatives in \eqref{eq:M_particle_plugged_in} appear in the combination $\partial_{n_m} - X_m$. This is where the explicit exponentials with the parameters~$\vect{r}$ in the ansatz~\eqref{eq:ansatz} come in. Since $\partial_{n_m} \bigl( \E^{\I \mspace{1mu} \vect{r}\cdot \vect{n}} \, F(\vect{n}) \bigr) = \E^{\I \mspace{1mu} \vect{r}\cdot \vect{n}} \bigl( \I \, r_m + \partial_{n_m} \bigr) F(\vect{n})$ let us choose $\I \, r_m = X_m$ to absorb all $X_m$ that come with a derivative. This relates $r_m = p_m - \tilde{p}_m$ to $p_m$. Using the Legendre relation it implies\,\footnote{\ Compare with (30) in \cite{Inozemtsev_1995} while noting that $p_\beta$ therein is closer to our $\tilde{p}_m$, defined in~\eqref{eq:quasiperiods}, than $p_m$.}
\begin{equation} \label{eq:p_requirement}
\tilde{p}_m = -\I \, \bar{\rho}_2\Bigl(\frac{\omega \, p_m}{2\,\pi}\Bigr) \, .
\end{equation}
Using this identification and the definition of $E'_\text{s}$ from \eqref{eq:M_particle_diff_eq_ell}, rearranging the terms in \eqref{eq:M_particle_plugged_in} results in the following equation:
\begin{equation} \label{eq:connection_to_eCS}
\begin{gathered}
\sum_{w \in S_M} \! \E^{\I \vect{r}\cdot \tilde{\vect{n}}} \Bigg(
\bigl( \widetilde{H}^{\ast} - \widetilde{E} \,\bigr) + \biggl( \widetilde{E} +\sum_{m=1}^M \left(\tfrac{1}{2} \, X_m^2 + Y_m -\frac{ \bar{\eta}_2}{\omega}\right) +E_\text{s}
\biggl) \Biggr) \widetilde{\Psi}_{\tilde{\vect{p}}}(\tilde{\vect{n}}) \\
= \sum_{m\neq m'}^M \sum_{w \in S_M} \!\! \frac{\Pi_{m\mspace{1mu}m'}(\tilde{\vect{n}})}{2\,\Delta(\tilde{\vect{n}})} \, \Bigl[ \E^{\I \vect{r}\cdot \tilde{\vect{n}}} \, \bigl(\partial_{\tilde{n}_m} - \partial_{\tilde{n}_{m'}} \bigr) 
\, \widetilde{\Phi}_{\tilde{\vect{p}}}(\tilde{\vect{n}}) \Bigr]_{\tilde{n}_m = \tilde{n}_{m'} } \, .
\end{gathered} 
\end{equation}
Here we recognised the Hamiltonian of the elliptic Calogero--Sutherland (eCS) model \cites{calogero1975exactly,sutherland1975exact} describing $M$ particles with unit mass moving on a circle with circumference $L$, whilst interacting pairwise via the (shifted) pair potential from~\eqref{eq:combs}:
\begin{equation} \label{eq:ham_eCS}
	\widetilde{H} \coloneqq -\frac{1}{2}\sum_{m=1}^M \! \frac{\partial^2}{\partial \, x_m^2} + 
	\beta \, (\beta-1) \! \sum_{m < m'}^M \biggl( \wp(x_m - x_{m'})+ \frac{\eta_2}{\omega} \biggr) \, .
\end{equation}
In \eqref{eq:connection_to_eCS} the asterisk~$\ast$ indicates that the reduced coupling is specialised to $\beta =2$ and the coordinates are `frozen' to the (equidistant) classical equilibrium positions $x_m = \tilde{n}_m \coloneqq n_{w(m)}$.

This is a very nice result. Let $\widetilde{\Psi}_{\tilde{\vect{p}}}$ be an eigenfunction of \eqref{eq:ham_eCS} with $\beta=2$ and energy $\widetilde{E}$,
\begin{equation} \label{eq:psi_tilde_eCS}
	\widetilde{H}\big|_{\beta = 2} \, \widetilde{\Psi}_{\tilde{\vect{p}}}(\vect{x}) = \widetilde{E} \, \widetilde{\Psi}_{\tilde{\vect{p}}}(\vect{x}) \, .
\end{equation} 
Then it inherits the double (quasi)periodicity \eqref{eq:quasiperiods},~\eqref{eq:periodicitygenM} and the analytic structure~\eqref{eq:psi_decomposition} from the pair potential in \eqref{eq:ham_eCS}\,%
\footnote{\ In fact, solutions will either have a simple pole or a double zero, corresponding to the two solutions $\beta=2$ and $\beta=-1$ of $\beta\,(\beta-1)=2$, respectively. We will briefly get back to this in \textsection\ref{sec:Lamé}.},
validating the computation of the sum $\Sigma_m$ that we started with. Of course \eqref{eq:psi_tilde_eCS} simplifies the first line of \eqref{eq:connection_to_eCS}. In \textsection\ref{app:derivative_numerator} we show that, moreover, the numerator $\widetilde{\Phi}_{\tilde{\vect{p}}}$ defined in \eqref{eq:psi_decomposition} is guaranteed to obey
\begin{equation}
\label{eq:F_requirement}
\Bigl[ \bigl(\partial_{x_m} - \partial_{x_{m'}} \bigr) \, 
\widetilde{\Phi}_{\tilde{\vect{p}}}(\vect{n}) \, \Bigr]_{x_{m} = x_{m'}} = 0 \, . 
\end{equation}
But this means that the right-hand side of \eqref{eq:connection_to_eCS} vanishes! Finally notice that
\begin{equation*}
	\tfrac{1}{2} \, X_m^2 + Y_m -\frac{ \bar{\eta}_2}{\omega} 
	= -\varepsilon_\text{s}(p_m) - \tfrac{1}{2} \, \tilde{p}_m^2 \, ,
\end{equation*}
where we recognised the dispersion~\eqref{eq:dispersion} and the square of \eqref{eq:p_requirement}. 

We should remark that the special case $\beta=2$ is quite special indeed. In the trigonometric limit (relevant for Haldane--Shastry) the corresponding eigenfunctions are spherical zonal polynomials, a special case of Jack polynomials related to (quaternionic) harmonic analysis~\cite{Mac_95}. In the hyperbolic limit, too, the eigenfunctions for $\beta=2$ are related to zonal spherical functions on a quaternionic symmetric space~\cite{OP_83,CV_90,inozemtsev1992extended}.
In general the cases $\beta \in \mathbb{Z}_{>0}$ are rather well understood. Note that $\beta=1$ is free, and not very interesting. The spectrum is much richer for $\beta \geq 2$, of which $\beta=2$ is the simplest case. For instance, when $M=2$ (see \textsection\ref{sec:Lamé}) there are simple explicit wave functions for any $\beta \in \mathbb{Z}_{>0}$, depending on auxiliary parameters that for $\beta>2$ must obey additional `Bethe-type' equations. 

\subsubsection{Result}\label{sec:result} Suppose that $\widetilde{\Psi}_{\tilde{\vect{p}}}$ is an eigenfunction of the elliptic Calogero--Sutherland model for $\beta=2$, solving \eqref{eq:psi_tilde_eCS}, with poles at coinciding arguments; as we have just seen it obeys the assumptions from \textsection\ref{sec:eCBA}. Then the ansatz~\eqref{eq:ansatz} yields a solution to the spin-chain eigenvalue equation~\eqref{eq:eigenvalue_equation} if the coordinates~$x_m$ are `frozen' to the spin-chain sites~$n_m \in \mathbb{Z}_L$ and the parameters $\tilde{p}_m$ are determined by $p_m$ through~\eqref{eq:p_requirement}. The corresponding energy has the `quasi-additive' form
\begin{equation} \label{eq:M_particle_energy'}
E_\text{s}
= \sum_{m=1}^M \varepsilon_\text{s}(p_m) - \widetilde{U} \, , \qquad \text{with $\widetilde{U}$ defined by} \qquad \widetilde{E} = \frac{1}{2}\sum_{m=1}^M \tilde{p}_m^2 + \widetilde{U} \, .
\end{equation}
Inozemtsev already found that the spin-chain energy is related to that of the elliptic Calogero--Sutherland model, but his relation is not quite as simple as \eqref{eq:M_particle_energy'}.

From the eCS viewpoint the relevant parameters are $\tilde{\vect{p}}$ and $\vect{\varphi}$, defined by the quasiperiodicity conditions \eqref{eq:quasiperiods}. These $2M$ parameters cannot all be independent; for example, when $M=2$ the eigenspace of \eqref{eq:ham_eCS} is two-dimensional (\textsection\ref{sec:M=2}).
For the spin chain the key parameter is $\vect{p}$, which is related to $\tilde{\vect{p}}$ via \eqref{eq:p_requirement}. For their interpretations see the paragraph preceding \textsection\ref{sec:mtm_lattices}, and \textsection\ref{sec:rapidity}, respectively. The \textsc{bae}~\eqref{eq:periodicitygenM} relate $\vect{p}$ and $\vect{\varphi}$.

Let us illustrate this by revisiting $M=1$. The ansatz \eqref{eq:ansatz} reads $\Psi_{\tilde{p},p}(n) = \widetilde{\Psi}_{\tilde{p}}(n)\,\E^{\I\mspace{1mu} (p-\tilde{p})\mspace{1mu}n}$. The assumptions further demand $\widetilde{\Psi}_{\tilde{p}}$ to be doubly quasiperiodic, $\widetilde{\Psi}_{\tilde{p}}(n+L) = \E^{\I\mspace{1mu} \tilde{p}\mspace{1mu}L+\I\mspace{1mu} \varphi} \, \widetilde{\Psi}_{\tilde{p}}(n)$, $\widetilde{\Psi}_{\tilde{p}}(n+\omega)=\E^{\I\mspace{1mu} \tilde{p}\mspace{1mu}\omega} \, \widetilde{\Psi}_{\tilde{p}}(n)$, and entire. The periodic boundary conditions for the spin chain finally mean that $p$ is of the form $L\, p = 2\pi\mspace{1mu}I +\varphi$ for $I\in\mathbb{Z}_L$. Now consider the wave equation $-\partial_x^2 \, \widetilde{\Psi}(x) = \widetilde{E} \, \widetilde{\Psi}(x)$. Its solution $\widetilde{\Psi}(x) = A \,\E^{\I B x}$ is indeed entire. Double quasiperiodicity fixes $B = \widetilde{p}$ and $\varphi=0$. Thus $\widetilde{\Psi}_{\tilde{p}}(x) = A \,\E^{\I\mspace{1mu}\tilde{p}\mspace{1mu}x}$, which has $\widetilde{E} = \tilde{p}^2/2$ so that $\widetilde{U}=0$. It follows that the spin-chain wave function is $\Psi_p(n) \coloneqq \Psi_{\tilde{p},p}(n) = A \, \E^{\I \mspace{1mu}p \mspace{1mu}n}$, with $p=2\pi I/L$ quantised as usual, and energy given by the dispersion~\eqref{eq:dispersion_normalised}. We thus recover the magnons from \textsection\ref{sec:M=1}.

Now we return to general $M$. Observe that the eCS Hamiltonian $\widetilde{H}$ is translationally invariant: it commutes with the \emph{total (continuum) momentum operator} $\widetilde{P} \coloneqq -\I \sum_m \partial_{x_m}$, so we can simultaneously diagonalise these two operators. Let us show that our double quasiperiodicity assumption~\eqref{eq:quasiperiods} fixes the $\widetilde{P}$-eigenvalue of $\widetilde{\Psi}_{\tilde{\vect{p}}}$ to
\begin{equation} \label{eq:ptilde_tot_varphi_sum}
\widetilde{p}_\text{tot} = \sum_{m=1}^M \widetilde{p}_m \in \mathbb{R}/2\pi\mathbb{Z} \, , \qquad \sum_{m=1}^M \varphi_m = 0 \ \mathrm{mod}\,2\pi \, ,
\end{equation}
where the latter is a consistency condition for the parameters $\vect{\varphi}$. Indeed, reality of the total momentum $\widetilde{p}_\text{tot}$ is guaranteed by hermiticity of $\widetilde{P}$. To find its value we exponentiate and pass to the \emph{(continuum) translation operator} 
\begin{equation*}
\widetilde{G}(a) \coloneqq \exp\bigl(\I \, a \mspace{1mu} \widetilde{P}\bigr) \, , \qquad \bigl(\widetilde{G}(a) \, \widetilde{\Psi}_{\tilde{\vect{p}}}\bigr)(\vect{x}) = \widetilde{\Psi}_{\tilde{\vect{p}}}(x_1+a,\cdots\mspace{-1mu},x_M+a) \, .
\end{equation*}
Since we view $\widetilde{\Psi}_{\tilde{\vect{p}}}$ as a function on $\mathbb{C}$ we can take $a=\omega$ to obtain the first relation in \eqref{eq:ptilde_tot_varphi_sum} thanks to \eqref{eq:quasiperiods}. By taking $a=L$ and using \eqref{eq:quasiperiods} we furthermore see that this value is only consistent if the second condition in \eqref{eq:ptilde_tot_varphi_sum} holds. 

From the spin-chain viewpoint the condition $\sum_m \varphi_m = 0$ allows us to identify $p_m$ as the contribution of the $m$th magnon to the total momentum~\eqref{eq:tot_mtm}. If it would furthermore be true that $\widetilde{U} = 0$ then the energy~\eqref{eq:M_particle_energy'} would become fully (functionally) additive, just like the Heisenberg spin chain, with contribution $\varepsilon_\text{s}(p_m)$ due to the $m$th magnon. Although we will see that this is not quite the case let us, perhaps by slight abuse of terminology, call $\vect{p}$ the \emph{quasimomenta}.

\subsection{Intermezzo: momentum lattices} \label{sec:mtm_lattices} Before we discuss the results of the Bethe-type analysis it will be useful to introduce reciprocal (momentum) lattices. These make the reality and positivity manifest for the dispersion~\eqref{eq:dispersion} and the relation \eqref{eq:p_requirement} between the (quasi)momenta, and are helpful for evaluating the Heisenberg and HS limits.

So far we have encountered two period lattices, $\mathbb{L} \coloneqq L \, \mathbb{Z} \oplus \omega \, \mathbb{Z}$ and $\!\!\bar{\,\,\mathbb{L}} \coloneqq \mathbb{Z} \oplus \omega \, \mathbb{Z}$. The former is naturally associated to the Hamiltonian. 
The latter originates from the (coordinate) shift symmetry appearing in the lattice Schr\"{o}dinger equation\,%
\footnote{\ From a computational point of view this shift symmetry originates in the periodicity of the `regularisations' \eqref{eq:W_definition_1}, \eqref{eq:W_definition_2} and \eqref{eq:Wm_definition} of the sums \eqref{eq:dispersion_gen} and \eqref{eq:M_particle_general_M_sum}.} and manifests itself in the dispersion relation \eqref{eq:dispersion} and the identification \eqref{eq:p_requirement}. In both of these equations the relevant variables are momenta, which appear in the combination $\omega\mspace{1mu}p/2\pi$. Since the momenta are defined mod $2\pi$ it is more natural to write these expressions on a lattice with real period $2\pi$. Let us therefore introduce the \emph{momentum} (reciprocal) lattices
\begin{equation}
\label{eq:momentum_lattice}
\begin{aligned}
\!\!\,\,\mathbb{L}^{\!\vee} &= \omega_1^\vee \, \mathbb{Z} \oplus \omega_2^\vee \, \mathbb{Z} \, , \qquad (\omega_1^\vee,\omega_2^\vee) \coloneqq (2\pi, -2\pi L /\omega) = (2\pi,2\mspace{1mu}\I\mspace{1mu}L\mspace{1mu}\kappa)\, , \\
\!\!\bar{\,\,\mathbb{L}}^{\!\vee} &= \bar{\omega}_1^\vee \, \mathbb{Z} \oplus \bar{\omega}_2^\vee \, \mathbb{Z} \, , \qquad (\bar{\omega}_1^\vee,\bar{\omega}_2^\vee) \coloneqq (2\pi, -2\pi /\omega) = (2\pi, 2\mspace{1mu}\I\mspace{1mu}\kappa) \, ,
\end{aligned}
\end{equation}
which are obtained from the \emph{coordinate} lattices $\mathbb{L}$ and $\!\!\bar{\,\, \mathbb{L}}$ by rescaling by $2\pi/\omega$. Note that we swapped the labels of the lattice generators so that $b=1$ ($b=2$) still corresponds to the real (imaginary) direction. We further included a minus sign for $b=2$ to ensure the imaginary part remains positive, so that the orientation is unchanged. Observe that the $\kappa$-dependence of the momentum lattices resides in the \emph{numerator} of the imaginary period. We already encountered the ratio $\momL{\omega}_2/\momL{\omega}_1$ in \eqref{eq:theta} as the lattice parameter~$\tau$ of $\theta_2$ (and $\omega_2/\omega_1$ for $\theta_1$).
The rescaling by $2\pi/\omega$ moreover takes care of the following. The special functions in \eqref{eq:rho_def} and \eqref{eq:F_def} all have analogues associated to the momentum lattices, which we will denote by $\momL{\rho}_b,\mom{\rho}_b$ and so on. Importantly, all of these functions have a \emph{homogeneity} property (\textsection\ref{sec:Weierstrass}) allowing us to pass between the coordinate and momentum lattices by rescaling the function and arguments. In particular,
\begin{subequations} \label{eq:rho_mtm_lattice}
\begin{gather}
	\momL{\zeta}(z) = \frac{\omega}{2\pi} \, \zeta\biggl(\frac{\omega\mspace{1mu}z}{2\pi}\biggr) \, , \qquad \text{so} \qquad \momL{\eta}_1 = \frac{\omega}{2\pi} \, \eta_2 \, , \quad 
	\momL{\eta}_2 = {-\frac{\omega}{2\pi}} \, \eta_1 \, , 
\checkedMma
\shortintertext{and therefore} 
	\momL{\rho}_a(z) = \frac{\omega}{2\pi} \, \rho_b\biggl(\frac{\omega\mspace{1mu}z}{2\pi}\biggr) \quad \text{and} \quad 
	\mom{\rho}_a(z) = \frac{\omega}{2\pi} \, \bar{\rho}_b\biggl(\frac{\omega\mspace{1mu}z}{2\pi}\biggr) \, , \qquad a \neq b \, .
\checkedMma
\end{gather}
\end{subequations}
These functions remain doubly (quasi)periodic, $\momL{\rho}_b(z + \momL{\omega}_b) = \momL{\rho}_b(z)$ and $\momL{\rho}_b(z + \momL{\omega}_a) = \momL{\rho}_b(z) + (-1)^b \, 2\pi/\momL{\omega}_b$ ($a\neq b$).{\checkedMma} 
Likewise, we have
\begin{equation} \label{eq:F_mtm_lattice}
	\momL{F}_a(z) = \biggl(\frac{\omega}{2\pi}\biggr)^{\!\!2} \, F_b\biggl(\frac{\omega\mspace{1mu}z}{2\pi}\biggr) \quad \text{and} \quad 
	\mom{F}_a(z) = \biggl(\frac{\omega}{2\pi}\biggr)^{\!\!2} \, \bar{F}_b\biggl(\frac{\omega\mspace{1mu}z}{2\pi}\biggr) \, , \qquad a \neq b \, .
	\checkedMma
\end{equation}

\subsubsection{Rapidity} \label{sec:rapidity}
In terms of this notation the identification \eqref{eq:p_requirement} acquires the simple form
\begin{equation} \label{eq:p_requirement_mom}
	\frac{\omega \mspace{1mu} \tilde{p}_m}{ 2 \pi \mspace{1mu} \I } = \frac{\tilde{p}_m}{ 2 \kappa} = \lambda_\text{u}(p_m) \coloneqq -\mom{\rho}_1( p_m) \, .
\checkedMma
\end{equation}
As $\mom{\rho}_1$ is real and strictly \emph{de}creasing on the interval $[0,2\pi]${\checkedMma}
the sign on the right-hand side ensures that $\tilde{p}_m$ \emph{in}creases with $p_m$. 

\begin{figure}[h]
	\centering
	\includegraphics[width=.95\textwidth]{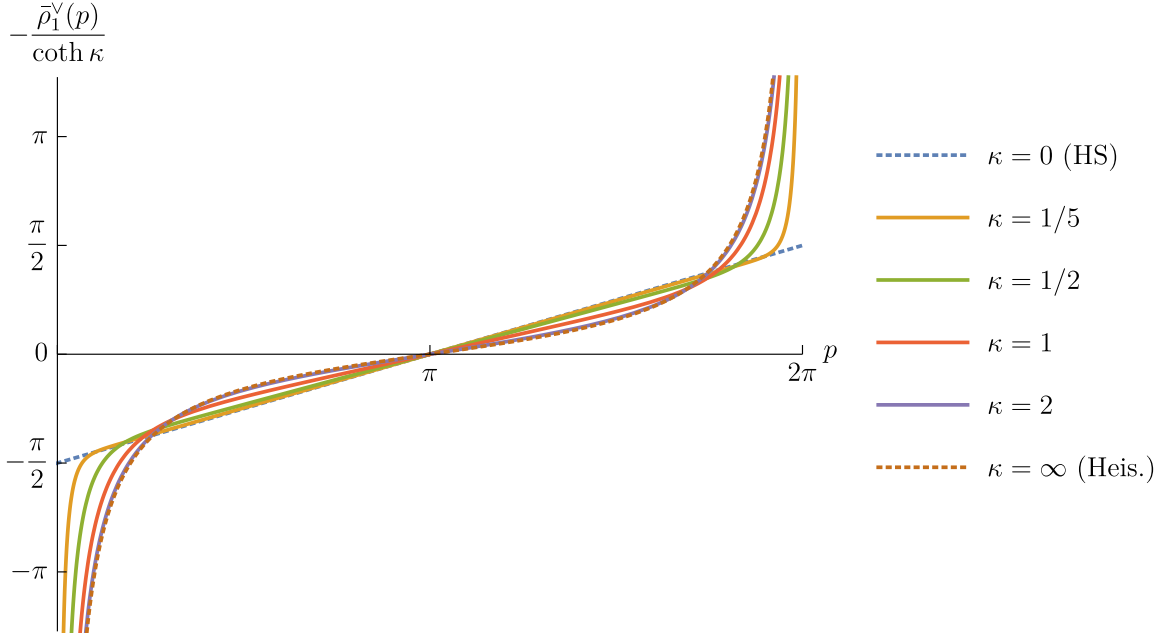}
	\caption{The (normalised) rapidity function~\eqref{eq:limits_rhobarcheck} for values of $\kappa$ interpolating between Haldane--Shastry and Heisenberg.}
	\label{fig:rapidity_kappa}
\end{figure}

To make sense of the HS limit we once more have to rescale a little. Let $n_\lambda(\kappa)$ be a function such that $\lim_{\kappa\to\infty} n_\lambda(\kappa) = 1$ while $n_\lambda(\kappa) = \kappa + O(\kappa^2)$,
\begin{equation} \label{eq:n_lambda}
\text{e.g.} \qquad n_\lambda(\kappa) = \tanh\kappa 
\, .
\end{equation}
Then for $\modulo{z}\neq 0$ we have\,%
\footnote{\ Note that $\modulo{{-}z} - \pi = \pi - \modulo{z}$ holds for $\modulo{z}\neq 0$. The value at $\modulo{z} = 0$, a pole of $\momL{\rho}_1$, is not determined by the limit.}
\begin{equation} \label{eq:limits_rhocheck}
\tikz[baseline={([yshift=-.5*10pt*0.6+5pt]current bounding box.center)},scale=0.6]{
	\matrix (m) [matrix of math nodes,row sep=0em,column sep=5em]{
		& \substack{\displaystyle \text{Inozemtsev} \\[.5em] \displaystyle 
		n_\lambda(L \, \kappa) \, \momL{\rho}_1(z) } & \\
		\substack{\displaystyle \text{Heisenberg} \\[.2em] \displaystyle \frac{1}{2} \cot \frac{z}{2} } & & \substack{\displaystyle \text{Haldane--Shastry} \\[.5em] \displaystyle \frac{\modulo{{-}z} - \pi}{2}  = \frac{\pi - \modulo{z}}{2} } \\
		& \displaystyle \frac{1}{2} \cot \frac{z}{2} & \\
	};
	\path[->] 
	(m-1-2) edge node [above left,xshift=.9cm,yshift=.3cm] {\footnotesize $\begin{array}{c} \kappa\to\infty \\ (\omega \to 0) \end{array}$} ([yshift=.1cm]m-2-1.east)
	(m-1-2) edge node [right] {\footnotesize $L\to\infty$ \ } (m-3-2)
	(m-1-2) edge node [above right,xshift=-.9cm,yshift=.3cm] {\footnotesize $\begin{array}{c} \kappa\to0 \\ (\omega \to \I\,\infty) \end{array}$} ([yshift=.2cm]m-2-3.west);
}
\checkedMma
\end{equation}
where we note that the real period~$2\pi$ survives in all limits, cf.~the notation~\eqref{eq:notation_mod2pi} for the \textsc{hs} limit, as is appropriate for a function of the momentum. At $L=1$ \eqref{eq:limits_rhocheck} yields, again for $\modulo{z}\neq 0$,
\begin{equation} \label{eq:limits_rhobarcheck}
\tikz[baseline={([yshift=-.5*10pt*0.6+5pt]current bounding box.center)},scale=0.6]{
	\matrix (m) [matrix of math nodes,row sep=0em,column sep=5em]{
		& \substack{\displaystyle \text{Inozemtsev} \\[.5em] \displaystyle 
		-n_\lambda(\kappa) \, \mom{\rho}_1(z) } & \\
		\substack{\displaystyle \text{Heisenberg} \\[.2em] \displaystyle -\frac{1}{2} \cot \frac{z}{2} } & & \substack{\displaystyle \text{Haldane--Shastry} \\[.5em] \displaystyle \frac{\pi-\modulo{{-}z}}{2} = \frac{\modulo{z} - \pi}{2} } \\
	};
	\path[->] 
	(m-1-2) edge node [above left,xshift=.9cm,yshift=.3cm] {\footnotesize $\begin{array}{c} \kappa\to\infty \\ (\omega \to 0) \end{array}$} ([yshift=.1cm]m-2-1.east)
	(m-1-2) edge node [above right,xshift=-.9cm,yshift=.3cm] {\footnotesize $\begin{array}{c} \kappa\to0 \\ (\omega \to \I\,\infty) \end{array}$} ([yshift=.2cm]m-2-3.west);
}
\checkedMma
\end{equation}
In the Heisenberg limit we obtain the familiar rapidity function. The HS limit gives $(z-\pi)/2$ shifted by a multiple of $\pi$ so as to take values in $(-\pi/2,\pi/2)$. See also Figure~\ref{fig:rapidity_kappa}.
Motivated by these limits we identify the combinations \eqref{eq:p_requirement_mom} and 
\begin{equation}
\lambda_\text{n}(p) \coloneqq n_\lambda(\kappa) \, \lambda_\text{u}(p) = -n_\lambda(\kappa) \, \mom{\rho}_1(p) 
\end{equation}
as the \emph{rapidity} from the spin-chain point of view. The sign in the Heisenberg limit is perhaps slightly unusual, but does not affect the denominator-free version of the Bethe-ansatz equations. It ensures that the rapidity is an increasing function (on the reals). Moreover, if (e.g.\ on $p\in[0,2\pi]$) we invert $\tilde{p} = \lambda_\textsc{h}(p) \coloneqq -\cot(p/2)/2$ then $\mathrm{d}\mspace{1mu}p/\mathrm{d}\mspace{1mu}\tilde{p}\,|_{\tilde{p} = \lambda_\textsc{h}(p)} = 4\, \sin^2(p/2)$ matches \eqref{eq:dispersion_limits}, as usual for the Heisenberg spin chain.\,%
\footnote{\ This relation, which can be understood as a consequence of the `trace identities' expressing the Heisenberg Hamiltonians as logarithmic derivatives of the transfer matrix, breaks down away from the Heisenberg limit. Indeed, numerical evaluation of $\mathrm{d}\mspace{1mu}p/\mathrm{d}\mspace{1mu}\tilde{p}\,|_{\tilde{p} = \lambda_\text{n}(p)} = -\coth\kappa \, ((\mom{\rho}_1)^{-1})'(\mom{\rho}_1(p))$ differs from $\varepsilon_\text{n}(p)$ for finite $\kappa$. (This is clear in the HS limit, where $\tilde{p}=(p-\pi)/2$ if $p\in[0,2\pi]$ so $\mathrm{d}\mspace{1mu}p/\mathrm{d}\mspace{1mu}\tilde{p} = 2$ is constant.)}
Conversely, this gives an interesting interpretation of the \emph{rapidity as the momentum of the excitations in the eCS model}.

In view of our definition of the quasimomenta $\vect{p}$ and the fact that their sum \eqref{eq:tot_mtm} is finite for any value of $\kappa$ it seems natural to assume that the individual $p_m$ are finite for any $\kappa$ too. We will see that this is indeed the case. By the \textsc{bae}~\eqref{eq:periodicitygenM} the same holds for $\vect{\varphi}$. The auxiliary parameters $r_m$ are no longer useful from the spin-chain point of view, and do not seem to have meaningful limits. 

\subsubsection{Limits of the dispersion} \label{sec:dispersion_limit}
In the same way we can now write the dispersion \eqref{eq:dispersion} as
\begin{equation} \label{eq:dispersion_in_check}
	\varepsilon_\text{s}(p) = \frac{1}{2} \, \biggl(\frac{2\pi}{\omega}\biggr)^{\!2} \, \mom{F}_1(p)  = {-2} \, \kappa^2 \mspace{1mu} \mom{F}_1(p) \, , \qquad
	\mom{F}_1(p) = \mom{\rho}_1{}'(p) + \mom{\rho}_1(p)^2 + \frac{3 \,\mom{\eta}_1}{2\pi} \, .
\checkedMma
\end{equation}
This form makes it more clear that the dispersion is real and $2\pi$-periodic and allows for a simple proof of its positivity.\,%
\footnote{\label{fn:positive_dispersion} 
Note that $\mom{F}_1{}'(p) = \mom{\rho}_1{}'(p) \, c(p)$ for $c(p) \coloneqq \mom{\rho}_1{}''(p)/\mom{\rho}_1{}'(p) +2\, \mom{\rho}_1(p)$. Thus $\mom{F}_1{}'(p)$ vanishes if $c(p)$ does. The function $c'(p)$ is elliptic of order two and thus has no more than two zeroes on $[0,2\pi)$. By Rolle's theorem, this implies that $c(p)$ has at most three zeroes in this interval. Since $\mom{\rho}_1{}'(p)$  is nonzero on $[0,2\pi)$ this implies $\mom{F}_1{}'(p)$ has at most three zeroes on $[0,2\pi)$, of which two are at $p=0$ and $p=\pi$. Since $\mom{F}_1{}'(p) = - \mom{F}_1{}'(2 \pi - p)$ we see that if there were a third zero $p^*$ then there would be another at $2 \pi - p^*$, yielding a contradiction. So $\bar{F}_1{}'(p)$ has precisely two zeroes. Since $\mom{F}_1$ vanishes at $p=0$ and $p=2\pi$ and $\mom{F}_1 (p) = \mom{F}_1(\pi-p)$ the above implies that $\mom{F}_1$ does not have any further zeroes and is therefore either nonnegative or nonpositive. Since $\mom{F}_1(\pi) <0$ it follows that it is nonpositive, proving that the dispersion \eqref{eq:dispersion_in_check} is nonnegative.}
Using \eqref{eq:limits_rhobarcheck} we can evaluate the Heisenberg and HS limits to obtain \eqref{eq:dispersion_limits}. Indeed, the quantity 
\begin{equation} \label{eq:limits_etabarcheck}
\tikz[baseline={([yshift=-.5*10pt*0.6+5pt]current bounding box.center)},scale=0.6]{
	\matrix (m) [matrix of math nodes,row sep=0em,column sep=5em]{
		& \displaystyle n_\lambda(\kappa)^2 \, \frac{3\,\mom{\eta}_1}{2\pi} & \\
		\displaystyle \frac{1}{4} & &  \displaystyle -\frac{\pi^2}{4} \\
	};
	\path[->] 
	(m-1-2) edge node [above left,xshift=.7cm,yshift=.1cm] {\footnotesize $\begin{array}{c} \kappa\to\infty \\ (\omega \to 0) \end{array}$} ([yshift=.1cm]m-2-1.east)
	(m-1-2) edge node [above right,xshift=-.7cm,yshift=.1cm] {\footnotesize $\begin{array}{c} \kappa\to0 \\ (\omega \to \I\,\infty) \end{array}$} ([yshift=.2cm]m-2-3.west);
}
\checkedMma
\end{equation}
has well-defined limits. Let us therefore write the normalised dispersion relation as
\begin{equation} \label{eq:dispersion_via_rapidity}
	\varepsilon_\text{n}(p) = {-2} \, n_H(\kappa)\,\kappa^2 \, \mom{F}_1(p) = \frac{2\,n_H(\kappa)\,\kappa^2}{n_\lambda(\kappa)} \, \lambda_\text{n}'(p) - \frac{2\,n_H(\kappa)\,\kappa^2}{n_\lambda(\kappa)^2} \, \biggl( \lambda_\text{n}(p)^2 + n_\lambda(\kappa)^2 \, \frac{3\,\mom{\eta}_1}{2\pi} \biggr) \, .
\end{equation}
Since $n_H(\kappa)\,\kappa^2/n_\lambda(\kappa) \in O(\kappa)$ the first term vanishes while $n_H(\kappa)\,\kappa^2/n_\lambda(\kappa)^2 \to1$ as $\kappa\to0$ and we obtain the HS limit given in \eqref{eq:dispersion_limits}. 

The Heisenberg limit is more subtle: for $\kappa\to\infty$ the leading terms cancel so we have to go one term further in the expansion in $q \coloneqq \E^{-\kappa}$. The Fourier expansion of $\zeta$ \cite[\textsection 23.8]{DLMF} implies
\begin{equation*}
	\mom{\rho}_1(p) = \frac{1}{2}\cot\frac{p}{2} + 2\, q^2 \sin p + \mathcal{O}(q^4) \, , \qquad
	\frac{3\,\mom{\eta}_1}{2\pi} = \frac{1}{4} - 6\, q^2 + \mathcal{O}(q^4) \, .
\end{equation*}
One can readily find the subleading term of the rapidity and constant in \eqref{eq:dispersion_via_rapidity}, but here it is simpler to use the expression of $\mom{F}_1$ in terms of $\mom{\rho}_1$ rather than $\lambda_\text{n}$. In either case we obtain the Heisenberg dispersion: 
\vspace{-10pt}
\begin{equation*}
	\begin{aligned}
	\varepsilon_\text{n}(p) = {} & {-}\frac{q^{-2} \, \bigl( 1+ O(q^2) \bigr)}{2} \Biggl( \frac{1}{4} \biggl(\,\overbrace{\! \frac{-1}{\sin^2(p/2)} + \cot^2\frac{p}{2} + 1\!}^{\displaystyle \mathrlap{ =0 } }\,\biggr) \\
	& \hphantom{{-}\frac{q^{-2} \, \bigl( 1+ O(q^2) \bigr)}{2} \Biggl(} 
	+ 2\,q^2 \Bigl(\,\underbrace{\!\cos p + \cot\frac{p}{2} \sin p -3 \!}_{\displaystyle \mathrlap{ = -4\sin^2(p/2) } }\,\Bigr) + \mathcal{O}(q^4) \Biggr) \\[-2ex]
	=\  {} &  4\sin^2(p/2) + O(q^2) \, . 
	\end{aligned}
\end{equation*}
This establishes the limits given in \eqref{eq:dispersion_limits}.

\section{The two-particle sector} \label{sec:M=2}

\subsection{From Lamé to two magnons} Let us apply the results of \textsection\ref{sec:connection}.

\subsubsection{Lamé equation} \label{sec:Lamé} For $M=2$ the elliptic Calogero--Sutherland Hamiltonian~\eqref{eq:ham_eCS} reads
\begin{equation}
	\widetilde{H}\big|_{M=2} 
	= -\biggl(\frac{\partial^2}{\partial \mspace{1mu} X^2} + \frac{\partial^2}{\partial \mspace{1mu} x^2}\biggr) + \beta\,(\beta-1) \, \biggl( \wp(x) + \frac{\eta_2}{\omega} \biggr) \, ,
\end{equation}
where we passed to coordinates $X \coloneqq x_1 + x_2$ and $x \coloneqq x_1 - x_2$. Separating variables as
\begin{equation*}
\widetilde{\Psi}_{\tilde{\vect{p}}}(x_1,x_2) = \Phi(X) \, \phi(x) 
\end{equation*}
we obtain free centre-of-mass motion,
\begin{equation*}
	\Phi(X) = \E^{\I \mspace{1mu} \tilde{p}_\text{tot} X/2} \, .
\end{equation*}
The constant $\tilde{p}_\text{tot}$ can be identified as the total momentum, i.e.\ the eigenvalue of the total momentum operator $\widetilde{P} = -\I \, (\partial_{x_1} + \partial_{x_2}) = -2 \, \I \, \partial_X$, reflecting the model's translational symmetry.
	
The relative motion obeys (the Weierstrass form of) the \emph{Lamé equation} \cite{lame1834memoire} 
\begin{equation} \label{eq:Lame}
	{-\phi''}(x) + \beta\,(\beta-1) \, \wp(x) \, \phi(x) = \bar{E} \, \phi(x) \, , \qquad\qquad
	\bar{E} \coloneqq \widetilde{E}_{M=2} -\frac{\tilde{p}_\text{tot}^2}{4} - \beta\,(\beta-1)\,\frac{\eta_2}{\omega} \, .
\end{equation}
This ordinary differential equation admits an exact solution that
was found by Hermite \cite[p.\,573]{whittaker1904course}.
For $\beta \in \mathbb{Z}_{\geq 2}$ the eigenfunctions of \eqref{eq:Lame} can be written as a product
\begin{equation*}
	\phi(x) = \prod_{j=1}^{\beta-1} \! h_{\gamma_j}(x)\, , \qquad\quad \text{with eigenvalue} \quad \bar{E} = -\sum_{j=1}^{\beta-1} \wp(\gamma_j) \, ,
\end{equation*}
where
\begin{equation} \label{eq:hermite}
	h_{\gamma}(x) \coloneqq \E^{-\zeta(\gamma)\,x} \, \frac{\sigma(x+\gamma)}{\sigma(x)\,\sigma(\gamma)} \, ,
	\qquad\quad 
	h_\gamma(x + \omega_b) = \E^{-\rho_b(\gamma) \, \omega_b} \, h_\gamma(x) \, , \quad b=1,2\,,
\checkedMma
\end{equation}
and the parameters~$\gamma_j$ must satisfy the conditions
\begin{equation*} 
	\sum_{k (\neq j)}^{\beta-1} \Bigl( \zeta(\gamma_k) - \zeta(\gamma_k - \gamma_j) - \zeta(\gamma_j) \Bigr) = 0 \, , \qquad 1 \leq j \leq \beta-1\, .
\end{equation*}
As an aside we mention that these transcendental equations can be recast in polynomial form by a change of variables, yielding a system equivalent to those discussed in \cite{Mai_07}.

Now we specialise the reduced coupling to $\beta=2$, the case that we are interested in. The functions \eqref{eq:hermite} and $h_{-\gamma}(x) = -h_\gamma(-x)$ span the eigenspace of the Lam\'{e} equation with eigenvalue $\bar{E} = -\wp(\gamma)$.{\checkedMma}
In this case $\gamma$ is free: the above condition is empty. The full two-particle wave function for the eCS model thus reads
\begin{equation} \label{eq:M=2_eCS_wave_fn}
\widetilde{\Psi}_{\tilde{\vect{p}}}(x_1,x_2) = \E^{\I \mspace{1mu} \tilde{p}_\text{tot} X/2} \, h_\gamma(x) \, .
\checkedMma 
\end{equation}
As promised in \textsection\ref{sec:connection} this function is doubly quasiperiodic. The quasiparameters $\tilde{\vect{p}},\vect{\varphi}$ from the assumption~\eqref{eq:quasiperiods} are related to $\tilde{p}_\text{tot}$ and $\gamma$ via
\begin{equation} \label{eq:quasiperiods_M=2}
\tilde{p}_1+\tilde{p}_2 = \tilde{p}_\text{tot}\, , \quad
\tilde{p}_1 - \tilde{p}_2 = 2 \mspace{1mu} \I \mspace{1mu} \rho_2(\gamma)\, , \qquad 
\varphi_1 + \varphi_2 = 0\, , \quad
\varphi_1 - \varphi_2 = -4 \mspace{1mu} \pi \mspace{1mu} \gamma/\omega \, ,
\checkedMma
\end{equation}
The second relation allows us to rewrite \eqref{eq:M=2_eCS_wave_fn} in the Bethe-ansatz-like form
\begin{equation} \label{eq:M=2_eCS_wave_fn_Bethe-like}
\widetilde{\Psi}_{\tilde{\vect{p}}}(x_1,x_2) = \E^{\I\,\tilde{p}_1 \,x_1 + \I \,\tilde{p}_2 \,x_2} \, \chi_2(x_1-x_2,\gamma) \, ,
\checkedMma
\end{equation} 
where we define the Kronecker elliptic functions\,%
\footnote{\ These functions are symmetric, $\chi_b(t,z) = \chi_b(z,t)$, and doubly (quasi)periodic, $\chi_b(z+\omega_b,t) = \chi_b(z,t)$ and $\chi_b(z+\omega_a,t) = \E^{(-1)^b \, 2\,\pi \I \,t/\omega_b} \,\chi_b(z,t)$ for $a\neq b$. 
See also Footnote~\ref{fn:theta_prime} on p.\,\pageref{fn:theta_prime}. These are the unique meromorphic functions of $z$ regular on $\mathbb{C}\setminus\mathbb{L}$ with a simple pole with residue one at the origin and the aforementioned double (quasi)periodicity.
Note that \cites{Inozemtsev:2002vb} uses $\chi_1(z,t) = \E^{-2\,\kappa \, z\,t/L} \, \chi_2(z,t)$, therein denoted by $-\tilde{\sigma}_z(-t)$. We will work with $\chi_2$, whose quasiperiods fit with \eqref{eq:quasiperiods} to yield the relation \eqref{eq:p_requirement_mom} with the eCS momenta~$\tilde{p}_m$.}
\begin{equation} \label{eq:chi_def}
\chi_b(z,t) \coloneqq 
\theta_b'(0)\,\frac{\theta_b(z+t)}{\theta_b(z)\,\theta_b(t)} \, = \, 
\E^{-\eta_b\, z\,t/\omega_b} \, \frac{\sigma(z+t)}{\sigma(z)\,\sigma(t)} \, , \qquad b = 1,2 \, .
\checkedMma
\end{equation}

The function \eqref{eq:M=2_eCS_wave_fn_Bethe-like} further has the desired analytic structure~\eqref{eq:psi_decomposition}: it is meromorphic with a simple pole at $x_1 = x_2$. With the help of the first two relations in \eqref{eq:quasiperiods_M=2} we can rewrite the energy as \eqref{eq:M_particle_energy'}, where the potential energy is given by
\begin{equation} \label{eq:U_tilde}
\widetilde{U} = -\frac{\tilde{p}_1^2 + \tilde{p}_2^2}{2} + \frac{\tilde{p}_\text{tot}^2}{4} -\wp(\gamma) + \frac{2\,\eta_2}{\omega} = F_2(\gamma) \, , \qquad F_b(z) \coloneqq \rho_b '(z) + \rho_b(z)^2 + \frac{3\,\eta_b}{\omega_b} \, .
\end{equation}
Note that the latter is a variant of the function that we encountered in the dispersion \eqref{eq:dispersion}.

The trigonometric and hyperbolic (infinite-length) limits of the preceding straightforwardly yields a description of the two-particle eigenfunctions of the corresponding Calogero--Sutherland Hamiltonians, respectively, with eigenfunctions that inherit the poles at the origin; cf.~Example~1 in \cite{CV_90}. The contact limit ($\omega\to 0$), however, is much more subtle. (For the spin chain, see e.g.\ the potential in \eqref{eq:limits_potential} and especially Figure~\ref{fig:potential_kappa}, this limit is regularised by the lattice spacing, providing a natural `ultraviolet cutoff'.)

Note that the double pole of the potential in the Lamé equation~\eqref{eq:Lame} can only be balanced if $\widetilde{\Psi}(x) = \text{cst}\, x^n + \text{higher-order monomials}$, with either $n=\beta$ or $n=1-\beta$. Although the assumption~\eqref{eq:psi_decomposition} required for the analysis in \textsection\ref{sec:connection} demands wave functions with a single pole at the origin, it is possible to construct wave functions with a double zero instead. Indeed, the eigenspace with energy $-\wp(\gamma)$ contains the even eigenfunction $\phi(x) = h_{\gamma}(x) + h_{\gamma}(-x) = h_{\gamma}(x) - h_{-\gamma}(x)$, which has a double zero at $x=0$ (rather than a simple pole: $\text{Res}_{x=0} \, h_\gamma = 1$ for any $\gamma$),{\checkedMma}
just like the usual eigenfunctions of the trigonometric Calogero--Sutherland model with $\beta=2$.
These eigenfunctions should limit to the usual description of the trigonometric CS spectrum in terms of Jack polynomials; this is particularly clear in the elliptic CS eigenfunctions of~\cite{langmann2000anyons}, and will yield a more transparent connection to the usual Haldane--Shastry eigenfunctions. We plan to report on this in a future publication.

\subsubsection{Two-magnon wave functions} \label{sec:M=2_wave_fns} Now we pass to the spin chain. Restricting (`freezing') the coordinates to $x_m = n_m \in \mathbb{Z}_L$ and plugging \eqref{eq:M=2_eCS_wave_fn_Bethe-like} into the extended \textsc{cba}~\eqref{eq:ansatz} we obtain the wave function found by Inozemtsev~\cite{Inozemtsev:1989yq}\,%
\footnote{\ The main difference with (26) therein is a shift of $\I \eta_2\gamma/\omega$ in our $p_m$ compared to \cite{Inozemtsev:1989yq}, yielding the factors $\E^{\pm \eta_2(n_1-n_2)\gamma/\omega}$ in \eqref{eq:M=2_wave_fn}.}
\begin{equation} \label{eq:M=2_wave_fn}
	\begin{aligned}
	\Psi_{\vect{p}}(n_1,n_2) \coloneqq \Psi_{\tilde{\vect{p}},\vect{p}}(n_1,n_2)
	= {} &  \chi_2(n_1-n_2,\gamma)\, \E^{\I\, p_1\,n_1 + \I\, p_2 \,n_2} + \chi_2(n_2-n_1,\gamma)\, \E^{\I\, p_1\,n_2 + \I\, p_2 \,n_1} \\
	= {} & 
		\frac{\sigma(n_1 - n_2 + \gamma)}{\sigma(n_1 - n_2)\,\sigma(\gamma)}\, \E^{-\eta_2\,(n_1 - n_2)\,\gamma/\omega} \, \E^{\I\, p_1\,n_1 + \I\, p_2 \,n_2} \\
	& + \frac{\sigma(n_1 - n_2 - \gamma)}{\sigma(n_1 - n_2)\,\sigma(\gamma)}\, \E^{\eta_2\,(n_1 - n_2)\,\gamma/\omega}\, \E^{\I \,p_1\,n_2 + \I \,p_2 \,n_1} \, .
	\end{aligned}
\end{equation}
The only remaining parameters are the quasimomenta $p_m = \widetilde{p}_m + r_m$ and~$\gamma$. In view of the last two relations in \eqref{eq:quasiperiods_M=2} we set 
\begin{equation}\label{eq:varphi_vs_gamma}
\varphi \coloneqq \varphi_1 = -\varphi_2 = -\frac{2\pi}{\omega} \, \gamma = 2\,\I\, \kappa\,\gamma \,.
\end{equation}
Let us recall that $\chi_2$ is $\omega$-periodic while $\chi_2(z+L,\gamma) = \E^{2\pi\I\gamma/\omega} \, \chi_2(z,\gamma)$.{\checkedMma}
In particular, \eqref{eq:M=2_wave_fn} is $2\pi$-periodic in $\varphi$.{\checkedMma}
For the spin chain this parameter is no longer free: the identification \eqref{eq:p_requirement_mom}, needed to solve the two-particle difference equation, turns the second relation in~\eqref{eq:quasiperiods_M=2} into the constraint
\begin{equation} \label{eq:M=2_constraint}
	2 \, \momL{\rho}_1(\varphi) = \mom{\rho}_1(p_1) - \mom{\rho}_1(p_2) \, . 
\end{equation}
This relation can be viewed (on a suitably restricted domain) as an implicit definition of $\varphi$ as a function of $\vect{p}$, cf.~\eqref{eq:M=2_constraint_normalised}, or for one of the $p_m$ in terms of $p_\text{tot}$ and $\varphi$. Generically the wave function is antisymmetric in $p_1,p_2$ when we take into account that by \eqref{eq:M=2_constraint} $\varphi$ acquires a sign if $p_1 \leftrightarrow p_2$.
Finally we impose the periodic boundary conditions: the wave function is $L$-periodic in each argument separately provided {\checkedMma}%
the \emph{Bethe-ansatz equations}~\eqref{eq:periodicitygenM} hold, i.e.\,%
$\vphantom{\,}$\footnote{\ Let us stress once more that, as we mentioned just after \eqref{eq:periodicitygenM}, in \cite{Inozemtsev:1989yq} the relation \eqref{eq:M=2_BAE} is implemented from the start, rather than a condition for the parameters $p_m$.}
\begin{equation} \label{eq:M=2_BAE}
L\,p_1 = 2\pi \mspace{1mu}I_1 + \varphi \, , \qquad 
L\,p_2 = 2\pi \mspace{1mu}I_2 - \varphi \, , \qquad I_m \in \mathbb{Z}_L \, .
\end{equation}
We will refer to \eqref{eq:M=2_constraint}--\eqref{eq:M=2_BAE}, which together determine the values of $\vect{p}$ and $\varphi$ given a suitable choice of $\vect{I}$, as the \emph{Bethe-ansatz system}; we will study it in detail below.

Each (suitable, see below) solution completely determines a two-particle wave function~\eqref{eq:M=2_wave_fn} of the Inozemtsev spin chain with energy\,%
\footnote{\ Compared to \cite{Inozemtsev:1989yq} the shift in our quasimomenta~$p_m$ leads to a simpler energy, built directly from the dispersion relation at the quasimomenta and the eCS potential energy~$\tilde{U}$.}
\begin{subequations} \label{eq:M=2_energy}
\begin{gather}
E_\text{n}(\vect{p}) = \varepsilon_\text{n}(p_1) + \varepsilon_\text{n}(p_2) + U_\text{n}(p_1,p_2) \, ,
\checkedMma
\intertext{where we view}
U_\text{n}(\vect{p}) \coloneqq -n_H(\kappa) \, \widetilde{U} = 4 \, n_H(\kappa) \, \kappa^2 \, \momL{F}_1(\varphi)
\checkedMma
\end{gather}
\end{subequations}
as a function of the quasimomenta by the constraint \eqref{eq:M=2_constraint}. 
Here the normalising prefactor is as in \eqref{eq:ham_normalised}, and we switched to the momentum lattice using \eqref{eq:F_mtm_lattice}. Note the similarity between the functions in the potential energy and the dispersion~\eqref{eq:dispersion_in_check}. The potential energy spoils (functional) additivity of the two-particle energy\,---\,but see \textsection\ref{sec:potential} for the limits. 
Moreover, \emph{on shell}, i.e.\ on solutions of \eqref{eq:M=2_constraint}--\eqref{eq:M=2_BAE},
the energy is \emph{elliptic} in $\varphi$;
we will prove this in \textsection\ref{sec:M=2_energy_rat}.

\subsubsection{\textit{S}-matrix and scattering phase} \label{sec:S-matrix} 
The wave function \eqref{eq:M=2_wave_fn} has the form of the extended coordinate Bethe ansatz~\eqref{eq:ansatz} also in terms of the quasimomenta~$p_m$:
\begin{subequations} \label{eq:CBA-like_M=2}
	\begin{gather}
	\Psi_{\vect{p}}(\vect{n}) = A_e(\vect{p};\vect{n}) \, \E^{\I \, \vect{p} \,\cdot\, \vect{n}} + A_\tau(\vect{p};\vect{n}) \, \E^{\I \, \vect{p} \, \cdot\, \vect{n}_\tau} \, ,
\intertext{where $\tau \in S_2$ denotes the transposition. Here we think of the coefficients}
	A_w(\vect{p};\vect{n}) = \E^{-\I \, \tilde{\vect{p}} \, \cdot\, \vect{n}_w} \, \widetilde{\Psi}_{\tilde{\vect{p}}}(\vect{n}_w) = \chi_2(n_{w(1)} - n_{w(2)},\gamma)
	\end{gather}
\end{subequations}
as depending on $\vect{p}$ through \eqref{eq:M=2_constraint}. The difference with the usual \textsc{cba} is their dependence on $\vect{n}$.

In the standard treatment of the Heisenberg spin chain one defines the (two-body) \textit{S}-matrix as the ratio of the coefficients in \eqref{eq:CBA-like_M=2}. At present, however, that ratio depends on the position, which seems to be an undesirable property for the \textit{S}-matrix. We have to be more careful. For the Heisenberg spin chain the coefficients in \eqref{eq:CBA-like_M=2} are independent of $\vect{n}$, so in that case plugging \eqref{eq:CBA-like_M=2} into the cyclicity condition~\eqref{eq:cyclicity_gen} yields an equation for each exponential in \eqref{eq:CBA-like_M=2}. Although presently the coefficients \emph{do} depend on $\vect{n}$ let us still try to ask for the (now a priori stronger) conditions obtained by equating the coefficients of those exponentials: 
\begin{equation} \label{eq:eBAE_M=2}
\E^{\I \mspace{1mu} L \mspace{1mu} p_1} = \frac{A_e(\vect{p};n_1,n_2)}{A_\tau(\vect{p};n_2,n_1+L)} \, , \qquad  \E^{\I \mspace{1mu} L \mspace{1mu} p_2} =  \frac{A_\tau(\vect{p};n_1,n_2)}{A_e(\vect{p};n_2,n_1+L)} \, .
\checkedMma
\end{equation}
Interestingly, the quasiperiodicity \eqref{eq:quasiperiods} of the extended \textsc{cba} wave function \eqref{eq:ansatz}\,---\,reflected in $\chi_2(n+L,\gamma) = \E^{2\pi \I \gamma/\omega} \, \chi_2(n,\gamma)${\checkedMma}
with $2\pi \I\mspace{1mu} \gamma/\omega = 2\mspace{1mu}\kappa\mspace{1mu}\gamma = -\I\,\varphi$\,---\,allows us to simplify the ratios on the right-hand sides in \eqref{eq:eBAE_M=2} to $\E^{\I \, \varphi}$ and $\E^{-\I \, \varphi}$, respectively.{\checkedMma}
The dependence on the positions drops out! Let us therefore define the (two-magnon) \emph{\textit{S}-matrix} as
\begin{equation} \label{eq:S-matrix_def}
S(\vect{p}) \coloneqq \frac{A_e(\vect{p};n_1,n_2)}{A_\tau(\vect{p};n_2,n_1+L)} = \frac{A_e(\vect{p};n_2,n_1+L)}{A_\tau(\vect{p};n_1,n_2)} = \E^{\I \, \varphi} \, ,
\checkedMma
\end{equation}
generalising the definition for the Heisenberg spin chain to the extended coordinate Bethe ansatz. Its interpretation as \textit{S}-matrix will be justified in \textsection\ref{sec:large_L}. From \eqref{eq:M=2_constraint} we deduce the usual properties
\begin{equation} \label{eq:S-mat}
S(p_1,p_2)^{-1} = S(p_2,p_1) \, , \qquad S(p_1,p_2)\big|_{p_1=p_2\neq 0} = -1 \, . 
\end{equation}
Let us point out that this function is not continuous at the origin. Indeed, $p_1=0$ is a pole of the constraint~\eqref{eq:M=2_constraint}, which can only be balanced if $\varphi=0$ (even though the residues on the two sides do not match). It follows that we should interpret $S(0,p_2) = 1$; this will be further justified when we consider descendants in \textsection\ref{sec:symms_trivial_roots}. Thus the limit of the \textit{S}-matrix at the origin depends on whether we approach it along an axis~$p_m=0$ or the diagonal $p_1=p_2$. The situation is exactly the same for Heisenberg, cf.~\eqref{eq:S_mat_Heis} below.

Plugging the \textit{S}-matrix into \eqref{eq:eBAE_M=2} we reobtain (the exponential form of)~\eqref{eq:periodicitygenM}, now in the familiar form
\begin{equation} \label{eq:M=2_BAE_exp}
\E^{\I \mspace{1mu} L \mspace{1mu} p_1} = S(p_1,p_2) \, , \qquad  \E^{\I \mspace{1mu} L \mspace{1mu} p_2} =  S(p_2,p_1) \, ,
\end{equation}
this time moreover with the standard interpretation of
the right-hand side of the \textsc{bae} as the phase acquired by the $m$th magnon as it is translated once around the chain. In particular we see that one may think of $\varphi$ as the two-particle \emph{scattering phase} up to normalisation. Let us emphasise that, unlike for Heisenberg, neither the \textit{S}-matrix nor the scattering phase features explicitly in the extended \textsc{cba} wave function.

The function $\momL{\rho}_1$ is invertible on a region $D \subset \mathbb{C}/\,\mathbb{L}^{\!\vee}$ slightly smaller than the fundamental domain \cite[App.~A]{Klabbers:2016cdd}. In terms of the properly normalised functions \eqref{eq:limits_rhocheck}--\eqref{eq:limits_rhobarcheck} we see that, like for Heisenberg, the scattering phase depends on the \emph{difference} of the rapidities:
\begin{equation} \label{eq:M=2_constraint_normalised}
\varphi = \Theta_\text{n}(\lambda_1 - \lambda_2) \, , \qquad\qquad
\begin{aligned}
\lambda_m \coloneqq {} & \lambda_\text{n}(p_m) =
{-n_\lambda(\kappa)} \, \mom{\rho}_1(p_m) \, , \\ 
\Theta_\text{n}(\lambda) \coloneqq {} & \Bigl(
n_\lambda(L\,\kappa) \, \momL{\rho}_1 \big|_D \Bigr)^{\!\!-1} \! \biggl( -
\frac{n_\lambda(L\,\kappa)}{2 \, n_\lambda(\kappa)} \, \lambda\biggr) \, .
\end{aligned} 
\end{equation}
This also provides a formal expression for the \textit{S}-matrix \eqref{eq:S-matrix_def}. We have not been able to find an explicit expression for $\Theta_\text{n}$ away from the limits. 

The definition \eqref{eq:S-matrix_def} of the \textit{S}-matrix extends to the $M$-magnon sector: if $w,w'\in S_M$ differ by a transposition, $w'= w \, \tau$ with $\tau= (m,m')$ for $m <m'$, the quasiperiodicity \eqref{eq:quasiperiods} and the fact that $\tilde{\Psi}$ is totally symmetric in positions immediately imply that the coefficients $A_w(\vect{p};\vect{n}) \coloneqq \E^{-\I \,\tilde{\vect{p}}\cdot \vect{n}_w} \, \widetilde{\Psi}_{\tilde{\vect{p}}}(\vect{n}_w)$ obey 
\begin{equation}
	\frac{A_{w\tau}(\vect{p}; \vect{n})}{A_{w}(\vect{p}; \vect{n}_{\tau} + L \, \hat{e}_{m'})} = \E^{ \I \, \varphi_{w^{-1}(m')}}\, . 
\end{equation}
As before the position dependence on the left-hand side implies we cannot write the coefficients $A_w$ as a product of (position-independent) \textit{S}-matrices.

\subsubsection{Symmetries and trivial roots} \label{sec:symms_trivial_roots} 
First of all we note that, as usual, the Bethe-ansatz equations~\eqref{eq:M=2_BAE} are invariant under simultaneous exchange $I_1 \leftrightarrow I_2$, $p_1 \leftrightarrow p_2$ and, cf.~\eqref{eq:M=2_constraint}, $\varphi\mapsto -\varphi$. Note that the energy is symmetric in $p_1,p_2$ while the wave function~\eqref{eq:M=2_wave_fn} is antisymmetric when we take into account the sign for $\varphi$. Thus we may without loss of generality order $0\leq I_1 \, \leq I_2 \leq L-1$. We will show that out of these $\binom{L+1}{2}$ pairs precisely $\binom{L}{2}$ yield a (different) solution, to account for all $M=2$ vectors.

By isotropy $\binom{L}{1}=L$ of the eigenvectors in the two-particle sector arise as descendants of $\ket{\uparrow\!\cdots\!\uparrow}$ ($M=0$) and the magnons with $p_\text{tot} \neq 0 \, \mathrm{mod}\,2\pi$ ($M=1$). Although the derivation of the $M$-particle difference equation in \textsection\ref{sec:M-part_diff_eq} assumes that the eigenvectors are highest weight, these descendants fit in as follows. If $I_1=0$ then \eqref{eq:M=2_BAE} is solved by taking $p_1=\varphi=0$ and $p \coloneqq p_2 = 2\pi I_2/L$. The constraint~\eqref{eq:M=2_constraint} is singular at this point, just as for the Heisenberg model. Nevertheless everything works fine at the level of the energy and wave function. Since $\mom{F}_1(0) = \momL{F}_1(0) = 0$ the energy $E_\text{n}(0,p) = \varepsilon_\text{n}(p)$ is as for $M=1$, in accordance with isotropy. Up to the vanishing but position-independent factor $1/\sigma(\gamma)$ the wave function \eqref{eq:M=2_wave_fn} becomes
\begin{equation} \label{eq:magnon_desc}
	\Psi_{0,p}(\vect{n}) \propto \E^{\I\mspace{1mu} p \mspace{1mu}n_1} + \E^{\I\mspace{1mu} p\mspace{1mu} n_2} = \cbra{n_1,n_2} \, S^- \sum_{n=1}^L \E^{\I \, p\, n} \cket{n} \, .
\end{equation} 

Our task is to find the remaining $\binom{L}{2} - L$ highest-weight vectors in the two-particle sector by solving \eqref{eq:M=2_BAE}--\eqref{eq:M=2_constraint} for (suitable) $1\leq I_1 \leq I_2 \leq L-1$. As we will see the structure of these solutions is very similar to that for the Heisenberg limit (\textsection\ref{sec:Heis_recap_M=2}).

Define
\begin{equation}
	I_\text{tot} \coloneqq I_1 + I_2 \, \mathrm{mod} \, L \ \in \{0,1,\cdots\mspace{-1mu},L-1\} \, .
\end{equation}
Translational invariance tells us how many eigenvectors there are at each $I_\text{tot}$ by a generalisation of the argument yielding the magnons for $M=1$ (\textsection\ref{sec:M=1}). By a translation every $M=2$ coordinate-basis vector can be brought to the form $\cket{1,1+n}$ for $1\leq n\leq \lfloor L/2 \rfloor$. Each of these can be `boosted' to a $G$-eigenvector $\sum_{k=1}^L \E^{\I \,p_\text{tot} k} \, \cket{k,k+n}$ with $p_\text{tot} = 2\pi I_\text{tot}/L$ (so linearly independent) --- except, when $L$ is even, the vector with $n=L/2$. Indeed, for the latter $k=1+L/2$ is equivalent to $k=1$ so the boosted vector vanishes if{f} $I_\text{tot}$ is odd. This shows that when $I_\text{tot}$ is even (odd) there are $\lceil (L-1)/2\rceil$ (resp.\ $\lfloor (L-1)/2\rfloor$) eigenvectors that span the two-particle sector at $p_\text{tot} = 2\pi\,I_\text{tot}/L$. We already obtained one $H$-eigenvector for each $I_\text{tot}$ amongst the descendants, so it remains to find $\lceil (L-3)/2\rceil$ eigenvectors at each even $I_\text{tot}$ and $\lfloor (L-3)/2\rfloor$ at each odd $I_\text{tot}$. In \textsection\ref{sec:counting} we will prove that the ansatz \eqref{eq:ansatz} does indeed yield this many eigenvectors.

Let us also comment on a discrete symmetry of the Inozemtsev spin chain: \emph{parity}, acting by a spatial reflection ($n\mapsto L-n-1$), which clearly preserves the Hamiltonian. Conjugation by the parity operator inverts the translation operator, so $p_\text{tot} \mapsto {-}p_\text{tot} \, \mathrm{mod} \, 2\pi$ under parity. As a consequence $H$-eigenvectors are either invariant under parity, with $p_\text{tot} \in \{0,\pi\}$, or come in parity-conjugate pairs with total momenta $p_\text{tot},2\pi-p_\text{tot}$. For $M=1$ this symmetry pairs up the magnons with $p_\text{tot} \notin \{0,\pi\}$, which do indeed have equal energy. In the present context it shows up as follows: if $p_1$, $p_2$, $\varphi$ solve \eqref{eq:M=2_BAE}--\eqref{eq:M=2_constraint} for a given choice of $I_1$, $I_2$ then $p'_1 \coloneqq 2\pi -p_2$, $p'_2 \coloneqq 2\pi - p_1$, $\varphi'\coloneqq \varphi$ do so for $I'_1 \coloneqq L - I_2$, $I'_2 \coloneqq L - I_1$ (still with $1\leq I'_1 \leq I'_2 \leq L-1$) and have $E_\text{n}(\vect{p}') = E_\text{n}(\vect{p})$, $p'_\text{tot}= 2\pi-p_\text{tot}$ and $\Psi_{\vect{p}'}(n_1,n_2) = \Psi_{\vect{p}}(-n_1,-n_2) = \Psi_{\vect{p}}(L-n_2,L-n_1)$. For example, when $L=6$ we will see that $\vect{I}=(1,5)$ admits a solution, which will yield a parity-invariant vector ($p_\text{tot} = 2\pi \, I_\text{tot}/L = 0$), while $\vect{I}=(1,1)$ will produce another solution, along with its parity conjugate at $\vect{I}'=(5,5)$ (with $p_\text{tot} = \pi/2$ and $p'_\text{tot} = 3\pi/2$, respectively).

\begin{figure}[h]
	\centering
	\begin{tikzpicture}
		\node at (0,0) {\includegraphics[width=\textwidth]{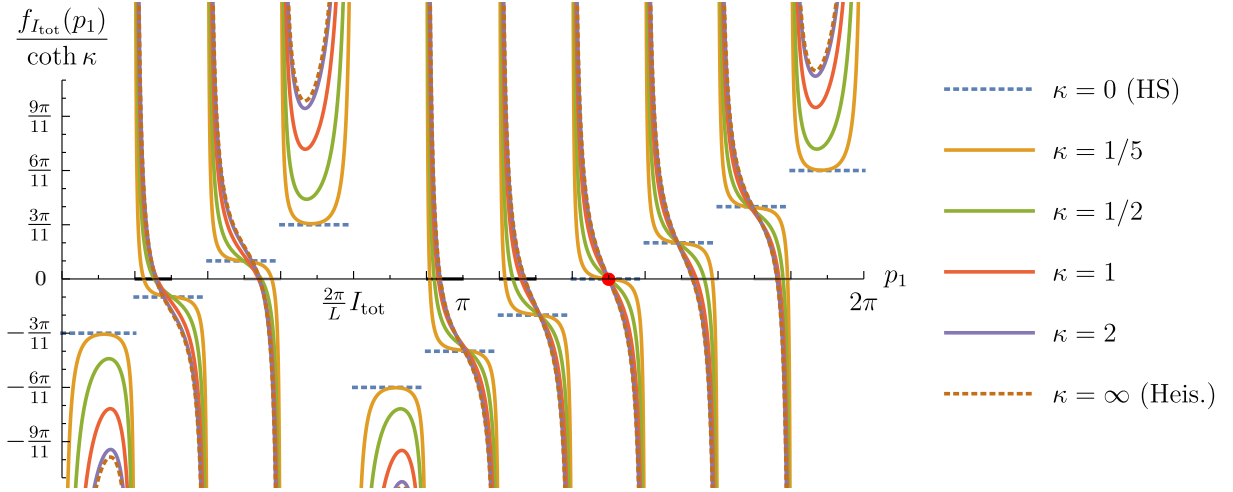}};
		\node at (-7.5,-.46) {\scriptsize $0$};
	\end{tikzpicture}
	\caption{Plots of the (normalised) function~\eqref{eq:f_constraint} with $L=11$ and $I_\text{tot} = 4$ for $p_1 \in [0,2\pi]$ and values of $\kappa$ interpolating between Haldane--Shastry, where the constraint becomes piecewise constant, and Heisenberg, see \eqref{eq:M=2_constraint_Heis}. The red dot indicates a trivial root\,---\,the second in \eqref{eq:trivial_roots}\,---\,which is a zero for any $\kappa$. The nontrivial roots may be labelled by $\vect{I} = (1,3), (5,10), (6,9)$ and come in pairs $p_1 \in \frac{2\pi}{L}[I_1,I_1+\frac{1}{2})$ along with $(p_2=)\,\frac{2\pi}{L}(I_1+I_2) - p_1 \in \frac{2\pi}{L}(I_2-\frac{1}{2},I_2]$; these intervals are indicated in black and grey, respectively. 
	The picture is similar for any $L,I_\text{tot}$. There are no nontrivial roots in $\frac{2\pi}{L} \, [0,1) \cup (I_\text{tot}-1,I_\text{tot}+1) \cup (L-1,L]$. The black and grey intervals are separated by a trivial root if present, and enumerated by $\vect{I}$ with $2\leq I_1+1< I_2-1\leq L-2$, first for $I_1+I_2 = I_\text{tot}$ and then $I_1+I_2 = I_\text{tot} + L$.} 
	\label{fig:constraint_real_kappa}
\end{figure}

Let us examine the structure of the constraint in more detail. Eliminating the scattering phase~$\varphi$ in favour of $p_1$, say, \eqref{eq:M=2_constraint} amounts to the vanishing of
\begin{equation} \label{eq:f_constraint}
	f_{I_\text{tot}}(p_1) \coloneqq 2 \, \momL{\rho}_1(L\,p_1) - \mom{\rho}_1(p_1) + \mom{\rho}_1(2\pi I_\text{tot}/L - p_1) \, .
\end{equation}
A new feature for Inozemtsev is that the double quasiperiodicity of $\momL{\rho}_1$ and $\mom{\rho}_1$ conspire to make \eqref{eq:f_constraint} \emph{elliptic}, with the periods $2\pi$ and $2\I \kappa$ of the momentum lattice $\!\!\bar{\,\,\mathbb{L}}^{\!\vee}$. The wave function and energy, similarly expressed in terms of $p_1$, are so too.
It therefore suffices to seek zeroes in the fundamental domain $\mathbb{C}/\!\bar{\,\,\mathbb{L}}^{\!\vee} \cong [0,2\pi) \times [0,2\I \kappa)$. Figure~\ref{fig:constraint_real_kappa} shows an example for real $p_1$.

In fact it suffices to consider roots in the lower half of the fundamental domain. Indeed, complex conjugate solutions 
are physically equivalent. For Inozemtsev the quasimomenta are moreover $2\I \kappa$-periodic, so that roots in the upper half of the fundamental domain are equivalent to those in the lower half. 

Certain roots are nonphysical and have to be excluded. These can be found explicitly. Since $\eta_b = 2\,\zeta(\omega_b/2)$ the functions $\rho_b$ have simple values at half the lattice periods: $\rho_b(\omega_b/2) = 0$ and, by the Legendre relation, $\rho_b(\omega_a/2) = (-1)^b \, \I \, \pi/\omega_b$ for $a\neq b$. Another way to see this is to exploit the quasiperiodicity and oddness to compute $\rho_b(\omega_a/2) = \rho_b(-\omega_a/2 + \omega_a) = -\rho_b(\omega_a/2) + \delta_{a,2-b} \, (-1)^b \, 2\, \pi\I/\omega_b$, which can be solved to reproduce the preceding exact values for $\rho_b(\omega_a/2)$. There is one more point in the fundamental domain where we can use this trick:
\begin{equation*}
	\rho_b\biggl(\frac{L+\omega}{2}\biggr) 
	= \rho_b\biggl(-\frac{L+\omega}{2}+L+\omega\biggr) 
	= -\rho_b\biggl(\frac{L+\omega}{2}\biggr) + (-1)^b\,\frac{2\pi \I}{\omega_b}\, .
\end{equation*}
For the special functions featuring in \eqref{eq:f_constraint} we thus obtain the following exact values: $\momL{\rho}_1(\pi) = 0$, $\momL{\rho}_1(\I\,L \, \kappa) = \momL{\rho}_1(\pi+\I\,L\,\kappa) = -\I/2$, and (take $L=1$) $\mom{\rho}_1(\pi) = 0$, $\mom{\rho}_1(\I\,\kappa) = \mom{\rho}_1(\pi+\I\,\kappa) = -\I/2$. As a consequence the constraint is solved, for given $I_\text{tot}$, by 
\begin{subequations} \label{eq:trivial_roots}
\begin{gather}
	p_1 = \pi\,I_\text{tot}/L + \delta \, , \qquad 
	p_2 = \pi\,I_\text{tot}/L - \delta \, \mathrm{mod} \, 2\pi \, ,
\intertext{provided $\delta \in \, \!\!\bar{\,\,\mathbb{L}}^{\!\vee}\!/2 \ \cap \ \mathbb{C}/\!\bar{\,\,\mathbb{L}}^{\!\vee}$ is given by}
	\begin{aligned}
	\delta & = 0 && \text{iff $I_\text{tot}$ is odd,} \\
	\delta & = \pi && \text{iff $L-I_\text{tot}$ is odd (i.e.\ exactly one of $L,I_\text{tot}$ is even),}  \\
	\delta & = \I \kappa \, , && \\
	\delta & = \pi + \I \kappa \, . &&
	\end{aligned}
\end{gather}
\end{subequations}
(Note that $p_2 = p_1^* \, \mathrm{mod} \, 2\pi$ in each case.) 
Indeed, writing $p_1 = p_\text{tot}/2 +\delta$ we have
\begin{equation}
f_{I_\text{tot}}(p_\text{tot}/2 + \delta) = 2 \, \momL{\rho}_1(L\,p_\text{tot}/2 + L\,\delta) - \mom{\rho}_1(p_\text{tot}/2 + \delta) + \mom{\rho}_1(p_\text{tot}/2 - \delta) \, .
\end{equation}
Then $\delta=0$ gives a zero provided $I_1+I_2$ is odd to ensure we're at $\momL{\rho}_1(\pi)=0$ (rather than the pole at the origin). Likewise, $\delta=\pi$ yields a root when $I_\text{tot}+L$ is odd by $2\pi$-periodicity of $\mom{\rho}_1$. The remaining two cases use $L\,p_\text{tot}/2 = \pi I_\text{tot}$ and the quasiperiodicity $\mom{\rho}_1(z - 2\I \kappa) = \mom{\rho}_1(z) + \I$. 
A straightforward check teaches us that for any of these roots the associated two-particle wave function vanishes, see \textsection\ref{app:trivialsols} for the details. That is, \eqref{eq:trivial_roots} are \emph{trivial roots}. One of them can be seen in Figure~\ref{fig:constraint_real_kappa}.

Parity provides a little more structure in the preceding, organising the trivial roots into pairs. Indeed, the first trivial root in \eqref{eq:trivial_roots} corresponds to
\begin{equation*}
	p_{1,2} = \frac{2\pi I_{1,2} \pm \pi (I_2 - I_1)}{L} = \frac{\pi (I_1+I_2)}{L} \, . 
\end{equation*}
The parity-conjugate integers $I'_1 = L-I_2$ and $I'_2 = L-I_1$ likewise correspond to
\begin{equation*}
	p'_{1,2} = \frac{2\pi I'_{1,2} \pm \pi (I'_2 - I'_1)}{L} = \frac{\pi (I'_1+I'_2)}{L} = \frac{2\pi L - \pi(I_1 + I_2)}{L} = \frac{\pi (L- (I_1+I_2))}{L} + \pi \, .
\end{equation*}
The latter may be reinterpreted as the second trivial root of \eqref{eq:trivial_roots} where $I_1+I_2$ is replaced by $L- (I_1+I_2)$, the representative of the parity conjugate $-(I_1+I_2) \, \mathrm{mod}\, L$ for $I_m$ with $2\leq I_1+I_2 \leq L$, cf.~Figure~\ref{fig:integers}. A similar argument relates the third and fourth trivial root in \eqref{eq:trivial_roots}. So there are essentially only two different trivial roots. 

In conclusion we seek $\binom{L}{2} - \binom{L}{1} = L\,(L-3)/2$ zeroes of \eqref{eq:f_constraint}: $\lceil (L-3)/2\rceil$ for every even $I_\text{tot}$ and $\lfloor (L-3)/2\rfloor$ for each odd $I_\text{tot}$. By ellipticity the domain is the fundamental rectangle $\mathbb{C}/\!\!\bar{\,\,\mathbb{L}}^{\!\vee} \cong [0,2\pi) \times [0,2\I \kappa)$, and we are to exclude the trivial roots~\eqref{eq:trivial_roots}. Although $f_{I_\text{tot}}$ only depends on the sum of the $I_m$ we will see that the $I_m$ provide a convenient way to label (count) the roots, just as for the Heisenberg spin chain.

\subsection{Heisenberg regime} \label{sec:Heis_regime} When $\kappa$ is large\,---\,in practice: $\kappa\gtrsim2$, cf.~Figures \ref{fig:potential_kappa} (p.\,\pageref{fig:potential_kappa}), \ref{fig:dispersion_kappa} (p.\,\pageref{fig:dispersion_kappa}) and \ref{fig:rapidity_kappa} (p.\,\pageref{fig:rapidity_kappa})\,---\,the interaction range is small and Inozemtsev spin chain differs only little from the Heisenberg chain. Since the latter is probably the most familiar for the reader, and the natural arena for the coordinate Bethe ansatz, we start our investigations in the limit.

\subsubsection{Heisenberg limit} \label{sec:Heis_limit_M=2} We let $\kappa\to\infty$, so that the imaginary quasiperiod $\omega \to \I \, 0^+$. Although this is the most singular limit we have already done the hard preparatory work. The limiting constraint readily follows from \eqref{eq:limits_rhocheck}--\eqref{eq:limits_rhobarcheck}:
\begin{equation} \label{eq:M=2_constraint_Heis}
2 \, \lambda_\textsc{h}(\varphi) 
= \lambda_\textsc{h}(p_1) - \lambda_\textsc{h}(p_2)\, , \qquad\quad \lambda_\textsc{h}(p) \coloneqq \lim_{\kappa\to\infty} \lambda_\text{n}(p) = -\frac{1}{2}\cot\frac{p}{2} \, ,
\end{equation}
where the appearance of $\lambda_\textsc{h}$ on the left-hand side is a coincidence for Heisenberg. This is precisely the relation between the scattering phase and momenta found by Bethe~\cite{bethe1931theorie}. In particular this shows that $\varphi$ remains finite in the Heisenberg limit. We will get back to the `exponential form' of the \textsc{bae} at the end of this subsection.

The coincidence also facilitates the evaluation of the Heisenberg limit of our \textit{S}-matrix using the formal expression~\eqref{eq:M=2_constraint_normalised}. As $\kappa\to\infty$ we can perform the inverse explicitly to recover the familiar Heisenberg \textit{S}-matrix
\begin{subequations} \label{eq:S_mat_Heis}
	\begin{gather}
	S_\textsc{h}(p_1,p_2) \coloneqq \lim_{\kappa\to\infty} S(p_1,p_2) = -\frac{\E^{\I\mspace{1mu}(p_1+p_2)} - 2 \, \E^{\I\mspace{1mu} p_1} +1}{\E^{\I\mspace{1mu}(p_1+p_2)} - 2 \, \E^{\I\mspace{1mu} p_2} +1} = \frac{\lambda_\textsc{h}(p_1) - \lambda_\textsc{h}(p_2) - \I}{\lambda_\textsc{h}(p_1) - \lambda_\textsc{h}(p_2) + \I} \, ,
\checkedMma
\intertext{where the signs of $\pm \I$ are due to the sign of the rapidity function. On shell, i.e.\ if the $p_m$ solve the constraint~\eqref{eq:M=2_constraint_Heis}, we reobtain \eqref{eq:S-matrix_def}:}
	S_\textsc{h}(p_1,p_2) = \E^{\I \mspace{1mu}\varphi} \, .
\checkedMma
	\end{gather}
\end{subequations}

Let us next show that the extended coordinate Bethe ansatz reduces to the usual \textsc{cba} in the Heisenberg limit. Recall that, rather than the \textit{S}-matrix~\eqref{eq:S-matrix_def}, the wave function~\eqref{eq:CBA-like_M=2} features the ($\vect{n}$-dependent) ratio
\begin{equation} \label{eq:A^e_tau}
A^e_\tau(\vect{p};\vect{n}) \coloneqq \frac{A_e(\vect{p};\vect{n})}{A_\tau(\vect{p};\vect{n})} 
= \E^{-2\,\eta_2\,(n_1-n_2)\,\gamma/\omega} \, \frac{\sigma(n_1-n_2+\gamma)}{\sigma(n_1-n_2-\gamma)} \, .
\checkedMma
\end{equation}
Since for suitable constants $c$ and $z_0$
\begin{equation*}
\sigma(z) = c \, \exp\biggl(  \int_{z_0}^z \zeta(t) \, \mathrm{d}t \biggr)
\checkedMma
\end{equation*}
we find the elegant expression 
\begin{equation*}
A^e_\tau(\vect{p};\vect{n}) = \exp\biggl(\int_{-\gamma}^{\gamma} \rho_2(n_1-n_2+t) \, \mathrm{d}t \biggr) \, .
\checkedMma
\end{equation*}

Since $\kappa^{-1} \, \rho_2 = -1 + O(\kappa^{-1})$ for all relevant values of $n_m$ and $\gamma${\checkedMma} its leading behaviour for large~$\kappa$ coincides with our \textit{S}-matrix~\eqref{eq:S-matrix_def}:
\begin{equation}
\label{eq:S-mat_via_kappa_gamma}
A^e_\tau(\vect{p};\vect{n}) \sim 
\E^{-2\kappa\gamma} = \E^{\I \mspace{1mu} \varphi} \, , \qquad \kappa \rightarrow \infty \, .
\end{equation}
In fact we can do better and compute the limit of the wave function itself. To switch from the parameter~$\gamma$ to the physically meaningful~$\varphi$, which will stay finite in the Heisenberg limit (cf.~the \textsc{bae}), we pass to the momentum lattice. Like in \textsection\ref{sec:mtm_lattices} set
\begin{subequations}
\label{eq:chi_on_momentum}
	\begin{gather}
	\chi^\vee_b(z,t) \coloneqq (\theta_b^\vee)'(0) \, \frac{\theta_b^\vee(z+t)}{\theta_b^\vee(z)\,\theta_b^\vee(t)} = \E^{-\eta_b^\vee\,z\,t/\omega_b^\vee} \, \frac{\sigma^\vee(z+t)}{\sigma^\vee(z)\,\sigma^\vee(t)} \, ,
\checkedMma
\intertext{which is related to \eqref{eq:chi_def} as} 
	\chi^\vee_b(z,t) = \frac{\omega}{2\pi} \, \chi_a\biggl(\frac{\omega\mspace{1mu}z}{2\pi},\frac{\omega\mspace{1mu}t}{2\pi}\biggr) \, , \quad a\neq b \, .
\checkedMma
	\end{gather}
\end{subequations}
The coefficients of the wave function are of the form $\chi_2(z,\gamma) 
= 2\mspace{1mu}\I\mspace{1mu}\kappa \, \momL{\chi}_1(2\mspace{1mu}\I\mspace{1mu}\kappa\,z,\varphi)$.{\checkedMma}
The momentum-lattice function has a well-defined limit,
$2\,\chi_1^\vee(z,\varphi) \to \cot(z/2) + \cot(\varphi/2)$ as $\kappa\to\infty$,{\checkedMma}
whence
\begin{equation} \label{eq:chi_2_heis_limit}
\begin{aligned}
\frac{\chi_2(z,\gamma)}{\I\,\kappa} = 2 \, \momL{\chi}_1(2\mspace{1mu}\I\mspace{1mu}\kappa\,z,\varphi) \sim {} & \cot(\varphi/2)  -\I \coth(\kappa\,z) && \kappa\gg1 \\[-1ex]
\to \!{} & \cot(\varphi/2) -\I\,\mathrm{sgn}(\Re z) = \frac{\E^{-\I \,\mathrm{sgn}(\Re z)\,\varphi/2}}{\sin(\varphi/2)} \, , \qquad && \kappa\to\infty \, .
\end{aligned}
\checkedMma
\end{equation}
We conclude that in the Heisenberg limit Inozemtsev's wave function~\eqref{eq:M=2_wave_fn} reduces, upon a simple rescaling, to the Bethe wave function~\cite{bethe1931theorie} 
\begin{equation} \label{eq:twoparticleansatz_Heis}
\lim_{\kappa\to\infty} \frac{\Psi_{\vect{p}}(\vect{n})}{\I\,\kappa} \, = \, \frac{\E^{\I\mspace{1mu}(\vect{p}\cdot\vect{n} \, + \, \varphi/2)} + \E^{\I\mspace{1mu}(\vect{p}\cdot\vect{n}_\tau \, - \, \varphi/2)}}{\sin(\varphi/2)} \, , \qquad n_1<n_2 \, .
\checkedMma
\end{equation}
In particular this again implies \eqref{eq:S-mat_via_kappa_gamma}. 

Next we turn to the energy~\eqref{eq:M=2_energy}. In the Heisenberg limit we find $\E^{2L\kappa} \, \momL{F}_1(\varphi) \to -8\,\sin(\varphi/2)^2${\checkedMma} 
so the contribution of the potential energy (without the exponential prefactor) vanishes rapidly, yielding the (functionally) additive result
\begin{equation} \label{eq:M=2_energy_Heis}
E_\textsc{h}(\vect{p}) = \varepsilon_\textsc{h}(p_1) + \varepsilon_\textsc{h}(p_2) \, , \qquad \quad \varepsilon_\textsc{h}(p_m) = 4\sin^2 \frac{p_m}{2} = \frac{1}{\lambda_\textsc{h}(p_m)^2 + 1/4} \, .
\end{equation}
We use the adjective `functionally' to stress that, unlike for the `strictly' additive Haldane--Shastry spin chain (\textsection\ref{sec:HS_recap_M=2}), it is not true in general that the actual values of the two-particle energies are the sum of two energies occurring for $M=1$. Indeed, the interactions cause $\varphi\neq 0$, which modify the allowed values of the the $p_m$ through the \textsc{bae}~\eqref{eq:M=2_BAE} and constraint~\eqref{eq:M=2_constraint_Heis}.

To conclude it is instructive to make the connection with the usual exponential form of the \textsc{bae} in terms of rapidities~\eqref{eq:M=2_BAE_exp}. The Heisenberg rapidity function obeys the identities
\begin{equation} \label{eq:exp_iLp}
\E^{\I\mspace{1mu}L\mspace{1mu}p_m} = \frac{\lambda_\textsc{h}(L\,p_m) -\I/2}{\lambda_\textsc{h}(L\,p_m) + \I/2} = \biggl(\frac{\lambda_\textsc{h}(p_m) -\I/2}{\lambda_\textsc{h}(p_m) + \I/2}\biggr)^{\!\!L} \, .
\checkedMma
\end{equation}
When we use the second equality in the \textsc{bae} for $\lambda_\textsc{h}(p_1)$, which reads
\begin{equation} \label{eq:M=2_constraint_Heis_exp}
\left(\frac{ \lambda_\textsc{h}(p_1)-\I/2}{\lambda_\textsc{h}(p_1)+\I/2}\right)^{\!\!L} \! = \frac{\lambda_\textsc{h}(p_1) -\lambda_\textsc{h}(p_2) -\I}{ \lambda_\textsc{h}(p_1) -\lambda_\textsc{h}(p_2) + \I} \, ,
\end{equation}
and remove the denominators we retrieve \eqref{eq:M=2_constraint_Heis} since $\lambda_\textsc{h}(L\,p_1) = \lambda_\textsc{h}(2\pi I_1 + \varphi) = \lambda_\textsc{h}(\varphi)$.{\checkedMma}
Note that the latter is not quite equivalent to \eqref{eq:M=2_constraint_Heis_exp}, since now solutions for which the right-hand side of \eqref{eq:M=2_constraint_Heis_exp} diverges are allowed. These `exceptional' solutions do occur \cite{bethe1931theorie}, as we will review shortly (\textsection\ref{sec:Heis_recap_M=2}). In this respect the denominator-free form is the better version of the \textsc{bae}. Importantly the second equality in \eqref{eq:exp_iLp} relies on general multi-angle formulae for the cotangent. However, in the elliptic case there is no such formula: one can show that no ratio of degree-$L$ polynomials in $\mom{\rho}_1$ can equal $\momL{\rho}_1(L \, p)$\,---\,but see \eqref{eq:rho_identity} for a related identity. 

\begin{figure}[h]
	\centering
	\includegraphics[width=\textwidth]{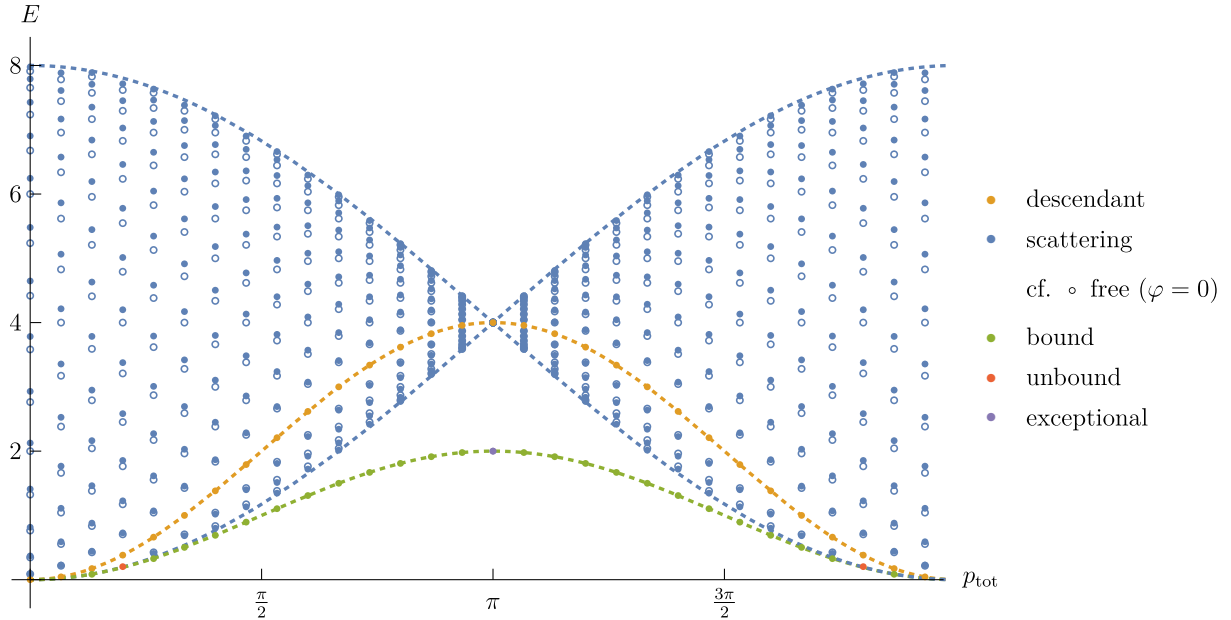}
	\caption{The two-particle spectrum of the Heisenberg spin chain with $L=30$ as described in \textsection\ref{sec:Heis_recap_M=2}; cf.~Figures~\ref{fig:Ep_M2_Ino} and \ref{fig:Ep_M2_HS}. Each dot represents an eigenspace. The descendants lie on the dispersion $\varepsilon_\textsc{h}(p)$ (dotted curve), shown in orange. The rest are highest-weight vectors. Scattering states are shown in blue, with energy close to that of two free magnons (open circles), and bounded by the energy of two identical magnons (dotted curve, two-valued as $p_\text{tot} = 2\,p \,\mathrm{mod}\, 2\pi$, $E(p,p) = 2\,\varepsilon_\textsc{h}(p)$). Bound states have $E \approx \varepsilon_\textsc{h}(p_\text{tot})/2$ (dotted curve) and come in three kinds: $p_{1,2} = \pi/2 \pm \I \infty$ exceptional (purple, once), $p_2^* = p_1$  complex conjugate (green) and $p_1 < p_2$ `unbound' (red, twice).}
	\label{fig:Ep_M2_Heis}
\end{figure}

\subsubsection{Recap of Heisenberg quasimomenta at $M=2$} \label{sec:Heis_recap_M=2} To complete the description of the two-particle spectrum of the Heisenberg spin chain one has to solve the Bethe-ansatz system \eqref{eq:M=2_BAE}, \eqref{eq:M=2_constraint_Heis}. Let us summarise the structure of its solutions \cite{bethe1931theorie,Essler1992,Deguchi:2015tfa,Cau18u}, see also \cite{KarbachMueller1998} for an introduction. The result is the two-particle spectrum depicted in Figure~\ref{fig:Ep_M2_Heis}. Recall from \textsection\ref{sec:symms_trivial_roots} that we seek $L\,(L-3)/2$ highest-weight vectors by finding equally many roots to \eqref{eq:M=2_constraint}--\eqref{eq:M=2_BAE}. Of these, $\lceil (L-3)/2\rceil$ occur per even value of $I_\text{tot}$ and $\lfloor (L-3)/2\rfloor$ per odd $I_\text{tot}$. Since $\kappa\to\infty$ the domain has become an infinite strip $\mathbb{C}/2\pi \mathbb{Z} \cong [0,2\pi) \times \I \, \mathbb{R}$. Recall our notation~\eqref{eq:notation_mod2pi} for the representatives. Depending on $L$ and $I_\text{tot}$ the two real trivial roots from \eqref{eq:trivial_roots} 
are to be discarded. As $p_m \neq 0 \, \mathrm{mod}\,2\pi$ (for all $m$) is a highest-weight condition~\cite[\textsection1.1.3]{Gau83}, see also \textsection\ref{sec:hw}, we exclude such quasimomenta. Then $S_\textsc{h}(p_1,p_1) = -1$, see \eqref{eq:S-mat}, so we finally require $p_1 \neq p_2$.

By total momentum quantisation the sum $p_1 + p_2$ is real. Reality of the energy a priori allows for three types of solutions: both $p_m$ are real (and without loss of generality $0\leq p_1 < p_2 < 2\pi$), or both $p_m$ are complex, with either $p_2 = p_1^*$ complex conjugate (and without loss of generality $\Im p_1 >0$) or $p_2 = \pi - p_1 \, \mathrm{mod} \, 2\pi$. 
As we will see the first two options exhaust all solutions. The roots can be labelled by (a subset of) the Bethe integers $I_1 \leq I_2 \leq L-1$, which pin down the approximate location of $\Re p_m$, ensuring that $p_1 < p_2$ for real solutions, and avoiding trivial roots. This works as follows, see also Figure~\ref{fig:integers}.

\DeclareRobustCommand\firstbox{\begin{tikzpicture}[scale=.27] \fill[pattern=crosshatch] (0,0) rectangle (1,1); \end{tikzpicture}}
\DeclareRobustCommand\secondbox{\begin{tikzpicture}[scale=.27] \fill[pattern=north east lines] (0,0) rectangle (1,1); \end{tikzpicture}}

\begin{figure}[t]
	\begin{tikzpicture}[scale=.39]
	\node at (3,2.5) {$L=6$};
	\node at (.5,1.5) {$I_2$}; \foreach \x in {0,...,5} \node at (\x+.5,.5) {$\x$};
	\node at (-1.3,-.5) {$I_1$}; \foreach \y in {0,...,5} \node at (-.5,-\y-.5) {$\y$};
	\draw[very thick]
	(0,0) -| (6,-1) -| cycle
	(1,-1) |- (2,-2) |- (3,-3) |- (4,-4) |- (5,-5) |- (6,-6) -- (6,-1)
	(3,-1) |- (4,-2) |- (5,-3) |- (6,-4);
	\foreach \x in {1,...,5} \draw[ultra thin] (\x,0) |- (6,-\x);
	\draw[dashed] (2.5,-4.5) -- (6,-1) -- (5.5,-1.5);
	\foreach \x in {1,4} \draw[dotted] (\x,-\x-2) -- (\x+1,-\x-1);
	\node at (3,-.5) [inner sep=.5pt,fill=white] {descendants};
	\node at (4.5,-1.5) [inner sep=.5pt,fill=white] {scatt.$\!$};
	\node at (1.5,-1.5) {b};
	\node at (2.5,-1.5) {e\vphantom{t}};
	\fill[pattern=north east lines] (2,-2) rectangle (3,-3);
	\fill[pattern=crosshatch] (3,-2) rectangle (4,-3);
	\fill[pattern=crosshatch] (3,-3) rectangle (4,-4);
	\fill[pattern=crosshatch] (4,-3) rectangle (5,-4);
	\fill[pattern=north east lines] (4,-4) rectangle (5,-5);
	\node at (5.5,-4.5) {e\vphantom{t}};
	\node at (5.5,-5.5) {b};
	\node at (0,-9) {}; 
	\end{tikzpicture}
	\
	\begin{tikzpicture}[scale=.39]
	\node at (3.5,2.5) {$L=7$};
	\node at (.5,1.5) {$I_2$}; \foreach \x in {0,...,6} \node at (\x+.5,.5) {$\x$};
	\node at (-1.3,-.5) {$I_1$}; \foreach \y in {0,...,6} \node at (-.5,-\y-.5) {$\y$};
	\draw[very thick]
	(0,0) -| (7,-1) -| cycle
	(1,-1) |- (2,-2) |- (3,-3) |- (4,-4) |- (5,-5) |- (6,-6) |- (7,-7) -- (7,-1)
	(3,-1) |- (4,-2) |- (5,-3) |- (6,-4) |- (7,-5);
	\foreach \x in {1,...,6} \draw[ultra thin] (\x,0) |- (7,-\x);
	\draw[dashed] (3,-5) -- (7,-1);
	\draw[dotted] (1.25,-3.25) -- (2,-2.5) (4.75,-6.75) -- (5.5,-6);
	\node at (3.5,-.5) [inner sep=.5pt,fill=white] {descendants};
	\node at (5.5,-2) [inner sep=.5pt,fill=white] {scatt.$\!$};
	\node at (1.5,-1.5) {b};
	\node at (2.5,-1.5) {b};
	\fill[pattern=north east lines] (2,-2) rectangle (3,-3);
	\fill[pattern=north east lines] (3,-2) rectangle (4,-3);
	\fill[pattern=crosshatch] (3,-3) rectangle (4,-4);
	\fill[pattern=crosshatch] (4,-3) rectangle (5,-4);
	\fill[pattern=crosshatch] (4,-4) rectangle (5,-5);
	\fill[pattern=north east lines] (5,-4) rectangle (6,-5);
	\fill[pattern=north east lines] (5,-5) rectangle (6,-6);
	\node at (6.5,-5.5) {b};
	\node at (6.5,-6.5) {b};
	\node at (0,-9) {}; 
	\end{tikzpicture}
	\
	\begin{tikzpicture}[scale=.39]
	\node at (4,2.5) {$L=8$};
	\node at (.5,1.5) {$I_2$}; \foreach \x in {0,...,7} \node at (\x+.5,.5) {$\x$};
	\node at (-1.3,-.5) {$I_1$}; \foreach \y in {0,...,7} \node at (-.5,-\y-.5) {$\y$};
	\draw[very thick]
	(0,0) -| (8,-1) -| cycle
	(1,-1) |- (2,-2) |- (3,-3) |- (4,-4) |- (5,-5) |- (6,-6) |- (7,-7) |- (8,-8) -- (8,-1)
	(3,-1) |- (4,-2) |- (5,-3) |- (6,-4) |- (7,-5) |- (8,-6);
	\foreach \x in {1,...,7} \draw[ultra thin] (\x,0) |- (8,-\x);
	\draw[dashed] (3.5,-5.5)  -- (8,-1);
	\foreach \x in {2,6} \draw[dotted] (\x-.5,-\x-1.5) -- (\x,-\x-1);
	\node at (4,-.5) [inner sep=.5pt,fill=white] {descendants};
	\node at (6.5,-2.5) [xshift=-.5,inner sep=.5pt,fill=white] {scatt.$\!$};
	\node at (1.5,-1.5) {b};
	\node at (2.5,-1.5) {b};
	\node at (2.5,-2.5) {e\vphantom{t}};
	\fill[pattern=north east lines] (3,-2) rectangle (4,-3);
	\fill[pattern=north east lines] (3,-3) rectangle (4,-4);
	\fill[pattern=crosshatch] (4,-3) rectangle (5,-4);
	\fill[pattern=crosshatch] (4,-4) rectangle (5,-5);
	\fill[pattern=crosshatch] (5,-4) rectangle (6,-5);
	\fill[pattern=north east lines] (5,-5) rectangle (6,-6);
	\fill[pattern=north east lines] (6,-5) rectangle (7,-6);
	\node at (6.5,-6.5) {e\vphantom{t}};
	\node at (7.5,-6.5) {b};
	\node at (7.5,-7.5) {b};
	\node at (0,-9) {}; 
	\end{tikzpicture}
	\
	\begin{tikzpicture}[scale=.39]
	\node at (4.5,2.5) {$L=9$};
	\node at (.5,1.5) {$I_2$}; \foreach \x in {0,...,8} \node at (\x+.5,.5) {$\x$};
	\node at (-1.3,-.5) {$I_1$}; \foreach \y in {0,...,8} \node at (-.5,-\y-.5) {$\y$};
	\draw[very thick]
	(0,0) -| (9,-1) -| cycle
	(1,-1) |- (2,-2) |- (3,-3) |- (4,-4) |- (5,-5) |- (6,-6) |- (7,-7) |- (8,-8) |- (9,-9) -- (9,-1)
	(3,-1) |- (4,-2) |- (5,-3) |- (6,-4) |- (7,-5) |- (8,-6) |- (9,-7);
	\foreach \x in {1,...,8} \draw[ultra thin] (\x,0) |- (9,-\x);
	\draw[dashed] (4,-6) -- (9,-1);
	\draw[dotted] (1.75,-3.75) -- (2.5,-3) (6.25,-8.25) -- (7,-7.5);
	\node at (4.5,-.5) [inner sep=.5pt,fill=white] {descendants};
	\node at (7,-3) [xshift=-.5,inner sep=.5pt,fill=white] {scatt.$\!$};
	\node at (1.5,-1.5) {b};
	\node at (2.5,-1.5) {b};
	\node at (2.5,-2.5) {b};
	\fill[pattern=north east lines] (3,-2) rectangle (4,-3);
	\fill[pattern=north east lines] (3,-3) rectangle (4,-4);
	\fill[pattern=north east lines] (4,-3) rectangle (5,-4);
	\fill[pattern=crosshatch] (4,-4) rectangle (5,-5);
	\fill[pattern=crosshatch] (5,-4) rectangle (6,-5);
	\fill[pattern=crosshatch] (5,-5) rectangle (6,-6);
	\fill[pattern=north east lines] (6,-5) rectangle (7,-6);
	\fill[pattern=north east lines] (6,-6) rectangle (7,-7);
	\fill[pattern=north east lines] (7,-6) rectangle (8,-7);
	\node at (7.5,-7.5) {b};
	\node at (8.5,-7.5) {b};
	\node at (8.5,-8.5) {b};
	\end{tikzpicture}
	\caption{The structure of the Bethe integers $0\leq I_1 \leq I_2\leq L-1$: descendants ($I_1=0$), scattering states ($I_2 \geq I_1 +2$) and bound states ($I_2 \in \{I_1,I_1+1\}$). $I_\text{tot} = I_1 + I_2 \, \mathrm{mod}\, L$ is constant along antidiagonals. Parity acts by reflection about the dashed antidiagonal ($I_\text{tot}=0$)\,---\,or, if $I_1=0$, about $I_2 =L/2$. Scattering states along the dashed antidiagonal admit explicit roots.\textsuperscript{\ref{fn:exact_selfconj_roots}} Complex roots occur for $I_2 \in \{I_1,I_1+1\}$, for which the $L-3$ preferred choices are marked by `b' for bound states and `e' for the exceptional root (equivalent, so choose one). The $L$ remaining choices approximately between the small dotted antidiagonals ($I_\text{tot} = L/2$) are excluded: `\firstbox' ($I_\text{tot} = 0,1,L-1$), `\secondbox' (scattering phase $\varphi$ not minimal) and, for $L$ even, one of the `e'.} 
	\label{fig:integers} 
\end{figure}
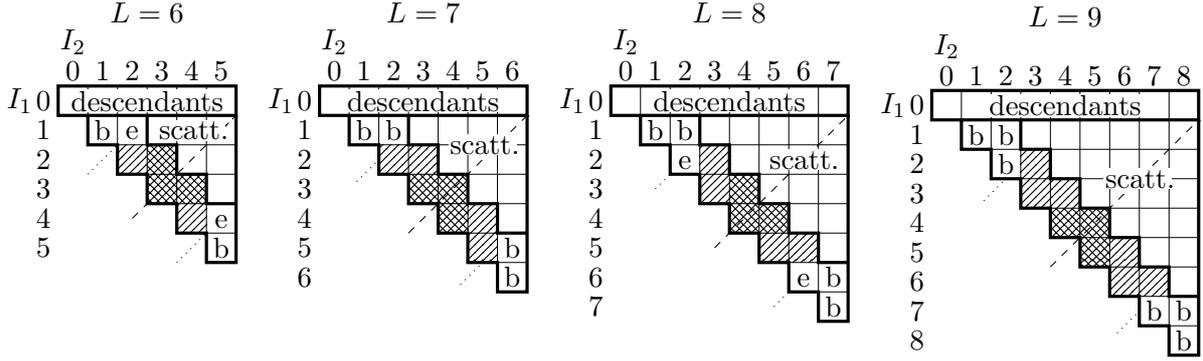

\begin{itemize}
	\item For completeness we start by recalling that the $L$ choices with $I_1 = 0$ enumerate the descendants of the $M=1$ vectors, see \textsection\ref{sec:symms_trivial_roots}. We will assume $I_1 \geq 1$.
	\item For each of the $\binom{L-2}{2}$ choices with $I_2 \geq I_1+2$ there exists a unique real root $\varphi \in (0,\pi)$, so $p_1 \in \frac{2\pi}{L}(I_1\,, I_1+\frac12)$, $p_2 \in \frac{2\pi}{L}(I_2 - \frac12, I_2)$.%
	\footnote{\ A priori the $I_m$ single out a interval of length $2\pi$ at the corresponding $I_\text{tot}$. We observe that the uniqueness holds for $\varphi \in [0,2\pi)$, with the actual root lying in $(0,\pi)$. This is natural from the structure illustrated in Figure~\ref{fig:constraint_real_kappa} and readily verified numerically for all roots in this class for $L\leq 30$. For the extra real roots present for $L\geq 22$ (see below) uniqueness only holds in the smaller interval $(0,\pi)$ just below the trivial root.}%
	$\hphantom{\,}$\footnote{\ \label{fn:exact_selfconj_roots} The parity self-conjugate cases $\vect{I} = (i,L-i)$, with $1\leq i \leq \lfloor L/2\rfloor -1$, admit explicit roots: $\varphi = 2\pi i/(L-1)$.{\checkedMma}}
	See Figure~\ref{fig:constraint_real_kappa} for a typical example.
	Nearly all real roots arise in this way, and for $L \leq 21$ all do (see below).
	The eigenvectors can be interpreted as \emph{scattering states}: the quasimomenta are close to their free values ($\varphi=0$) and the wave function is broad.
	As $L$ becomes large these energies form a continuum above the dispersion relation, cf.~Figure~\ref{fig:Ep_M2_Heis}.
	
	\item The $L-3$ remaining roots have $I_2 \in \{I_1,I_1+1\}$; we seek one per $2\leq I_\text{tot} \leq L-2$. Although each $2\leq I_\text{tot} \leq L-2$ admits two $\vect{I}$ that give rise to the same root, there is a preferred choice with $\varphi$ minimal in the sense that $\Re\varphi \in [0,2\pi)$.
	\begin{itemize}
		\item For each $I_2 = I_1 \in \{1,\cdots\mspace{-1mu},\lfloor(L-1)/4\rfloor\}$
		there exists a unique root with $\varphi \in \I\, \mathbb{R}_{>0}$ (so $p_2 = p_1^*$). By parity the same holds for each $I_2 = I_1 \in \{\lceil (3L+1)/4\rceil, \cdots\mspace{-1mu}, L-1\}$. 
		These eigenvectors represent bound states, with magnons that are close to each other. 
		Their energy is close to $\varepsilon_{\textsc{h}}(p_\text{tot})/2$,\,%
		\footnote{\ This bound-state curve follows from $p_{1,2} = p_\text{tot}/2 \mp \I \log[\cos(p_\text{tot}/2)]$, the exact solution for $L\to \infty$.}
		forming a band below the scattering states, see Figure~\ref{fig:Ep_M2_Heis}.
		
		\item When $L$ is even there is an \emph{exceptional root} at $I_1 = \lfloor L/4 \rfloor$, $I_2 = \lceil L/4 \rceil$, so $p_\text{tot} = \pi$. This root can be exactly given: $\varphi = \pi(I_2-I_1) + \I\,\infty$, $p_{1,2} = \pi/2 \pm \I\, \infty$. Another exceptional root occurs at the parity-conjugate $I_1 = \lfloor 3L/4 \rfloor$, $I_2 = \lceil 3L/4 \rceil$, with the same $p_\text{tot},\varphi$ but $p_{1,2} = 3\pi/2 \pm \I\,\infty$. Both solve \eqref{eq:M=2_constraint}, with real part fixed by the \textsc{bae}~\eqref{eq:M=2_BAE}. In terms of rapidities both solutions correspond to the exact two-string $\lambda_1 = \I/2$, $\lambda_2 = -\I/2$, in particular showing that they are equivalent.\,%
		\footnote{\ \label{fn:exceptional_Heis} In fact there is a whole one-parameter family of physically equivalent roots at $\I\,\infty$. For definiteness take $I_1 = \lfloor L/4 \rfloor$, $I_2 = \lceil L/4 \rceil$. The Bethe equations~\eqref{eq:M=2_BAE} are solved by $p_1 = u + \I v$, $p_2 = \pi - p_1$ (allowed by the reality of $p_\text{tot},E$) and $\varphi = L \,u - 2\pi \lfloor L/4\rfloor + \I \,L\, v$. When $v \to \infty$ the constraint~\eqref{eq:M=2_constraint_Heis} is obeyed for \emph{any} value of $u \in \mathbb{R}/2\pi\mathbb{Z}$. The choices $u = \pi/2$ and $u = -\pi/2 \equiv 3\pi/2$ are preferred in that $p_2=p_1^* \,\mathrm{mod}\,2\pi$ and $\Re \varphi \in [0,\pi]$ is small (if we switch to $I_1 = \lfloor 3L/4 \rfloor$, $I_2 = \lceil 3L/4 \rceil$ for $u = 3\pi/2$), as for all other roots.}
		The energy $E=2$ and wave function require some regularisation.\,%
		\footnote{\ As the pole in \eqref{eq:M=2_energy_Heis} signals, the naive value of the energy~\eqref{eq:M=2_energy_Heis} via $\vect{p}$ is incorrect. Regularise $\lambda_{1,2} = \pm\I/2 - \epsilon$ so that $\epsilon \to 0^+ \, (0^-)$ gives $p_1 \to \pi/2 + \I\,\infty$ ($p_1 \to 3\pi/2 + \I\,\infty$) and $p_2 = p_1^*$, yielding the correct value $E = 2$. (More generally the $p_m$ from Footnote~\ref{fn:exceptional_Heis} arise as $\epsilon \to \E^{\I (u-\pi/2)}\,0^+$ with $u \in [-\pi,\pi)$.) The \textit{S}-matrix~\eqref{eq:S_mat_Heis} is still singular. Following \cite{avdeev1985exceptional} keep $\lambda_1 = \I/2 - \epsilon$ but take $\lambda_2 = -\I/2 - \epsilon - 2\mspace{1mu}\I\, (\I\mspace{1mu}\epsilon)^L$. This doesn't affect the limits of $\vect{p},E$ yet allows the wave function to be renormalised: 	$\epsilon^{(L-2)/2} \, \Psi_{\vect{p}}(n_1,n_2) \to \pm [(-1)^{n_2}\,\delta_{n_2,n_1+1} - \delta_{n_1,1}\,\delta_{n_2,L}]$ as $\epsilon \to 0$ for $1\leq n_1 < n_2\leq L$, with sign depending on $L$ and the direction from which $\epsilon$ approaches zero.} 
		This is the most tightly bound state: the magnons are always adjacent.
		
		\item For each $I_2 = I_1 +1 \in \{2,\cdots\mspace{-1mu},\lfloor(L+1)/4\rfloor\}$, and by parity each $I_2 = I_1 +1 \in \{\lceil(L+3)/4\rceil,\cdots\mspace{-1mu},L-1\}$, there is a unique solution. To understand its behaviour we fix $\vect{I} = (i,i+1)$, $i\in\mathbb{Z}_{>0}$, set $n = I_1 + I_2 = 2i+1$ and vary $L>2\,n$.\,%
		\footnote{\ \label{fn:bound_state_starts_as_exceptional} If we ignore that $\vect{I}$ might first occur as the parity-conjugate of some lower $\vect{I}'$, any such root starts out as exceptional, $\varphi = \pi + \I\,\infty$, at $L=2\,n$.}
		For sufficiently low $L$ the root is of the form $\varphi \in \pi + \I \, \mathbb{R}_{>0}$, where $\Re\varphi=\pi$ ensures $p_2 = p_1^*$. As $L$ increases $\Im \varphi$ becomes smaller, until $p_\text{tot} \approx 0 \,\mathrm{mod}\,2\pi$ and the energy gap between the bound-state curve and the $M=2$ scattering-state continuum closes, see again Figure~\ref{fig:Ep_M2_Heis}. At some `critical length' $L_\text{cr}^{(n)}$ the $p_m$ collide on the real axis ($\varphi = \pi + 0\,\I$)\,---\,precisely at the trivial root $\delta=0$ from \eqref{eq:trivial_roots}. For $L \geq \lceil L_\text{cr}^{(n)} \rceil$ the root becomes real \cite{bethe1931theorie}. These extra real roots have $\varphi \in [0,\pi)$ so that $p_1 < p_2$ just as for scattering states. This collision is shown in Figure~\ref{fig:Lcr_pm}. 
		
		The value of $L_\text{cr}^{(n)} \in \mathbb{R}_{>0}$ where the roots for $\vect{I}=(i,i+1)$ and its parity-conjugate $\vect{I}'=(L-i-1,L-i)$ become real is determined uniquely by \cite{Essler1992}
		\begin{equation} \label{eq:L_crit}
			\arctan\sqrt{L_\text{cr}^{(n)} - 1} = \frac{\pi}{2} \biggl( 1 - \frac{n}{L_\text{cr}^{(n)}} \biggr) \, , \qquad n = 2\,i+1 \, .
		\end{equation}
		To very good approximation $L_\text{cr}^{(n)} \approx (\pi \, n/2)^2 - 1/3$, with first few values (for $n=3,5,\dots$) approximately equal to $21.9, 61.3, 120.6, 199.5, 298.2, 416.7$. \\[-1ex]
	\end{itemize}
	Together these bound states can be labelled by $I_1 = \lfloor n/2 \rfloor, I_2 = \lceil n/2 \rceil$ where for each $I_\text{tot} = n \, \mathrm{mod} \, L$ with $I_\text{tot} \neq L/2$ the preferred choices are $n \in \{ 2,\dots, \lfloor L/2 \rfloor \}$ plus their parity-conjugate values $n'=2 \, (L-1)-n$. For the exceptional root at even $L$ pick either of $n\in \{L/2,3L/2\}$.
\end{itemize}

\begin{figure}[h]
	\centering
	\includegraphics[width=\textwidth]{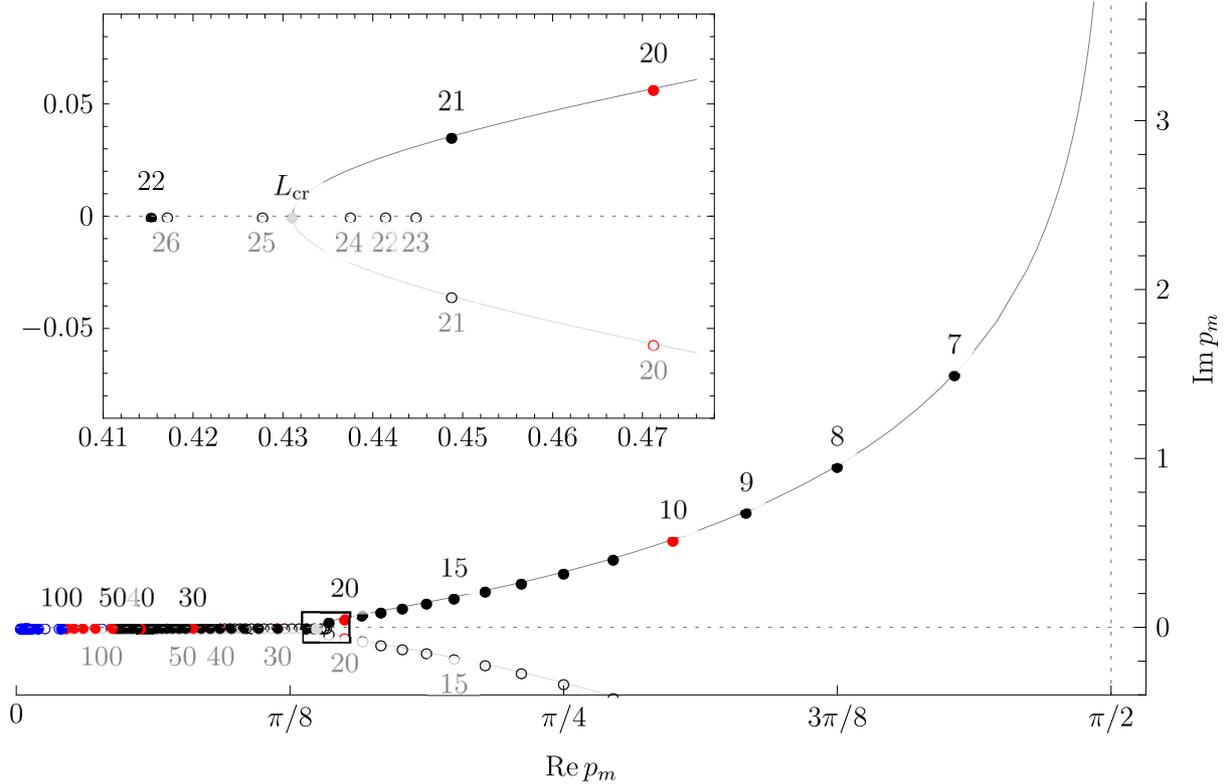}
	\caption{The `unbinding' of bound-state quasimomenta for the Heisenberg spin chain with $\vect{I} = (1,2)$ and varying system size. Here $p_1$ ($p_2$) is represented by $\bullet$ ($\circ$), blue for $L$ a multiple of 100, red for $L=10\,n$ ($1 \leq n \leq 9$) and grey for the formal value $L_\text{cr}^{(3)} \approx 21.9$. The (formal) trajectories of $p_{1,2} \to \pi/2 \pm \I \infty$ as $L\to 6^+$ are indicated too. The inset shows what happens near $L_\text{cr}^{(3)} \approx 21.9$.}
	\label{fig:Lcr_pm}
\end{figure}

In summary the counting for the two-particle sector of the Heisenberg spin chain is as follows. Let $N_L \coloneqq 2 \, \#\{L^{(n)}_\text{cr} \leq L , \, n \in 1 + 2\,\mathbb{Z}_{>0} \} = \lfloor (2/\pi) \, \sqrt{L+1/3} - 1 \rfloor$, so $N_L = 0$ for $4\leq L\leq 21$, $N_L = 2$ if $22\leq L\leq 61$, and so on. Then the Bethe-ansatz system~\eqref{eq:M=2_BAE}, \eqref{eq:M=2_constraint_Heis} admits $\binom{L-2}{2} + N_L$ real roots, $\varphi \in [0,2\pi)$, and $L-3-N_L$ complex roots, $\varphi \in \pi \, (I_2-I_1) + \I \, \mathbb{R}_{>0}$ so that $p_2^*=p_1$ with $\Re p_m = \pi\,(I_1+I_2)/L$. The Bethe integers~$\vect{I}$ enumerating these roots are illustrated in Figure~\ref{fig:integers}. Together these give all $\binom{L-1}{2} - L$ highest-weight vectors, yielding the two-particle spectrum shown in Figure~\ref{fig:Ep_M2_Heis}.

\subsection{Inozemtsev regime}  \label{sec:Ino_regime} For sufficiently high values of $\kappa$ the spectrum will be close to that of Heisenberg. Let us therefore use the classification in terms of descendants, scattering states and bound states, labelled by $\vect{I}$ like in \textsection\ref{sec:Heis_recap_M=2}. Before we go into details let us preview what happens as we flow down from Heisenberg ($\kappa = \infty$) to Haldane--Shastry ($\kappa =0$). We focus on the intermediate regime, $\kappa \sim 1$; we will discuss the HS limit in \textsection\ref{sec:HS_regime}.

\begin{figure}[h]
	\centering
	\includegraphics[width=\textwidth]{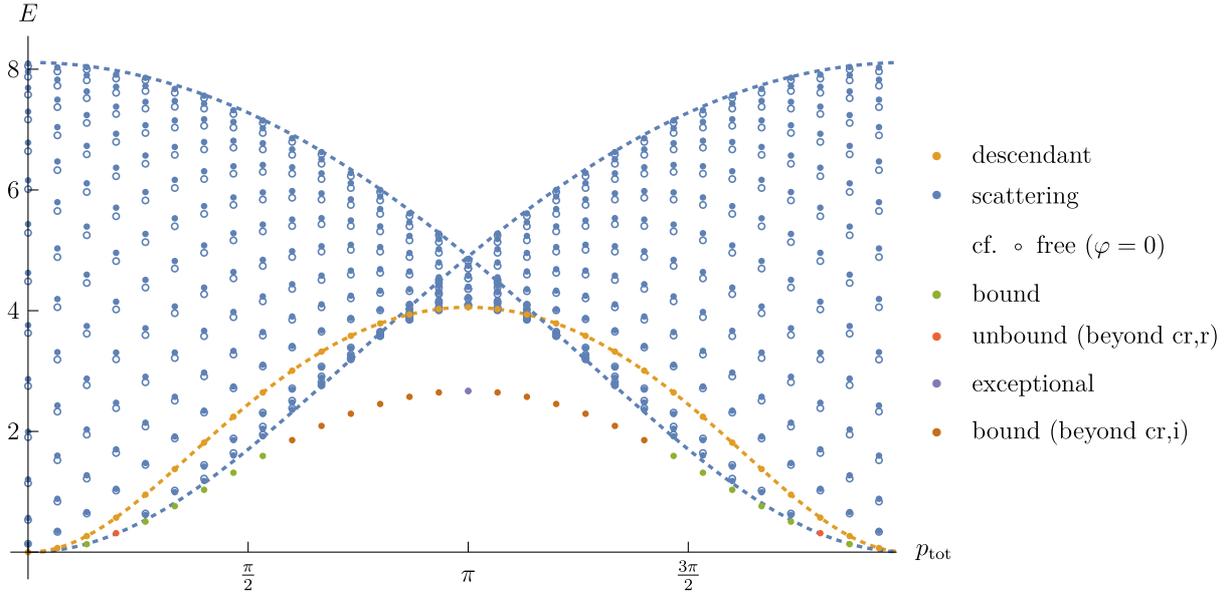}
	\caption{The two-particle spectrum of the Inozemtsev spin chain with $L=30$, $\kappa =1$, in between Figures~\ref{fig:Ep_M2_Heis} and \ref{fig:Ep_M2_HS}. Each dot represents an eigenspace. The descendants lie on the dispersion $\varepsilon_\text{n}(p_\text{tot})$ (dotted curve), shown in orange. The rest are highest-weight vectors. Scattering states are shown in blue, with energy rather close to that of two free magnons (open circles), and bounded from above by the energy of two identical magnons (dotted curve, two-valued as $p_\text{tot} = 2\,p \,\mathrm{mod} \,2\pi$, $E(p,p) = 2\,\varepsilon_\text{n}(p)$). Bound states are closer to the other states than for Heisenberg and now come in four kinds: $p_1 = 0 + \I \kappa, p_2 = \pi - \I\kappa$ exceptional (purple, once), $p_2^* = p_1$  complex conjugate (green), $p_m \in \mathbb{R}$ `unbound' (red, twice) as well as $\Re p_1 < \Re p_2$ with $\Im p_m = \kappa$ (brown, around $p_\text{tot} =\pi$).}
	\label{fig:Ep_M2_Ino}
\end{figure}

\subsubsection{Qualitative description} \label{sec:qualitative} The descendants ($I_1=0$) are completely determined by homogeneity, with wave function \eqref{eq:magnon_desc}; only their eigenvalue~\eqref{eq:dispersion} varies.

Scattering states behave as for Heisenberg. Each $\vect{I}$ with $1\leq I_m\leq L-1$ subject to $I_2 \geq I_1 +2$ singles out a unique real root $\varphi \in (0,\pi)$ for any value of $\kappa>0$.\,%
\footnote{\ The self-conjugate cases do no longer admit simple exact roots for finite $\kappa$, cf.~Footnote~\ref{fn:exact_selfconj_roots} and Tables~\ref{tb:spec_L=4}--\ref{tb:spec_L=6}.} 
Although the value of $\varphi$ may initially increase a little as $\kappa$ is lowered\,%
\footnote{\ This first occurs at $L=9$, for $\vect{I}=(1,3)$ and its parity conjugate $\vect{I}'=(6,8)$, with $\varphi|_{\kappa=3/2} \approx 2.02$ vs $\varphi|_{\kappa=\infty} \approx 2.00$.{\checkedMma} If one looks closely it is visible for the left-most pair of zeroes, with $\vect{I} = (1,3)$, in Figure~\ref{fig:constraint_real_kappa}.}
it eventually decreases to $\varphi|_{\kappa=0} = 0$. See Figure~\ref{fig:constraint_real_kappa} for an illustration.

The $L-3$ remaining roots are enumerated by $I_2 \in \{I_1,I_1+1\}$. When $L$ is even this includes the exceptional root at $I_\text{tot} = L/2$. This root admits an explicit expression for any $\kappa$: we find that $p_1 = u + \I \kappa$ and $p_2 = \pi - p_1$ 
solves the constraint for $u\in\{0,\pi/2,\pi,3\pi/2\}$. 
Although the choices $u \in\{\pi/2,3\pi/2\}$ seem to be the natural $\kappa$-deformations of the exceptional roots given in \textsection\ref{sec:Heis_recap_M=2}, the corresponding wave function vanishes: these $p_m$ are amongst the trivial roots from \eqref{eq:trivial_roots}. Hence, by exchanging the roles of $p_{1,2}$ if necessary, we pick $u = 0$.\,%
\footnote{\ \label{fn:exceptional_Ino} Let us make contact with Footnote~\ref{fn:exceptional_Heis}. For Inozemtsev $v = \Im p_1 = \kappa$ ensures that the imaginary part of \eqref{eq:f_constraint} is zero, after which the real part of the constraint requires $u \in\{0,\pi/2,\pi,3\pi/2\}$. As $\kappa \to \infty$ the constraint flattens out, vanishing identically to give rise to the one-parameter family of exceptional roots from Footnote~\ref{fn:exceptional_Heis}. As we argued there $u \in \{\pi/2, 3\pi/2\}$ are preferred for Heisenberg. Inozemtsev requires $u\in\{0,\pi\}$, corresponding to `macroscopic' $\Re\varphi \sim L/2$. See the end of \textsection\ref{sec:crit} for a way to understand this shift of $u$.}
Using the $\omega$-periodicity in~$\gamma$ of the wave function \eqref{eq:M=2_wave_fn} one may take $\gamma = L/2$. Using the $L$-quasiperiodicity of $\chi_2$ and the Legendre relation the wave function assumes the simple form
\begin{subequations} \label{eq:exceptional_wave_fn}
	\begin{gather}
	\Psi_{\I \kappa,\pi-\I\kappa}(n_1,n_2) = \bigl( \E^{\I \, \pi \,n_1} - \E^{\I \, \pi \,n_2} \bigr) \, \chi_1(n_1-n_2,L/2) \, .
\checkedMma
\intertext{The prefactor demands the $n_m$ to have opposite parity, forcing the magnons to be separated by an even number of sites. 
with finite limits for the renormalised wave function}
	n_H(\kappa)^{1/2} \ \Psi_{\I \kappa,\pi-\I\kappa}(n_1,n_2)
	\end{gather}
\end{subequations}
with prefactor possessing the properties of the square root of that in \eqref{eq:ham_normalised}.
This function is real at physical values, $n_1 \neq n_2 \in \mathbb{Z}$, of the coordinates. For general $\kappa$ it is nonzero if the magnons are at distance $d_L(n_1,n_2) \in \{1,3,\cdots\mspace{-1mu},2\lfloor L/4\rfloor-1\}$. Away from the Haldane--Shastry regime (so for $\kappa \gtrsim 1$) it is exponentially suppressed, with nonzero values of \eqref{eq:exceptional_wave_fn} behaving like $2\,\E^{-\kappa \, (d_L(n_1,n_2)-1)}$. The Heisenberg limit yields the nearest-neighbour wave function from \textsection\ref{sec:Heis_recap_M=2}.

For all other bound states $\Im \varphi$ initially \emph{in}creases, so both $p_1$ and $p_2 = p_1^*$ move away from the real axis. Let us use the periodicity in the imaginary direction to think of $p_2 = p_1^* + 2\I \kappa$ as lying in the upper half of the fundamental domain. From this perspective the $p_m$ move towards each other, to collide with each other (and a trivial root) at $\Im p_m = \kappa_\text{cr}$ for some `critical' value $\kappa_\text{cr}$ that depends on $L$ and $I_1 + I_2$: see \textsection\ref{sec:crit}. As our notation suggests this phenomenon is closely related to that with $L_\text{cr}$ for Heisenberg (\textsection\ref{sec:Heis_recap_M=2}). The main differences are that here the real period~$L$ is fixed while the imaginary period $\omega = \I \pi/\kappa$ increases, that $\Im \varphi|_{\kappa = \kappa_\text{cr}} \neq 0$, and that this new phenomenon happens for \emph{all} bound states (see Footnote~\ref{fn:exceptional_Ino} for the exceptional root). Since at $\kappa_\text{cr}$ the $p_m$ coincide with a trivial root the wave function vanishes. (The same holds for $L_\text{cr}$, but that value is non integral so does not occur for any physical wave function.) It should be possible
to renormalise $\Psi_{\vect{p}}$ by division by a function whose zero determines $\kappa_\text{cr}$ (see \textsection\ref{sec:crit}) and viewing its value as a limit $\kappa \to \kappa_\text{cr}$. When we further lower $\kappa \ (< \kappa_\text{cr}$) the roots stay at $\Im p_m = \kappa$ while $\Re \varphi$ starts to vary. This type of root ($p_2 \neq p_1^* \, \mathrm{mod}\,2\I \kappa$, $p_2 \neq \pi - p_1 \, \mathrm{mod}\,2\I \kappa$) is new for Inozemtsev; the total momentum is real $\mathrm{mod}\,2\I \kappa$, and the energy is real. In analogy with the phenomena of $L_\text{cr}$ we will take $\varphi|_{\kappa < \kappa_\text{cr}}$ to be such that $\Re p_1 < \Re p_2$. In the HS limit $\kappa \to 0$ the $p_m$ hit the real axis; we will determine their limiting value in \textsection\ref{sec:HS_limit}.

\subsubsection{Critical equation} \label{sec:crit} In fact the Inozemtsev spin chain allows for even richer scenarios of bound-state quasimomenta than what we have just described. To understand this we need to analyse the behaviour of the bound states as a function of the periodicity parameters $L,\kappa$ in more detail. We wish to determine the values of these parameters where the $p_m$ collide.\,%
\footnote{\ We thank Istv\'{a}n M.\ Sz\'{e}cs\'{e}nyi for a suggestion that led to the following approach.} 
Fix $I_1,I_2$. Then the vanishing of 
\begin{equation*}
	g_{I_1,I_2}(\varphi;L,\kappa) \coloneqq L\, f_{I_1 + I_2}\Bigl(\frac{2\pi I_1 +\varphi}{L}\Bigr) = 2 \, L\, \momL{\rho}_1(\varphi) - L \, \mom{\rho}_1\Bigl(\frac{2\pi I_1 +\varphi}{L}\Bigr) + L \, \mom{\rho}_1\Bigl(\frac{2\pi I_2 - \varphi}{L}\Bigr)
\end{equation*}
implicitly characterises the scattering phase as a function of the quasiperiodicity parameters~$L,\kappa$ of $\momL{\rho}_1,\mom{\rho}_1$. More precisely, given $L_0,\kappa_0$ let $\varphi_0$ be a root of $g_{I_1,I_2}$. If we formally extend the real period to values $L \in \mathbb{R}_{>0}$ the function $g_{I_1,I_2}$ is analytic in $\varphi$ as well as $L,\kappa$. By the implicit function theorem $g_{I_1,I_2}(\varphi;L,\kappa) = 0$ has a unique analytic solution $\varphi(L,\kappa)$ extending $\varphi_0 = \varphi(L_0,\kappa_0)$ to a neighbourhood of $L_0,\kappa_0$ as long as
\begin{equation*} 
	\frac{\partial \, g_{I_1,I_2}}{\partial \mspace{1mu} \varphi} = 2 \, L \, \momL{\rho}_1{}'(\varphi) - \mom{\rho}_1{}'\Bigl(\frac{2\pi I_1 +\varphi}{L}\Bigr) - \mom{\rho}_1{}'\Bigl(\frac{2\pi I_2 - \varphi}{L}\Bigr)
\end{equation*}
is nonzero at $\varphi_0;L_0,\kappa_0$. We are looking for roots that collide as we vary the parameters. At such a point $L_0,\kappa_0$ there is no \emph{unique} extension, so \eqref{eq:crit} must vanish at that point. The `critical equation' can thus be written as
\begin{equation} \label{eq:crit}
	\frac{1}{2}\,\mom{\rho}_1{}'(p_1) + \frac{1}{2}\,\mom{\rho}_1{}'(p_2) = L \, \momL{\rho}_1{}'(\varphi) \, .
\end{equation}
For colliding quasimomenta the terms on the left-hand side are equal. The functions in \eqref{eq:crit} are just $\mom{\wp},\momL{\wp}$ up to some constants: recall that $-\rho'_b = \wp + \eta_b/\omega_b$.

\begin{figure}[h]
	\centering
	\includegraphics[width=.95\textwidth]{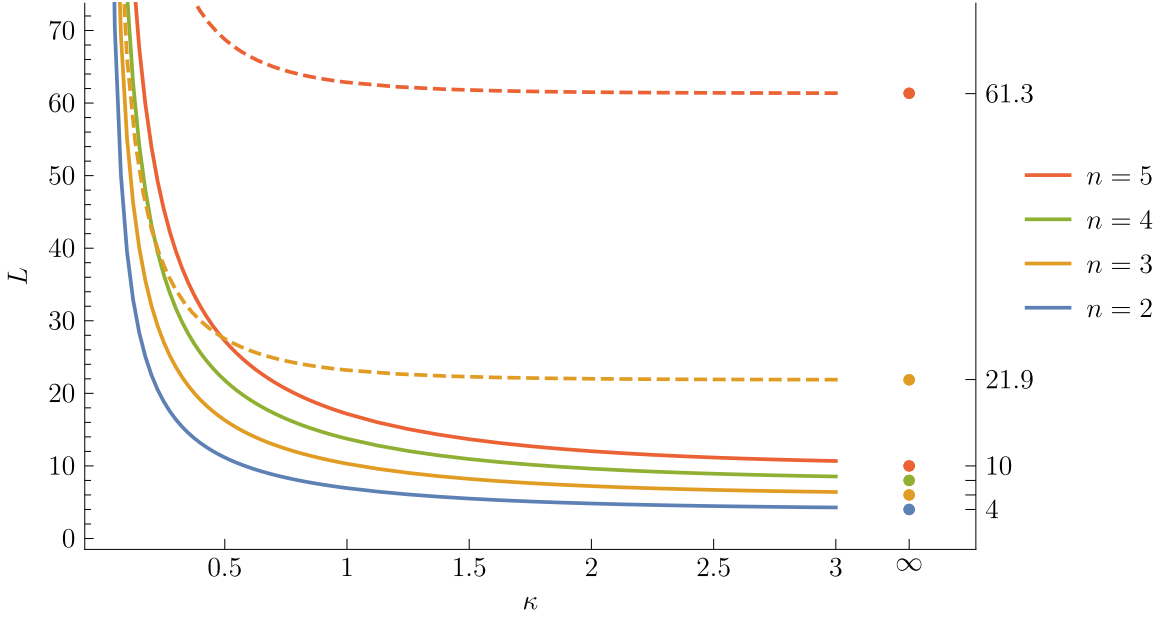}
	\caption{Critical loci of the Inozemtsev spin chain, characterising, for the first few values of $n = I_1 + I_2 \geq 2$, the parameters at which the quasimomenta collide at the imaginary half-period ($L_\text{cr,i},\kappa_\text{cr,i}$, solid curves) or the real axis ($L_\text{cr,r},\kappa_\text{cr,r}$, dashed curve, only for $n$ odd). For fixed $\kappa$ note that $L_\text{cr,i}^{(n)} < L_\text{cr,i}^{(n+1)}$ and, for $n$ odd, $L_\text{cr,r}^{(n)} \gg L_\text{cr,i}^{(n)}$ away from the Haldane--Shastry regime. For fixed $L$, coming from Heisenberg, the bound state with $n$ closest to $L/2$ (along with its parity conjugate, not indicated) is the first to exhibit colliding quasimomenta. Note that $L_\text{cr,i}^{(n)} = 2n$ as $\kappa\to\infty$, i.e.\ for $L$ even $\kappa_\text{cr,i}^{(L/2)} = \infty$.}
	\label{fig:crit_locus}
\end{figure}

Now we use $p_1 = p_2$. Then $\varphi = \pi \, (I_2-I_1)$ by the \textsc{bae}~\eqref{eq:M=2_BAE}. This occurs precisely at the first trivial solution from \eqref{eq:trivial_roots}, present if $I_1+I_2$ is odd. By $2\pi$-periodicity we obtain
\begin{equation*} 
	\mom{\rho}_1{}'\biggl(\frac{\pi\,(I_1 + I_2)}{L}\biggr) = L \, \momL{\rho}_1{}'\bigl(\pi\,(I_1 + I_2)\bigr) \, ,
\end{equation*}
where the left-hand side features $I_1 + I_2$ \emph{without} the mod $L$. The equation is the same for $I_1+I_2$ and the parity-conjugate $I'_1+I'_2 = 2\,L-(I_1+I_2)$. When $I_1 + I_2$ (and $I'_1+I'_2$) is odd the pole of $\momL{\rho}_1$ is avoided. 
Focus on $\vect{I}=(i,i+1)$,\,%
\footnote{\ Note that the critical equation does not single out this choice, as, by $2\pi$-periodicity, it only depends on $I_1+I_2$. Numerics confirm that colliding momenta occur only for choices of the $I_m$ that correspond to Heisenberg bound states; the roots for scattering states only vary in a small real interval, never getting near any trivial roots.}
with critical equation
\begin{equation} \label{eq:crit_r}
	\text{cr, r}: \qquad \mom{\rho}_1{}'\biggl(\frac{\pi\,n}{L}\biggr) = L \, \momL{\rho}_1{}'(\pi) \, , \qquad n = 2\, i + 1 \, .
\end{equation}
(The meaning of `cr, r' will become clear soon.) Since $\mom{\rho}_1{}'(z),\momL{\rho}_1{}'(z) \to [2\sin(z/2)]^{-2}$ in the Heisenberg limit, the condition reduces there to
\begin{equation} \label{eq:crit_r_Heis}
	\sin^{-2}\Bigl(\frac{\pi\,n}{2L}\Bigr) = L \, , \qquad \text{i.e.} \qquad \tan^{-2} \Bigl(\frac{\pi\,n}{2L}\Bigr) = L-1 \, .
\end{equation}
Take the square root and use $\arctan(1/z) = \pi/2 - \arctan(z)$ for $z>0$ to retrieve \eqref{eq:L_crit}. In general \eqref{eq:crit_r} determines the \emph{cricital locus} $\{(L,\kappa)\} \in \mathbb{R}_{>0}^2$ at which the quasimomenta collide on the real axis (whence the `r') for $\vect{I} = (i,i+1)$ and $\vect{I}' = (L-i-1,L-i)$ separately. This curve is illustrated in Figure~\ref{fig:crit_locus}. In particular, \eqref{eq:crit_r} gives the critical length $L_\text{cr,r}^{(n)} = L_\text{cr,r}^{(n)}(\kappa)$ for which the quasimomenta become real for the Inozemtsev spin chain with fixed parameter~$\kappa$.\,%
\footnote{\ More precisely, $L_\text{cr,r}^{(n)}$ is the \emph{largest} real root of \eqref{eq:crit_r}. Just like for \eqref{eq:crit_r_Heis} there are further roots $L$ with $0<L \lesssim n$. In the Heisenberg limit these `unphysical' roots disappear when we take square roots: $\cot[\pi\,n/2L] = -\sqrt{L-1}$ accounts for all unphysical roots, while $L_\text{cr}^{(n)}$ is the unique positive solution to $\cot[\pi\,n/2L] = \sqrt{L-1}$, which is equivalent to \eqref{eq:L_crit}. We have not been able to obtain the elliptic version of the latter equation.} 
For $\kappa = \infty$ this gives what we denoted by $L_\text{cr}^{(n)}$ or $L_\text{cr}$ in \textsection\ref{sec:Heis_recap_M=2} and \textsection\ref{sec:qualitative}. Numerics suggest that $L_\text{cr,r}^{(n)}(\kappa)$ increases monotonically when $\kappa$ is lowered, cf.~Figure~\ref{fig:crit_locus}. We will return to the Haldane--Shastry limit in \textsection\ref{sec:HS_limit}.

The imaginary period of the quasimomentum lattice for Inozemtsev allows for another root that corresponds to coinciding quasimomenta: $\varphi = \pi \, (I_2-I_1) + \I L \kappa$ yields $p_2 = p_1^* = p_1 - 2\I \kappa = p_1 \,\mathrm{mod}\,2\I\kappa$. This, too, is a trivial root\,---\,the third in \eqref{eq:trivial_roots}. The critical equation reads
\begin{equation} \label{eq:crit_i}
	\text{cr, i}: \qquad \mom{\rho}_1{}'\Bigl(\frac{\pi \, n}{L} +\I \kappa \Bigr) = L \, \momL{\rho}_1{}'\bigl(\pi \, n + \I L \kappa \bigr) \, , \qquad n=I_1+I_2 \, ,
\end{equation}
where $\vect{I} = (\lfloor n/2 \rfloor, \lceil n/2 \rceil)$ and the values of $n$ are as in \textsection\ref{sec:Heis_recap_M=2}. This time the imaginary shift avoids the pole at the origin, so both odd and even $n$ are allowed. This equation determines the critical locus at which the quasimomenta for $\vect{I}$, or $\vect{I}'$ with $I'_1 + I'_2 = 2L-n$, collide at the imaginary half-period (whence the `i'). In particular, if we keep $L$ fixed and pick $2\leq I_\text{tot} \leq L-2$ it gives the value of $\kappa_\text{cr,i}^{(n)} = \kappa_\text{cr,i}^{(n)}(L)$ for which the quasimomenta of the bound state with $n=I_\text{tot}$ and go from complex conjugate (for $\kappa>\kappa_\text{cr,i}^{(n)}$) to having $\Im p_m = \kappa$ but $\Re p_1 \neq \Re p_2$ ($\kappa>\kappa_\text{cr,i}^{(n)}$). The same is true for the parity-conjugate quasimomenta at $n'=2\,L - I_\text{tot}$. This is what we denoted by $\kappa_\text{cr}$ in \textsection\ref{sec:qualitative}. We observe numerically that $\kappa_\text{cr,i}^{(n)} = \infty$ for the exceptional root ($L$ even and $n=L/2, 3L/2$), and that the largest finite $\kappa_\text{cr,i}^{(n)}$ occurs for $n \approx L/2, 3L/2$ close to the exceptional root. See again Figure~\ref{fig:crit_locus}.

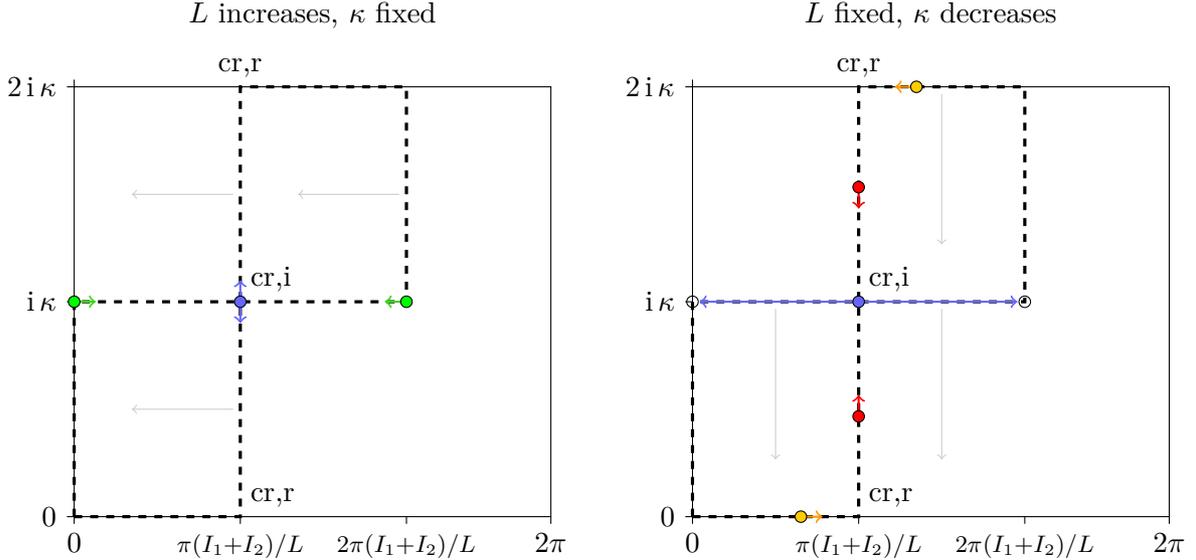
\begin{figure}[h]
	\centering
	\begin{tikzpicture}[scale=.95]
		\node at (3.3,7) {$L$ increases, $\kappa$ fixed};
		\draw (0,0) rectangle (6.6,6);
		\draw (0,0) -- (-.1,0) node [left] {$0$};
		\draw (0,3) -- (-.1,3) node [left] {$\I \,\kappa$};
		\draw (0,6.1) |- (-.1,6) node [left] {$2\,\I \,\kappa$};
		\draw (0,0) -- (0,-.1) node [below] {$0$};
		\draw (2.3,0) -- (2.3,-.1) node [below] {\footnotesize$\pi(I_1{+}I_2)/L$};
		\draw (4.6,0) -- (4.6,-.1) node [below] {\footnotesize$2\pi(I_1{+}I_2)/L$};
		\draw (6.6,0) -- (6.6,-.1) node [below] {$2\pi$};
		\node at (2.3,3) [above right] {cr,i};
		\draw [very thick, dashed] (0,0) -- (0,3) -| (4.6,6); 
		\draw [very thick, dashed] (0,0) -- (2.3,0) |- (4.6,6);
		\node at (2.3,0) [above right] {cr,r}; 
		\node at (2.3,6) [above] {cr,r};
		\draw[thick, green!70!brown,->] (0,3) -- (.3,3); 
		\draw [fill=green] (0,3) circle (.08); 
		\draw[thick, green!70!brown,->] (4.6,3) -- (4.3,3); 
		\draw [fill=green] (4.6,3) circle (.08); 
		\draw[thick,blue!60!white,->] (2.3,3) -- (2.3,2.7); \draw[thick,blue!60!white,->] (2.3,3) -- (2.3,3.3); \draw [fill=blue!60!white] (2.3,3) circle (.08); 
		\draw[gray!40!white,->] (2.2,4.5) -- (.8,4.5); \draw[gray!40!white,->] (2.2,1.5) -- (.8,1.5); \draw[gray!40!white,->] (4.5,4.5) -- (3.1,4.5);
	\end{tikzpicture}
	\quad
	\begin{tikzpicture}[scale=.95]
		\node at (3.3,7) {$L$ fixed, $\kappa$ decreases};
		\draw (0,0) rectangle (6.6,6);
		\draw (0,0) -- (-.1,0) node [left] {$0$};
		\draw (0,3) -- (-.1,3) node [left] {$\I \,\kappa$};
		\draw (0,6.1) |- (-.1,6) node [left] {$2\,\I \,\kappa$};
		\draw (0,0) -- (0,-.1) node [below] {$0$};
		\draw (2.3,0) -- (2.3,-.1) node [below] {\footnotesize$\pi(I_1{+}I_2)/L$};
		\draw (4.6,0) -- (4.6,-.1) node [below] {\footnotesize$2\pi(I_1{+}I_2)/L$};
		\draw (6.6,0) -- (6.6,-.1) node [below] {$2\pi$};
		\node at (2.3,3) [above right] {cr,i};
		\draw [very thick, dashed] (0,0) -- (0,3) -| (4.6,6); 
		\draw [very thick, dashed] (0,0) -- (2.3,0) |- (4.6,6);
		\node at (2.3,0) [above right] {cr,r}; 
		\node at (2.3,6) [above] {cr,r};
		\draw (0,3) circle (.08); 
		\draw (4.6,3) circle (.08); 
		\draw[thick,blue!60!white,->] (2.3,3) -- (.1,3); \draw[thick,blue!60!white,->] (2.3,3) -- (4.5,3); 
		\draw [fill=blue!60!white] (2.3,3) circle (.08); 
		\draw[thick, red,->] (2.3,1.4) -- (2.3,1.7); \draw [fill=red] (2.3,1.4) circle (.08); 
		\draw[thick, red,->] (2.3,4.6) -- (2.3,4.3); \draw [fill=red] (2.3,4.6) circle (.08); 
		\draw[thick,red!40!yellow,->] (1.5,0) -- (1.8,0); \draw [fill=red!20!yellow] (1.5,0) circle (.08);  
		\draw[thick,red!40!yellow,->] (3.1,6) -- (2.8,6); \draw [fill=red!20!yellow] (3.1,6) circle (.08); 
		\draw[gray!40!white,->] (1.15,2.9) -- (1.15,.8); \draw[gray!40!white,->] (3.45,5.9) -- (3.45,3.8); \draw[gray!40!white,->] (3.45,2.9) -- (3.45,.8);
	\end{tikzpicture}
	\caption{Cartoon of the different trajectories of bound-state quasimomenta in the fundamental domain, in a `rest frame' where we pretend that $L,\kappa$ are both fixed. In reality, some of the lines move as indicated by the gray arrows. The upper half of the fundamental rectangle is to be identified with its $2\I\kappa$-translate in the lower half plane (not shown). \textsc{Left.} When $\kappa$ is fixed we start for $L=2\,n$ (cf.~Footnote~\ref{fn:bound_state_starts_as_exceptional}) at $\Im p_m =\pm\kappa$: at the green circles if $\kappa \in \mathbb{R}_{>0}$ and at blue if $\kappa = \infty$ (cf.~Footnote~\ref{fn:exceptional_Heis}). As $L$ grows the $p_m$ move towards `cr,i' and then `cr,r'. For odd~$n$ the $p_m$ collide at `cr,r' and then move apart (cf.\ the inset of Figure~\ref{fig:Lcr_pm}), while for even~$n$ `cr,r' is only reached as $L\to \infty$, when everything (save $2\pi$) collapses to the imaginary axis; in either case $p_m \to 0 \, \mathrm{mod}\,2\I\kappa$. \textsc{Right.} When $L$ is fixed the $p_m$ start for $\kappa=\infty$ at the orange circles if both $n$ is odd and $L> L_\text{cr,r}^{(n)}(\kappa=\infty)$, at red else. As $\kappa$ decreases the $p_m$ collide at (possibly `cr,r' and then) `cr,i'; the open circles are reached as $\kappa \to 0$, when everything collapses to the real axis. In the exceptional case $n=L/2$ the $p_m$ start at blue, jumping to the open circles as soon as $\kappa<\infty$. }
	\label{fig:pm_cartoon}
\end{figure}

Altogether the two critical equations for Inozemtsev allow for five different scenarios for the corresponding quasimomenta for a given $\vect{I} = (\lfloor n/2\rfloor, \lceil n/2\rceil)$ with $n \geq 2$ like in \textsection\ref{sec:Heis_recap_M=2}. These correspond to the five qualitatively different horizontal and vertical slices possible in Figure~\ref{fig:crit_locus}.
\begin{itemize}
	\item Fix $\kappa \in (0,\infty)$ and increase $L$ (cf.~Footnote~\ref{fn:bound_state_starts_as_exceptional} on p.\,\pageref{fn:bound_state_starts_as_exceptional}).
	\begin{itemize}
		\item When $n$ is even we start with the exceptional solution $\vect{p} = (0+\I \kappa,\pi -\I \kappa)$ at $L=2n$. As $L$ increases $\Im p_{1,2} = \pm \I \kappa$ is constant while the real parts vary, with $\Re p_1$ ($\Re p_2$) increasing (decreasing) towards the trivial solution $\pi n/L + \I \kappa$. The $p_m$ collide ($\mathrm{mod}\,2\I \kappa$) at that point when (formally) $L=L_\text{cr,i}^{(n)}$.\,%
		\footnote{\ For example, at $\kappa =1$ the first few values are $L_\text{cr,i}^{(2)} \approx 6.9$, $L_\text{cr,i}^{(4)} \approx 13.7$, $L_\text{cr,i}^{(6)} \approx 21.6$, $L_\text{cr,i}^{(8)} \approx 27.5$, $L_\text{cr,i}^{(10)} \approx 34.4$.} 
		For $L > L_\text{cr,i}^{(n)}$ the quasimomenta are complex conjugate, with $\Re p_m =\pi n/L$ ($\Re \varphi = 0$) while $\Im p_1 = -\!\Im p_2 \in (0,\kappa)$ decreases as $L$ grows. The $p_m$ hit the origin as $L\to \infty$.
		\item If $n$ is odd we start out as before, but this time there is a second critical length, $L_\text{cr,r}^{(n)} > L_\text{cr,i}^{(n)}$,\,%
		\footnote{\ At $\kappa =1$ the first few values are $L_\text{cr,i}^{(3)} \approx 10.3$, $L_\text{cr,i}^{(5)} \approx 17.2$, $L_\text{cr,i}^{(7)} \approx 24.1$, $L_\text{cr,i}^{(9)} \approx 30.9$ versus $L_\text{cr,r}^{(3)} \approx 23.2$, $L_\text{cr,r}^{(5)} \approx 62.9$, $L_\text{cr,r}^{(7)} \approx 122.1$, $L_\text{cr,r}^{(9)} \approx 201.1$. Compare the latter with the values for $\kappa=\infty$ given below \eqref{eq:L_crit}.}
		for which the quasimomenta collide on the real axis at the trivial solution $p_m = \pi n/L_\text{cr,r}^{(n)}$. When $L$ increases $>L_\text{cr,r}^{(n)}$ the $p_m$ stay real and move apart. This end is as for Heisenberg (\textsection\ref{sec:Heis_recap_M=2}, Figure~\ref{fig:Lcr_pm}). 
	\end{itemize}
	\item Fix $L \geq 2n$ and vary $\kappa \in [0,\infty]$ to interpolate from Heisenberg to Haldane--Shastry.
	\begin{itemize}
		\item If $L$ is even $n=L/2$ is the exceptional case. (See also Footnotes~\ref{fn:exceptional_Heis} and \ref{fn:exceptional_Ino}.) For $\kappa = \infty$ one may take $p_{1,2} = \pi/2 \pm \I\infty$; note that $\pi/2 = n\,\pi/L$. As soon as we lower $\kappa$ the exceptional solution jumps to $p_1 = 0+\I \kappa, p_2= \pi - \I \kappa$. This change in real part is perhaps not too surprising from the observation that $\lim_{\kappa\to\infty} L_\text{cr,i}^{(n)} = 2n$ so $\kappa_\text{cr,i}^{(n)} = \infty$ for $L=2n$. The $p_m$ approach the real axis as $\kappa$ decreases. (Even if $n$ is odd $L = 2n$ is below $\min_\kappa L_\text{cr,r}^{(n)}(\kappa) = L_\text{cr,r}^{(n)}(\kappa = \infty) \sim (\pi n/2)^2$ so there is no $\kappa_\text{cr,r}^{(n)}$.)
		\item The typical scenario is as follows. We start for Heisenberg with a bound state at $\Re p_m = n\pi/L$ and finite $\Im p_1 = -\!\Im p_2$. As $\kappa$ is lowered the quasimomenta move away from the real axis, with $\Re p_m = n\pi/L$ (i.e.\ $\varphi = \pi \,(I_2-I_1)$) and increasing $\Im p_1 = -\!\Im p_2 \, (<\kappa)$. At $\kappa = \kappa_\text{cr,i}^{(n)}$ the quasimomenta collide ($\mathrm{mod}\,2\I\kappa$) at the trivial solution $p_m = \pi n/L +\I \kappa$. When $\kappa$ decreases further the quasimomenta stay at $\Im p_{1,2} = \pm \kappa$ while moving apart in the real direction.
		\item In case $n$ is odd and $L > \min_\kappa L_\text{cr,r}^{(n)}(\kappa) = L_\text{cr,r}^{(n)}(\kappa = \infty)$ we instead start with two real (`unbound') quasimomenta $p_1 < \pi n/L < p_2$ for Heisenberg (\textsection\ref{sec:Heis_recap_M=2}). As $\kappa$ is lowered the $p_m$ move towards each other, meeting at the trivial solution $p_m = \pi n/L$ for $\kappa = \kappa_\text{cr,r}^{(n)}$. When $\kappa$ diminishes further the $p_m$ move away from the real axis while keeping $\Re p_m = \pi n/L$. The rest is as in the typical scenario. 
	\end{itemize}
\end{itemize}
The trajectories for the different scenarios are illustrated in Figure~\ref{fig:pm_cartoon}. In \textsection\ref{sec:HS_limit} we will see that all three scenarios for fixed $L$ lead to the same result in the Haldane--Shastry limit.

Finally let us mention that, another, more heuristic way to derive the critical equation \eqref{eq:crit} is to consider the map $\vect{p} \mapsto \vect{I}(\vect{p})$ defined from \eqref{eq:M=2_BAE}, where we formally view $\vect{I} \in \mathbb{R}^M$. The derivative of this map\,---\,which for $L\to \infty$ is the density of states\,---\,has vanishing determinant
whenever a two-particle state trivialises. The equation that governs this vanishing is precisely \eqref{eq:crit}.

\subsubsection{Large-\textit{L} asymptotics} \label{sec:large_L} Before we proceed to Haldane--Shastry let us consider the limit that is the easiest at the level of the various functions, but perhaps least familiar: the asymptotic regime $L \to \infty$ in which the pair potential of the Hamiltonian becomes hyperbolic.

For the wave function we find $\sigma(z) \to \kappa^{-1}\, \E^{-\kappa^2\,z^2/6}\,\sinh(\kappa z)${\checkedMma} 
and $\eta_2/\omega \to -\kappa^2/3${\checkedMma} 
whence
\begin{equation} \label{eq:chi_2_L_limit}
	\chi_2(z,\gamma) \to \kappa\,\frac{\sinh[\kappa\,(z+\gamma)]}{\sinh(\kappa\, z)\,\sinh(\kappa\,\gamma)} = \kappa \, [\coth(\kappa\, z) + \coth(\kappa\, \gamma)] \, , \qquad L \to \infty \, .
\checkedMma
\end{equation}
In particular the ratio~\eqref{eq:A^e_tau} of the extended Bethe-ansatz coefficients becomes
\newcommand{\dn}{\delta n}
\begin{equation} \label{eq:A^e_tau_hyp}
A_{\tau}^e(\vect{p};\vect{n}) \to \frac{\sinh [\kappa\, (n_1 - n_2+\gamma)]}{\sinh [\kappa\,(n_1 - n_2-\gamma)]} \, , \qquad L\to \infty \, ,
\checkedMma
\end{equation}
reproducing Inozemtsev's $M=2$ solution for his hyperbolic ($L\to\infty$) spin chain \cite{Inozemtsev:1989yq}. At $L=\infty$ our definition of the \textit{S}-matrix does not make sense. Factorised scattering, i.e.\ the factorisation of the extended Bethe-ansatz coefficients for $M>2$,
instead suggests defining the \textit{S}-matrix as the limit of \eqref{eq:A^e_tau_hyp} as $n_1 - n_2 \to -\infty$ 
\cite{inozemtsev1992extended,klabbers2015inozemtsev}, which yields 
$S_\text{hyp}(\vect{p}) = \E^{-2\kappa \gamma}= \E^{\I \varphi}$ in line with \eqref{eq:S-matrix_def}. For a more explicit expression we turn to the constraint.

For the constraint~\eqref{eq:M=2_constraint} only the left-hand side depends on $L$. Using \eqref{eq:limits_rhobarcheck} its limit is
\begin{equation} \label{eq:M=2_constraint_hyp}
\cot (\varphi/2) = \mom{\rho}_1(p_1) - \mom{\rho}_1(p_2) \, ,
\end{equation}
which matches \cite{Inozemtsev:1989yq} after switching back to 
$\gamma = -\omega\,\varphi/2\pi = -\I \varphi/2\kappa$. In the limit we lose (total) momentum quantisation and the Bethe-ansatz equations~\eqref{eq:M=2_BAE}. 
For large but finite $L$ the latter become the asymptotic \textsc{bae} in terms of the hyperbolic \textit{S}-matrix, cf.\ \eqref{eq:S_mat_Heis},
\begin{equation}
\label{eq:S_hyp}
	S_\text{hyp}(p_1,p_2) 
	= \frac{\lambda_\text{u}(p_1) - \lambda_\text{u}(p_2) - \I}{\lambda_\text{u}(p_1) - \lambda_\text{u}(p_2) + \I} = \frac{\lambda_\text{n}(p_1) - \lambda_\text{n}(p_2) - \I\,  n_\lambda(\kappa)}{\lambda_\text{n}(p_1) - \lambda_\text{n}(p_2) + \I\, n_\lambda(\kappa)} = \E^{\I \mspace{1mu}\varphi} \, ,
\end{equation}
with $n_\lambda$ behaving as \eqref{eq:n_lambda}.

The dispersion relation~\eqref{eq:dispersion} was independent of $L$, so applies here as well. The potential energy from~\eqref{eq:M=2_energy} vanishes when $L\to \infty$, yielding a (functionally) additive expression for the two-particle energy.
We thus see that although the pair potential~\eqref{eq:limits_potential} and wave functions degenerate to hyperbolic functions, the dispersion, rapidity, and \textit{S}-matrix remember the elliptic origin of the model.

Finally we turn to the implications of the limit $L\to\infty$ for the discussion in \textsection\ref{sec:qualitative}--\ref{sec:crit}. 

We claim that a sequence of bound states with fixed $p_{\text{tot}}$ can only limit to a bound state if the $I_m$ and $\Im\varphi$ diverge as $L\to \infty$. To see this consider a sequence of bound-state solutions of the constraint~\eqref{eq:M=2_constraint} indexed by $L$ such that $\vect{p} = (2\pi I_1 + \varphi,2\pi I_2 - \varphi)/L$ remains finite as $L\to \infty$. Recall that for bound states $I_1$ and $I_2$ differ at most by one. If both the $I_m$ are fixed, or, more generally, remain finite as $L$ grows then the critical equation~\eqref{eq:crit} stays valid. Indeed, even in the latter case there must be some $L'<\infty$ such that the $I_m$ are constant for $L>L'$. This corresponds to the first bullet point from \textsection\ref{sec:crit} and does not yield a bound state. Next, if $\Im\varphi$ were to remain finite and the quasimomenta were to become real as $L\to \infty$, the limit would yield a scattering state with $\varphi$ and the $p_m$ related by \eqref{eq:M=2_constraint_hyp}. In order to end up with a bound state the sequence of bound-state solutions must therefore have $\Im \varphi \to \infty$. 

Letting $\Im \varphi \to \infty$ in the constraint~\eqref{eq:M=2_constraint} we recover the bound-state constraint derived directly from the hyperbolic extended Bethe ansatz~\eqref{eq:A^e_tau_hyp} by Dittrich and Inozemtsev~\cite{dittrich1997two}, see (15) therein:
\begin{equation}
{-\I} = \mom{\rho}_1(p_1) - \mom{\rho}_1(p_2)\, .
\end{equation}
Its solutions either have (\textsc{i})~$p_1^* = p_2$ or (\textsc{ii}) $p_1^* \neq p_2$.\footnote{\ See \cite{klabbers2015inozemtsev} for another perspective: one can understand the existence of these two types of bound states from the non-injectivity of $\mom{\rho}_1(z)$ ($=\phi(z)/(2\I \kappa)$ therein).} At least qualitatively, solutions of type~(\textsc{i}) should arise from a sequence of finite-size bound states for which the $I_m$ increase in such a way that $L_{\text{cr,i}}^{(n)} < L < L_{\text{cr,r}}^{(n)}$ eventually, whereas type~(\textsc{ii}) solutions should be limits of sequences with $I_m$ increasing such that $L < L_{\text{cr,i}}^{(n)}$ throughout. To see this, we study the large-$L$ asymptotics of the critical equations \eqref{eq:crit_i} and \eqref{eq:crit_r} at constant $p_\text{tot}$ (implying that $n = L \, p_\text{tot} /2\pi$ increases with $L$), which is given by
\begin{equation}
\text{cr, i}: \quad \mom{\rho}_1{}'(p_{\text{tot}}/2+\I \kappa) = 0, \qquad \text{cr, r}: \quad \mom{\rho}_1{}'(p_{\text{tot}}/2) \to \infty\, . 
\end{equation}
The former\,---\,which can be recognised in (19) in \cite{dittrich1997two} and is more explicitly given in (22) in \cite{klabbers2015inozemtsev}\,---\,has a unique solution $p_{\text{tot}}=p_{\text{cr}}\in [0,2\pi)$, whereas the latter is solved by $p_{\text{tot}}=0$. Keeping in mind the scenarios from the first bullet point in \textsection\ref{sec:crit} we can interpret these limits as follows. If $L_{\text{cr,i}}^{(n)}$ grows faster than $L$, in the limit $L\to \infty$ the resulting bound state has $p_{\text{tot}}> p_{\text{cr}}$ and is of type~(\textsc{ii}), with $\vect{p} = (p' + \I \kappa,p'' - \I \kappa)$ such that $\mom{\rho}_1(p'+\I \kappa) = \mom{\rho}_1(p''+\I \kappa)$. If, on the other hand $L$ surpasses $L_{\text{cr,i}}^{(n)}$ then the two momenta will move through the trivial solution $p_{\text{tot}}/2 +\I \kappa$ and become complex conjugate. In case $L_{\text{cr,r}}^{(n)}$ grows more rapidly than $L$ the resulting bound state is of type~(\textsc{i}) and has $0 < p_{\text{tot}} < p_{\text{cr}}$, while if $L$ also grows past $L_{\text{cr,r}}^{(n)}$ then the bound states unbinds to become a scattering state that eventually limits to $\vect{p} = 0$, resulting in a descendant of the vacuum. 

\subsection{Haldane--Shastry regime} \label{sec:HS_regime} In the Haldane--Shastry limit the imaginary period goes to zero, so the fundamental domain collapses and all quasimomenta will become real as $\kappa\to0$. This limit is more subtle than the Heisenberg limit. A notable exception is the case when we approach HS from the regime of asymptotically large~$L$~\cite{Serban2021}.
Indeed, consider the Inozemtsev spin chain with large but finite $L$ and write the $M$-particle \textsc{bae} in the asymptotic form $L \, p_m = 2\pi \, I'_m -\I \sum_{m'(\neq m)}^M \log S_\text{hyp}(p_m,p_{m'})$, where $I'_m$ are the `asymptotic Bethe integers'\,\footnote{\ Note that these are not just the $I_m$: in order to follow a given solution to the \textsc{bae} (at fixed $p_\text{tot}$) in a meaningful way as $L$ increases one has to increase the $I_m$ in the appropriate way, cf.~the end of \textsection\ref{sec:large_L}.} and the asymptotic \textit{S}-matrix was given in \eqref{eq:S_hyp}. For real quasimomenta $p_m \in [0,2\pi)$ the asymptotic scattering phase simplifies drastically in the HS limit:
\begin{equation} 
	\begin{aligned}
	-\I \log \frac{\lambda_{\text{n}}(p_m) - \lambda_{\text{n}}(p_{m'}) - \I \, n_\lambda(\kappa)}{\lambda_{\text{n}}(p_m) - \lambda_{\text{n}}(p_{m'}) + \I \, n_\lambda(\kappa) } \ \sim \ \, & {-\I} \log\frac{p_m - p_{m'} - 2\I\kappa}{p_m - p_{m'} +2\I\kappa} \\
	\ \to \ & {-\pi} + \pi\,\mathrm{sgn}(p_m- p_{m'}) \, , 
	\end{aligned} 
	\qquad \kappa \to 0 \, , \ L\gg 1 \, ,
\end{equation}
where we used \eqref{eq:limits_rhobarcheck} together with $\lim_{\kappa\to 0} \log[(x+2\I \kappa)/(x-2\I \kappa)] = \pi \, \I \,(1 - \mathrm{sgn}\,x)$ on the real line, as follows from a careful inspection of the branch cut.
Thus the HS limit of the asymptotic \textsc{bae} for Inozemtsev yields~\cite{Serban2021} Haldane's Bethe-ansatz-like equations \cite{Hal_91a}\,\footnote{\ To the best of our knowledge, Haldane's equation~\eqref{eq:BAE_like_HS} does not arise from requiring ($L$-periodicity for) a Bethe ansatz\,---\,the HS wave functions are manifestly $L$-periodic, see \eqref{eq:HS_wavefn_M=2}\,---\,but is of \textsc{bae} form and, by construction, has the desired solutions. Note that Haldane~\cite{Hal_91a} works with (distinct, chosen in ascending order) quantum numbers $I''_m \coloneqq I'_m + (M-1)/2 - (m-1) \in \{(M+1)/2,(M+1)/2 + 1, \dots, L-(M+1)/2\}$. Here the shift over $(M-1)/2$ absorbs the constant in~\eqref{eq:BAE_like_HS}.} 
\begin{equation} \label{eq:BAE_like_HS}
	L \, p_m = 2 \pi \, I'_m - (M-1)\,\pi + \pi \!\!\!\! \sum_{m'(\neq m)}^M \!\!\! \mathrm{sgn}(p_m- p_{m'}) \, , \quad I'_m \in \mathbb{Z}_L \, , \qquad 1\leq m\leq M \, .
\end{equation}
If we order $p_1 < \dots < p_M$ then this simple combinatorial equation dictates its solution:
\begin{equation} \label{eq:BAE_like_HS_sol}
	\frac{L \, p_m}{2\pi} = I'_m - M + m \eqqcolon \mu_m \, , \quad 1\leq \mu_m \leq L-1 \, , \qquad 1\leq m\leq M \, ,
\end{equation}
and we exclude $\mu_m = 0$ because we are considering highest-weight vectors (\textsection\ref{sec:hw}). For $I'_1 < \dots < I'_M$ \cite{Hal_91a} the quasimomenta are quantised exactly, subject to the `Pauli exclusion principle' (cf.~\cite{Hal_91b}) $\mu_{m+1} > \mu_m +1$. The quantity $\vect{\mu} = (L/2\pi)\,\vect{p}$, consisting of an $M$-tuple of non-consecutive integers $\vect{\mu}=(\mu_1,\cdots\mspace{-1mu},\mu_M) \subset \{1,\cdots\mspace{-1mu},L-1\}$, is known as a `motif'~\cite{HH+_92}. One remarkable fact about the HS spin chain is that the asymptotic \textsc{bae} solution~\eqref{eq:BAE_like_HS_sol} remains exact for \emph{finite} system size $L$. 

The goal of the remainder of this section is to analyse the HS limit for the two-particle sector of the Inozemtsev spin chain at finite size. Since the exact eigenvectors in the two-particle sector of the Haldane--Shastry spin chain might not be as familiar we start by reviewing the latter.

\begin{figure}[h]
	\centering
	\includegraphics[width=\textwidth]{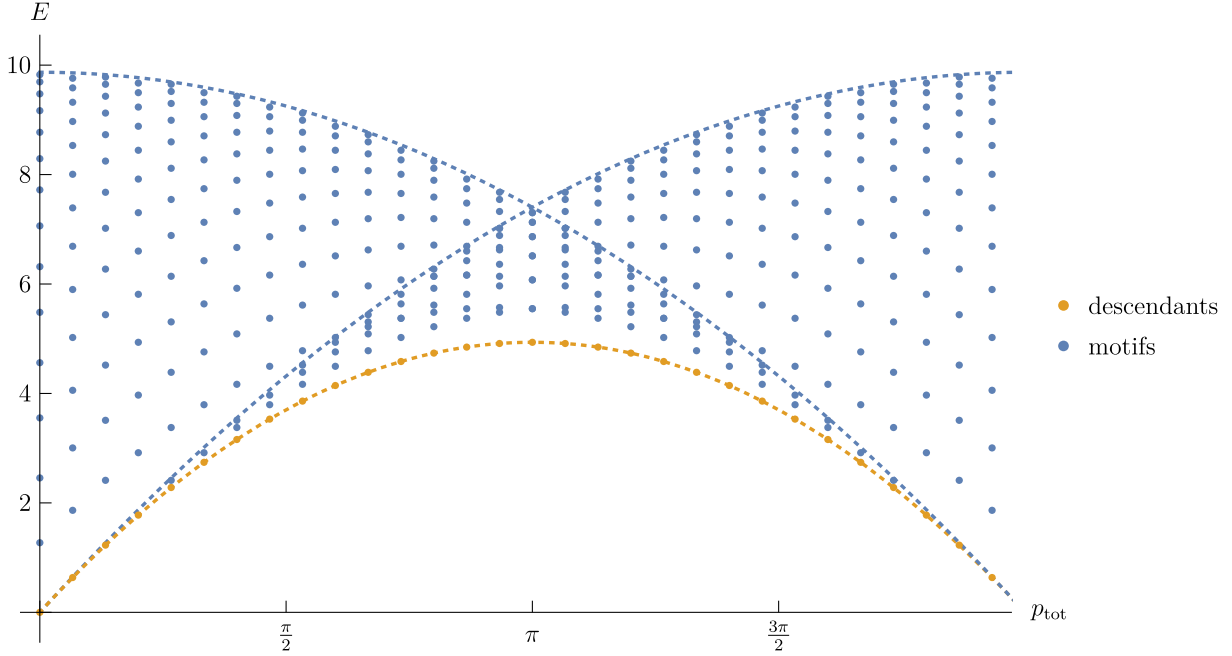}
	\caption{The two-particle spectrum of the Haldane--Shastry spin chain with $L=30$, cf.~Figures~\ref{fig:Ep_M2_Heis} and \ref{fig:Ep_M2_Ino}. Each dot represents an eigenspace. The (Yangian) descendants lie on the dispersion $\varepsilon_\textsc{hs}(p_\text{tot})$ (dotted curve), shown in orange. The rest are highest-weight vectors, shown in blue. Their energy is equal to that of two free magnons, with energy an integer multiple of $2\pi^2/L^2$, resulting in a particularly regular plot. The energy is bounded from below by that of two identical magnons (dotted curve, two-valued as $p_\text{tot} = 2\,p \,\mathrm{mod} \, 2\pi$, $E(p,p) = 2\,\varepsilon_\textsc{hs}(p)$). The occurring energy values can be labelled by motifs.}
	\label{fig:Ep_M2_HS}
\end{figure}

\subsubsection{Recap Haldane--Shastry spectrum at $M=2$} \label{sec:HS_recap_M=2} For the Haldane--Shastry model isotropy is enhanced to Yangian symmetry~\cite{HH+_92,bernard1993yang}. First we consider the (Yangian) highest-weight vectors, which are labelled by the motifs~\cite{HH+_92} that we encountered above. For $M=2$ a \emph{motif} is just a pair $1\leq \mu_1 , \mu_2 \leq L-1$ with $\mu_2 \geq \mu_1 +2$. Therefore $\nu(\vect{\mu}) = (\nu_1,\nu_2) \coloneqq (\mu_2 - 2,\mu_1)$ is a partition of length two, $\nu_1\geq \nu_2 \geq 1$. The corresponding highest-weight wave function is known completely explicitly:
\begin{equation} \label{eq:HS_wavefn_M=2}
\Psi^\textsc{hs}_{\vect{\mu}}(n_1,n_2) \coloneqq \cbraket{n_1,n_2}{\vect{\mu}} = \mathrm{ev}\Bigl[ (z_{n_1}-z_{n_2})^2 \, P_{\nu(\vect{\mu})}(z_{n_1},z_{n_2}) \Bigr] \, , \qquad  \mathrm{ev} \colon z_j \mapsto \E^{2\pi\I j/L} \, .
\end{equation}
There are no free parameters. Let us unravel this expression. The \emph{evaluation} sends the `coordinate'~$z_j$ to the $j$th power of the primitive $L$th root of unity, which we can view as the position of the $j$th site of the chain $\mathbb{Z}_L \subset S^1 \subset \mathbb{C}$. The wave function is the evaluation of a symmetric polynomial in two variables: the square of the Vandermonde factor times a Jack polynomial $P_\nu \coloneqq P^{(1/2)}_\nu$ (in the monic `P-normalisation') with parameter $\alpha = \beta^{-1} = 1/2$, i.e.\ a zonal spherical polynomial. These symmetric polynomials have been studied extensively~\cite{Sta_89,Mac_95}. The condition $\nu_2 \geq 1$, which ensures that $\nu$ has degree at least one in each variable in view of
\begin{equation}\label{eq:Jack_shift}
P_{(\nu_1,\nu_2)}(z_1,z_2) = (z_1\,z_2)^{\nu_2} \, P_{(\nu_1-\nu_2,0)}(z_1,z_2) \, , 
\end{equation}
is a Yangian highest-weight condition \cite{bernard1995one,LPS_20u}. Jack polynomials in two variables admit a simple explicit expansion in terms of Schur polynomials:
\begin{equation} \label{eq:Jack_M=2}
P_{(I,0)}(z_1,z_2) = \! \sum_{\substack{\lambda \, \vdash \mspace{1mu} I \\ \ell(\lambda)\leq 2}} \!\!\! \frac{\lambda_1 - \lambda_2 +1}{I+1} \, s_\lambda(z_1,z_2) = \sum_{i=0}^{\lfloor I/2 \rfloor} \frac{I-2\,i+1}{I+1} \, s_{(I-i,i)}(z_1,z_2) \, ,
\end{equation}
where the first sum ranges over all partitions $\lambda$ of $I$ of length at most two. In  turn, Schur polynomials are defined as \cite{Mac_95}
\begin{equation} \label{eq:Schur_M=2}
s_\lambda(z_1,z_2) \coloneqq 
 \frac{z_1^{\lambda_1 +1} \, z_2^{\lambda_2} - z_1^{\lambda_2}\,z_2^{\lambda_1 +1}}{z_1 - z_2} 
= \sum_{k=0}^{\lambda_1 - \lambda_2} \!\! z_1^{\lambda_1 - k} \, z_2^{\lambda_2 + k} \,.
\end{equation}
We find that \eqref{eq:Jack_M=2} becomes
\begin{equation*}
P_{(I,0)}(z_1,z_2) = \sum_{k=0}^{I} \frac{(I+1-k)(k+1)}{I+1} \, z_1^{I-k} \, z_2^k \, ,
\end{equation*}
and, remarkably,
\begin{equation*}
(z_1-z_2)^2 \, P_{(I,0)}(z_1,z_2) 
= s_{(I+2,0)}(z_1,z_2) - \frac{I+3}{I+1} \, s_{(I+1,1)}(z_1,z_2) \, .
\end{equation*}
Thus the two-particle highest-weight wave function~\eqref{eq:HS_wavefn_M=2} is the evaluation of
\begin{equation} \label{eq:HS_wavefn_M=2_rewrite}
\begin{aligned}
(z_{n_1}-z_{n_2})^2 \, P_{\nu(\mu)}(z_{n_1},z_{n_2}) & = s_{(\mu_2,\mu_1)}(z_{n_1},z_{n_2}) - \frac{\mu_2-\mu_1+1}{\mu_2-\mu_1-1} \, s_{(\mu_2-1,\mu_1+1)}(z_{n_1},z_{n_2}) \, .
\end{aligned}
\checkedMma
\end{equation}
The motif condition guarantees that $\mu_2-\mu_1-1>0$ is nonzero, and $\mu_2 - 1\geq \mu_1+1$. The corresponding energy is
\begin{equation} \label{eq:M=2_energy_HS}
E_{\vect{\mu}}^\textsc{hs} = 
\frac{2\,\pi^2}{L^2} \, \bigl( \mu_1 \, (L - \mu_1) + \mu_2 \, (L - \mu_2) \bigr) \, .
\checkedMma
\end{equation}
Note that the HS energy is `strictly additive'. Indeed, \eqref{eq:M=2_energy_HS} is a sum like in \eqref{eq:M=2_energy_Heis} where each term is of the form $\varepsilon_\textsc{hs}(p) = p\,(2\pi - p)/2 = (2\,\pi^2/L^2) \, I \, (L - I)$ as in \eqref{eq:dispersion_limits}. Moreover, the `quasimomenta' $2\pi \, \mu_m/L$ take the \emph{same} values as for free magnons: if we think of $\mu_m$ as playing the role of the $I_m$ in the \textsc{bae}~\eqref{eq:M=2_BAE} then $\varphi =0$, i.e.\ there are only `1-strings'. That is, for HS magnons interact solely through their (exclusion) statistics, $\mu_{m+1} > \mu_1 + 1$. 

Next we turn to the descendants. Recall from \textsection\ref{sec:M=1} that the one-particle eigenvectors are fixed by translational invariance. In terms of the notation from \eqref{eq:HS_wavefn_M=2} these magnons can be written as $\sum_n \mathrm{ev}[z_n^I] \, \cket{n}$, where we recognise $z_n^I = s_{(I)}(z_n) = P_{(I)}(z_n)$, and we may pick representatives for $I \in \mathbb{Z}_L$ in $\{0,\cdots\mspace{-1mu},L-1\}$ by virtue of the evaluation. (In terms of motifs: a one-particle motif is just $\mu=(I)$ with $1\leq I \leq L-1$, while the empty motif $\mu = 0$ characterises the ferromagnetic pseudovacuum.)
The lowering operator for $\mathfrak{sl}_2 \subset Y(\mathfrak{sl}_2)$ just adds a variable: \eqref{eq:magnon_desc} can be rewritten as
\begin{subequations} \label{eq:magnon_desc_via_ev}
\begin{gather} 
\cbra{n_1,n_2}\,S^-\sum_{n=1}^L \mathrm{ev}\bigl[z_n^I\bigr] \, \cket{n} = \mathrm{ev}\bigl[ P_{(I)}(z_{n_1},z_{n_2}) \bigr] \, ,
\checkedMma
\shortintertext{where}
P_{(I)}(z_1,z_2) = s_{(I)}(z_1,z_2) = z_1^I + z_2^I = s_{(I,0)}(z_1,z_2) - s_{(I-1,1)}(z_1,z_2) \, .
\checkedMma
\end{gather}
\end{subequations}

In view of its Yangian symmetry the HS spin chain has additional descendants. The `affine lowering operator' is $Q^- 
= \I \sum_{k<l}^L \mathrm{ev}\, \frac{z_k + z_l}{z_k - z_l} \, (\sigma^-_k \, \sigma^z_l - \sigma^z_k \, \sigma^-_l)$ \cite{HH+_92}. For magnons we find that the corresponding affine descendant has wave function\,%
\footnote{\,The first equality hinges on the evaluation: split the sum on the right-hand side of the first line into two, one sum for each summand, and use $\sum_{k(\neq n)}^L \mathrm{ev}[ (z_n + z_k)/(z_n - z_k)] = 0$, valid for any $n \in \{1,\dots\mspace{-1mu},L\}$.}
\begin{equation} \label{eq:magnon_desc_Yangian}
\begin{aligned}
\cbra{n_1,n_2}\,Q^-\sum_{n=1}^L \mathrm{ev}\bigl[z_n^I\bigr] \, \cket{n} 
& = 2 \,\I \!\!\!\!\! \sum_{k(\neq n_1,n_2)}^L \!\!\!\!\! \mathrm{ev}\biggl[ z_{n_1}^I \, \frac{z_{n_2} + z_k}{z_{n_2} - z_k} +  \frac{z_{n_1} + z_k}{z_{n_1} - z_k} \, z_{n_2}^I \biggr] \\
& = 2\,\I \ \, \mathrm{ev}\Bigl[ s_{(I,0)}(z_{n_1},z_{n_2}) + s_{(I-1,1)}(z_{n_1},z_{n_2}) \Bigr] \, , \qquad n_1 < n_2 \, .
\end{aligned}
\checkedMma
\end{equation}
Note that the final polynomial is \emph{not} a spherical zonal polynomial, which are after all determined by \eqref{eq:Jack_shift}--\eqref{eq:Jack_M=2}. For $I=0$ (i.e.\ the $p=0$ magnon $S^- \, \ket{\uparrow\cdots\uparrow}$) the sum~\eqref{eq:magnon_desc_Yangian} vanishes. When $I \in \{1,L-1\}$ the function~\eqref{eq:magnon_desc_Yangian} equals $2\,\I$~times \eqref{eq:magnon_desc_via_ev}. 
In the remaining cases, $2\leq I \leq L-2$,
the two Yangian descendants are linearly independent. There is one subtlety, however. The affine lowering operator $Q^-$ does \emph{not} commute with $S^+$, so \eqref{eq:magnon_desc_Yangian} is not guaranteed to have $\mathfrak{sl}_2$ highest weight. We find that, up to normalisation, the unique linear combination of the ordinary and affine descendant that is annihilated by $S^+$ is $\bigl( Q^- + 2\mspace{1mu}\I \, \frac{L-2\,I}{L-2} \, S^- \bigr) \sum_{n=1}^L \mathrm{ev}\bigl[z_n^I\bigr] \, \cket{n}$. By \eqref{eq:magnon_desc_via_ev}--\eqref{eq:magnon_desc_Yangian} its wave function can be written as 
\begin{equation} \label{eq:magnon_desc_Yangian_sl2_hw}
2\,\I\, \frac{2}{L-2} \, \mathrm{ev}\Bigl[(I-1) \, s_{(I,0)}(z_{n_1},z_{n_2}) + (L-1 - I) \, s_{(I-1,1)}(z_{n_1},z_{n_2}) \Bigr] \, .
\end{equation}
In particular, when $L$ is even the case $I=L/2$ yields the particularly simple wave function
\begin{equation} \label{eq:magnon_desc_Yangian_excep}
	2\,\I\, \mathrm{ev}\Bigl[(z_{n_1} + z_{n_2}) \, s_{(I-1,0)}(z_{n_1},z_{n_2}) \Bigr] \, .
\end{equation}
The $\mathfrak{sl}_2$ highest-weight property only holds upon evaluation.

All in all the two-particle spectrum of Haldane--Shastry consists of $L$ $\mathfrak{sl}_2$-descendants, $\binom{L-2}{2}$ Yangian highest-weight vectors, plus $L-3$ affine descendants (or their versions with highest weight for $\mathfrak{sl}_2$). The astute reader will have noticed that this breakdown of the $\binom{L-1}{2}$ two-particle eigenvectors parallels that for Heisenberg (\textsection\ref{sec:Heis_recap_M=2}): as we will show now, this is no coincidence.

\subsubsection{Haldane--Shastry limit} \label{sec:HS_limit} It remains to understand the limit of the two-particle spectrum of the Inozemtsev spin chain in which $\kappa\to 0$, so that the imaginary period $\omega \to \I\,\infty$ disappears. Of course the $\mathfrak{sl}_2$ descendants are independent of $\kappa$, like the magnons from $M=1$ that they descend from. For the $\mathfrak{sl}_2$  highest-weight vectors, however, this is the trickiest limit, in part because it is not a priori clear how the $\mathfrak{sl}_2$ highest-weight wave functions \eqref{eq:HS_wavefn_M=2_rewrite} and \eqref{eq:magnon_desc_Yangian_sl2_hw} are related to the extended \textsc{cba}~\eqref{eq:CBA-like_M=2}. To compare the two in more detail consider the two-particle Inozemtsev wave function~\eqref{eq:M=2_wave_fn}. In the trigonometric limit the Weierstrass sigma function degenerates to $\sigma(z) \to (L/\pi)\,\exp(\pi^2\,z^2/6L^2) \, \sin(\pi\,z/L)${\checkedMma},
whilst $\eta_2/\omega \to \pi^2/3L^2${\checkedMma}.
Thus the coefficients in \eqref{eq:M=2_wave_fn} limit to [cf.~\eqref{eq:chi_2_L_limit}]
\begin{equation} \label{eq:chi_2_HS_limit}
	\begin{aligned}
		\chi_2(n_1-n_2,\gamma) &\approx {} \frac{\pi}{L} \,  \E^{2(n_1 - n_2) \kappa \gamma/L } \, \biggl( \I \, \cot\frac{\pi(n_1-n_2)}{L} + \cot\frac{\pi\gamma}{L}  \biggr) \quad\qquad \\ 
		&\!\hspace{-1.2pt}\!\stackrel{\kappa\to0}{=} \frac{\pi}{L} \, \biggl( \I \, \cot\frac{\pi(n_1-n_2)}{L} + \cot\frac{\pi\gamma}{L} \biggr)
		 = \frac{\pi}{L} \, \mathrm{ev}\biggl[ \I \, \frac{z_{n_1}+z_{n_2}}{z_{n_1}-z_{n_2}}+\cot\frac{\pi\gamma}{L} \biggr] \, ,
	\end{aligned}	
\end{equation}
where the first line includes the first-order correction to the strict $\kappa \to0$ limit. The first expression on the second line is close to that in \eqref{eq:chi_2_heis_limit} when we switch periods, so $\pi/L \leftrightarrow \pi/\omega = -\I \, \kappa$. The plane waves in \eqref{eq:M=2_wave_fn} can be rewritten via $\E^{\I\,p\,n_m} = \mathrm{ev} [z_m^{L\,p/2\pi}]$, with exponent  $L\,p_{1,2}/2\pi = I_{1,2} \pm \varphi/2\pi$ by \eqref{eq:M=2_BAE}.

To make contact with the HS wave functions we need to solve the constraint~\eqref{eq:M=2_constraint} for the scattering phase $\varphi = 2\I \kappa \gamma$ in the limit $\kappa\to0$. In order to get a polynomial in the $z_{n_m}$ the limiting value of $\varphi$ better be an integer multiple of $2\pi$. There are two possible scenarios: either $\gamma$ has a finite limit, in which case $\varphi \propto \kappa \,\gamma$ tends to zero, or $\gamma$ diverges in such a way that $\varphi$ can be nonzero while $\cot(\pi\gamma/L)$ in \eqref{eq:chi_2_HS_limit} has a well-defined limiting value. Let us for a moment assume that these limits exist and, given some choice of Bethe integers~$\vect{I}$, write
\begin{equation} \label{eq:def_j_xi}
	j \coloneqq \lim_{\kappa \to 0} \frac{\varphi}{2\pi} \, , \qquad \xi \coloneqq {-\I}\, \lim_{\kappa\to0} \cot\frac{\pi\gamma}{L}
\end{equation}
for the asymptotic solution to the corresponding Bethe-ansatz system~\eqref{eq:M=2_constraint}--\eqref{eq:M=2_BAE} as $\kappa\to 0$. By \eqref{eq:chi_2_HS_limit} the limiting wave function is $\I \pi/L$ times the evaluation of
\begin{equation} \label{eq:M=2_wavefn_HS_limit_pre}
	\begin{aligned}
	& \biggl( \frac{z_{n_1}+z_{n_2}}{z_{n_1}-z_{n_2}} + \xi \biggr) \, z_{n_1}^{I_1 + j} \, z_{n_2}^{I_2 - j} \, + \, (n_1 \leftrightarrow n_2) \\
	& = (\xi+1) \, \frac{z_{n_1}^{I_1 +j+1} \, z_{n_2}^{I_2-j} \, - \, (n_1 \leftrightarrow n_2)}{z_{n_1}-z_{n_2}} - (\xi-1)\, \frac{z_{n_1}^{I_1 + j} \, z_{n_2}^{I_2-j+1} \, - \, (n_1 \leftrightarrow n_2) }{z_{n_1}-z_{n_2}} \, ,
	\end{aligned}
\end{equation}
where we use the shorthand `$(n_1 \leftrightarrow n_2)$' for the same terms with $n_1$ and $n_2$ reversed. Provided $j$ is an integer in the range $-I_1 \leq j \leq I_2$ the divisions by $z_{n_1}-z_{n_2}$ result in Schur polynomials, cf.~\eqref{eq:Schur_M=2}, and is of the form we are looking for, cf.~\eqref{eq:HS_wavefn_M=2_rewrite} and \eqref{eq:magnon_desc_Yangian_sl2_hw}. Motivated by these observations we turn to the constraint and limits in \eqref{eq:def_j_xi}.

At leading order the constraint becomes piecewise constant: by \eqref{eq:limits_rhocheck}--\eqref{eq:limits_rhobarcheck} the normalised version of the function~\eqref{eq:f_constraint} limits to 
\begin{equation} \label{eq:M=2_constraint_HS_limit_naive}
	\begin{aligned}
	n_\lambda(\kappa) \, f_{I_\text{tot}}(p_1) \to {} & \frac{\pi-\modulo{L\,p_1\!}}{L} + \frac{\modulo{p_1\!} - \modulo{2\pi I_\text{tot}/L - p_1\!}}{2} 
	\\
	& = \frac{\pi}{L} \, \Bigl( 2 \, \lfloor L \, p_1/2\pi \rfloor - L \, \lfloor p_1/2\pi \rfloor + L \, \lfloor I_\text{tot}/L - p_1/2\pi \rfloor + 1 - I_\text{tot} \Bigr) \, , 
	\end{aligned}
\end{equation}
where $p_1$ has become real since the imaginary period $\kappa \to 0$, and the equality uses \eqref{eq:notation_mod2pi}. These are the plateaus shown in Figure~\ref{fig:constraint_real_kappa}. 
As the latter illustrates \eqref{eq:M=2_constraint_HS_limit_naive} is not sufficient to determine the limiting values of the $p_m$ (and therefore $j$): the zeroes of \eqref{eq:M=2_constraint_HS_limit_naive} are given by intervals around the trivial roots, while missing the roots that we are after.\,%
\footnote{\ Namely, let $x \coloneqq L\,p_1/2\pi$. We may take $p_1 \in [0,2\pi)$ so $x \in [0,L)$, and $\lfloor x/L\rfloor = 0$. There are two cases to consider. If $x\leq I_\text{tot}$ then $\lfloor (I_\text{tot}-x)/L\rfloor = 0$ and \eqref{eq:M=2_constraint_HS_limit_naive} vanishes if $\lfloor x\rfloor = (I_\text{tot} -1)/2$, i.e.\ $2\,x \in [I_\text{tot} - 1, I_\text{tot} + 1)$. Both $p_m$ then lie in the interval of length $2\pi/L$ around the first trivial root from~\eqref{eq:trivial_roots}. In case $x>I_\text{tot}$ (possible when $I_1+I_2>L$) instead $\lfloor (I_\text{tot}-x)/L\rfloor = -1$ and we obtain $\lfloor x\rfloor = (I_\text{tot} + L -1)/2$. This time both $p_m$ lie in the interval around the \emph{second} trivial root from~\eqref{eq:trivial_roots}. The latter case is shown in Figure~\ref{fig:constraint_real_kappa}.}
The roots for HS should rather be understood by continuity in $\kappa$, as limits of the roots when we let $\kappa \to 0$.

First we consider the case where $\gamma$ has a well-defined, finite limit as $\kappa\to 0$. By \eqref{eq:varphi_vs_gamma} the first limit in~\eqref{eq:def_j_xi} then is $j = 0$. Let us compute the second limit in~\eqref{eq:def_j_xi} from the constraint. Since $\gamma$ remains finite we should write the (normalised) constraint as
\begin{equation} \label{eq:constraint_for_HS_limit}
	-2 \, n_\lambda(\kappa)\,\momL{\rho}_1(2\pi \gamma/\omega) = n_\lambda(\kappa) \, \mom{\rho}_1(p_1) - n_\lambda(\kappa) \, \mom{\rho}_1(p_2) \, .
\end{equation} 
To extract the dependence on $\kappa$ in the left-hand side we pass to the coordinate lattice via~\eqref{eq:rho_mtm_lattice}. The coordinate-lattice version of the limit $\kappa\to 0$ from \eqref{eq:limits_rhocheck} is 
\begin{equation}
	\label{eq:rho_HS_lim_coord}
	\lim_{\kappa \to 0} \, n_\lambda(L \, \kappa) \, \frac{\omega}{2\pi} \, \rho_2(\gamma) 
	= -\frac{\pi\,\xi}{2} \, .
\end{equation}
The limit of the right-hand side of \eqref{eq:constraint_for_HS_limit} follows from \eqref{eq:limits_rhobarcheck} as usual. Thus \eqref{eq:constraint_for_HS_limit} tends to
\begin{equation*}
	\frac{\pi \,\xi}{L} = \frac{\pi - \modulo{p_1}}{2} - \frac{\pi - \modulo{p_2}}{2} \, , \qquad \text{whence} \qquad \xi = \frac{L}{2\pi}\,(p_2-p_1) = I_2 - I_1 \quad \text{for} \ j=0 \, .
\end{equation*}
Now let us define 
\begin{subequations}
	\begin{gather}
	\mu_m \coloneqq \lim_{\kappa\to 0} \frac{L\,p_m}{2\pi} = I_m \, .
\intertext{Then $\vect{\mu}$ is a motif provided $2\leq I_1 +1 < I_2 - 1\leq L-2$ corresponded to a scattering state. Numerics corroborates that $\gamma$ remains finite for such $I_m$. The identification of the motif in terms of $\vect{I}$ is supported by \eqref{eq:M=2_energy_HS}. Using \eqref{eq:M=2_wavefn_HS_limit_pre} we conclude that the on-shell Inozemtsev wave function~\eqref{eq:M=2_wave_fn} limits to a simple multiple of~\eqref{eq:HS_wavefn_M=2_rewrite}:}
	\Psi_{\vect{p}}(n_1,n_2) \to (\mu_2-\mu_1-1) \, \frac{\I\mspace{1mu} \pi}{L} \, \Psi^\textsc{hs}_{\vect{\mu}}(n_1,n_2) \, , \qquad \kappa \to 0 \, .
	\end{gather}
\end{subequations}
As a special case this includes (for $L$ even) the closed-form exceptional wave function~\eqref{eq:exceptional_wave_fn}, which limits to \eqref{eq:magnon_desc_Yangian_excep}. That is, \emph{the scattering states in the two-particle sector of the Inozemtsev spin chain limit to Yangian highest-weight vectors of the Haldane--Shastry spin chain}.

It remains to consider the bound states, labelled by (some choices of) $I_2 \in \{ I_1,I_1 + 1\}$ with $1\leq I_m \leq L-1$. We have not been able to determine the limiting value of $j \neq 0$ analytically, except for the exceptional case ($I_\text{tot} = L/2$) where the explicit solutions $p_1 = 0+\I \kappa$, $p_2 = \pi -\I \kappa$, valid for all $\kappa \in \mathbb{R}_{>0}$, limit to $p_1 \to 0$ and $p_2 \to \pi$, so that $j=-I_1$. From numerics we observe that $j \in \{-I_1,I_2\}$ for all bound states, whence $L\,\vect{p}/2\pi \to (0,I_\text{tot}) \equiv (I_\text{tot},L)$. The above analysis now has to be adjusted as one of the arguments on the right-hand side of \eqref{eq:constraint_for_HS_limit} vanishes, which is a pole of $\mom{\rho}_1$. For definiteness consider $j = -I_1$, so that $p_1 < p_2$ as usual. Since $p_1\propto \kappa$ we should pass to the coordinate lattice to evaluate the limit of the first term on the right-hand side of \eqref{eq:constraint_for_HS_limit} too. By \eqref{eq:rho_HS_lim_coord} we have $n_\lambda(\kappa) \, \mom{\rho}_1(p_1) \to  -\pi \,\xi/2$ as $\kappa \to 0$. Therefore \eqref{eq:constraint_for_HS_limit} now tends to
\begin{equation*}
\frac{\pi \,\xi}{L} = \frac{\pi \,\xi}{2} - \frac{\pi - p_2}{2} \, , \qquad \text{whence} \qquad \xi = \frac{ L-2\, I_{\text{tot}}}{L-2} \quad \text{for} \ j= -I_1 \, .
\end{equation*}
The limit of $\Psi_{\vect{p}}(n_1,n_2)$ again follows from \eqref{eq:M=2_wavefn_HS_limit_pre}, yielding $-\pi/2L$ times \eqref{eq:magnon_desc_Yangian_sl2_hw} with $I = I_\text{tot}$. We conclude that \emph{the two-particle bound states of the Inozemtsev spin chain limit to (the $\mathfrak{sl}_2$ highest-weight versions of) the affine descendants of the magnons (with $2\leq I \leq L-2$) of the Haldane--Shastry spin chain}. This is true for all bound states, despite the different scenarios for the trajectories of the $p_m$ as a function of $\kappa$ described at the end of \textsection\ref{sec:crit}.

\subsubsection{Potential energy and additivity} \label{sec:potential}
The potential energy vanishes in the Haldane--Shastry limit. It may come as a surprise that the way in which this happens is somewhat subtle. 
The potential energy \emph{function}~\eqref{eq:M=2_energy} does \emph{not} vanish in the limit. Indeed, $\mom{\rho}_1{}'$ disappears exponentially fast for $\kappa \to 0$, mirroring the behaviour of the coordinate-lattice function $\rho$ in the Heisenberg limit from \textsection\ref{sec:dispersion_limit}. Therefore 
\begin{equation*}
	U_\text{n} = 4\, n_H(\kappa) \, \kappa^2 \, \momL{F}_1(\varphi) 
	\approx 4 \, \biggl( 0 + \Bigl(\! -n_\lambda(\kappa)\momL{\rho}_1(\varphi)\Bigr)^{\!2} + n_\lambda(\kappa)^2 \, \frac{3\,\momL{\eta}_1}{2\pi}  \biggr) \, .
\end{equation*}
With the help of the limits \eqref{eq:limits_rhocheck} and \eqref{eq:limits_etabarcheck} we find
\begin{equation*}
	U_\text{n} \to \frac{4}{L^2} \, \biggl( \frac{(\modulo{\varphi}-\pi)^2}{4} - \frac{\pi^2}{4} \biggr) = -\frac{1}{L^2} \, \modulo{\varphi}\,(2\pi -\modulo{\varphi}) \, , \qquad \kappa\to 0 \, ,
\end{equation*}
where we use the notation \eqref{eq:notation_mod2pi}.
Happily, we are saved by the constraint. In \textsection\ref{sec:HS_limit} we have seen that $\varphi \in 2\pi \mathbb{Z}$ in the Haldane--Shastry limit, so the potential energy is zero \emph{on shell}, i.e.\ for $\varphi,p_m$ solving the Bethe-ansatz system. This is how the (strictly) additive energies~\eqref{eq:M=2_energy_HS} of the Haldane--Shastry spin chain arise as a limit of the not even functionally additive energy~\eqref{eq:M=2_energy} Inozemtsev spin chain.

\subsection{Rationalisation and completeness} \label{sec:completeness}
To conclude our discussion of the two-particle sector we show that the extended coordinate Bethe ansatz is complete for $M=2$, elaborating on \cite{inozemtsev1993Hermitelike}. To this end we will `rationalise' the Bethe-ansatz system~\eqref{eq:M=2_constraint}--\eqref{eq:M=2_BAE}, turning it into a polynomial equation for which we can count the number of solutions. This polynomial form is furthermore of great practical benefit, allowing for much more efficient numerical solutions.

We will use the \textsc{bae}~\eqref{eq:M=2_BAE} to express the $p_m$ in terms of $\varphi$. By \eqref{eq:M=2_BAE} the latter naturally takes values in the torus $\mathbb{C}/L\, \!\!\bar{\,\,\mathbb{L}}^{\!\vee}$. To accommodate for this, we will use
\newcommand{\momphi}[1]{\hat{#1}^{\vee}}
\begin{equation}
\label{eq:momphi_lattice}
\!\!\hat{\,\,\mathbb{L}}^{\mspace{-4mu}\vee} \coloneqq L\, \!\!\bar{\,\,\mathbb{L}}^{\!\vee} = \momphi{\omega}_1 \, \mathbb{Z} \oplus \momphi{\omega}_2 \, \mathbb{Z} \, , \qquad (\momphi{\omega}_1,\momphi{\omega}_2) \coloneqq (2\pi L, - 2\pi L/\omega) = (2\pi L, 2\,\I\,L\, \kappa)\, .
\end{equation} 
from \textsection\ref{sec:elliptic_curve_M=2} onwards. As the notation shows, \eqref{eq:momphi_lattice} is the reciprocal (momentum version) of another lattice, $\hat{\mathbb{L}\!\!}$\,\,, which was used by Inozemtsev~\cite{inozemtsev1993Hermitelike} and appears in \textsection\ref{app:ts}--\ref{app:trivialsols}.

\subsubsection{Highest-weight property}
\label{sec:hw}
First of all we show that solutions for which both quasimomenta are nonzero correspond to $\mathfrak{sl}_2$ highest-weight vectors. Our task is to demonstrate that if $p_{1,2} \neq 0$ then the wave function of the descendant,
\begin{equation*}
\cbra{n}\,S^+ \!\! \sum_{n_1<n_2}^L \!\! \Psi(n_1,n_2) \,\cket{n_1,n_2} = \sum_{n'(<n)} \!\! \Psi(n',n) + \!\sum_{n'(>n)}^L \!\! \Psi(n,n') = \sum_{n'(\neq n)}^L \!\! \Psi(n,n') \, ,
\end{equation*}
vanishes. The second equality uses that, without loss of generality, $\Psi$ inherits the (formal) symmetry of the coordinate basis.  We adapt the proof for the Heisenberg limit given in \cite[\textsection1.1.3]{Gau83}, see also below. For Inozemtsev's wave function~\eqref{eq:M=2_wave_fn} the problem is to calculate
\begin{equation} \label{eq:hw_sum_to_compute}
\begin{aligned}
\sum_{n'(\neq n)}^L \!\! \Psi_{\vect{p}}(n,n') 
& = \sum_{n'(\neq n)}^L \bigl( \E^{\I\, p_1\,n' + \I\, p_2 \,n} \, \chi_2(n'-n,\gamma) + \E^{\I\, p_1\,n + \I\, p_2 \,n'} \, \chi_2(n-n',\gamma) \bigr) \\ 
& = \E^{\I\, (p_1 + p_2)\,n} \sum_{n'=1}^{L-1} \bigl( \E^{\I\, p_1\,n'} \, \chi_2(n',\gamma) - \E^{\I\, p_2\,n'} \, \chi_2(n',-\gamma) \bigr) \, .
\end{aligned}
\end{equation}
In \textsection\ref{app:hw_sum} we use the \textsc{bae}~\eqref{eq:M=2_BAE} to show that
\begin{equation} \label{eq:hw_sum}
\sum_{n'=1}^{L-1} \! \E^{\I\, p\, n'} \chi_2(n',t) = -\rho_2(t)-\bar{\rho}_2\Bigl(\frac{\omega\,p}{2\pi}\Bigr) = -\frac{2\pi}{\omega}\biggl( \momL{\rho}_1\Bigl(\frac{2\pi\,t}{\omega}\Bigl) + \mom{\rho}_1(p) \biggr) \, , \qquad p \neq 0 \, .
{\checkedMma}
\end{equation}
As long as the $p_m \neq 0$ we can use this for both terms in \eqref{eq:hw_sum_to_compute}. By \eqref{eq:varphi_vs_gamma} the result is proportional to the constraint~\eqref{eq:M=2_constraint}:
\begin{equation*}
\sum_{n'(\neq n)}^L \!\! \Psi_{\vect{p}}(n,n') = \frac{2\pi}{\omega} \, \E^{\I\, (p_1 + p_2)\,n} \, \bigl( 2 \, \momL{\rho}_1(\varphi)-\mom{\rho}_1(p_1) + \mom{\rho}_1(p_2) \bigr) \, , \qquad p_m\neq 0 \, .
\end{equation*}
On shell, i.e.~upon the constraint, the two-particle eigenvectors therefore have highest weight.

It is instructive to consider the Heisenberg limit. Since the Bethe-ansatz coefficients become independent of the coordinates~$n,n'$, see \eqref{eq:chi_2_heis_limit}, in this case \eqref{eq:hw_sum} reduces to a simple geometric sum:
\begin{equation*}
	\begin{aligned}
	\lim_{\kappa\to\infty} \, \sum_{n'=1}^{L-1} \E^{\I\, p\, n'} \, \frac{\chi_2(n',\pm\gamma)}{\I \kappa} \, & = \pm\frac{\E^{\mp\I \,\varphi/2}}{\sin(\varphi/2)} \sum_{n'=1}^{L-1} \E^{\I\, p\, n'} \\
	& = \pm\frac{\E^{\mp\I \,\varphi/2}}{\sin(\varphi/2)} \, \frac{\E^{\I\, L \,p} - \E^{\I\, p}}{\E^{\I\, p} - 1} \, , \qquad p \neq 0 \, ,
	\end{aligned}
\end{equation*}
where $p=p_1$ ($p=p_2$) is associated with the upper (lower) sign. The Bethe-ansatz equations~\eqref{eq:periodicitygenM} further yield
\begin{equation*}
	\pm \frac{\E^{\mp\I \varphi/2}}{\sin(\varphi/2)} \, \frac{\E^{\pm \I \varphi} - \E^{\I p_m}}{\E^{\I p_m} - 1} = \pm\frac{\sin[(p_m \mp \varphi)/2]}{\sin(p_m/2)\sin(\varphi/2)} = \cot(p_m/2) \mp \cot(\varphi/2) \, ,
\end{equation*}
which is indeed the Heisenberg limit of the right-hand side of \eqref{eq:hw_sum}, cf.~\eqref{eq:limits_rhocheck}--\eqref{eq:limits_rhobarcheck}. If both quasimomenta are nonzero the result vanishes on solutions of the Heisenberg constraint~\eqref{eq:M=2_constraint_Heis}, reproducing the proof in \cite[\textsection1.1.3]{Gau83}. When instead $p_1=0$ we have $\varphi = 0$, so to regularise we multiply the wave function by $\sin(\varphi/2)$ to remove the denominator. Rather than a geometric progression the first sum then just gives $L-1$, while the geometric sum with $p = p_2$ yields $\cos(\varphi/2) = 1$. Including the overall exponential in \eqref{eq:hw_sum_to_compute} we thus obtain the nonzero result $L\,\E^{\I \,p \,n}$.
In particular this implies that the (regularised) result of \eqref{eq:hw_sum_to_compute} has to be nonzero when $p_1 = 0$ for generic values of $\kappa$. We conclude that, conversely, a two-particle vector of the Inozemtsev spin chain with $p_m = 0$ does \emph{not} have highest weight, in accordance with \textsection\ref{sec:symms_trivial_roots}.

\subsubsection{Onto the elliptic curve}\label{sec:elliptic_curve_M=2} We now wish to find a parametrisation such that the constraint becomes an algebraic equation. We will work in terms of the scattering phase, since that will generalise to general~$M$, as we will show in a future publication. Such an algebraic equation allows for a much more efficient way of solving the Bethe-ansatz system~\eqref{eq:M=2_BAE}--\eqref{eq:M=2_constraint} numerically, and will allow us to prove that the extended coordinate Bethe ansatz produces all highest-weight eigenvectors for $M=2$ (\textsection\ref{sec:completeness}). In \textsection\ref{sec:symms_trivial_roots} we saw that the constraint~\eqref{eq:f_constraint} only depends on the $I_m$ through the sum $I_\text{tot} = I_1+I_2 \, \mathrm{mod}\,L$. In terms of the scattering phase this can be seen by using the \textsc{bae}~\eqref{eq:M=2_BAE} and passing to the shifted phase $\varphi' \coloneqq \varphi - 2\pi \, I_2 $ to recast \eqref{eq:M=2_constraint} in the form 
\begin{equation} \label{eq:M=2varphi'}
2\,\momL{\rho}_1(\varphi') =\mom{\rho}_1\biggl( \frac{2\,\pi I_\text{tot} + \varphi'}{L}\biggr)+\mom{\rho}_1\biggl( \frac{\varphi'}{L}\biggr) \, ,
\end{equation}
where we used the $2\pi$-periodicity of $\momL{\rho}_1$. Since both $\chi_2$ and $\momL{F}_1$ are also $2\pi$-periodic in $\varphi$ this shift does not affect the wave function or energy. We will from now on drop the primes. 

We can rewrite \eqref{eq:M=2varphi'} using
\begin{equation}
\label{eq:rho_identity}
\momL{\rho}_1( z ) =  \sum_{n=0}^{L-1} \momphi{\rho}_1(z + 2\pi \, n )
\end{equation}
and the homogeneity property $\mom{\rho}_1(z) = L \, \momphi{\rho}_1(L\,z)$ to get
\begin{equation} \label{eq:M=2_constraints_rewrite_2}
2 \sum_{n=0}^{L-1} \momphi{\rho}_1 (\varphi + 2\pi \, n ) = L \, \momphi{\rho}_1 ( \varphi +2\pi \, I_\text{tot}) + L \, \momphi{\rho}_1 (\varphi) \, ,
\end{equation}
where we introduced the seemingly unnecessary signs in the arguments to make the final signs come out nicer. Use the addition rule \eqref{eq:doubling} with $u= \varphi + \pi\,I_\text{tot}$ and $v=2\pi\, n \, -\pi\,I_\text{tot}$ ($v= \mp \pi\,I_\text{tot}$) on the left-hand (right-hand) side of \eqref{eq:M=2_constraints_rewrite_2}, respectively. Since $\momphi{\wp}$ is even and $\momphi{\wp}{}'$ odd the result can be rearranged as
\begin{equation*}
\begin{aligned}
& 2 \, \sum_{n=0}^{L-1} \left( \momphi{\rho}_1(2\pi \, n -\pi\,I_\text{tot}) +\frac{1}{2} \, \frac{\momphi{\wp}{}'(\varphi + \pi\,I_\text{tot}) - \momphi{\wp}{}'(2\pi \, n-\pi\,I_\text{tot})}{\momphi{\wp}(\varphi+\pi\,I_\text{tot}) - \momphi{\wp}(2\pi \, n-\pi\,I_\text{tot})} \right) \\
& \qquad\qquad = \frac{L}{2} \, \frac{2 \, \momphi{\wp}{}'(\varphi + \pi\,I_\text{tot})}{\momphi{\wp}(\varphi+\pi\,I_\text{tot}) - \momphi{\wp}(\pi\,I_\text{tot})} \, .
\end{aligned}
\end{equation*}
For even $I_\text{tot}$ the singular terms at $n=I_\text{tot}/2$ cancel and are to be omitted from the sum. Moreover, the parity of $\rho_2$, $\momphi{\wp}$ and $\momphi{\wp}{}'$ together with their $2\pi L$-periodicity implies
\begin{equation*}
\sum_{\substack{n=0 \\ (\neq I_\text{tot}/2)}}^{L-1} \!\! \momphi{\rho}_1 (2\pi \, (n-I_\text{tot}/2)) = 0 \, , \qquad \sum_{\substack{n=0 \\ (\neq I_\text{tot}/2)}}^{L-1} \! \frac{\momphi{\wp}{}'(2\pi \, (n-I_\text{tot}/2))}{\momphi{\wp}(\varphi+\pi\,I_\text{tot}) - \momphi{\wp}(2\pi \, (n-I_\text{tot}/2))} = 0 \, .
\end{equation*}
The first of these identities leaves us with an expression that solely features the elliptic Weierstrass function and its first derivative. Introduce 
\begin{equation}
\label{eq:curve_coordinates}
\begin{aligned}
x_{\varphi} & \coloneqq \momphi{\wp}(\varphi + \pi\,I_\text{tot}) \, , \\
y_{\varphi} &\coloneqq \momphi{\wp}{}'(\varphi + \pi\,I_\text{tot}) \, , 
\end{aligned} 
\qquad\quad
\text{and}
\qquad\quad
\begin{aligned} 
x_n &\coloneqq \momphi{\wp}(2\pi \, (n-I_\text{tot}/2)) \, , \\ 
y_n &\coloneqq \momphi{\wp}{}'(2\pi \, (n-I_\text{tot}/2)) \, , 
\end{aligned} \qquad n \in \mathbb{Z}_L \, ,
\end{equation}
which parametrise our variable~$\varphi$ and a family of points on the elliptic curve $y^2 = 4\,x^3 - \momphi{g}_2 \, x -\momphi{g}_3$. Then we are left with
\begin{equation} \label{eq:ell_curve}
\sum_{\substack{n=0 \\ (\neq I_\text{tot}/2)}}^{L-1} \!\! \frac{y_{\varphi}}{x_{\varphi} - x_n} = \frac{L \, y_{\varphi}}{x_{\varphi} - x_0} \, . 
\end{equation}
This is an equation for $\varphi$ given a choice of $I_\text{tot}$. In fact, it holds for $I_\text{tot}=0$ as well when the right-hand side of \eqref{eq:ell_curve} is treated as zero since both denominators diverge there. Indeed, the preceding argument carries through provided at the intermediate stage one drops the two terms that for $I_\text{tot}=0$ diverge but cancel, and for $I_\text{tot}\neq 0$ give rise to the right-hand side of \eqref{eq:ell_curve}.

A neat property of this form of the constraint is that it allows us to factorise out the trivial roots listed in \eqref{eq:trivial_roots}. Indeed, $y_\varphi = 0$ has three solutions corresponding to the last three roots in \eqref{eq:trivial_roots}, whereas the pole on the right-hand side of \eqref{eq:ell_curve} corresponds to the first root. This useful feature is caused by the inclusion of a shift by $\pi\,I_\text{tot}$ in \eqref{eq:curve_coordinates} and is the unique choice for elliptic coordinates with this property. 

\textit{Nontrivial roots.} Our aim is to get rid of the trivial roots. Dividing by $y_{\varphi}$ we get
\begin{equation} \label{eq:before_removing_singular_root}
{-}\frac{L}{x_{\varphi} - x_0} + \sum_{\substack{n=0 \\ (\neq I_\text{tot}/2)}}^{L-1} \!\! \frac{1}{x_{\varphi} - x_n} = 0 \, .
\end{equation}
To remove the singular root we will proceed slightly differently for either parity of $I_\text{tot}$, but the result will be the same. 
For odd $I_\text{tot}$ the left-hand side can be rewritten as
\begin{equation} \label{eq:rewrite}
\sum_{n=0}^{L-1} \left( -\frac{ 1}{x_\varphi - x_0} + \frac{1}{x_\varphi - x_n}  \right) = \sum_{n=0}^{L-1} \frac{x_n - x_0}{(x_\varphi - x_0)(x_\varphi - x_n)} \, . 
\end{equation}
Now we can remove the factor in the denominator corresponding to the singular root. Omitting the vanishing term at $n=0$ we end up with
\begin{equation*}
\sum_{n=1}^{L-1} \frac{x_{0} - x_n}{x_{\varphi} - x_n} = 0 \, . 
\end{equation*}
For even $I_\text{tot}$ the sum has one less term, so \eqref{eq:rewrite} only accounts for $L-1$ times $-1/(x_{\varphi} - x_0)$. Compensating for this \eqref{eq:before_removing_singular_root} becomes
\begin{equation*}
{-}\frac{ 1}{x_{\varphi} - x_0} + \sum_{\substack{n=0 \\ (\neq I_\text{tot}/2)}}^{L-1} \!\! \frac{x_n - x_{0}}{(x_{\varphi} - x_{0})(x_{\varphi} - x_n)} = 0 \, .
\end{equation*}
Again removing the factor responsible for the singular root and omitting $n=0$ from the sum, for which the summand vanishes identically, we obtain 
\begin{equation*}
{-1} + \! \sum_{\substack{n=1 \\ (\neq I_\text{tot}/2)}}^{L-1} \!\! \frac{x_n - x_{0}}{x_{\varphi} - x_n} = 0 \, .
\end{equation*}
By interpreting the first term as the limit of the summand as $n \to I_\text{tot}/2$ we can include it in the sum, again yielding the pleasingly simple result
\begin{equation*}
\sum_{n=1}^{L-1} \frac{x_0 - x_n}{x_{\varphi} - x_n} = 0 \, . 
\end{equation*}

Note that this equation holds for any $I_\text{tot}$ after properly accounting for divergencies: not only do we need to be careful to interpret the summand correctly at $n=I_\text{tot}/2$, but there is another divergence at $I_\text{tot} = 0$ since $x_0$ now features in the numerator. Dividing by $x_0$ we finally obtain
\begin{equation}
\label{eq:M=2curveequations}
\sum_{n=1}^{L-1} \frac{1 - x_n/x_0}{x_\varphi - x_n} = 0 \, . 
\end{equation}
This equation, with suitable interpretation of the singularities, yields the roots that we are after for all $I_\text{tot} \in \mathbb{Z}_L$. In practice it is easiest to turn \eqref{eq:M=2curveequations} into a polynomial equation for $x_\varphi$, 
\begin{equation} \label{eq:M=2curveequations_polynomial}
\sum_{n=1 }^{L-1}  A_n^{(I_\text{tot})} \!\!\!\!\!\! \prod_{\substack{ k =1 \\ (\neq \, n,\,I_\text{tot}/2)}}^{L-1} \!\!\!\!\!\!\! (x_\varphi - x_{k}) = 0 \, ,  \qquad\quad
A_n^{(I_\text{tot})} = \begin{cases} 1 & \text{if} \ I_\text{tot}=0 \, , \\ 1/x_0 & \text{if} \ n=I_\text{tot}/2 \, , \\ 1 - x_n/x_0 & \text{else}\,. \end{cases}
\end{equation}
Solving this equation numerically is straightforward and very fast. To obtain the momenta that parametrise the solution one can (numerically) invert \eqref{eq:curve_coordinates} to compute $\varphi$ and subsequently \eqref{eq:M=2_BAE} to determine the $p_m$. Since the polynomial equation only determines $x_\varphi$ (and not $y_\varphi$ as well), this inversion is not unique, but rather requires us to choose between the two preimages in the fundamental parallelogram of the lattice $\!\!\hat{\,\,\mathbb{L}}^{\mspace{-4mu}\vee}$ from \eqref{eq:momphi_lattice}. As we will discuss in the next section (\textsection\ref{sec:triv_sym}), though, this choice is immaterial as both options parametrise the same energy and wave function.

Finally, it is instructive to compare this to the usual rational \textsc{bae} of the isotropic Heisenberg spin chain. (The HS limit is much more singular.) After a renormalisation we can take the Heisenberg limit of the elliptic coordinates \eqref{eq:curve_coordinates} to obtain
\begin{equation}
\label{eq:curve_coordinates_heis}
\begin{aligned}
x_{\textsc{h},\varphi} & = \hat{\lambda}_{\textsc{h}}'(\varphi+\pi\, I_{\text{tot}}), \quad & x_{\textsc{h},n} =\hat{\lambda}_{\textsc{h}}'\bigl(2\pi\,( n -I_{\text{tot}}/2)\bigr)\, , \\
y_{\textsc{h},\varphi} & = \hat{\lambda}_{\textsc{h}}''(\varphi+I_{\text{tot}}), \quad & y_{\textsc{h},n} =\hat{\lambda}_{\textsc{h}}''\bigl(2\pi\, (n -I_{\text{tot}}/2)\bigr)\, ,
\end{aligned}
\end{equation}
where we defined $\hat{\lambda}_{\textsc{h}}(z) \coloneqq\lambda_{\textsc{h}}(z/L)/L$ with $\lambda_{\textsc{h}}$ the Heisenberg \textsc{xxx} rapidity function~\eqref{eq:M=2_constraint_Heis}. These variables satisfy
\begin{equation*}
y^2 = 4 \, \biggl(x-\frac{1}{12 \, L^2}\biggr)^{\!3} - \frac{1}{12 \, L^4} \biggl(x-\frac{1}{12 \, L^2}\biggr) - \frac{1}{216 \, L^6} = 4 \, x^3 - \frac{1}{L^2} \, x^2\, ,
\end{equation*}
so we see that in the Heisenberg limit the elliptic curve degenerates. Following the same steps as above, which in particular involves a Heisenberg analogue of~\eqref{eq:rho_identity}, viz.\ 
\begin{equation*}
\lambda_{\textsc{h}}(z) = \sum_{n=0}^{L-1} \hat{\lambda}_{\textsc{h}}(z + 2\pi n)  \, ,
\end{equation*}
we can write the Heisenberg constraint \eqref{eq:M=2_constraint_Heis} in the form \eqref{eq:ell_curve} using the coordinates \eqref{eq:curve_coordinates_heis}. In practice, this means one could use \eqref{eq:M=2curveequations_polynomial} in the single variable $x_\varphi$ to find the complete $M=2$ spectrum of the Heisenberg \textsc{xxx} spin chain, by solving the polynomial equation of degree~$L$ for a choice of $I_\text{tot} \in \mathbb{Z}_L$. In contrast, the usual rational \textsc{bae} in terms of the two rapidities $\lambda_{\textsc{h}}(p_m)$\,---\,see \eqref{eq:M=2_constraint_Heis_exp} and its analogue with $p_1\leftrightarrow p_2$\,---\, involves \emph{two} coupled (rational, or equivalently polynomial) equations in \emph{two} variables, but without specifying $I_{\text{tot}}$. For a fixed $I_\text{tot}$ one can eliminate $p_2 =  2\pi I_\text{tot}/L-p_1$ in \eqref{eq:M=2_constraint_Heis_exp} (or vice versa), using addition laws for $\lambda_{\textsc{h}}$ to obtain a polynomial of degree $L+1$ in $\lambda_{\textsc{h}}(p_1)$, which differs from the Heisenberg limit of \eqref{eq:M=2curveequations_polynomial}. 

\subsubsection{Trivial symmetry} 
\label{sec:triv_sym}
In order to investigate the effect of choosing one $\varphi$ out of the two preimages that correspond to $x_\varphi$, let us back track a little to make contact with Inozemtsev's original description of the spectrum~\cite{inozemtsev1993Hermitelike}. We study the constraint \eqref{eq:M=2varphi'} on $\!\!\hat{\,\,\mathbb{L}}^{\mspace{-4mu}\vee}$, i.e. 
\begin{equation}
	\label{eq:num_constraint}
	\frac{2}{L} \, \momL{\rho}_1(\varphi) - \momphi{\rho}_1(2\pi I_\text{tot} + \varphi) - \momphi{\rho}_1(\varphi) = 0 \, . 
\end{equation}
Inozemtsev observed that if $\varphi$ solves \eqref{eq:num_constraint} then so does\,%
\footnote{\ We have reformulated Inozemtsev's original statement for the lattice $\!\!\hat{\,\,\mathbb{L}}^{\mspace{-4mu}\vee}$ and fixed a slight inconsistency that we discuss in more detail in \textsection\ref{app:ts}.}
\begin{subequations} \label{eq:trivialsymmetry_in_text}
	\begin{gather}
		\momphi{R}_x(\varphi) \coloneqq -\varphi -2\pi \,I_\text{tot} + \frac{2\pi L}{\omega} \,\mathrm{sgn}(\Im\varphi) + \,  2\pi L \, \theta\mspace{-1mu}\bigl(2\pi \, I_\text{tot} + \Re\varphi \bigr) \,  
\intertext{with $\theta$ the Heaviside step function at the convention $\theta(0)=0$. Inozemtsev moreover observed that these solutions are physically equivalent, with the same wave function and energy. That is, $\momphi{R}_x$ is a \emph{trivial symmetry} of \eqref{eq:num_constraint}, which should be removed from the spectrum. In \textsection\ref{app:ts} we derive the preceding formula from the observation that}
		\momphi{R}_x(\varphi) =  -\varphi - 2\pi \,I_\text{tot} \, \ \mathrm{mod } \hat{\,\,\mathbb{L}}^{\mspace{-4mu}\vee} \, ,
\intertext{which can be lifted to the elliptic curve as} 
		(x_{\momphi{R}_x(\varphi)}, y_{\momphi{R}_x(\varphi)}) = (x_\varphi, -y_\varphi) \, .
	\end{gather}
\end{subequations}
In other words, $\momphi{R}_x$ is just a reflection in the $x$-axis, so a morphism of the elliptic curve. 
This equivalence between solutions suggests that the entire spectral problem can be formulated in terms of $x_\gamma$ alone, which is what we started doing by the reformulation of the constraint in the previous section. 
We will finish this reformulation in \textsection\ref{sec:M=2_energy_rat} by deriving such an expression for the two-particle energy in terms of $x_\varphi$, thereby showing that the whole $M=2$ spectral problem can be solved in terms of $x_\varphi$ alone. The choice between preimages from \textsection\ref{sec:elliptic_curve_M=2} is therefore truly trivial. This moreover shows that in order to count the number of independent states parametrised by solutions of the constraint one only needs to count the roots of \eqref{eq:M=2curveequations}. 

\subsubsection{Completeness}
\label{sec:counting}
Let us count how many roots \eqref{eq:M=2curveequations} has. First we need to find the number of values that occur twice amongst the $x_n$. This can be found from the following table:
\begin{equation*} \label{table:xnvalues}
	\begin{tabular}{lcccccc}
	\midrule
	& & $I_\text{tot} = 0$  & & even $I_\text{tot} \neq 0$  & & odd $I_\text{tot}$ \\ \hline
	even $L$ & & $2 \, \frac{\strut L-2}{2} + 1 = L-1$ & & $2 \, (\frac{L-2}{2}-1) + 2 = L-2$ & & $2 \, \frac{L-2}{2} + 1 =L-1$  \\
	odd $L$ & & $2 \, \frac{\strut L-1}{2} + 0 = L-1$ & & $2 \, (\frac{L-1}{2}-1) + 1 = L-2$ & & $2 \, (\frac{L-1}{2}-1) +2 = L-1$  \\
	\bottomrule
	\hline
	\end{tabular}
\end{equation*}
Here each entry contains, as a check, on the right-hand side the total number of $x_n$ (for even $I_\text{tot} \neq 0$ omitting $n=I_\text{tot}/2$). On the left-hand side we have refined the enumeration by counting, with multiplicities, distinct values of $x_n$. For example, $I_\text{tot} = 0$ gives $x_{L-n} = x_n$ and thus doubly occurring values, except for a single value at $n=L/2$ if $L$ is even. If $I_\text{tot}$ is even but nonzero then $n=I/2$ is excluded, so one of the preceding double occurrences becomes a single value. Finally, when $I_\text{tot}$ is odd the only singly occurring values are at $n=I_\text{tot}$ and, if $L$ is odd, $n=(L+I_\text{tot})/2$.

To count the number of solutions we turn to the polynomial form \eqref{eq:M=2curveequations_polynomial} of the constraint, keeping in mind that we require $x_\varphi \neq x_k$. We examine the different cases for $I_\text{tot}$ separately.
\begin{itemize}
	\item When $I_\text{tot}$ is even there are $\lceil\frac{L-3}{2}\rceil$ roots:
	\begin{itemize} 
		\item For $I_\text{tot}=0$ the left-hand side of \eqref{eq:M=2curveequations_polynomial} has degree $L-2$ in $x_\varphi$. However, every doubly occurring value $x_n$ will feature at least once in every summand, to become a root that was not allowed. So there are $L-2 - \lceil\frac{L-2}{2}\rceil$ solutions for $x_\varphi$. 
		\item If $I_\text{tot} \neq 0$ is even 
		the degree in $x_\varphi$ is $L-3$. Again accounting for the disallowed roots due to doubly occurring points we get $L-3 - \lceil\frac{L-2}{2}-1\rceil$
		roots as well.
	\end{itemize}
	\item When $I_\text{tot}$ is odd there are $\lfloor\frac{L-3}{2}\rfloor$ roots. Indeed, there is one $n$ such that $x_n = x_0$, causing one term from the sum to vanish. Since $k=I_\text{tot}$ has to be excluded too the degree is again $L-3$. Once more accounting for the disallowed roots yields $L-3 - \lfloor \frac{L-2}{2}\rfloor$
	solutions. 
\end{itemize}
Together this means that for any $L$, as we vary $I_\text{tot} \in \mathbb{Z}_L$, we find $L\,(L-3)/2$ roots $x_\gamma$ in total, distributed per $I_\text{tot}$ as foreseen in \textsection\ref{sec:symms_trivial_roots}. We will next prove that the two-particle energy only depends on $\varphi$ through $x_\varphi$, implying that we need exactly one out of the two available preimages $\varphi$ for every $x_\varphi$ we counted above. This then yields the $\binom{L}{2} - \binom{L}{1}$ highest-weight vectors in the two-particle sector and shows, in accordance with the results in \cite{inozemtsev1993Hermitelike}, that the ansatz for $M=2$ is complete.

\subsubsection{Rationalising the \textit{M} = 2 energy} 
\label{sec:M=2_energy_rat}
To complete the proof that the entire spectral problem for the two-particle sector can be rationalised we need to express the two-particle energy \eqref{eq:M=2_energy} as a rational function in $x_\varphi$ as well. This is reminiscent of the rational expressions for the Haldane--Shastry spin chain (in terms of momenta) and the Heisenberg model (in terms of rapidities), although the latter are not straightforward limits of the formula that we shall now derive.
On shell, i.e.\ on solutions of the constraint \eqref{eq:ell_curve}, we may rewrite the two-particle energy as
\begin{equation}
\label{eq:M=2_on_curve_1}
\begin{aligned}
\frac{E(\varphi)}{-4\,n_H(\kappa)\,\kappa^2}\bigg|_{\text{on-shell}} = {} & L^2 \Big( \momphi{F}_1( \varphi +2 \pi I_\text{tot}) + \momphi{F}_1 (\varphi ) \Big) - 2 \, \momL{F}_1(\varphi ) \\
& + 2 \, L \, \momphi{\rho}_1(\varphi + \pi I_\text{tot}) \,  y_{\varphi} \Biggl( \, \frac{L}{x_{\varphi} - x_0} - \! \sum_{\substack{n=0 \\ (\neq I_\text{tot}/2)}}^{L-1} \!\! \frac{1}{x_{\varphi} - x_n}  \Biggr) \, ,
\end{aligned}
\end{equation}
where the first line is the two-particle energy from \eqref{eq:M=2_energy} and the second line is a judicious multiple of \eqref{eq:ell_curve}, which we are free to add. By construction this function is elliptic in $\varphi$, as can be checked using $\momL{F}_1( z+ \momL{\omega}_2) = \momL{F}_1(z)-2\,\I\, \momL{\rho}_1(z)-1$. In particular it equals any other elliptic function with the same pole structure on this lattice up to a constant $C$, including 
\begin{equation*}
C - 2 \, L \Biggl( \, (L-1)\, \frac{\momphi{\rho}_1(\pi \,I_\text{tot}) \, y_{0}}{x_{\varphi} - x_0} - \!\! \sum_{\substack{n=0 \\ (\neq I_\text{tot}/2)}}^{L-1} \!\!\!\! \frac{\momphi{\rho}_1((2n-I_\text{tot})\pi) \, y_{n}}{x_{\varphi}- x_n}  \Biggr) \, .
\end{equation*}
Computing the residues show that its pole structure matches that of \eqref{eq:M=2_on_curve_1}. To determine $C$ we evaluate both expressions at $\varphi = -\pi \,I_\text{tot}$ (not a pole). This yields the following result for the two-particle energy:
\begin{equation}
\label{eq:M=2_on_curve_3}
\begin{aligned}
\frac{E(\varphi)}{-8\,n_H(\kappa)\,\kappa^2}\bigg|_{\text{on-shell}} = {} & L^2 \, \momphi{F}_1( \pi \,I_\text{tot}) - \momL{F}_1(\pi \,I_\text{tot}) \\
& - L\,(L-1)\, \frac{\momphi{\rho}_1(\pi \,I_\text{tot}) \, y_{0}}{x_{\varphi} - x_0} + L \!\!\!\! \sum_{\substack{n=0 \\ (\neq I_\text{tot}/2)}}^{L-1} \!\!\!\! \frac{\momphi{\rho}_1\bigl((2\,n-I_\text{tot})\pi\bigr) \, y_{n}}{x_{\varphi}- x_n} \, .
\end{aligned}
\end{equation}
The point here is not that this expression would be more elegant than the original~\eqref{eq:M=2_energy}, but above all is a proof of principle to show that the energy depends rationally on the curve coordinate~$x_\varphi$ along with various constants, depending on $I_\text{tot}$ but not $\varphi$. 

The fact that we managed to write the energy without any dependence on $y_\varphi$ once more verifies that when counting solutions of the constraint~\eqref{eq:ell_curve} we rightfully identified solutions related by the reflection $R_x$: such solutions have the same image under $x_\varphi$. Since $y_\varphi$ obeys $y_{R_x(\varphi)} = -y_\varphi$ any true dependence of the energy on $y_\varphi$ would spoil such an identification.

\subsection{Summary} \label{sec:M=2_summary} Let us sum up the highlights of our journey through the two-particle sector of the Inozemtsev spin chain. The key ingredients are the wave function~\eqref{eq:M=2_wave_fn}, Bethe-ansatz system \eqref{eq:M=2_constraint}--\eqref{eq:M=2_BAE} determining the quasimomentum parameters, and energy \eqref{eq:M=2_energy} given in \textsection\ref{sec:M=2_wave_fns}, with dispersion~\eqref{eq:dispersion_in_check}. In \textsection\ref{sec:S-matrix} we defined a two-body \textit{S}-matrix~\eqref{eq:S-mat} that is independent of the positions and features in the exponential form~\eqref{eq:M=2_BAE_exp} of the Bethe-ansatz equations. The wave function can be written in terms of this \textit{S}-matrix in the large-$L$ regime for magnons that are far apart (\textsection\ref{sec:large_L}) and in the Heisenberg limit (\textsection\ref{sec:Heis_limit_M=2}). For $L$ even the exceptional bound state admits a closed-form wave function~\eqref{eq:exceptional_wave_fn}.

\subsubsection{Spectrum} 
One of our main results is the correspondence between the different classes of states of the Heisenberg and Haldane--Shastry spin chains, which we were able to study at the level of the quasimomenta and the wave function. 
\begin{itemize}
	\item Since the magnons (highest weight at $M=1$ for $p\neq 0$) are independent of $\kappa$ (\textsection\ref{sec:M=1}) their $\mathfrak{sl}_2$-descendants are also the same for Heisenberg, Inozemtsev and Haldane--Shastry. All other $M=2$ vectors are highest weight for $\mathfrak{sl}_2$ (\textsection\ref{sec:hw}). 
	\item The scattering states for the Heisenberg spin chain, characterised by real quasimomenta $p_m \approx 2\pi I_m/L$ and enumerated by $2 \leq I_1 + 1 \leq I_2 - 1 \leq L$ (\textsection\ref{sec:Heis_recap_M=2}), retain these characteristics for the Inozemtsev spin chain (\textsection\ref{sec:qualitative}). In the Haldane--Shastry limit these eigenvectors become Yangian highest-weight vectors with motif~$\vect{\mu} = \vect{I}$ (\textsection\ref{sec:HS_limit}).\,\footnote{\ It is remarkable that even though the elliptic wave functions have a simple pole at coinciding arguments, see \eqref{eq:psi_decomposition} and \eqref{eq:M=2_wave_fn}, in the Haldane--Shastry limit the pole disappears: the numerator becomes (the evaluation of) an antisymmetric polynomial that is divisible by the denominator, see \eqref{eq:M=2_wavefn_HS_limit_pre}. Upon evaluation the result is equal to a polynomial with a double zero rather than a simple pole, see \eqref{eq:HS_wavefn_M=2_rewrite}.}
	\item The bound states for the Heisenberg spin chain typically have complex conjugate quasimomenta and are enumerated by $L-3$ pairs $1 \leq I_1 \leq L-1$ with $I_2 \in \{I_1,I_1+1\}$, one pair for each value of $I_\text{tot} = I_1 + I_2 \, \mathrm{mod} \, L$ in the range $2\leq I_\text{tot} \leq L-2$ (\textsection\ref{sec:Heis_recap_M=2}). When $L$ is sufficiently large, see \eqref{eq:L_crit}, the quasimomenta may start out as real, to collide and become complex conjugate at some $\kappa_\text{cr,r}$ determined by \eqref{eq:crit_r}, see also Figure~\ref{fig:Lcr_pm}. In either case the imaginary period for Inozemtsev allows one to think of both $p_m$ as lying in the fundamental rectangle $[0,2\pi) \times [0,2\I\kappa)$, with $p_1$ ($p_2$) in the lower (upper) half, respectively. As $\kappa$ is lowered the roots move towards each other to collide at the imaginary halfperiod $\Im p_m = \I \kappa$ for some value $\kappa_\text{cr,i}$ determined by \eqref{eq:crit_i} (\textsection\ref{sec:crit}). Below this (possibly second) critical value of $\kappa$ the quasimomenta are $p_1 = \pi I_\text{tot}/L - u +\I \kappa$, $p_2 = \pi I_\text{tot}/L + u - \I \kappa$ decreasing from $u|_{\kappa = \kappa_\text{cr,i}} = \pi I_\text{tot}/L$ to $u|_{\kappa =0}$ in the Haldane--Shastry limit (\textsection\ref{sec:HS_limit}). This should be compared to the `squeezing' of \cite{ha1993squeezed}. The corresponding eigenvectors become ($\mathfrak{sl}_2$-highest weight version of) affine descendants of the magnons with momentum $p_\text{tot} = 2\pi I_\text{tot}/L$, present due to the enhanced (Yangian) symmetry in that limit. (For $L\to\infty$ our critical equation reduces to that of \cite{dittrich1997two}, see \textsection\ref{sec:large_L}.)
\end{itemize}

At least for the two-particle sector we thus find the elegant correspondence
\begin{equation} \label{eq:correspondence}
	\begin{tabular}{lcccl} 
		$(I_1,I_2)$ & Heisenberg & & Haldane--Shastry & $\vect{\mu}$ \\[1ex] 
		$I_2 > I_1 + 1$ &
		scattering state & $\longleftrightarrow$ & Yangian highest-weight vector & $(I_1,I_2)$ 
		\\ 
		$I_2 \in \{I_1, I_1+1 \}$ & 
		bound state & $\longleftrightarrow$ & affine descendant 
		& $(I_\text{tot})$ 
		\\
		$I_1 = 0 $ & 
		($\mathfrak{sl}_2$-)descendant & $\longleftrightarrow$ & (non-affine) descendant 
		& $(I_2)$ 
	\end{tabular}
\end{equation}
where, more precisely, $I_1>0$ for the first two classes, and bound states correspond to  affine descendants with $\mathfrak{sl}_2$-highest weight.
Together these cover all $\binom{L-2}{2} + (L-3) + L = \binom{L}{2}$ vectors of the two-particle sector. The corresponding spectra are illustrated in Figures~\ref{fig:Ep_M2_Heis}, \ref{fig:Ep_M2_Ino} and~\ref{fig:Ep_M2_HS} for Heisenberg, Inozemtsev and Haldane--Shastry, respectively. Tables \ref{tb:spec_L=4}--\ref{tb:spec_L=6} contain the full two-particle spectrum in the Inozemtsev regime ($\kappa =1$) for small system size.

\begin{table}[h]
	\begin{tabular}{cccccl} \toprule
		$\vect{I}$ & $p_1$ & $p_2$ & $p_\text{tot}$ & $E_\text{n}$ & type \\ \midrule
		$(0,0)$ & $0$ & $0$ & $0$ & $0$ & descendant of $\ket{\uparrow\uparrow\uparrow\uparrow}$ \\ \midrule
		$(0,1)$ & $0$ & $\pi/2$ & $\pi/2$ & $2.44814$ & \\
		$(0,2)$ & $0$ & $\pi$ & $\pi$ & $4.05607$ & descendants of magnons \\
		$(0,3)$ & $0$ & $3\pi/2$ & $3\pi/2$ & $2.44814$ & \\
		\midrule
		$(1,3)$ & $1.90871$ & $4.37448$ & $0$ & $6.08410$ & scattering state \\ \midrule
		$(1,1)$ & $0+\I\kappa$ & $\pi-\I\kappa$ & $\pi$ & $2.86825$ & exceptional bound state \\
		\bottomrule
		\hline
	\end{tabular}
	\caption{The two-particle spectrum and quasimomenta of the $\kappa = 1$ Inozemtsev spin chain with $L=4$. $\vect{I} =(1,1)$ may be replaced by $\vect{I} = (3,3)$. \vspace{-\baselineskip}}
	\label{tb:spec_L=4}
\end{table}

\begin{table}[h]
	\begin{tabular}{cccccl} \toprule
		$\vect{I}$ & $p_1$ & $p_2$ & $p_\text{tot}$ & $E_\text{n}$ & type \\ \midrule
		$(0,0)$ & $0$ & $0$ & $0$ & $0$ & descendant of $\ket{\uparrow\uparrow\uparrow\uparrow\uparrow}$ \\ \midrule
		$(0,1)$ & $0$ & $2\pi/5$ & $2\pi/5$ & $1.81426$ & \multirow{4}{*}{descendants of magnons} \\
		$(0,2)$ & $0$ & $4\pi/5$ & $4\pi/5$ & $3.78899$ & \\
		$(0,3)$ & $0$ & $6\pi/5$ & $6\pi/5$ & $3.78899$ & \\
		$(0,4)$ & $0$ & $8\pi/5$ & $8\pi/5$ & $1.81426$ & \\
		\midrule
		$(1,3)$ & $1.57985$ & $3.44670$ & $8\pi/5$ & $6.45733$ & \multirow{3}{*}{scattering states} \\
		$(1,4)$ & $1.46531$ & $4.81787$ & $0$ & $4.48260$ & \\
		$(2,4)$ & $2.83649$ & $4.70333$ & $2\pi/5$ & $6.45733$ & \\ \midrule
		$(1,1)$ & $0.25028 +\I\kappa$ & $2.26299-\I\kappa$ & $4\pi/5$ & $2.50786$ & \multirow{2}{*}{bound states} \\
		$(4,4)$ & $4.02020 +\I\kappa$ & $6.03290 -\I\kappa$ & $6\pi/5$ & $2.50786$ & \\
		\bottomrule
		\hline
	\end{tabular}
	\caption{The two-particle spectrum and quasimomenta of $\kappa = 1$ Inozemtsev with $L=5$. The bound states show that we are below $\kappa^{(2)}_\text{cr,i}(L=5) \approx 1.83$. \vspace{-\baselineskip}}
	\label{tb:spec_L=5}
\end{table}

\begin{table}[h]
	\begin{tabular}{cccccl} \toprule
		$\vect{I}$ & $p_1$ & $p_2$ & $p_\text{tot}$ & $E_\text{n}$ & type \\ \midrule
		$(0,0)$ & $0$ & $0$ & $0$ & $0$ & descendant of $\ket{\uparrow\uparrow\uparrow\uparrow\uparrow\uparrow}$ \\ \midrule
		$(0,1)$ & $0$ & $\pi/3$ & $\pi/3$ & $1.37585$ & \multirow{5}{*}{descendants of magnons} \\
		$(0,2)$ & $0$ & $2\pi/3$ & $2\pi/3$ & $3.32131$ & \\
		$(0,3)$ & $0$ & $\pi$ & $\pi$ & $4.05607$ & \\
		$(0,4)$ & $0$ & $4\pi/3$ & $4\pi/3$ & $3.32131$ & \\
		$(0,5)$ & $0$ & $5\pi/3$ & $5\pi/3$ & $1.37585$ & \\
		\midrule
		$(1,3)$ & $1.34732$ & $2.84147$ & $4\pi/3$ & $5.99702$ & \multirow{6}{*}{scattering states} \\
		$(1,4)$ & $1.25028$ & $3.98571$ & $5\pi/3$ & $5.37678$ & \\
		$(1,5)$ & $1.19556$ & $5.08763$ & $0$ & $3.37286$ & \\
		$(2,4)$ & $2.40586$ & $3.87732$ & $0$ & $7.38088$ & \\
		$(2,5)$ & $2.29748$ & $5.03290$ & $\pi/3$ & $5.37678$ & \\
		$(3,5)$ & $3.44171$ & $4.93587$ & $2\pi/3$ & $5.99702$ & \\ \midrule
		$(1,2)$ & $0 +\I\kappa$ & $\pi -\I\kappa$ & $\pi$ & $2.69665$ & exceptional bound state \\ \midrule
		$(1,1)$ & $0.45587 + \I\kappa$ & $1.63853 - \I\kappa$ & $\pi/3$ & $2.07650$ & \multirow{2}{*}{other bound states} \\
		$(5,5)$ & $4.64466+\I\kappa$ & $5.82732 -\I\kappa$ & $5\pi/3$ & $2.07650$ & \\
		\bottomrule
		\hline
	\end{tabular}
	\caption{The two-particle spectrum and quasimomenta of $\kappa = 1$ Inozemtsev with $L=6$. $\vect{I} =(1,2)$ may be replaced by $\vect{I} = (4,5)$, cf.~Figure~\ref{fig:integers}. The other bound states attest that we are below $\kappa^{(2)}_\text{cr,i}(L=6) \approx 1.28$. \vspace{-\baselineskip}}
	\label{tb:spec_L=6}
\end{table}

\textit{Conjecture.} The correspondence~\eqref{eq:correspondence} extends to general $M$. Namely, scattering states for Heisenberg will correspond one to one, via Inozemtsev, to Yangian highest-weight vectors for Haldane--Shastry with $\vect{\mu} = \vect{I}$. Moreover, bound states for Heisenberg, with complex $p_m,p_{m+1}$ associated to $\vect{I}$ with $I_{m+1} \in \{I_m,I_m+1\}$, will flow to ($\mathfrak{sl}_2$-highest-weight versions of) affine descendants for Haldane--Shastry, with motif obtained by 'squeezing' $\vect{I} = (\dots,I_m,I_{m+1},\dots) \mapsto (\dots,I_m + I_{m+1},\dots) = \vect{\mu}$. 

\begin{figure}[b]
	\centering
	\vspace{-30pt}
	\begin{tikzpicture}[font=\scriptsize]
		\node at (0,0) {\includegraphics[width=.9\textwidth]{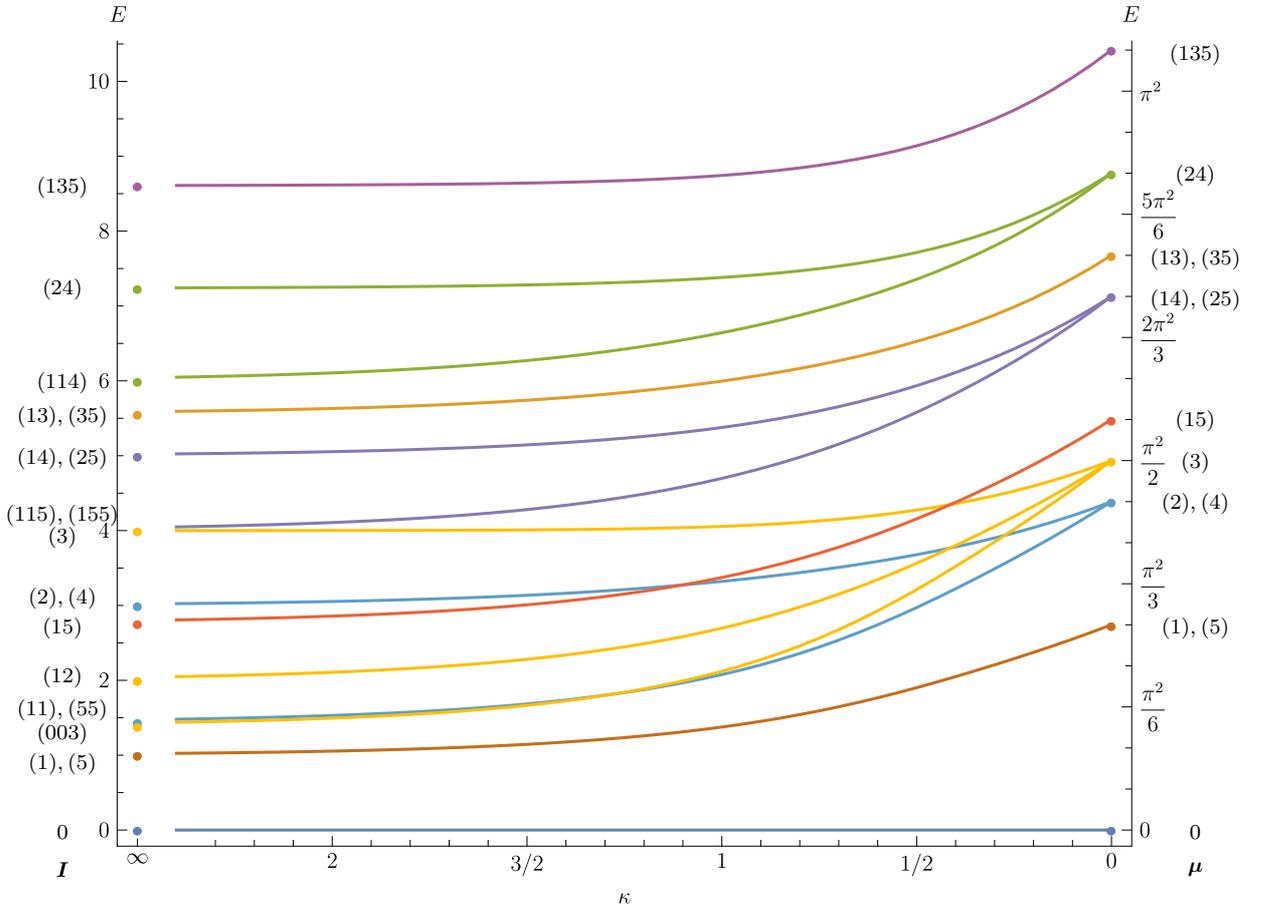}};
		\node at (-7.5,-5.5) {$\vect{I}$};
		\node at (-7.5,-5) {$0$};
		\node at (-7.5,-4.1) {$(1),(5)$};
		\node at (-7.5,-3.7) {$(003)$};
		\node at (-7.5,-3.38) {$(11),(55)$};
		\node at (-7.5,-2.95) {$(12)$};
		\node at (-7.5,-2.3) {$(15)$};
		\node at (-7.5,-1.9) {$(2),(4)$};
		\node at (-7.5,-1.1) {$(3)$};
		\node at (-7.5,-.8) {$(115),(155)$};
		\node at (-7.5,-.05) {$(14),(25)$};
		\node at (-7.5,.5) {$(13),(35)$};
		\node at (-7.5,.95) {$(114)$};
		\node at (-7.5,2.2) {$(2 4)$};
		\node at (-7.5,3.55) {$(135)$};
		\node at (7.4,-5.5) {$\vect{\mu}$};
		\node at (7.4,-5) {$0$};
		\node at (7.4,-2.3) {$(1),(5)$};
		\node at (7.4,-.65) {$(2),(4)$};
		\node at (7.4,-.1) {$(3)$};
		\node at (7.4,.45) {$(15)$};
		\node at (7.4,2.05) {$(14),(25)$};
		\node at (7.4,2.6) {$(13),(35)$};
		\node at (7.4,3.7) {$(24)$};
		\node at (7.4,5.3) {$(135)$};
	\end{tikzpicture}
	\caption{The full spectrum of the $L=6$ Inozemtsev spin chain as a function of $\kappa$. For Heisenberg ($\kappa=\infty$) the energies and momenta are labelled by $\vect{I}=(I_1,\cdots\mspace{-1mu},I_M)$, with eigenspaces of dimension~$L-2M+1$ that are either parity invariant or come in parity-conjugate pairs. (The exceptional cases $(1,2) \equiv (4,5)$ and $(1,1,4) \equiv (2,5,5)$ are in fact self conjugate.) The degeneracy for $(3)$ and $(1,1,5),(1,5,5)$ is a coincidence that is lifted by $p_\text{tot} = 2\pi I_\text{tot}/L$. Intermediate values of $\kappa$ show various crossing energy levels. In the Haldane--Shastry limit ($\kappa=0$) the (Yangian highest-weight) eigenspaces are labelled by motifs $\vect{\mu}=(\mu_1,\cdots\mspace{-1mu},\mu_M)$, with energies in multiples of $2\pi^2/L^2$. Bound states, with $I_{m+1} \in \{ I_m , I_m +1\}$, have become affine descendants with $\mu_{m'} = I_m +I_{m+1} \, \text{mod}\,L$; the magnon with $\vect{\mu} = (3)$ has two affine descendants.}
	\label{fig:E_L=6_kappa}
\end{figure}

Our conjecture clearly works at the level of counting the number of different states in each class. Preliminary numerical investigations for $M=3$ suggest that it does indeed generalise. To illustrate this the full spectrum for $L=6$ is shown as a function of $\kappa$ in Figure~\ref{fig:E_L=6_kappa}. Note the level crossing, which provide strong hints of the model's quantum integrability, indicating the existence of higher conserved charges that protect the states from mixing at coinciding energies. One could imagine that these level crossings are related to the critical phenomena from \textsection\ref{sec:crit}, but we have not been able to connect the two; for instance, the critical value(s) of $\kappa$ only depend on the label~$n$ associated to a single energy, while \emph{both} levels that cross affect the value of $\kappa$ at which this occurs. 

\subsubsection{Rationalisation and completeness} Another main result is our finding that after a suitable change of variables the entire two-particle spectral problem can be rationalised, i.e.\ turned into a system of rational equations. More precisely:
\begin{itemize}
\item For given $I_m$ the constraint \eqref{eq:f_constraint} is an elliptic function in $p_1$ (or equivalently $p_2$) on $\bar{\mathbb{L}\!\!}^{\,\vee}$, and so are the wave function and the two-particle energy. Using the Bethe-ansatz equations \eqref{eq:M=2_BAE} we can also express the constraint in terms of the scattering phase $\varphi$ alone and in this form it becomes an elliptic function on the lattice $\!\!\hat{\,\,\mathbb{L}}$. 
\item The constraint can be written in terms of the elliptic curve coordinates \eqref{eq:curve_coordinates}, allowing for an effective factoring into trivial and non-trivial roots. The result is a polynomial equation~\eqref{eq:M=2curveequations_polynomial} that can be solved very efficiently numerically. The lift of the constraint to the elliptic curve also provides a geometric meaning to the trivial symmetry~\eqref{eq:trivialsymmetry_in_text} observed by Inozemtsev and automatically removes it from the spectrum.
\item It is straightforward to count the solutions to the polynomial constraint \eqref{eq:M=2curveequations_polynomial}, directly verifying the completeness of the ansatz \eqref{eq:ansatz} for the two-particle sector. 
\item The explicit expression \eqref{eq:M=2_on_curve_3} for the two-particle energy in terms of the elliptic curve coordinate $x_\varphi$ concludes the proof that the entire spectral problem for $M=2$ can be lifted to the elliptic curve to become fully rational.
\end{itemize}

\subsubsection{Summary of limits} For easy reference we collect the most important limits of various quantities of interest. The total momentum $p_\text{tot} = 2\pi(I_1+I_2)/L$ is discrete, so independent of~$\kappa$. As we have just summarised the quasimomenta~$p_m$ vary with $\kappa$ but have well defined limits. The limits of the corresponding rapidities are given in \eqref{eq:limits_rhobarcheck}, supplemented by \eqref{eq:rho_HS_lim_coord} with $L\rightsquigarrow 1$ for the bound-state rapidities in the Haldane--Shastry limit. As for the energy, the limit of the dispersion was given in \eqref{eq:dispersion_limits} (see \textsection\ref{sec:dispersion_limit}), whereas the (normalised) potential energy~\eqref{eq:M=2_energy} behaves as (see \textsection\ref{sec:Heis_limit_M=2}, \ref{sec:large_L}, \ref{sec:potential}) 
\begin{equation} \label{eq:limits_potential_energy}
	\tikz[baseline={([yshift=-.5*10pt*0.6+5pt]current bounding box.center)},scale=0.6]{
		\matrix (m) [matrix of math nodes,row sep=0em,column sep=5em]{
			& \substack{\displaystyle \text{Inozemtsev} \\[.5em] \displaystyle 4\,n_H(\kappa)\,\kappa^2 \, \momL{F}_1(\varphi) } & \\
			\substack{\displaystyle \text{Heisenberg} \\[.2em] \displaystyle 0 } & & \substack{\displaystyle \text{Haldane--Shastry} \\[.5em] \displaystyle -\frac{\modulo{\varphi} \, (2\pi-\modulo{\varphi})}{2 \, L^2} } \\
			& \displaystyle 0 & \\
		};
		\path[->] 
		(m-1-2) edge node [above left,xshift=.9cm,yshift=.2cm] {\footnotesize $\begin{array}{c} \kappa\to\infty \\ (\omega \to 0) \end{array}$} ([yshift=.1cm]m-2-1.east)
		(m-1-2) edge node [left] {\footnotesize $L\to\infty$ \ } (m-3-2)
		(m-1-2) edge node [above right,xshift=-.9cm,yshift=.2cm] {\footnotesize $\begin{array}{c} \kappa\to0 \\ (\omega \to \I\,\infty) \end{array}$} ([yshift=.2cm]m-2-3.west);
	}
	\checkedMma
\end{equation}
For general values of $\varphi$ Haldane--Shastry is the only limit that is not functionally additive. On shell, however, it is, and the energy is even \emph{strictly} additive (\textsection\ref{sec:HS_recap_M=2}).

Finally we turn to the wave function. The building block is the function $\chi_2$ defined in~\eqref{eq:chi_def}. It depends on the parameter~$\gamma$, which is the coordinate-lattice counterpart of the scattering phase~$\varphi$, see \eqref{eq:varphi_vs_gamma}. By the \textsc{bae}~\eqref{eq:M=2_BAE} the quasimomenta can also be expressed via $\varphi$. Conversely, the latter has the usual \emph{difference property} in that it is a function $\varphi = \Theta_\text{n}(\lambda_1 - \lambda_2)$ of the difference of rapidities, see \eqref{eq:M=2_constraint_normalised}. 
By \eqref{eq:chi_2_heis_limit}, \eqref{eq:chi_2_L_limit} and~\eqref{eq:chi_2_HS_limit} the extended Bethe-ansatz coefficient has the following limits:
\begin{equation} \label{eq:limits_chi2}
	\tikz[baseline={([yshift=-.5*10pt*0.6+5pt]current bounding box.center)},scale=0.6]{
	\matrix (m) [matrix of math nodes,row sep=0em,column sep=5em]{
		& \substack{\displaystyle \text{Inozemtsev} \\[.5em] \displaystyle n_\Psi(\kappa)\,\chi_2(z,\gamma) } & \\
		& & \substack{ \displaystyle \text{Haldane--Shastry} \\[.5em] \displaystyle \frac{\pi}{L} \, \frac{\sin[\pi(z+\gamma)/L]}{\sin(\pi\,z/L)\sin(\pi\,\gamma/L)} } \\
		\substack{ \displaystyle \text{Heisenberg} \\[.2em] \displaystyle \frac{\I\,\E^{-\I \,\mathrm{sgn}(\Re z)\,\varphi/2}}{\sin(\varphi/2)} } & \displaystyle n_\Psi(\kappa) \, \kappa \, \frac{\sinh[\kappa(z+\gamma)]}{\sinh(\kappa\,z) \, \sinh(\kappa\,\gamma)} & \\
		& & \displaystyle \frac{z+\gamma}{z\,\gamma} \\
	};
	\path[->] 
		(m-1-2) edge node [above left,xshift=.3cm,yshift=.1cm] {\footnotesize $\begin{array}{c} \kappa\to\infty \\ (\omega \to 0) \end{array}$} ([yshift=.3cm]m-3-1.east)
		(m-1-2) edge node [right] {\footnotesize $L\to\infty$ \ } (m-3-2)
		(m-1-2) edge node [above right,xshift=-.9cm,yshift=.2cm] {\footnotesize $\begin{array}{c} \kappa\to0 \\ (\omega \to \I\,\infty) \end{array}$} ([yshift=.2cm]m-2-3.west)
		(m-2-3) edge node [right] {\footnotesize $L\to\infty$ \ } (m-4-3)
		(m-3-2.west) edge node [below right,xshift=-.8cm,yshift=0cm] {\footnotesize $\begin{array}{c} \kappa\to\infty \\ (\omega \to 0) \end{array}$} (m-3-1.east) 
		(m-3-2.east) edge node [below right,xshift=-.9cm,yshift=-.2cm] {\footnotesize $\begin{array}{c} \kappa\to0 \\ (\omega \to \I\,\infty) \end{array}$} ([yshift=.2cm]m-4-3.west);
	}
\checkedMma
\end{equation}
Here the normalising factor must obey $\lim_{\kappa\to0} n_\Psi = 1$ and vanish as $n_\Psi \sim 1/\kappa$ when $\kappa\to\infty$; in the context of this work it seems more natural to take
\begin{equation*}
	\text{e.g.} \quad n_\Psi(\kappa) = \kappa \, \coth \kappa
\end{equation*}
than $1/(1+\kappa)$. 
Note that the hyperbolic, trigonometric and rational limits admit alternative expressions using $\sin(x+y)/(\sin x\sin y) = \cot x + \cot y$. 
In the Heisenberg limit one may similarly rewrite $\I\,\E^{-\I \,\mathrm{sgn}(\Re z)\,\varphi/2}/\sin(\varphi/2) = \I\,\cot(\varphi/2) + \mathrm{sgn}(\Re z)$, see \eqref{eq:chi_2_heis_limit}.

\section{Outlook} \label{sec:conclusions}
\noindent
By modifying Inozemtsev's extended coordinate Bethe ansatz (\textsection\ref{sec:eCBA}) we have shed new light on the exact spectrum of Inozemtsev's elliptic spin chain, obtaining new results in general (summarised in \textsection\ref{sec:result}) and for the two-particle sector in particular (summarised in \textsection\ref{sec:M=2_summary}). 

Although our proposed reparametrisation yields a new, much better structured formula for the $M$-particle energy we have not explored the consequences beyond $M=2$ much in the present paper. To find out whether the relations and structures we found are general, analysing our findings for higher $M$ is an important step. One possible way forward is to follow Inozemtsev~\cite{Inozemtsev_2000} and use the eigenfunctions of the elliptic CS model described in \cite{Felder:1995iv} to describe the spin chain spectrum. These eigenfunctions depend on $N\coloneqq M(M-1)/2$ auxiliary parameters~$t_\alpha$ related to the momenta through a system of $N$ constraints that may be viewed as Bethe-ansatz-like equations for eCS. For the spin chain these relations depend on the quasimomenta~$p_m$ through the identification~\eqref{eq:p_requirement} [or \eqref{eq:p_requirement_mom}]; let us denote this system by $\{\Xi_\alpha(t_1,\cdots\mspace{-1mu},t_N) = 0\}_{\alpha=1}^N$. 
Periodicity further give rise to the \textsc{bae} as in \textsection\ref{sec:Bethe_analysis} and fixes the scattering phases~$\varphi_m$ in terms of the $t_\alpha$. In this way one can express $\{\Xi_\alpha = 0\}_{\alpha=1}^N$ in terms of $t_1,\cdots\mspace{-1mu},t_N$ and $I_1,\cdots\mspace{-1mu},I_M$. Viewed as a function of any single $t_\alpha$ the result is elliptic on the lattice $\!\!\hat{\mathbb{\,\,L}}$ from \textsection\ref{sec:completeness}, implying that the system can be rewritten in terms of elliptic curve coordinates. The associated $M$-particle energy $E_\text{n}$ can also be written as a function of the $t_\alpha$~\cite{Felder:1995iv,Inozemtsev_2000}. We find that under translations by the lattice periods~$\hat{\omega}_b$ this function satisfies
\begin{equation}
	\begin{aligned}
	E_\text{n}(t_1,\cdots\mspace{-1mu},t_{\alpha} + \hat{\omega}_1,\cdots\mspace{-1mu},t_N) & = E_\text{n}(t_1,\cdots\mspace{-1mu},t_N) \, , \\
	E_\text{n}(t_1,\cdots\mspace{-1mu},t_{\alpha} + \hat{\omega}_2,\cdots\mspace{-1mu},t_N) & = E_\text{n}(t_1,\cdots\mspace{-1mu},t_N) + \text{cst}\times \Xi_\alpha(t_1,\cdots\mspace{-1mu},t_N) \, .
	\end{aligned}
\end{equation}
That is, \emph{the $M$-particle energy is elliptic in the $t_\alpha$ on shell}, i.e.\ on solutions of $\{\Xi_\alpha = 0\}_{\alpha=1}^N$. In particular it is possible to rewrite $E_\text{n}$ in terms of elliptic coordinates, just as we did for the two-particle energy in \textsection\ref{sec:M=2_energy_rat}, yielding a fully rational $M$-particle spectral problem in terms of the eCS solutions from \cite{Felder:1995iv}. This is reminiscent of how the Heisenberg (or six- or eight-vertex) \textsc{bae} can be obtained from analytic properties of the $M$-particle energy. In the present case ellipticity of 
$E_\text{n}(t_1,\cdots\mspace{-1mu},t_N)$ similarly gives rise to $\{\Xi_\alpha = 0\}_{\alpha=1}^N$. We will elaborate on this, as well as our conjecture that the correspondence \eqref{eq:correspondence} generalises to the $M$-particle sector, in a future publication. 

One of the key issues that remains open is whether the Inozemtsev spin chain is \emph{quantum integrable}: is it possible to construct a hierarchy of commuting Hamiltonians that includes the translation operator~\eqref{eq:shift_op} and the spin-chain Hamiltonian from \textsection\ref{sec:model}, and to uncover an underlying quantum-algebraic structure (\textit{R}-matrix, algebraic Bethe ansatz)? Besides the various works mentioned in \textsection\ref{sec:intro} that hint in this direction, this is already expected from the level crossings (Figure~\ref{fig:E_L=6_kappa}) and, of course, the model's exact solvability. By clarifying various aspects of the latter we hope that the present work will prove a stepping stone for addressing this question.

Based on recent results for the classical continuum limit of the Haldane-Shastry spin chain and the large-$L$ limit of the Inozemtsev spin chain \cite{berntson2020multi}, one could expect that the relationship could be searched for in a `parent' field theory, which would be interesting in itself as well. 

Another natural question is what happens beyond the isotropic level. Recently the spectrum of the (partially isotropic or) $q$-deformed Haldane--Shastry spin chain was obtained \cite{LPS_20u}, building on work of Uglov~\cite{Ugl_95u} and one of us~\cite{Lam_18}. One naturally wonders whether the Inozemtsev spin chain admits a similar \textit{q}-deformation that interpolates between the $q$-deformed Haldane--Shastry spin chain and (a version of) the \textsc{xxz} spin chain. Proceeding like in \textsection\ref{sec:limits_combs} we expect its potential to be the following point splitting of Inozemtsev's $\wp$ pair potential from \textsection\ref{sec:model}:
\begin{equation} \label{eq:q-def_potential}
	\frac{1}{2\gamma}\,\partial_x \log \frac{\theta_2(x-\gamma)}{\theta_2(x+\gamma)} = -\frac{\rho_2(x+\gamma) - \rho_2(x-\gamma)}{2\gamma} = -\frac{\zeta(x+\gamma) - \zeta(x-\gamma)}{2\gamma} + \frac{\eta_2}{\omega} \, ,
\end{equation} 
where $\gamma$ is the anisotropy parameter. Since \eqref{eq:q-def_potential} is a discrete derivative the isotropic limit $\gamma\to0$ readily gives the shifted pair potential $-\rho'_2(x) = \wp(x) +\eta_2/\omega$ from~\eqref{eq:ham_shifted}. Including the normalisation $n_H(\kappa,\gamma) = 2\,\gamma \sinh^2(\kappa)/[\kappa \sinh(2 \kappa \gamma)]$ the trigonometric limit of this potential is
\begin{equation*}
	-\frac{\frac{\pi}{L}\cot\frac{\pi}{L}(x+\gamma) - \frac{\pi}{L}\cot\frac{\pi}{L}(x-\gamma)}{2\gamma} = \frac{\sin(2\pi \gamma/L)}{2\pi \gamma/L} \, \frac{(\pi/L)^2}{\sin[\pi(x+\gamma)/L]\,\sin[\pi(x-\gamma)/L]} \, .
\end{equation*}
which equals the potential from \cite{Lam_18} up to the prefactor, and we can identify $q = \E^{\I \pi\gamma/L}$. The large-$L$ (hyperbolic) limit gives $n_H(\kappa,\gamma)$ times the same potential with $\pi/L \rightsquigarrow \pi/\omega = -\I\,\kappa$ as usual. The rational limit of both is $1/[(x-\gamma)(x+\gamma)]$, and conversely gives rise to the trigonometric, hyperbolic and elliptic potentials using the comb trick from \textsection\ref{sec:limits_combs}. The normalisation $n_H(\kappa,\gamma)$ ensures that in the contact limit the potential \eqref{eq:q-def_potential} reduces to the usual nearest-neighbour potential $\delta_{d_L(z),1}$ from \eqref{eq:limits_potential}. The spin interactions of the \textit{q}-deformed Inozemtsev spin chain will be rather more complicated and involve multi-particle interactions, as this is already the case for \textit{q}-deformed Haldane--Shastry. This will make it much more complicated to work with the model in the coordinate basis and investigate the key point\,---\,whether the exact solvability persists at the partially isotropic level. We intend to return to this question in the future.

\appendix

\section{Notation compared with the literature} 
\label{app:notation_comparison}

\noindent In this work we have made some changes with respect to Inozemtsev's extended Bethe ansatz, see \textsection\ref{sec:eCBA}. To facilitate comparison with the literature we have summarised the notation used for important quantities here and in the literature in Table~\ref{table:notations}. When applicable the relation to our notation is indicated in square brackets; for example, the $\vect{p}$ of \cite{Inozemtsev_1995} coincides with our $\widetilde{\vect{p}} - \vect{\varphi}$.

\begin{table}[h] 
	\begin{tabular}{lccc}
		\toprule
		&  & Inozemtsev & Inozemtsev \\
		quantity & this work & \cites{Inozemtsev_1995,Inozemtsev_2000} &  \cite{Inozemtsev:1989yq}
		\\
		& & general $M$ & $M=2$ \\
		\midrule
		(spin-chain) length & $L$ & $N$ & $N$
		\\
		magnon number & $M$ & $M$ & -- \\
		lattice positions & $\vect{n}$ & $\vect{n}$ & $\vect{k}$ \\
		continuum positions & $\vect{x}$ & $\vect{x}$ & -- \\
		quasimomenta & $\vect{p}$  & $\vect{p}$ & $\vect{p} \ 
		[= \vect{p} - \I \frac{\eta_2}{2\pi} \, \vect{\varphi} ]$ \\
		continuum momenta & $\widetilde{\vect{p}}$ & $\vect{p} \ [= \widetilde{\vect{p}} - \vect{\varphi} ]$ & -- \\
		quasiperiodicty parameters & $\vect{\varphi}$ & $ \vect{q} \ [ = \frac{\omega}{2\pi \, L} \,\vect{\varphi} ] $& $\gamma \ [ = \frac{\omega}{2\pi} \varphi ]$ \\[.2ex]
		spin-chain wave function & $\Psi_{\vect{p}} = \Psi_{\tilde{\vect{p}},\vect{p}}$ & $\psi_M$ & $\psi^{(N)}_{\vect{p}}$ \\
		continuum wave fn. & $\widetilde{\Psi}_{\tilde{\vect{p}}}$ & $\chi^{(p)}_M$ & -- \\
		\qquad building block & $\chi_2(z,t)$ & $\tilde{\sigma}_w(t) \ [= \chi_1(w,t)]$ & --  \\
		spin-chain energy & $E_\text{s}$ & $\varepsilon_M$ & $\varepsilon_{\vect{p}}$ \\
		\qquad unshifted & $E_\text{u}$ & $\mathcal{E}_M$  \cite{Inozemtsev_1995} & -- \\
		\qquad normalised & $E_\text{n}$ & -- & -- \\
		\qquad dispersion & $\varepsilon_s$ & $\varepsilon$ & $\varepsilon^{(1)}(p)$ \\ 
		continuum energy  & $\widetilde{E}$ & $\mathsf{E}_M$ & --
		\\
		\bottomrule \hline
	\end{tabular}
	\caption{Comparison of notations used in this work and Inozemtsev's papers.} \label{table:notations}
\end{table} 

\section{Elliptic functions}
\noindent
\label{app:elliptic_fns}

\noindent In this work we make use of six sets of Weierstrass elliptic functions associated to six different lattices depending on the spin chain length $L \in \mathbb{N}$ and the interpolation parameter $\kappa \in \mathbb{R}_+$ setting $\omega = \I \pi/\kappa$. The six lattices can be divided into three `coordinate lattices' along with three (reciprocal) `momentum lattices' that differ by a rescaling by $-2\pi/\omega$ (see \textsection\ref{sec:mtm_lattices}). The Hamiltonian and wave function are defined on the coordinate lattice $\mathbb{L}$ with periods $(\omega_1,\omega_2) = (L,\omega)$. The elliptic sums discussed in \textsection\ref{app:Inozemtsev_trick} are naturally defined on the lattice $\!\!\bar{\,\,\mathbb{L}}$ with periods $(\bar{\omega}_1,\bar{\omega}_2) = (1,\omega)$, whereas the coordinate-version $\gamma$ of the scattering phase $\varphi$ in e.g.~\eqref{eq:M=2_eCS_wave_fn} behaves most naturally on the lattice~$\!\!\hat{\,\, \mathbb{L}}$ with periods $(\hat{\omega}_1,\hat{\omega}_2) = ( L, L\,\omega)$, which appears in \textsection\ref{app:ts}--\ref{app:trivialsols}. The momentum analogues are decorated with a `${}^\vee$'. The lattices $\mathbb{L}^{\!\vee}$ with periods $(\omega_1^\vee,\omega_2^\vee) \coloneqq (2\pi, -2\pi L /\omega)$ and $\!\!\bar{\,\,\mathbb{L}}^{\!\vee}$ with $(\bar{\omega}_1^\vee,\bar{\omega}_2^\vee) \coloneqq (2\pi, -2\pi /\omega)$ are discussed at length in \textsection\ref{sec:mtm_lattices}, providing the natural setting for expressions involving the quasimomenta. The lattice $\!\!\hat{\,\,\mathbb{L}}^{\mspace{-4mu}\vee}$ with periods $(\momphi{\omega}_1,\momphi{\omega}_2) \coloneqq (2\pi L, -2\pi L/\omega)$ is the natural choice for the scattering phase $\varphi$ and therefore the lattice that defines the elliptic curve on which the spectral problem gets rationalised in \textsection\ref{sec:completeness}.

Now we turn to the associated elliptic functions. Useful references are \cite{DLMF,abramowitz1948handbook,whittaker1904course}.

\subsection{Weierstrass functions} \label{sec:Weierstrass}
The Weierstrass elliptic function with periods $(\omega_1,\omega_2)$ is 
\begin{equation}
\wp(z) = \frac{1}{z^2} + \sum_{j,k \in \mathbb{Z}} {}^{\!\!\!^{\scriptstyle *} }\left(\frac{1}{(z+j \, \omega_1 + k\, \omega_2 )^2} - \frac{1}{(j\, \omega_1 + k\, \omega_2)^2}\right) \, ,
\end{equation} 
where the asterisk indicates that the term with $j=k=0$ is to be omitted from the sum. This (even) function is related to the other two (odd) Weierstrass functions by
\begin{equation}
\wp(z) =  -\zeta'(z) \, , \qquad\qquad \zeta(z) = (\log \sigma(z))' = \frac{\sigma'(z)}{\sigma(z)} \, .
\end{equation}
These two functions are doubly quasiperiodic with quasiperiods $(\omega_1,\omega_2)$. Denote (twice) the values of $\zeta$ at half its quasiperiods by
\begin{subequations}
	\begin{gather}
	\label{eq:eta12}
	\eta_b \coloneqq 2\,\zeta(\omega_1/2) \, , \qquad b=1,2\, .
\intertext{These constants obey the Legendre relation}
	\label{eq:Legendre relation}
	\omega_2 \, \eta_1 - \omega_1 \, \eta_2 = 2 \pi\I \, .
	\end{gather}
\end{subequations}
Then the (arithmetic) quasiperiodicity of $\zeta$ reads $\zeta(z+\omega_b)=\zeta(z) + \eta_b$ for $b=1,2$, while $\sigma$ is doubly (geometric) quasiperiodic, $\sigma(z+\omega_b) = {-}\E^{(z+\omega_b/2)\,\eta_b }\,\sigma(z)$. 

We make frequent use of the functions
\begin{equation} \label{eq:rho}
	\rho_b(z) \coloneqq \zeta(z) - \frac{\eta_b}{\omega_b} \, z \, , \qquad b=1,2\,.
\end{equation} 
These functions are odd, $\rho_b(-z) = -\rho_b(z)$, periodic in one direction, $\rho_b(z+\omega_b) = \rho_b(z)$, and (arithmetically) quasiperiodic in the other, $\rho_b(z+\omega_a) = \rho_b(z) + (-1)^a\, 2\pi\I/\omega_a$ for $a\neq b$. Due to the Legendre relation they differ by $\rho_2(z) - \rho_1(z) = (2\pi\I)/(L\,\omega)\,z$.

We will need the doubling formulae, valid as long as $u,v,u+v \notin \mathbb{L}$:
\begin{subequations}
	\label{eq:doubling}
	\begin{gather}
		\wp(u+v) = \mathrm{P}(u,v)^2 -\wp(u) - \wp(v) \, , \qquad \rho_b(u+v) =  \mathrm{P}(u,v) + \rho_b(u) + \rho_b(v)  \, , 
\shortintertext{where, to make the structure of the equations more clear, we write}
		\mathrm{P}(u,v) \coloneqq \frac{1}{2} \frac{\wp'(u) - \wp'(v)}{\wp(u) - \wp(v)} \, .
	\end{gather}
\end{subequations}

\subsection{Jacobi theta functions} 
The odd Jacobi theta function $\vartheta$ is defined as
\begin{equation}
\label{eq:vartheta}
\vartheta(z | \tau ) \coloneqq 2 \sum_{n=0}^\infty (-1)^n \, q^{(n+1/2)^2}\sin( (2n+1)\,z)\, , \qquad q = \E^{\I \pi \tau}\,  
\end{equation}
It is double quasiperiodic, $\vartheta(z +\pi | \tau )  = -\vartheta(z | \tau ) $ and $\vartheta(z + \pi \,\tau | \tau ) = - q^{-1} \, \E^{-2\I z} \, \vartheta(z | \tau )$, and entire with a simple zero at the origin and its lattice shifts. We will use two variants of this function, which we denote by
\begin{equation} \label{eq:theta_1,2}
	\theta_1(z) \coloneqq \vartheta\bigl(\pi \,z/L\bigm|\omega/L\bigr) \, , \qquad \theta_2(z)\coloneqq \vartheta\bigl(\pi\,z/\omega\bigm|{-L}/\omega\bigr) \, ,
\checkedMma
\end{equation}
and are related by $\theta_1(z) = \I \, (\I \, L/\omega)^{1/2} \, \E^{-\I \pi\,z^2/(L\omega)} \, \theta_2(z) = \I \, (L \, \kappa/\pi)^{1/2} \, \E^{-\kappa \, z^2/L} \, \theta_2(z)$. These functions inherit the double quasiperiodicy in the form $\theta_b(z+\omega_b) = -\theta_b(z)$ along with $\theta_1(z+\omega) = -\E^{-2\, \pi\I z/L -\pi\I \, \omega/L} \theta_1(z)$ and $\theta_2(z+L) = -\E^{2\, \pi\I z/\omega +\pi\I \, L/\omega} \theta_2(z)$. In particular these functions are closely related to Weierstrass $\sigma$. We have
\begin{equation}
\sigma(z) = \exp\biggl(\frac{\eta_b \, z^2}{2\,\omega_b}\biggr) \, \frac{\theta_b(z)}{\theta_b'(0)} \, , \qquad \zeta(z) = \bigl(\log \theta_b(z)\bigr)' + \frac{\eta_b}{\omega_b}\,z \, , \qquad \wp(z) = -\bigl(\log \theta_b(z)\bigr)'' - \frac{\eta_b}{\omega_b} \, .
\end{equation}
Notice that \eqref{eq:rho} naturally arises as
\begin{subequations}
	\begin{gather}
	\rho_b(z) = \bigl(\log \theta_b(z)\bigr)' = \frac{\theta'_b(z)}{\theta_b(z)} \, .
\intertext{This is a close cousin of the Jacobi zeta function. Its derivative}
	\rho'_b(z) = -\wp(z)- \frac{\eta_b}{\omega_b}
	\end{gather}
\end{subequations}
appears in \textsection\ref{sec:limits_combs} as the pair potential (for $b=2$). That is, the Weierstrass elliptic functions $\wp : \zeta : \sigma$ have Jacobi counterparts $V_{\text{s}} : \rho_2 : \theta_2$ (along with a version for $b=1$). Let us stress once more that, despite their subscripts, both of \eqref{eq:theta_1,2} are versions of the \emph{odd} Jacobi theta function~\eqref{eq:vartheta}.

\section{Inozemtsev's trick for computing elliptic sums}
\label{app:Inozemtsev_trick}
\noindent
To compute the sums that appear in various places of the analysis of Inozemtsev's spin chain we often use the fact that two elliptic functions with the same pole structure must be equal up to a constant, as stipulated by Louiville's theorem. This allows us to find simple explicit quasiperiodic functions for complicated object of which we only know its poles and quasiperiodicity properties. This trick was used repeatedly by Inozemtsev~\cite{Inozemtsev:1989yq, inozemtsev1992extended,Inozemtsev_1995}.

Consider an analytic function $W$ that is doubly (quasi)periodic on the lattice $\!\!\bar{\,\,\mathbb{L}} = \mathbb{Z} \oplus \omega \, \mathbb{Z}$,
\begin{equation} \label{eq:W_double_quasiperiodicity}
W(z+1) = W(z) \, , \quad W(z+\omega) = \E^{\I \,\delta} \, W(z)\, ,
\end{equation}
where we assume $\delta \neq 0$ to ensure $W$ is not doubly periodic, and whose only pole in the fundamental parallelogram occurs at $z=0$, with Laurent expansion
\begin{equation}
\label{eq:WLaurent}
W(z) = \frac{c_{-2}}{z^2} + \frac{c_{-1}}{z} + c_0 + O(z) \, .
\end{equation}
For us the real periodicity of $W$ is typically true only on-shell, i.e.\ by virtue of the Bethe-ansatz equations~\eqref{eq:periodicitygenM}. 
In order to rewrite $W$ in terms of known elliptic functions we first engineer an elliptic function with first and second order poles at $z=0$. This goes as follows. The second-order pole is taken care of by  $\bar{\wp}(z)$, which is doubly periodic. For the first-order pole we use $\bar{\zeta}(z)$, which has the right expansion, but does not have the correct double quasiperiodicity properties. To remedy this we subtract another zeta function with argument shifted by some $s \neq 0$ to be determined: 
\begin{equation*}
\bar{\zeta}(z)-\bar{\zeta}(z+s) \, .
\end{equation*}
So far our ansatz therefore is
\begin{equation} \label{eq:W_ansatz_pre}
A\,\bar{\wp}(z) + B\,\Bigl(\bar{\zeta}(z)-\bar{\zeta}(z+s)\Bigr) \, ,
\end{equation}
which has the right double quasiperiodicity, but has a spurious pole at $z=-s$. 
To remove this spurious pole we multiply the ansatz by $\bar{\sigma}(z+s)$, and divide by $\bar{\sigma}(z-s)$ to compensate for the $z$-dependent behaviour under $z \mapsto z+\omega_b$,
\begin{equation*}
\frac{\bar{\sigma}(z+s)}{\bar{\sigma}(z-s)} \, \biggl( A\,\bar{\wp}(z) + B\,\Bigl(\bar{\zeta}(z)-\bar{\zeta}(z+s)\Bigr) \biggr). 
\end{equation*}
However, this prefactor introduces another spurious pole at $z=s$; to compensate for this we add a zero at that point by subtracting the value of \eqref{eq:W_ansatz_pre} at $z=s$:
\begin{equation*}
\frac{\bar{\sigma}(z+s)}{\bar{\sigma}(z-s)}\, \biggl(A\,\Bigl(\bar{\wp}(z)-\bar{\wp}(s)\Bigr) + B\,\Bigl(\bar{\zeta}(z)-\bar{\zeta}(z+s)-\bar{\zeta}(s)+\bar{\zeta}(2s)\Bigr) \biggr). 
\end{equation*}
To finish we note that the first factor has two quasiperiods, but only one quasiperiodicity parameter to tune. Multiplying the ansatz with the simplest doubly quasiperiodic, $\E^{q \, z}$, we arrive at the Weierstrass form of our function $W$:
\begin{equation}
\label{eq:Weierstrass_form_W_AB}
\E^{q \, z} \, \frac{\bar{\sigma}(z+s)}{\bar{\sigma}(z-s)} \, \biggl(A \, \Bigl(\bar{\wp}(z)-\bar{\wp}(s)\Bigr) + B\, \Bigl(\bar{\zeta}(z)-\bar{\zeta}(z+s)-\bar{\zeta}(s)+\bar{\zeta}(2s)\Bigr)\Bigr) \, . 
\end{equation}
This is the general form of a doubly quasiperiodic function with second and first-order poles at the origin. The parameters $q$ and $s$ are fixed by~\eqref{eq:W_double_quasiperiodicity}, yielding the system of linear equations
\begin{equation*}
q+ 2\, \bar{\eta}_1 \, s = 0 \, , \qquad q \, \omega +2 \, \bar{\eta}_2 \, s = \I\,\delta \, ,
\end{equation*}
with unique solution 
\begin{equation}
\label{eq:rdelta}
s= -\frac{\delta}{4 \pi} \, , \qquad q = \frac{\bar{\eta}_1}{2\pi} \, \delta \, . 
\end{equation}
The final step is to determine the residues~$A,B$. To this end we expand \eqref{eq:Weierstrass_form_W_AB} at $z=0$,
\begin{equation*}
-\frac{A}{z^2} + \frac{\bigl(q + 2\,\bar{\zeta}(s)\bigr) A - B}{z} + O(z^0) \, ,
\end{equation*}
which matches \eqref{eq:WLaurent} provided
\begin{equation}
A = -c_{-2} \, , \qquad B = c_{-2} \, \bigl(q + 2\,\bar{\zeta}(s)\bigr) - c_{-1} \, .
\end{equation}
Putting everything together we conclude that the function $W$ can be written as
\begin{equation}
\label{eq:Weierstrass_form_W}
\begin{aligned}
W(z) = \E^{q \, z} \, \frac{\bar{\sigma}(z+s)}{\bar{\sigma}(z-s)} \, \biggl( & {-}c_{-2} \,\Bigl(\bar{\wp}(z)-\bar{\wp}(s)\Bigr) \\
& + \Bigl(c_{-2}\,\bigl(q + 2\bar{\zeta}(s) \bigr) - c_{-1}\Bigr) \,\Bigl(\bar{\zeta}(z)-\bar{\zeta}(z+s)-\bar{\zeta}(s)+\bar{\zeta}(2s)\Bigr)\bigg). 
\end{aligned}
\end{equation}
supplemented by \eqref{eq:rdelta}.

Note that we did not discuss the possibility that the above expression for $W$ is off by a constant, since the quasiperiodicity (rather than periodicity) demands that such a constant is zero. This means in particular that we can express $c_0$, the zeroth order term in the Laurent expansion \eqref{eq:WLaurent} of $W$ in terms of the other expansion coefficients along with $s$ and $q$:
\begin{equation}
\label{eq:w0_Weierstrass_form}
c_0 = -c_{-2}\left(\bar{\wp}(s) +\tfrac{1}{2}q^2-2\,\bar{\zeta}(s)^2+q\,\bar{\zeta}(2s)+2\,\bar{\zeta}(s)\,\bar{\zeta}(2s) \right)+c_{-1}\bigl(q+\bar{\zeta}(2s)\bigr). 
\end{equation}
This will prove to be very useful for the computation of various sums, which we will now discuss in turn.

\subsection{Dispersion}
\label{app:dispersion}
Since there are some subtleties in the evaluation of the sums in \eqref{eq:dispersion_gen} needed to find the dispersion relation we include a detailed computation here. This also serves as a warm-up for the general-$M$ case in  \textsection\ref{app:general_M_derivation}. 
We wish to compute the sums
\begin{equation*}
\sum_{j=1}^{L-1} \wp(j) \, , \qquad \sum_{j=1}^{L-1} \E^{\I \, p \, j} \, \wp(j) \, .
\end{equation*}

We will start with the $p$-dependent sum, which is a variant of a sum computed in \cite{Inozemtsev:1989yq}. Following Inozemtsev let us consider the function
\begin{equation} \label{eq:W_definition_1} 
W(z) \coloneqq \sum_{j \in \mathbb{Z}_L} \E^{\I \mspace{1mu} p \mspace{1mu} (j+z)} \, \wp(j+z) \, .
\end{equation} 
The presence of the exponential can be exploited: it makes $W$ doubly quasiperiodic,
\begin{equation}
W(z+1) = W(z) \, , \qquad W(z+\omega) = \E^{\I \mspace{1mu} p_m \mspace{1mu} \omega} \, W_m(z) \, ,
\end{equation}
if{f} $\E^{\I \mspace{1mu} L \mspace{1mu} p} = 1$, which is required by the periodicity of the one-particle wave function anyway. Therefore $W$ can be expressed in the Weierstrass form \eqref{eq:Weierstrass_form_W} with the choices \eqref{eq:rdelta} with $\delta = p \, \omega$ for the two parameters $s$ and $q$:
\begin{equation}
s= -\frac{p \, \omega}{4 \pi}, \qquad q = \frac{\bar{\eta}_1}{2\pi} \, p \, \omega,
\end{equation}
including the now important restriction that we must have $s\neq 0$ to ensure $W$ is only quasiperiodic and not periodic. 
Since $W$ has only a second order pole at zero with residue $c_{-2} = 1$ we can use \eqref{eq:w0_Weierstrass_form} to find that
\begin{equation}
-c_0 = \bar{\wp}(s) + \tfrac{1}{2} \, q^2 - 2\, \bar{\zeta}(s)^2 + q \, \bar{\zeta}(2s) + 2\, \bar{\zeta}(s) \, \bar{\zeta}(2s) \, ,
\end{equation}
which by computing the zeroth-order term of the sum directly is seen to equal the sum
\begin{equation*}
\sum_{j=1}^{L-1} \wp(j) \, \E^{\I \, p \, j}
\end{equation*}
we are after. Plugging in $s$ and $q$ and using the doubling formulae for $\bar{\wp}$ and $\bar{\zeta}$ we conclude
\begin{equation} \label{eq:disp_sum_p}
\sum_{j=1}^{L-1} \wp(j) \, \E^{\I \, p \, j} = \frac{1}{2} \bar{\wp}\Bigl(\frac{\omega p}{2\,\pi}\Bigr) -\frac{1}{2} \left(   \bar{\zeta}\Bigl(\frac{\omega p}{2\,\pi}\Bigr) - \frac{\bar{\eta}_2}{\omega} \frac{\omega p}{2\,\pi} \right)^{\!2} \qquad \text{provided} \quad p \in \frac{2\pi}{L} \,(\mathbb{Z}_L \setminus \{0\} ) \, .
\end{equation}
This equality only holds when $p = 2\pi I/L$ with $I \in \mathbb{Z}_L$ quantised by the periodicity condition,\,%
\footnote{\ We stress that \eqref{eq:disp_sum_p} only holds `on shell'; for $p \notin  (2\pi/L) \, \mathbb{Z}_L$ the left-hand side typically is not even real. This not related to $\wp$ but rather a feature of the general sum \eqref{eq:dispersion_gen}, and already applies to Heisenberg.}
and moreover $s\neq 0$, which implies $I \neq 0 \text{ mod } L$.

It remains to compute the sum with $p=0$, which we also use to get the elliptic $M$-particle difference equation~\eqref{eq:M_particle_diff_eq_ell}.\,%
\footnote{\ Another way to compute this sum can be found in \cite{Finkel_2014}.} To this end we analyse the function 
\begin{equation} \label{eq:W_definition_2}
W(z) = \sum_{j=0}^{L-1} \wp(z+j),
\end{equation}
which is obviously doubly periodic with periods $(1,\omega)$ and has a double pole at the origin. This means that 
\begin{equation*}
W(z) = \bar{\wp}(z) +C,
\end{equation*}
with $C$ a constant. By performing the Laurent expansion we see that $C$ equals the sum we are after. We can find an expression for $C$ by analysing $W(z) - \bar{\wp}(z)$: as long as we pick $\tau$ such that we do not pass any poles we have
\begin{equation*}
	\begin{aligned}
	C \, \omega & = \int_{\tau -\omega/2}^{\tau +\omega/2} \bigl( W(z) - \bar{\wp}(z) \bigr) \mathrm{d}z \\
	& = \bar{\zeta}\Bigl(\tau  + \frac{\omega}{2}\Bigr) - \bar{\zeta}\Bigl(\tau - \frac{\omega}{2}\Bigr) - \sum_{j=0}^{L-1} \left( \bar{\zeta}\Bigl(\tau + j + \frac{\omega}{2}\Bigr) - \bar{\zeta}\Bigl(\tau + j - \frac{\omega}{2}\Bigr) \right) .
	\end{aligned}
\end{equation*}
Differentiation with respect to $\tau$ shows that this expression, and in fact each pair of $\zeta$s, is constant. This allows us to choose $\tau$ as we like. Sending $\tau \to 0$ (or to $-n$ for the summands) we find
\begin{equation*}
C \, \omega = 2 \left( \bar{\zeta}(\omega/2) - L \, \zeta(\omega/2)\right) \, .
\end{equation*}
We conclude that
\begin{equation} \label{eq:disp_sum_p=0}
\sum_{j=1}^{L-1} \wp(j) = \frac{\bar{\eta}_2 - L \, \eta_2}{\omega}  = \bar{\eta}_1 - \eta_1 \, . 
\end{equation}

\subsection{Highest-weight sum} \label{app:hw_sum} 
In order to prove the highest-weight property of the $M=2$ eigenfunctions we need to compute the sum 
\begin{equation}
	\sum_{n=1}^{L-1} \! \E^{\I\, p_1\,n} \, \chi_2(n,\gamma)\, .
\end{equation}
To compute it we first introduce the function
\begin{equation}
	W(z) = \sum_{n'= 0 }^{L-1} \E^{\I p_1 (n'+z)} \chi_2(n'+z,\gamma)\,  ,
\end{equation}
which, due to the periodicity properties of $\chi_2$ has the following periodicity properties:
\begin{equation}
	W(z+1) = W(z)\, , \quad  W(z+\omega) = \E^{\I p_1 \omega} W(z)\, ,
\end{equation}
where the first equation is only true on-shell, i.e. as long as $p_1$ and $\gamma$ satisfy \eqref{eq:M=2_BAE}. Further inspection shows that $W$ has only one pole in the fundamental parallelogram, of first order. This means that we can again postulate that $W$ is of the form \eqref{eq:Weierstrass_form_W}, where we can now choose $w_ {-2} = 0$ and find the parameters $q$ and $s$ from the quasiperiods (for which we must assume that $p_1 \neq 0$). To continue we first note that at $z=0$ $W$ expands as
\begin{equation}
	W(z) = \frac{1}{z} + \I p_1 + \rho_2(\gamma) +\sum_{n' (\neq 0) }^{L-1} \E^{\I p_1 n'} \chi_2(n',\gamma)+O(z)\, . 
\end{equation}
Using the relation between the Laurent coefficients given in \eqref{eq:w0_Weierstrass_form} we straightforwardly find
\begin{equation}
	\begin{aligned}
		\sum_{n' (\neq 0) }^{L-1} \E^{\I p_1 n'} \chi_2(n',\gamma) = - \rho_2(\gamma) -\I p_1  -\bar{\rho}_1\left(\frac{\omega\,p_1}{2\pi}\right) = -\rho_2(\gamma)-\bar{\rho}_2\left(\frac{\omega\,p_1}{2\pi}\right) \, .
	\end{aligned}
	{\checkedMma}
\end{equation}

\subsection{\textit{M}-particle sums}
\label{app:general_M_derivation}
Recall the short-hand $\vect{r} \coloneqq \vect{p} - \tilde{\vect{p}}$. We wish to compute the sum
\begin{equation}
\Sigma_m(\vect{n}) \coloneqq \sum_{j (\notin \vect{n})}^L \! \wp(n_{m}-j) \, [\widetilde{\Psi}_{\tilde{\vect{p}}}(\vect{n}) \, \E^{\I \vect{r}\cdot \vect{n}}]_{n_m \mapsto j} \, 
\end{equation} 
defined in \eqref{eq:M_particle_general_M_sum}.  
Finding an explicit expression for this sum is the main task in order to analyse the difference equation and follows the strategy of Inozemtsev~\cite{Inozemtsev_1995}. 

We view $\vect{n}$ as fixed and introduce the function 
\begin{equation}
\label{eq:Wm_definition}
W_m(z) \coloneqq \sum_{n' \in \mathbb{Z}_L} \wp(n_{m}- n' -z) \, \widetilde{\Psi}_{\vect{r}}(n_{1},\dots\mspace{-1mu},n' +z,\dots\mspace{-1mu}, n_{M}),
\end{equation}
which obeys the periodicity conditions 
\begin{equation}
\label{eq:W_periodicity}
W_m(z+1) = W_m(z), \quad W_m(z+\omega) = \E^{\I p_m \omega} \, W_m(z),
\end{equation}
using the Bethe-ansatz equations \eqref{eq:periodicitygenM}. 
Its only pole in the fundamental parallellogram occurs at zero with Laurent expansion matching \eqref{eq:WLaurent}. Therefore $W_m$ can be expressed as \eqref{eq:Weierstrass_form_W} with $q$ and $s$ determined by \eqref{eq:rdelta} with $\delta = p_m \, \omega$. Because of this we know that we can express the constant term $c_0$ of $W_m$ as in \eqref{eq:w0_Weierstrass_form}, which we will here sometimes abbreviate as $c_0 = Y_m \, c_{-2} + X_m \, c_{-1}$. On the other hand we can directly evaluate the Laurent coefficients $c_{-2}$, $c_{-1}$ and $c_0$ by performing the relevant residue integrals, see below. Recall that $\widetilde{\Phi}_{\tilde{p}}(\vect{n})$ denotes the numerator from \eqref{eq:psi_decomposition} and abbreviate $\widetilde{\Phi}_{\tilde{\vect{p}},\vect{p}}(\vect{n}) \coloneqq \widetilde{\Phi}_{\tilde{p}}(\vect{n}) \, \E^{\I \vect{r} \cdot \vect{n}}$. Then we claim that
\begin{equation}
\label{eq:w0_W}
\begin{aligned}
c_0 = {} & \Sigma_m(\vect{n}) + \frac{1}{2} \, \partial^2_{n_m} \widetilde{\Psi}_{\vect{r}}(\vect{n}) \\
& + \frac{
1}{\Delta(\vect{n})} \sum_{m' (\neq m)}^M \!\!\! \Pi_{mm'}(\vect{n}) \Bigg( \biggl(\frac{\wp'(n_{m'} - n_{m})}{\wp(n_{m'} - n_{m})} + \!\! \sum_{m''(\neq m,m')}^{M} \!\!\!\!\!\!\! \zeta(n_{m''} - n_{m'}) \biggr) \, \bigl[ \widetilde{\Phi}_{\tilde{\vect{p}},\vect{r}}(\vect{n})\bigr]_{n_m=n_{m'}} \\
& \hphantom{+ \frac{1}{\Delta(\vect{n})} \sum_{m' (\neq m)}^M \!\!\! \Pi_{mm'}(\vect{n}) \Bigg( } \ + \bigl[ \partial_{n_m} \widetilde{\Phi}_{\tilde{\vect{p}},\vect{r}}(\vect{n})\bigr]_{n_{m} = n_{m'}} \Bigg),
\end{aligned}
\end{equation}
where
\begin{equation}
\Pi_{m m'}(\vect{n}) \coloneqq \wp(n_m-n_{m'}) \, \sigma(n_{m}-n_{m'})\prod_{m''(\neq m,m')}^M \frac{\sigma(n_{m''} - n_{m})}{\sigma(n_{m''} - n_{m'})} \, .
\end{equation}
By equating this expression for $c_0$ with the expression \eqref{eq:w0_Weierstrass_form} containing $c_{-2}$ and $c_{-1}$ we find an expression for $\Sigma_m(\vect{n})$, which can be plugged into the $M$-particle difference equation \eqref{eq:M_particle_general_M}.

Using the definition \eqref{eq:Wm_definition} we can compute the Laurent coefficients of $W_m$ directly. 

\textit{Computing $c_{-2}$.} To find $c_{-2}$ we only need to consider the second-order poles in the sum, which is relatively easy. Since we assumed that the summands in the ansatz have no second-order poles they can only occur in the $\wp$ function and restricted to the $(1,\omega)$ lattice the only relevant one occurs when $n'=n_m$. Thus we find
\begin{equation}
c_{-2} = \widetilde{\Psi}_{\vect{\tilde{p}},\vect{r}}(\vect{n}). 
\end{equation}

\textit{Computing $c_{-1}$.} Things already get a little more interesting when we consider the first order poles, as there are many more of those. Nevertheless, let us start noting that there is also a contribution coming from the second-order pole at $n'=n_m$, yielding $\partial_{n_m} \widetilde{\Psi}_{\vect{\tilde{p}},\vect{r}}$. Further contributions from this pole combined with possible zeroes that compensate are impossible, since such a zero would mean that the entire eigenfunction is zero. Therefore all further contributions come from the first order poles in the eigenfunctions in the summands with $n'\neq n_{m}$. So let us set $n'=n_{m'}$ with $m'\neq m$, so that the summand reads
\begin{equation*}
 \wp(n_m-n_{m'}-z) \, \frac{\widetilde{\Phi}_{\tilde{\vect{p}},\vect{r}}(n_1,\ldots,n_{m'}+z,\ldots,n_{M})}{\Delta(n_1,\ldots,n_{m'}+z,\ldots,n_{M})} \, .
\end{equation*} 
As the pole comes from $1/\Delta$ we can set $z=0$ everywhere else in the expression, yielding 
\begin{equation*}
\begin{aligned}
 &\wp(n_m-n_{m'}) \, \widetilde{\Phi}_{\tilde{\vect{p}},\vect{r}}(n_1,\ldots,n_{m'},\ldots,n_{M}) \\
 & \times \Biggl(\, \prod_{\substack{m''< m''' \\ (\neq m)}}^M \!\!\!\! \sigma(n_{m''} - n_{m'''}) \prod_{m''=1}^{m-1} \!\! \sigma(n_{m''} - n_{m'}-z) \prod_{m'''(>m)}^M \!\!\!\!\! \sigma(n_{m'}+z - n_{m'''}) \Biggr)^{\!\! -1}.
 \end{aligned}
\end{equation*} 
If $m'<m$ the residue is
\begin{equation*}
\begin{aligned}
 & {-}\wp(n_m-n_{m'}) \, \widetilde{\Phi}_{\tilde{\vect{p}},\vect{r}}(n_1,\ldots,n_{m'},\ldots,n_{M}) \\
 & \times \Biggl(\, \prod_{\substack{m''< m''' \\ (\neq m)}}^M \!\!\!\! \sigma(n_{m''} - n_{m'''}) \prod_{m'' (\neq m')}^{m-1} \!\! \sigma(n_{m''} - n_{m'}) \prod_{n'''(>m)}^M \!\! \sigma(n_{m'} - n_{m'''}) \Biggr)^{\!\! -1} \\
 = {} & {-}\wp(n_m-n_{m'}) \, \widetilde{\Phi}_{\tilde{\vect{p}},\vect{r}}(n_1,\ldots,n_{m'},\ldots,n_{M}) \\
 & \times \Biggl(\,\prod_{m''< m''' }^M \!\!\!\! \sigma(n_{m''} - n_{m'''}) \prod_{m'' (\neq m')}^{m-1} \frac{\sigma(n_{m''} - n_{m'})}{\sigma(n_{m''} - n_{m})} \prod_{m'''(>m)}^M \frac{\sigma(n_{m'} - n_{m'''})}{\sigma(n_{m} - n_{m'''})} \Biggr)^{\!\! -1} \sigma(n_{m'}-n_{m}) \, ,
 \end{aligned}
\end{equation*} 
where we noted the minus sign explicitly coming from the minus in front of the $z$ that gives the pole and introduced extra factors of $\sigma$ to simplify the products. Namely, we see that the first product is basically $\Delta(n_1,\ldots,n_M)$.
So we find
\begin{equation*}
\wp(n_m-n_{m'}) \, \frac{ \widetilde{\Phi}_{\tilde{\vect{p}},\vect{r}} (n_1,\ldots,n_{m'},\ldots,n_{M}) }{ \Delta(n_1,\ldots,n_M) } \, \sigma(n_{m}-n_{m'})\prod_{m''(\neq m,m')}^M \frac{\sigma(n_{m''} - n_{m})}{\sigma(n_{m''} - n_{m'})} \, . 
\end{equation*}
Repeating the same argument for $m'>m$ one finds the same result, the lack of a minus sign being compensated for by the oddness of the separate $\sigma$ function. This expression is the same as that obtained by Inozemtsev in \cite{Inozemtsev_1995} up to a factor $\wp(n_m-n_{m'})$. So we finally have
\begin{equation}
\begin{aligned}
c_{-1} &= \partial_{n_m} \widetilde{\Psi}_{\vect{\tilde{p}},\vect{r}} \, + \, \frac{1}{\Delta(\vect{n})} \! \sum_{m'(\neq m)}^M \!\!\! \widetilde{\Phi}_{\tilde{\vect{p}},\vect{r}}(n_1,\ldots,n_{m'},\ldots,n_{M}) \, \Pi_{m m'}(\vect{n}) \, ,
\end{aligned}
\end{equation}
where we define\,\footnote{\ This slightly extends the definition of what Inozemtsev denotes by $T$ in the equation below (23c) in \cite{Inozemtsev_1995}.}
\begin{equation}
\Pi_{m m'}(\vect{n}) \coloneqq \wp(n_m-n_{m'}) \, \sigma(n_{m}-n_{m'})\prod_{m'' (\neq m,m')}^M \frac{\sigma(n_{m''} - n_{m})}{\sigma(n_{m''} - n_{m'})} \, .
\end{equation}

\textit{Computing $c_0$.} Of course now comes the most important coefficient, the one that contains the sum we are actually after. It has three contributions: the direct contribution from setting $z=0$ is exactly $\Sigma_m(\vect{n})$. The contribution from the second order pole is also straightforward, as it is simply $\partial_{n_m}^2 \widetilde{\Psi}_{\vect{r}}/2$. 
The contribution coming from the first order poles at $n'=n_{m'}$ is the most complicated, as it requires us to differentiate 
\begin{equation*}
\begin{aligned}
 &\wp(n_m-n_{m'}-z) \, \widetilde{\Phi}_{\tilde{\vect{p}},\vect{r}}(n_1,\ldots,n_{m'}+z,\ldots,n_{M}) \\
 & \times \Biggl(\, \prod_{\substack{m''< m''' \\ (\neq m)}}^M \!\!\!\! \sigma(n_{m''} - n_{m'''}) \prod_{m''=1}^{m-1} \!\! \sigma(n_{m''} - n_{m'}-z) \prod_{m'''(>m)}^M \!\! \sigma(n_{m'}+z - n_{m'''}) \Biggr)^{\!\! -1}
 \end{aligned}
\end{equation*} 
with respect to $z$ before setting it to zero. Doing this yields
\begin{equation*}
\frac{\Pi_{mm'}(\vect{n})}{\Delta(\vect{n})} \Biggl(\, \biggl(\, \frac{\wp'(n_{m'} - n_{m})}{\wp(n_{m'} - n_{m})} +\sum_{m''(\neq m,m')}^{M} \!\!\!\!\!\!\! \zeta(n_{m''} - n_{m'}) \biggr) F_{\vect{r}}(\vect{n})\big|_{n_m=n_{m'}} +\partial_{n_m} \widetilde{\Phi}_{\tilde{\vect{p}},\vect{r}}(\vect{n})\big|_{n_{m} = n_{m'}}  \Biggr) \, .
\end{equation*}
In this way we obtain \eqref{eq:w0_W}.

\section{Further proofs} \label{app:further_proofs}

\subsection{A simplification} \label{app:simplification}
Before we can plug \eqref{eq:w0_W} into the $M$-particle difference equation~\eqref{eq:M_particle_general_M} let us consider the action of the permutation $w \in S_M$ on the coordinates in $\Sigma(\vect{n}_w)$. In the first line of \eqref{eq:w0_W} we can simply replace $\vect{n} \to \vect{n}_w$, while in the second line the antisymmetry of the Vandermonde factor, $\Delta(\vect{n}_w) = \mathrm{sgn}(w) \, \Delta(\vect{n})$, yields a total antisymmetrisation. This allows the summands in that line to be simplified in pairs (whence the factor 1/2 in front) with permutations differing by the transposition $(m,m') \in S_M$. Indeed, define
\begin{equation}
\begin{aligned}
Z_{mm'}(\vect{n}) \coloneqq \Pi_{mm'}(\vect{n}) \Bigg( & \biggl(
\frac{\wp'(n_{m'} - n_{m})}{\wp(n_{m'} - n_{m})} -X_m +\sum_{\alpha(\neq m,m')}^{M} \zeta(n_{\alpha} - n_{m'}) \biggr)
\widetilde{\Phi}_{\tilde{\vect{p}}}(\vect{n}) \big|_{n_m=n_{m'}} \\
&+\partial_{n_m} 
\widetilde{\Phi}_{\tilde{\vect{p}}}(\vect{n}) \, \E^{\I \vect{r}\cdot \vect{n}} \big|_{n_{m} = n_{m'}} 
\Bigg).
\end{aligned}
\end{equation}
Then it follows that
\begin{equation} 
\label{eq:M_auxiliary_sum}
\frac{1}{\Delta(\vect{n})} \sum_{w\in S_M} \!\! \mathrm{sgn}(w) \!\! \sum_{m\neq m'}^M \!\! Z_{mm'}(\tilde{n}) = 
\frac{1}{2\,\Delta(\vect{n})}\sum_{w\in S_M} \!\! \mathrm{sgn}(w) \! \sum_{m\neq m'}^M \! \left( Z_{mm'}(\tilde{n}) - Z_{m'm}(\tilde{n}_{\tau})\right),
\end{equation}
where $\tau \coloneqq (m m')$ is a transposition. By noting that 
\begin{equation}
\begin{aligned}
\Pi_{mm'}(\vect{n}) &= \Pi_{m'm}(\vect{n}_{\tau}) \\
\widetilde{\Phi}_{\tilde{\vect{p}}}(\vect{n}) \, \E^{\I \vect{r}\cdot \vect{n}} \big|_{n_m = n_{m'}} & = 
\widetilde{\Phi}_{\tilde{\vect{p}}}(\vect{n}_\tau) \, \E^{\I \vect{r}\cdot \vect{n}_\tau} \big|_{n_{\tau(m')} = n_{\tau(m)}}, \\
\frac{\wp'(n_{m'} - n_{m})}{\wp(n_{m'} - n_{m})} &= \frac{\wp'(n_{\tau(m)} - n_{\tau(m')})}{\wp(n_{\tau(m)} - n_{\tau(m')})}, \\
\sum_{m''(\neq m,m')}^{M} \!\!\!\!\!\!\!\! \zeta(n_{m''} - n_{m'}) &= \sum_{m''(\neq m,m')}^{M} \!\!\!\!\!\!\!\! \zeta(n_{\tau(m'')} - n_{\tau(m)}),
\end{aligned}
\end{equation}
we can drastically simplify 
\begin{equation}
\label{eq:M_particle_rhs}
Z_{mm'}(\vect{n}) - Z_{m'm}(\vect{n}_{\tau}) = \Pi_{mm'}(\vect{n})\left( (\partial_{n_m}-X_m)  - (\partial_{n_{m'}}-X_{m'}) \right) 
\widetilde{\Phi}_{\tilde{\vect{p}}}(\vect{n}) \, \E^{\I \vect{r}\cdot \vect{n}} \, \big|_{n_{m} = n_{m'}}.
\end{equation}
Plugging \eqref{eq:w0_W} with $\vect{n} \to \vect{n}_w$ into the $M$-particle difference equation~\eqref{eq:M_particle_general_M} yields a left-hand side containing the sum \eqref{eq:M_auxiliary_sum}. Using \eqref{eq:M_particle_rhs} we obtain \eqref{eq:M_particle_plugged_in}. 

\subsection{Derivatives of the numerator} 
\label{app:derivative_numerator}
By assuming that $\widetilde{\Psi}_{\tilde{\vect{p}}}$ is an eigenfunction of the eCS Hamiltonian at $\beta=2$ we can deduce further information about its numerator $\widetilde{\Phi}_{\tilde{\vect{p}}} = \Delta \, \widetilde{\Psi}_{\tilde{\vect{p}}}$, see~\eqref{eq:psi_decomposition}, from the eCS eigenvalue equation. Carefully cancelling the denominator~$\Delta$ we find 
\begin{equation*}
\begin{gathered}
\sum_{m=1}^M \partial^2_{x_m} \widetilde{\Phi}_{\tilde{\vect{p}}}(\vect{x}) + \left( 2 \, \widetilde{E} - \sum_{m\neq m'}^M \!\! \wp(x_m-x_{m'}) + \sum_{m=1}^M \biggl(\, \sum_{m'(\neq m)}^M \!\!\!\! \zeta(x_m-x_{m'}) \biggr)^{\!\! 2} \ \right) \mspace{-1mu} \widetilde{\Phi}_{\tilde{\vect{p}}}(\vect{x}) \\
= \sum_{m\neq m'}^M \!\!\zeta(x_m -x_{m'}) \, (\partial_{m} - \partial_{m'}) \, \widetilde{\Phi}_{\tilde{\vect{p}}}(\vect{x}) \, .
\end{gathered}
\end{equation*}
This looks slightly different than equation (62b) in \cite{Inozemtsev:2002vb}
but luckily we can draw the same conclusions: the left-hand side of this equation is analytic, as every pole from $\wp$ gets cancelled by one from the square of $\zeta$ functions. Since the right-hand side has $\zeta$ functions as well (and therefore poles), we see that it must be true that
\begin{equation*}
(\partial_{m} - \partial_{m'}) \, \widetilde{\Phi}_{\tilde{\vect{p}}}(\vect{x}) \big|_{x_{m} = x_{m'}} = 0
\end{equation*}
to ensure the analyticity of the right-hand side.

\subsection{Trivial symmetry map} \label{app:ts} In this appendix we try to give some more explanation to one of the central tools Inozemtsev used in \cite{inozemtsev1993Hermitelike} to perform his version of the counting of the $M=2$ solutions as we have done in \textsection\ref{sec:counting}, because it turns out to be a nice piece of Weierstrass analysis that can be applicable in other situations. In order to facilitate comparison with Inozemtsev's work we will work on the lattice $\!\!\hat{\,\,\mathbb{L}}$ with periods $(L, L\,\omega)$, which is the coordinate version of the lattice $\!\!\momphi{\,\, \mathbb{L}}$ defined in \eqref{eq:momphi_lattice}. 

Let us first define the map $\hat{R}_x$ on the fundamental parallelogram of $\!\!\hat{\,\,\mathbb{L}}$ by requiring it to satisfy $x_{\hat{R}_x(\gamma)} = x_{\gamma}$, i.e.\ $\gamma$ and $\hat{R}_x(\gamma)$ belong to the same solution of \eqref{eq:M=2curveequations}. Here $\gamma$ relates to $\varphi$ as in \eqref{eq:varphi_vs_gamma}. Note that there is a unique $\gamma' \neq \gamma$ inside the fundamental parallelogram with the same image under $\wp$ since the latter is a second-order elliptic function, implying the map $\hat{R}_x$ is indeed well-defined. 

More precisely, let $\gamma$ be in the fundamental parallelogram and 
\begin{equation}
x_{\gamma} = \wp(\gamma -I_\text{tot} \, \omega/2) \in \mathbb{C}\, . 
\end{equation}
We can find this $\gamma'$ by congruence:
\begin{equation}
x_{\gamma} = \wp\bigl(-(\gamma- \omega \, I_\text{tot}/2)\bigr) = \wp((-\gamma+  \omega I_\text{tot} ) - I_\text{tot} \omega/2 ) \, ,
\end{equation}
which tells us that the second solution is
\begin{equation}
\gamma' = -\gamma+ \omega \, I_\text{tot} \,  \ \mathrm{mod }\,\,\!\!\hat{\,\,\mathbb{L}} \, ,
\end{equation}
i.e. we can define the map $\hat{R}_x$ by
\begin{equation}
\label{eq:TS_as_a_mod}
\hat{R}_x(\gamma) \coloneqq (-\gamma+ I_\text{tot} \, \omega)\, \mathrm{mod }\,\, \!\!\hat{\,\,\mathbb{L}}\, .
\end{equation}
It is not so hard to work out what the modulo means in practice and derive a formula for $\hat{R}_x(\gamma)$. By assumption $0\leq \Re\gamma < L$ and $0\leq \Im\,\gamma\leq L |\omega|$. First we consider the real part of $\gamma'$. If $\gamma$ is purely imaginary then so is $\gamma'$ and we do not have to adjust the real part, while if $\Re\gamma$ is positive then $\Re(I_\text{tot} \, \omega -\gamma)<0$ so we should add $L$ to land in the fundamental parallelogram. In total, then, the real shift can be written as $L \, \mathrm{sgn}(\Re\gamma)$, with the convention that $\mathrm{sgn}\,0 = 0$. To determine the imaginary shift there are three cases to consider. If $I_\text{tot} \, \Im\omega - \Im \gamma$ is positive we do not have to do anything, while if it is negative we can add $L\,\omega$ to land in the fundamental parallelogram, and when
\begin{equation}
\label{eq:equal_im_bound}
I_\text{tot} \, \Im\omega - \Im\gamma =0
\end{equation}
then $\gamma' = -\Re\gamma$ is real. This means only the real shift by $L$ is necessary in order for the solution to be in the fundamental parallelogram. We therefore find that 
\begin{equation}
\gamma '= \hat{R}_x(\gamma) = I_\text{tot} \, \omega - \gamma + L \, \mathrm{sgn}(\Re \gamma) + L \, \omega\, \theta(I_\text{tot}\, |\omega| -\Im\gamma )\, ,
\end{equation}
where $\theta$ is the Heaviside step function satisfying $\theta(0) = 0$. 

This expression looks a lot like the equation for $\gamma_0'$ below (14) in \cite{inozemtsev1993Hermitelike}, with the only difference sitting in the final term. 
Before exploring this difference in more details, let us first note that the map $\hat{R}_x$ has a nice geometric meaning. Indeed, $\gamma' = \hat{R}_x(\gamma)$ is the unique second solution to $x_{\gamma'} = x_\gamma$, which has $y_{\hat{R}_x(\gamma)} = -y_{\gamma}$. Therefore $\hat{R}_x$ maps a point $(x_{\gamma}, y_{\gamma})$ on the elliptic curve to the point $(x_{\gamma}, - y_{\gamma})$: it is nothing but a reflection in the $x$-axis. 

To investigate the difference between $\hat{R}_x$ and the expression for $\gamma_0'$ in \cite{inozemtsev1993Hermitelike}, let us first note that on the set of solutions to the constraints the maps are nearly identical, with the only difference occurring when \eqref{eq:equal_im_bound} holds. However, solutions like this are extremely rare: consider such a solution $\gamma = \gamma_\text{r} + I_\text{tot} \, \omega$ where $\gamma_\text{r} \in [0,L]$. Plugging in this form into the constraint yields
\begin{equation}
2 \, \rho_2(\gamma_\text{r} ) - \bar{\rho}_2\biggl(\frac{\gamma_\text{r}}{L}\biggr) - \bar{\rho}_2\biggl(\frac{\gamma_\text{r}+I_\text{tot} \, \omega}{L}\biggr) = 0\, , 
\end{equation}
where we used the periodicity of $\rho_2$ to omit the shift by $I_\text{tot} \, \omega$ in the first term. Now note that $\rho_2$ and $\bar{\rho}_2$ are analytic functions away from poles (i.e.\ meromorphic) and that there is only one term here that will generate a non-trivial imaginary part as long as $I_\text{tot} \not\cong 0 $ mod $L/2$.\,%
\footnote{\ One can show this by noting that the complex conjugate $\rho_2(\gamma_\text{r} +\I\, \gamma_\text{i})^* = \rho_2 (\gamma_\text{r} - \I \, \gamma_\text{i})$, thus the only way this term is real is when $\rho_2 (\gamma_\text{r} - \I\, \gamma_\text{i}) = \rho_2 (\gamma_\text{r} + \I\, \gamma_\text{i})$, which only happens if $\I \, \gamma_\text{i} = L \omega /2$.}

Thus this can happen only if $I_\text{tot} = 0,L/2$, but $I_\text{tot} = 0$ yields the trivial solution $\gamma = L/2$. Choosing $I_\text{tot} = L/2$ yields a unique non-trivial solution with $\gamma_\text{r} = L/2$, i.e.\ $\gamma = L/2 + L\omega/2$. This is not a trivial root: the two-particle wave function is nonzero for this $\gamma$. Now we see that $\hat{R}_x(L/2 + L\omega/2) = L/2$, while $\gamma'_0 = L/2 + L\omega/2$. So both expressions send this solution to a solution in the fundamental domain, but in the case discussed in \cite{inozemtsev1993Hermitelike} we have $\gamma'_0 = \gamma_0$. 

Let us finally note that this analysis can be performed on any of the lattices. For our purposes the lattice $\!\!\hat{\,\,\mathbb{L}}^{\mspace{-4mu}\vee}$, in terms of $\varphi$, is most convenient. Following the preceding steps one readily finds that on that lattice the action of the reflection~$\momphi{R}_x$ is given by
\begin{equation}
\momphi{R}_x(\varphi) \coloneqq -\varphi -2\pi \,I_\text{tot} + \frac{2\pi L}{\omega} \,\mathrm{sgn}(\Im\varphi) + \,  2\pi L \, \theta(2\pi \, I_\text{tot} + \Re\varphi) \, . 
\end{equation}

\subsection{Trivial roots are nonphysical}
\label{app:trivialsols}
In \textsection\ref{sec:M=2} we simplified \eqref{eq:ell_curve} by discarding the four simple explicit solutions~\eqref{eq:trivial_roots}. Let us show that this is allowed by analysing these trivial solutions. First, all the trivial solutions of \eqref{eq:ell_curve} can be identified with the solutions to\,%
\footnote{\ Inozemtsev first noticed this in \cite{inozemtsev1993Hermitelike}, but his formula includes the rare but nontrivial solution from \textsection\ref{app:ts}.}
\begin{equation}
\label{eq:TS_gamma}
\hat{R}_x(\gamma) = \gamma \, ,
\end{equation}
which on the elliptic curve is easy to see: checking that all our trivial solutions satisfy the above equation is straightforward, so let us consider the converse. If $x_{\hat{R}_x(\gamma)} = x_{\gamma} \in \mathbb{C}$, then we have $y_{\hat{R}_x(\gamma)} = -y_{\gamma}$, so in particular if $\hat{R}_x(\gamma) = \gamma$ we find $y_{\gamma} = -y_{\gamma}$, i.e.\ $y_{\gamma} = 0$. This yields the last three trivial solutions listed in \eqref{eq:trivial_roots}. The fourth trivial root at $\gamma = I_\text{tot} \, \omega/2$ is slightly more subtle: for this value $x_\gamma$ and $y_\gamma$ diverge, i.e. $x_\gamma \notin \mathbb{C}$. This implies we cannot follow the approach above as there is only one point for which $x_\gamma$ diverges in the fundamental parallelogram. Moreover, we technically have not proven the validity of \eqref{eq:TS_as_a_mod} for such $\gamma$. Nevertheless, by the previous we see that if $x_{\hat{R}_x(\gamma)} = x_\gamma \notin \mathbb{C}$ it must be that $\gamma$ lies at a pole, which can only be $\gamma = I_\text{tot} \, \omega/2$. 
Thus we see that the set of trivial solutions is in one-to-one correspondence with the solutions to \eqref{eq:TS_gamma}.

Now we would like to furthermore demonstrate that the two-particle wave function corresponding to a trivial root vanishes. On the lattice $\!\!\hat{\,\,\mathbb{L}}^{\mspace{-4mu}\vee}$ the roots in question are 
\begin{equation*}
\varphi = -I_\text{tot} \, \pi \, , \quad  \varphi = (L-I_\text{tot}) \, \pi\, , \quad \varphi = -I_\text{tot} \, \pi  + \I \,L\,\kappa \, , \quad \varphi = (L-I_\text{tot})\, \pi + \I \,L\,\kappa \, ,
\end{equation*}
which on the corresponding coordinate lattice $\!\!\hat{\,\,\mathbb{L}}$ yields
\begin{equation}
\label{eq:app_triv_roots}
\gamma = \frac{I_\text{tot} \, \omega}{2} \, ,\quad  \gamma = \frac{I_\text{tot} \, \omega  + L}{2}, \quad \gamma = \frac{(I_\text{tot}+L) \, \omega  }{2} \, , \quad \gamma = \frac{(I_\text{tot}+L) \, \omega +L  }{2} \, .
\end{equation}
The $M=2$ wave function can be written in terms of $\gamma$ as
\begin{equation}
\label{eq:M=2_wf_bargamma}
\Psi \sim \frac{2}{\sigma(\gamma)} \left(\exp\left(\frac{2\pi \I \, I_\text{tot} \, n_1 + 2 \, \eta_1 \gamma \, \delta n  }{L}\right) \sigma(\gamma-\delta n) +\exp\left(\frac{2\pi \I\, I_\text{tot} \,n_2}{L}\right)  \sigma(-\gamma-\delta n)  \right),
\end{equation}
where we have omitted irrelevant regular non-vanishing factors. The overall factor is finite as long as $\gamma$ does not lie on the lattice. The mechanism for vanishing is the same for all four roots: we check that at these roots the two occurring sigma functions are the same by applying the quasi-periodicity rule we defined just below \eqref{eq:Legendre relation} as often as necessary. The easiest one is the root $\gamma = I_\text{tot} \, \omega/2$: by applying the quasi-periodicity rule $I_\text{tot}$ times we obtain 
\begin{equation}
\sigma\left( \frac{I_\text{tot} \, \omega}{2} -\delta n\right) = (-1)^{I_\text{tot}} \, \E^{-\delta n \, I_\text{tot} \, \eta_2} \, \sigma\left( -\frac{I_\text{tot} \, \omega}{2} -\delta n\right),
\end{equation}
thus we can rewrite the first factor in the brackets of \eqref{eq:M=2_wf_bargamma} as  
\begin{equation*}
\begin{aligned}
&(-1)^{I_\text{tot}} \exp\biggl(\frac{2\pi \I \, n_1 + 2\gamma\, \eta_1 \delta n  }{L} -\delta n \, I_\text{tot} \, \eta_2 \biggr) \, \sigma\biggl( -\frac{I_\text{tot} \omega}{2} -\delta n \biggr) \\
&= (-1)^{I_\text{tot}} \exp\biggl(\frac{2\pi \I \, I_\text{tot} \, n_2}{L} \biggr) \, \sigma\biggl( -\frac{I_\text{tot} \, \omega}{2} -\delta n\biggr).
\end{aligned}
\end{equation*}
This directly implies that the wave function vanishes as long as $I_\text{tot}$ is odd. Note that the possible pole occurring due the first factor in \eqref{eq:M=2_wf_bargamma} is avoided due to this requirement on $I_\text{tot}$ as well. 

Performing a similar analysis on $\gamma = (I_\text{tot}\, \omega  + L)/2$ shows that we can rewrite the first term in the brackets of \eqref{eq:M=2_wf_bargamma} as
\begin{equation}
{-}\! \exp\biggl(\frac{2\pi \I \, I_\text{tot} \, n_2}{L} \biggr) \, \sigma\biggl( -\frac{I_\text{tot} \, \omega}{2}-\frac{L}{2} -\delta n\biggr),
\end{equation}
thus in this case the wave function vanishes without any restrictions.

The third and fourth root require another repeated application of the quasi-periodicity rule: for the third root (which is parity-dual to the first one) we find after another $L$ applications that
\begin{equation}
\sigma\biggl( \frac{(I_\text{tot}+L) \,\omega}{2} -\delta n\biggr) = (-1)^{I_\text{tot}+L} \, \E^{-\delta n (I_\text{tot}+L) \eta_2} \, \sigma\biggl( -\frac{(I_\text{tot}+L)\, \omega}{2} -\delta n\biggr)
\end{equation}
and when evaluated at the third root in \eqref{eq:app_triv_roots} the first term in the brackets of the $M=2$ wave function becomes
\begin{equation}
(-1)^{I_\text{tot}+L} \exp\biggl(\frac{2\pi \I \, I_\text{tot} \, n_2}{L} +2\pi \I \, \delta n \biggr) \, \sigma\biggl( -\frac{(I_\text{tot}+L) \, \omega}{2} -\delta n\biggr),
\end{equation}
which when evaluated for integer $\delta n$ vanishes against the second factor in the brackets of \eqref{eq:M=2_wf_bargamma} as long as $L+I_\text{tot}$ is odd, which is exactly what is necessary to avoid the pole coming from the overall factor. 

Finally, for the fourth root $\gamma = [(I_\text{tot}+L) \, \omega +L]/2$ we can follow the approach of its parity-dual, the second root,
and rewrite the first term to cancel the second term without any restrictions on $L$ and $I_\text{tot}$. 

Thus we see that the wave function vanishes precisely when these roots also occur as solutions to the constraint \eqref{eq:ell_curve} and can thus safely be removed from this equation to ultimately yield the constraint \eqref{eq:M=2curveequations}. It would be interesting to see this analysis for higher $M$ as it could reveal what would be suitable coordinates for the spectral problem in those sectors.


\bibliography{bibliography}

\end{document}